\documentclass[prl,twocolumn,showpacs,amsmath,amssymb,floatfix, superscriptaddress]{revtex4-2}
\usepackage{placeins}

\usepackage{amsmath,amssymb}
\usepackage{graphicx}
\usepackage{dcolumn}
\usepackage{bm}
\usepackage{mathtools}
\usepackage{braket}
\usepackage[dvipsnames]{xcolor}
\usepackage{ulem}
\graphicspath{{./figures/}}
\usepackage{listings}

\definecolor{codegreen}{rgb}{0,0.6,0}
\definecolor{codegray}{rgb}{0.5,0.5,0.5}
\definecolor{codepurple}{rgb}{0.58,0,0.82}
\definecolor{backcolour}{rgb}{0.95,0.95,0.92}
\lstdefinestyle{mystyle}{
  backgroundcolor=\color{backcolour},
  commentstyle=\color{codegreen},
  keywordstyle=\color{magenta},
  stringstyle=\color{codepurple},
  basicstyle=\ttfamily\footnotesize,
  breakatwhitespace=false,
  breaklines=true,
  captionpos=b,
  columns=fullflexible,
  keepspaces=false,
  numbers=none,
  numbersep=5pt,
  showspaces=false,
  showstringspaces=false,
  showtabs=false,
  tabsize=2,
  extendedchars=true
}
\usepackage[
 colorlinks = true,
  linkcolor = blue,
  urlcolor  = blue,
  citecolor = blue,
anchorcolor = blue]{hyperref}
\usepackage{natbib}
\usepackage[mathscr]{euscript}
\usepackage[T1]{fontenc}

\begin{document}
\title{Accelerating Crystal Structure Prediction Using Data-Derived Potentials: High-Pressure Binary Hydrides}

\author{Lewis J. Conway}
\email{lewisjohnconway@gmail.com}
\affiliation{Department of Materials Science and Metallurgy, University of Cambridge, 27 Charles Babbage Road, Cambridge CB30FS, UK}
\affiliation{Advanced Institute for Materials Research, Tohoku University, Sendai, 980-8577, Japan}
\author{Chris J. Pickard}
\email{cjp20@cam.ac.uk}
\affiliation{Department of Materials Science and Metallurgy, University of Cambridge, 27 Charles Babbage Road, Cambridge CB30FS, UK}
\affiliation{Advanced Institute for Materials Research, Tohoku University, Sendai, 980-8577, Japan}

\begin{abstract}

  Crystal structures can be predicted from first-principles using \textit{ab initio} random structure searching (\textsc{AIRSS}) and density functional theory (DFT).
  \textsc{AIRSS} provides a method to sample the potential energy landscape and DFT provides a robust and accurate description of that landscape.
  Classical interatomic potentials can describe energy landscapes at a significantly lower computational cost, typically at the expense of robustness and accuracy.
  Modern machine-learning interatomic potentials offer a compromise, with greater robustness and accuracy than classical potentials at a fraction of the computational cost of DFT.
  In this work, we use Ephemeral Data-Derived Potentials (\textsc{EDDP}s) to perform accelerated \textsc{AIRSS} calculations for the binary hydrides at 100\,GPa.
  Since the training data is generated iteratively using \textsc{AIRSS}, the searches can be performed with no prior knowledge of hydrides.
  These potentials allow for more diverse searches, sampling a wider range of compositions, larger unit cells, and orders-of-magnitude more structures.
  In addition to recovering many of the known structures, the searches reveal structures such as the hydrogen-rich phases of H$_{22}$(BrH), H$_{23}$Pb, and H$_{32}$Mg, supermolecular phases of H$_{25}$Cs and H$_{26}$Rn, and many `substoichiometric' variants of known hydrides.
  Our results indicate that using the current generation of pretrained universal MLIPs to search for novel high-pressure hydrides is less effective due to model instabilities or markedly slower inference speeds and highlight the necessity of generating new, targeted data to drive further discoveries.

\end{abstract}
\maketitle

First-principles crystal structure prediction is a powerful tool for the discovery of novel materials.
Coupled with density functional theory (DFT), it can be applied to a wide range of chemical environments.
These include high-temperature superconductors~\cite{Dolui2024}, battery materials~\cite{Lu2021}, and two-dimensional phases of ice~\cite{Chen2016}. It can also be used to predict structures at extremely high pressures, such as those found in planetary interiors~\cite{Conway2021, Martinez-Canales2012}.
Hydride synthesis results can often be better understood through computational structure prediction techniques~\cite{Duan2014,Duan2015,Liu2017}, representing complementary approaches to materials design.

The principle behind structure prediction calculations is to generate a great many sensible candidate structures which can then be optimised towards nearby local energy minima.
The optimisation is performed by computing the energy, forces, and stresses using an appropriate (and often computationally demanding) model.
With sufficient sampling, this approach should eventually find the global energy minimum.
Crucially, a structure prediction algorithm need only identify the appropriate energy basins rather than the precise minima, which can then be refined through subsequent geometry optimisations.
The deeper the initial structures lie within the energy basins, the shorter the geometry optimisations, and the more efficient the search. The versatility of any structure prediction algorithm --- such as random sampling~\cite{Pickard2006, Pickard2011}, particle swarm optimisation~\cite{Wang2012}, genetic algorithms~\cite{Glass2006, Lonie2011}, or generative machine-learning~\cite{Zeni2023,Collins2024} --- relies on the robustness and accuracy of the energy, forces, and stress calculations.
These are typically obtained using plane-wave DFT, incurring significant computational costs.
Advances in machine-learning interatomic potentials (MLIPs) have provided a cheaper alternative that is sufficiently accurate for many applications. However, they may not be robust enough for random structure search.

Several universal interatomic potentials have been developed in recent years, allowing for a wide range of calculations for elements across the periodic table at significantly lower computational cost than DFT~\cite{Batatia2024,Yang2024,Neumann2024,Merchant2023}.
However, at present, these models remain less transferable to extremely diverse chemical environments than DFT.
Typically, they do not provide sufficiently accurate estimates in `out-of-distribution' applications, such as random structure search, often resulting in pathological and unphysical structures.

Furthermore, universal potentials typically include millions of parameters in order to learn a wide range of chemical environments. For many applications involving only two or three elements, significantly smaller and faster models could be used.
One potential solution is to distil and fine-tune the pre-trained models to create faster, more accurate potentials~\cite{Wang2025}.
However, in this work we opted to train bespoke small models on data specifically designed for structure prediction, which we then use to predict new structures.

Ephemeral Data-Derived Potentials (\textsc{EDDP}s) are a class of small multi-layer perceptron models initially designed to accelerate first-principles crystal structure prediction~\cite{Pickard2022}.
Their efficacy can be attributed to an iterative training scheme using randomly generated data consisting of small diverse unit cells, to fast and trivially parallelisable CPU-based inference, and to uncertainty estimation using ensembles~\cite{Salzbrenner2023}.
Despite their apparent simplicity, \textsc{EDDP}s perform remarkably well for structure prediction.
One of the first applications of the \textsc{EDDP}s identified a stable structure of silane (SiH$_4$) containing 12 formula units~\cite{Pickard2022}, a structure which would not have been found in the small system sizes typically used in DFT-based structure prediction~\cite{Pickard2006}.
They have also been used to great effect in the prediction of stable Sc-H, Mg-Ir-H and Lu-N-H hydrides, continuing to outperform traditional DFT-based structure prediction techniques~\cite{Salzbrenner2023, Dolui2024, Ferreira2023}.

High-pressure binary hydrides present a fitting test for machine-learning accelerated structure prediction as they were among the first applications of \textit{ab initio} structure prediction methods~\cite{Pickard2006,Pickard2007a, Zurek2009}.
Most X-H binary hydride systems have since been studied using DFT-based techniques~\cite{Bi2019, Zurek2017, Boeri2022}, unveiling novel chemistry at high pressure~\cite{Hooper2012}, clathrate hydrides~\cite{Wang2012b, Liu2017}, and high-temperature superconductors~\cite{Duan2014, Liu2017}.
With increasing computing power, structure prediction efforts have shifted towards more complex compositions, including ternary and quaternary hydrides~\cite{DiCataldo2022,Lucrezi2022,Sun2022, Dolui2024}.
However, the binary hydrides have yet to be studied in detail with MLIP-accelerated structure prediction.

Computational crystal structure prediction is particularly important for the experimental characterisation of hydrides.
In X-ray diffraction experiments, where the hydrogen atoms are effectively invisible, diffraction peaks can only be attributed to the lattice formed by the heavy atoms.
The hydrogen content is typically inferred from the unit cell volume assuming a linear mixture of elements or is determined using computational methods such as structure prediction. This inference can be misleading as discrepancies exist between experimentally and computationally deduced hydrogen content.
These discrepancies may result from choices in computational methods --- exchange-correlation functional, finite-temperature effects or anharmonicity --- or (to a lesser extent) from experimental uncertainties in pressure measurements.
They could also stem from the physical processes involved in hydride synthesis.
Increasing evidence suggests that hydrides exhibit highly variable hydrogen compositions depending on pressure, temperature, hydrogen availability and time~\cite{Laniel2022}.
A notable example is the superconducting phase of H$_{10}$La which can be synthesised by laser heating a sample of H$_3$La in a hydrogen-rich environment~\cite{Drozdov2019}.

While not common practice, it is helpful for computational works to explicitly specify the spacegroup of the total crystal (including the positions of the hydrogen atoms) and the spacegroup of the heavy atom lattice.
This way, the symmetries observed in experiments and in calculations can be interpreted separately.
Since many structures are common across different chemical systems, in this work we will refer to a crystal structure type common to many species by the anonymised formula, H$_{m}$X$_{n}$, where X represents the heavy atom.

\begin{figure}[t]
  \includegraphics[width=\linewidth]{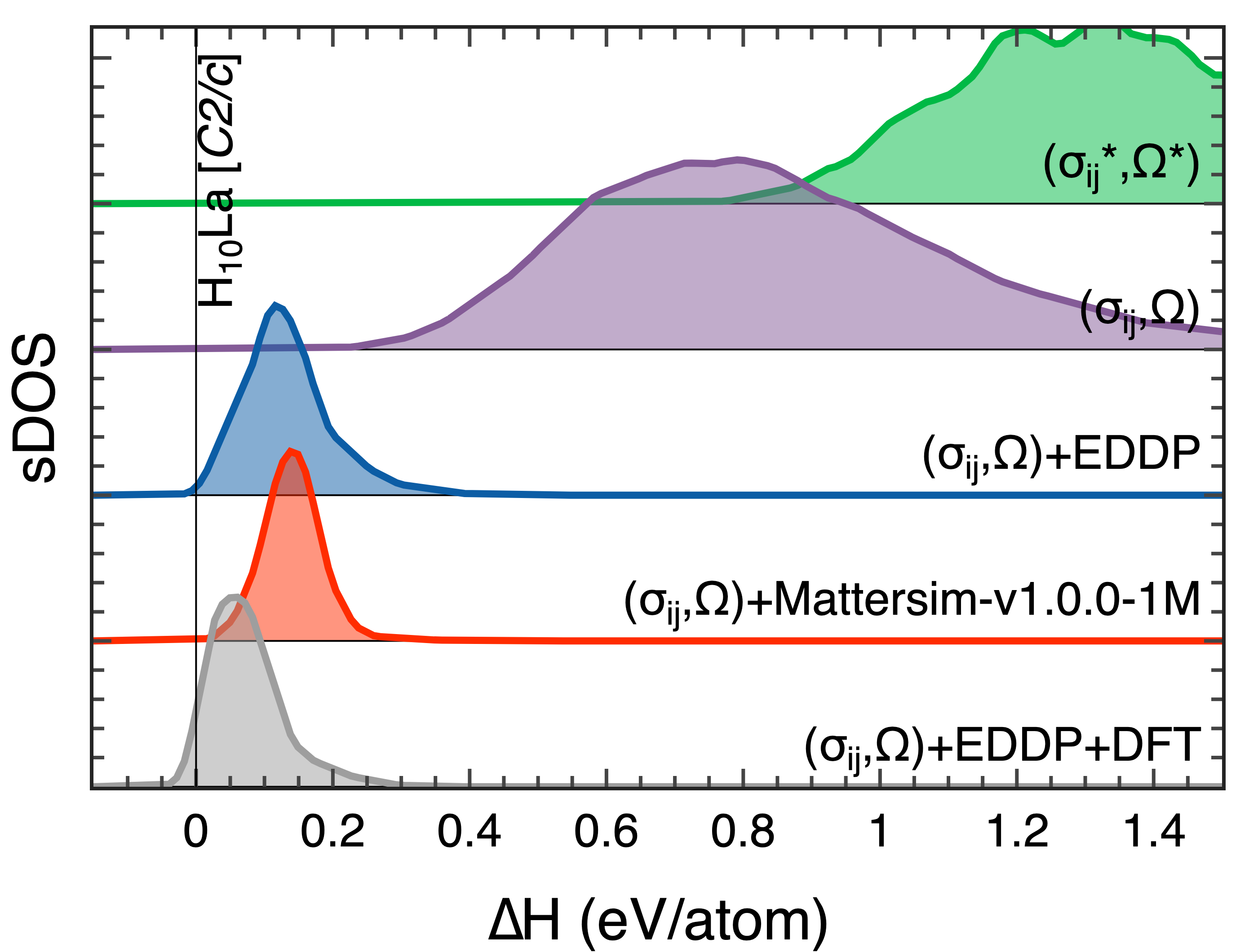}
  \caption{Structural density of states (sDOS) for searches of 2 formula units of LaH$_{10}$ at 100\,GPa with different optimisation schemes. Green curve; Structures generated with randomly allocated \texttt{MINSEP} and \texttt{TARGVOL} parameters ($\sigma_{ij}*$ and $\Omega*$, see text). Purple curve; structures generated using data-derived parameters ($\sigma_{ij}$, $\Omega$, see text). Blue curve: structures generated using data-derived parameters and optimised using an \textsc{EDDP}. Red curve: structures generated using data-derived parameters and optimised using Mattersim-v1.0.0-1M. Grey curve: structures optimised using data-derived parameters, \textsc{EDDP}, and then DFT. All enthalpies were then recalculated using single-point DFT calculations for comparison and plotted relative to the lowest enthalpy structure.}
  \label{fig:SDOS}
\end{figure}

\section{Methods}

The structure prediction code, \textsc{AIRSS} (\textit{Ab-initio} Random Structure Search)~\cite{Pickard2006, Pickard2011}, generates random sensible structures based on chemically informed parameters.
These structures form a distribution relatively close to (but not necessarily at) the local minima.
Distributions, or structural densities of states, of randomly generated structures are shown in green and purple in FIG.~\ref{fig:SDOS}.

In practice, these parameters are imposed by specifying the \texttt{MINSEP} and \texttt{TARGVOL} inputs.
\texttt{MINSEP} defines the pair-wise minimum separation, $\sigma_{ij}$, between atoms of species $i$ and $j$, while \texttt{TARGVOL} specifies the target volume, $\Omega$, for the structure.
The constraints are satisfied by a two-stage optimisation; first by iteratively pushing overlapping atoms apart until no more overlaps remain (within some tolerance threshold), and second by optimising the lattice and ionic degrees of freedom (with a fixed volume, $\Omega$) using a truncated Lennard-Jones model of the form
\begin{equation}
  E_{ij}(r) =
  \begin{cases}
    \left(\frac{\sigma_{ij}}{r}\right)^6 - \left(\frac{\sigma_{ij}}{r}\right)^{12} & r<\sigma_{ij} \\
    0      & r\ge\sigma_{ij}
  \end{cases}.
  \label{eq:LJ}
\end{equation}

\begin{figure}[t]
  \includegraphics[width=\linewidth]{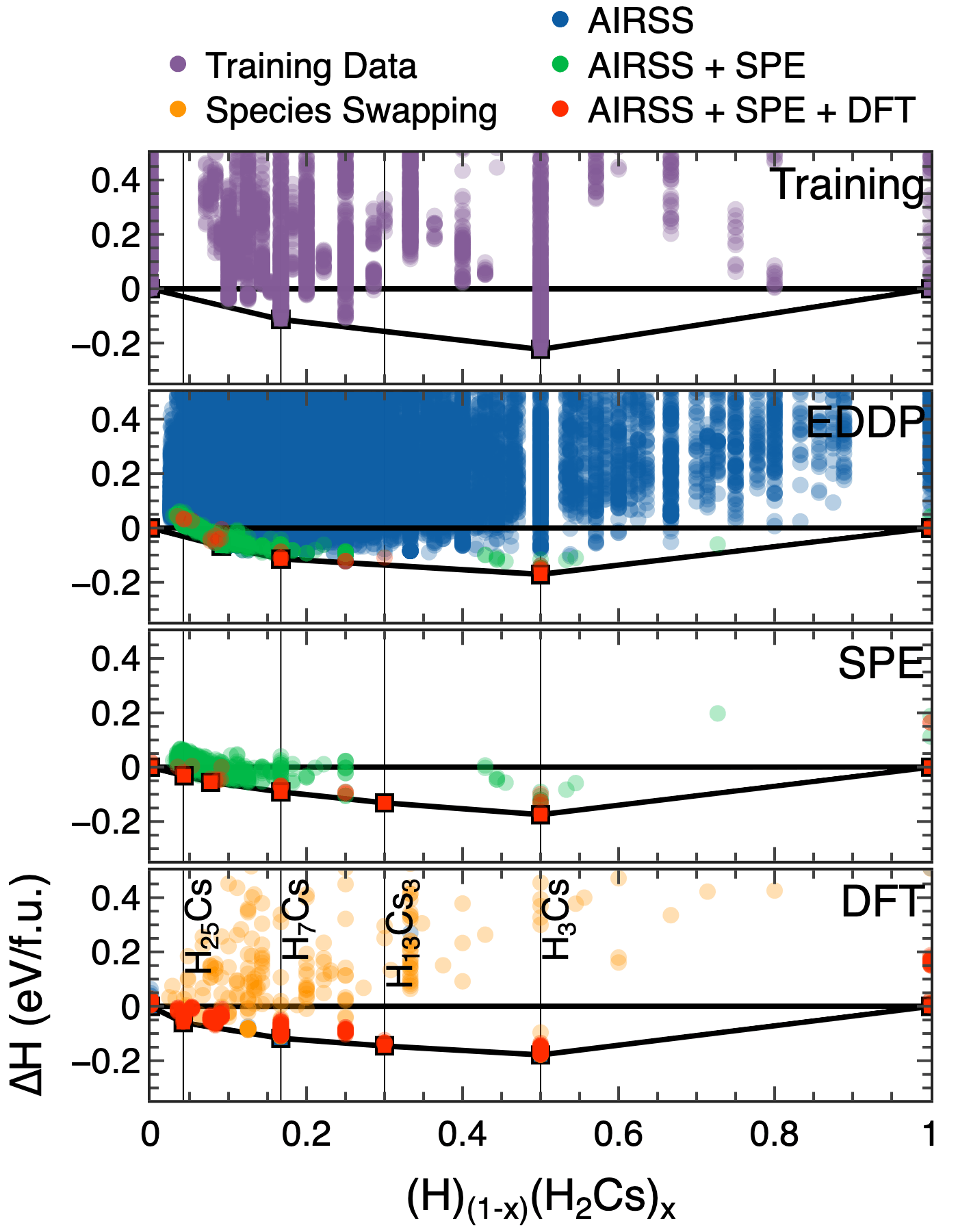}
  \caption{Convex hulls for the Cs-H datasets. The first panel is the training dataset, followed by the \textsc{EDDP} convex hull, SPE convex hull and final DFT-optimised convex hull. Symbols are coloured to indicate structures carried forward in the filtering process. Blue points were generated by the \textsc{EDDP} but filtered out. Green points were only carried forward from the \textsc{EDDP} dataset to the SPE stage. Red points were also carried forward to the final DFT optimisation step. Orange points were generated by species swapping in the final prototype dataset.}
  \label{fig:Cs}
\end{figure}

While the energy ranking from this simple potential is not physically meaningful, its role in generating sensible starting geometries is a potentially overlooked aspect of this method.
The parameters in this potential are often hand-crafted but they can be derived from data by first performing a more broad search with an initial guess of parameters, denoted here by $\sigma_{ij}^*$ and $\Omega^*$.
The approach used in this work is to randomly allocate values in $\sigma_{ij}^*$ to be between 1 and 3~\AA, and estimate a value for $\Omega^*$ by calculating the volume of packed spheres with tabulated atomic radii~\cite{Clementi1967}.
By default, \textsc{AIRSS} targets more open structures with a packing fraction of 34\%, but for high pressures, a value of around~74\% is more appropriate.
This search, while relatively inefficient, results in a dataset from which we can \textit{learn} the basic chemistry and from which more sensible $\sigma_{ij}$ and $\Omega$ values can be measured.

The difference between structure distributions generated with the default parameters, $(\sigma_{ij}^*,\Omega^*)$, and with data-derived parameters, $(\sigma_{ij}, \Omega)$, is demonstrated by comparing the DFT energies shown in FIG.~\ref{fig:SDOS}.
The enthalpy difference, $\Delta H$, between the stable structures and the randomly generated structures indicates how deep the they are within the energy basins, and hence how sensible they are.
Using the data-derived parameters lowers the peak of the structural density of states (sDOS) from 1.4\,eV to 0.9\,eV above the energy minimum.

\textsc{EDDP}s are MLIPs inspired by Lennard-Jones potentials, extended to include three-body interactions and non-linearity by way of a neural network.
The model architecture and data generation process are designed such that, just as the pair-potential defined in~Eq.\eqref{eq:LJ}, they can generate extremely sensible crystal structures with a low computational cost. By training on DFT energies, unlike~Eq.\eqref{eq:LJ}, these potentials give a physically meaningful ranking of their energies with typical Spearman correlations greater than 0.99.
After relaxing the structures with an \textsc{EDDP}, the peak in the sDOS in FIG.~\ref{fig:SDOS} is less than 0.2\,eV above the energy minimum.

We performed similar \textsc{AIRSS} calculations using several open-source pre-trained models --- MatterSim-v1.0.0-1M~\cite{Yang2024}, ORB-V2~\cite{Neumann2024}, Mace-MP-0 and Mace-MPA-0~\cite{Batatia2024} --- with the results detailed in the Supplementary Information.
The Orb model often failed to converge during geometry optimisations, likely due to the non-conservative implementation of the force calculations, although it eventually generated sensible structures after reaching the maximum number of optimisation steps.
The Mace models, however, exhibited pathological gaps in the potential energy landscape, and resulted in very few low-energy structures.
In contrast, MatterSim reliably gave chemically sensible structures (see FIG.~\ref{fig:SDOS}), although the difference in the width of the sDOS peak in comparison with the \textsc{EDDP} and DFT optimisations should be noted.

Of course, these models are typically not designed for this application and would likely require fine-tuning to work at high pressure.
However, the inference cost of these models should be considered.
The time to relax a structure using MatterSim on CPU is around 100 times slower than using an \textsc{EDDP}.
On a GPU, MatterSim is around 5 times slower as shown in Figures~S2, S3, S5, and S6 in the Supplementary Information.
Given the scaling challenges of global optimisation problems such as structure prediction, even modest throughput improvements can have significant benefits.

\begin{table*}[t!]
  \caption{Notable binary hydride prototypes at 100\,GPa, listed with anonymised formula, H$_n$X$_m$, and the elements, X, for which they are stable or metastable. Structures not on the convex hull are listed with their distance in meV/atom in parenthesis.}
  \label{tab:Structures}
  \centering
  \begin{tabular}{lllp{0.5\linewidth}l}
    \toprule
    Formula & Spacegroup & n.\ form & X ($\Delta H$ [meV/atom]) & Heavy atom Spacegroup \\
    H$_{11}$X$_{3}$                    & $P4/mmm$                           & 1                                  & Mn (3)                                                           & $Pm\overline{3}m$                  \\
    H$_{16}$X$_{3}$                    & $C2/m$                             & 1                                  & Na (6)                                                           & $Pm\overline{3}m$                  \\
    H$_{25}$X                          & $Fm\overline{3}c$                  & 2                                  & Cs                                                               & $Pm\overline{3}m$                  \\
    H$_{25}$X                          & $R\overline{3}$                    & 2                                  & Kr (8)                                                           & $Pm\overline{3}m$                  \\
    H$_{25}$X                          & $R3$                               & 2                                  & Rn (6)                                                           & $Pm\overline{3}m$                  \\
    H$_{26}$X                          & $Iba2$                             & 2                                  & Rn, Xe, Kr (3)                                                   & $Pm\overline{3}m$                  \\
    H$_{14}$X$_{5}$                    & $I4/m$                             & 1                                  & Lu, Sc, Ti                                                       & $Fm\overline{3}m$                  \\
    H$_{15}$X$_{4}$                    & $I\overline{4}3d$                  & 2                                  & Hf, Ti, Zr                                                       & $I\overline{4}3d$                  \\
    H$_{5}$X                           & $P\overline{1}$                    & 2                                  & Br                                                               & $P6/mmm$                           \\
    H$_{6}$X                           & $Pm$                               & 4                                  & Fe (5)                                                           & $P6/mmm$                           \\
    H$_{6}$X                           & $Imma$                             & 4                                  & Ru, Fe (6), Os (7)                                               & $P6/mmm$                           \\
    H$_{9}$X                           & $Cccm$                             & 2                                  & K, Rb (2), Cs (8)                                                & $P6/mmm$                           \\
    H$_{11}$X                          & $P\overline{6}m2$                  & 1                                  & Cl, Br (7)                                                       & $P6/mmm$                           \\
    H$_{13}$X                          & $P6_3/mmc$                         & 2                                  & Na                                                               & $P6/mmm$                           \\
    H$_{7}$X                           & $P\overline{3}m1$                  & 2                                  & Tc, Re (12)                                                      & $P\overline{3}m1$                  \\
    H$_{13}$X$_{3}$                    & $I4/mmm$                           & 1                                  & Fe                                                               & $I4/mmm$                           \\
    H$_{23}$X                          & $C2/m$                             & 1                                  & Ba, K, Rb, Sr (2), La (2)                                        & $I4/mmm$                           \\
    H$_{23}$X                          & $Cc$                               & 2                                  & Br, Cl (5)                                                   & $I4_1/amd$                         \\
    H$_{11}$X                          & $Cmc2_1$                           & 2                                  & K, Rb (7)                                                        & $Cmcm$                             \\
    H$_{32}$X                          & $P\overline{1}$                    & 1                                  & Na, H (5), Ca (5), Mg (7), Yb (7), Li (13), Tm (14)              & $C2/m$                             \\
    H$_{24}$X                          & $C2/c$                             & 2                                  & Ca, Sr, Yb, La (3), K (3), Tm (6), Y (7), Er (11)             & $C2/c$                             \\
    H$_{17}$X                          & $P\overline{1}$                    & 2                                  & Ar, Kr, Li (8), Na (11)                                          & $P\overline{1}$                    \\
    H$_{42}$X                          & $P\overline{1}$                    & 2                                  & Mg, Na (9), Yb (13), Li (14)                                 & $P\overline{1}$                    \\
    \hline \hline
  \end{tabular}
\end{table*}

To accelerate the prediction of high-pressure binary hydrides, we trained 81 different \textsc{EDDP}s for the X-H binaries where X is an element from the periodic table up to and including Rn~\footnote{The element palette includes Li, Be, Na, Mg, Al, Si, P, S, Cl, Ar, K, Ca, Sc, Ti, V, Cr, Mn, Fe, Co, Ni, Cu, Zn, Ga, Ge, As, Se, Br, Kr, Rb, Sr, Y, Zr, Nb, Mo, Tc, Ru, Rh, Pd, Ag, Cd, In, Sn, Sb, Te, I, Xe, Cs, Ba, Hf, Ta, W, Re, Os, Ir, Pt, Au, Hg, Tl, Pb, Bi, Po, At, Rn, La, Ce, Pr, Nd, Pm, Sm, Eu, Gd, Tb, Dy, Ho, Er, Tm, Yb and Lu.}.

For each binary, we train an ensemble of potentials with the following recipe.
We use \textsc{AIRSS} to generate random structures with composition X$_m$H$_n$ ($m=1$--$4$,~$n=0$--$20$), with all values of $\sigma_{ij}$ selected uniformly between 0.5 and 2\AA.
The atoms are placed in a unit cell of volume, $\Omega=m\Omega_X + n\Omega_H$ where $\Omega_X$ is randomly selected between 17 and 23\AA$^3$ and $\Omega_H$ is randomly selected between 2 and 3.5\AA$^3$.
These constraints are deliberately chosen to ensure that some high-energy close contacts of X-X, X-H and H-H are present in the training data, constraining the model to be repulsive at short distances.

In the initial training step, 10,000 random structures are generated with these constraints.
The single-point energy (SPE) of each of these structures is then calculated using \textsc{castep}~\cite{Clark2005}.
In all DFT calculations, we use \textsc{castep}'s C19 pseudopotentials, the PBE exchange-correlation functional, a 600\,eV plane-wave cut-off, and a $k$-point spacing of no more than $2\pi\times~0.03$\,\AA${}^{-1}$.

At this stage, the dataset contains no minima and consists of very diverse high-energy configurations.
The \textsc{EDDP} typically has a mean absolute error (MAE) of $30$--$40$\,meV/atom in the test set.
In the next step, we use the \textsc{EDDP}s, to optimise a new set of 5,000 randomly generated structures to local enthalpy minima at pressures randomly drawn uniformly between 25 and 150\,GPa.
In order to sample the minima and nearby configurations, we add shaken variants of the optimised structures, where the atoms are displaced randomly by up to 0.1\AA.
For each of these minima, 10 shaken structures are evaluated with a SPE calculation.
These energies are added to the dataset on which a new \textsc{EDDP} is trained.
After four more iterations of minima searching, the MAE on the test set is typically between 10 and 30\,meV/atom.
An example of the training data generated this way is shown in purple in FIG.~\ref{fig:Cs}.

For the structure searches, we continue generating random structures but with a wider range of stoichiometries, X$_m$H$_n$, $m=2$--$8$, $n=0$--$80$, and with between 2 and 48 symmetry operations.
Each of these structures is relaxed to a local enthalpy minimum at 100\,GPa.
For the first 20,000 structures, we use the initial guess parameters, $\sigma_{ij}^*$ and $\Omega^*$.
We then filter out any pathological structures by removing those with an ensemble standard deviation greater than 3~eV/atom.
From the filtered dataset, we measure the values for $\Omega$ and $\sigma_{ij}$ for the lowest energy configuration for each composition.
A second round of searching is then performed with the data-derived parameters to generate a further 50,000 structures.

We then employ several post-processing steps to filter for promising stable structures.
For each composition, we extract the structures within 50\,meV per atom of the convex hull at 100\,GPa.
Next, we identify metastable structures by linearly extrapolating the enthalpies down to 0\,GPa and up to 250\,GPa, as previously described by Pickard and Needs~\cite{Pickard2011}.
This process allows for slightly denser structures (which might be stabilised at higher pressures) or more open structures (which might be stabilised at lower pressures) to fall within our energy window.
It also provides some tolerance for structures where the volume has been under- or overestimated by the potential.
Structures identified at this stage (red and green points in FIG.~\ref{fig:Cs}) are recalculated using a \textsc{castep} SPE calculation.
We then recompute the convex hull using the SPE calculations.

Structures that remain within 10\,meV per atom of the convex hull after the \textsc{castep} calculations are then shaken and optimised using \textsc{castep}.
The shaking step helps account for overly smooth regions of the \textsc{EDDP} potential energy landscape around the energy minima, enabling structures to escape saddle-points.
The structures which remain at this stage are shown in red in FIG.~\ref{fig:Cs}.
As an example, FIG.~\ref{fig:Cs} shows data for Cs-H but the results of these steps for each of the other 76 binaries are plotted in the supplementary material.

\section{Results}

\begin{figure*}[t!]
  \centering
  \includegraphics[scale=0.35]{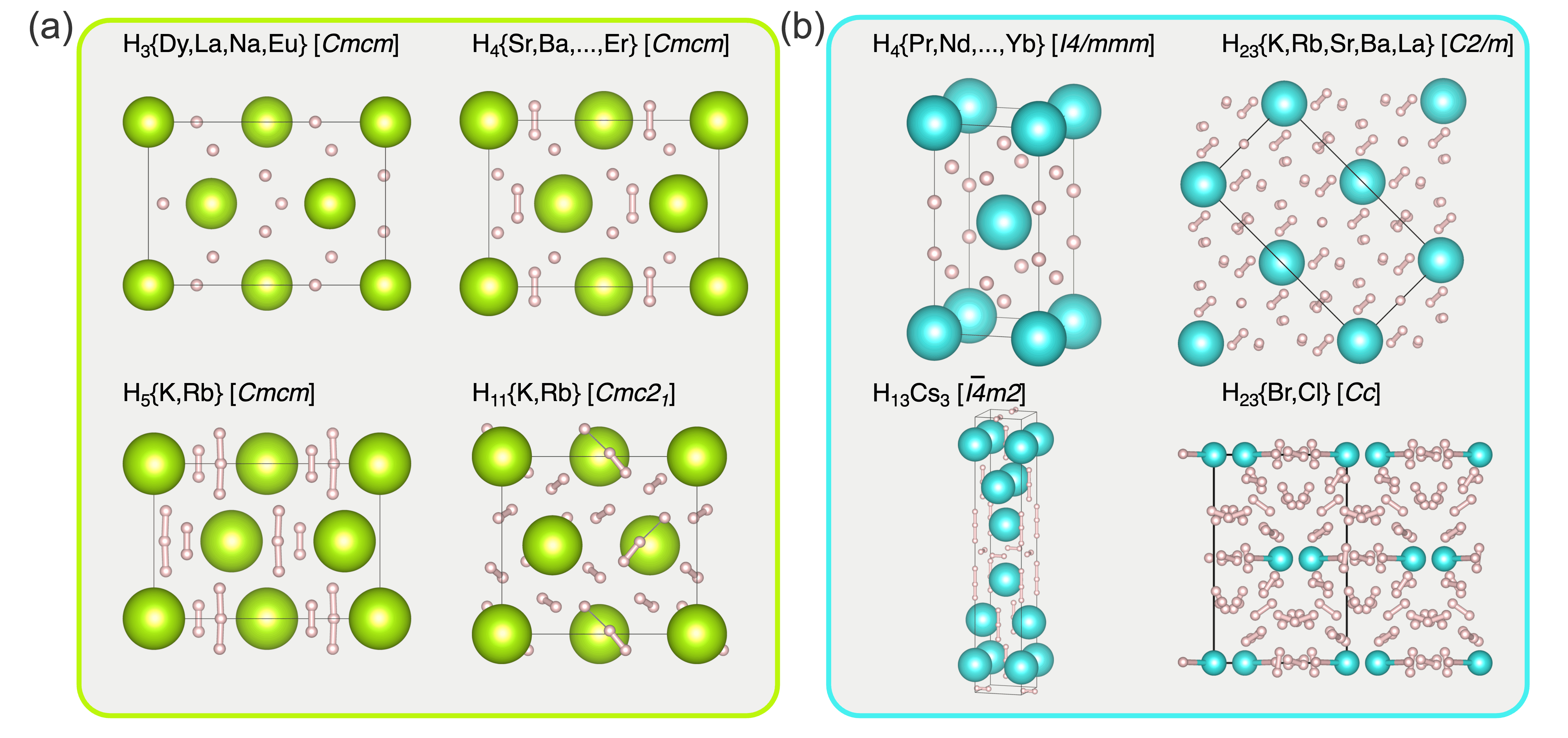}
  \caption{Structure prototypes grouped by their heavy-atom Bravais lattice. The elements in braces represent elements for which the prototype is within 10\,meV/atom of the convex hull. The full lists are tabulated in the Supplementary Materials. (a) Base-centred monoclinic. (b) Body-centred tetragonal}
  \label{fig:struc1}
\end{figure*}

The searches recovered most of the previously known hydride structures and uncovered several more stable structures.
To simplify the discussion, we categorise all structures into prototypes labelled by anonymised formula.
For example, the fcc trihydrides such as H$_3$Lu, H$_3$Y, H$_3$Sc can all be described by the prototype H$_3$X-$Fm\overline{3}m$.

In a final search step we employ species swapping.
For each of the structure prototypes, we iterate through the element palette, relax the geometry using \textsc{castep} and calculate the distances from the convex hull.
For example, we optimise the structure of the prototype H$_3$X-$Fm\overline{3}m$ with X being each of the 76 elements in our palette at 100\,GPa.
This revealed further stable structures that had not appeared in the original searches either because of the stochastic nature of \textsc{AIRSS} or errors in the \textsc{EDDP} stability estimation.
These structures appear in orange in FIG.~\ref{fig:Cs}.
For example, swapping the Tc atoms in H$_7$Tc-$P\overline{3}m1$ (FIG.~\ref{fig:struc2}c), which had been recovered by the \textsc{EDDP}, shows that isostructural compounds H$_7$Mn and H$_7$Re --- missed by the \textsc{EDDP} --- are also stable at 100\,GPa.
This highlights a strength in  high-throughput composition searches, where combining the results of each potential produces outcomes that exceed those of the individual searches.

TABLE~\ref{tab:Structures} presents noteworthy structure prototypes, sorted by the Bravais lattice formed by the heavy atoms.
The full sorted lists of prototypes form TABLE~S1-S14 in the supplementary material.
Sorting in this manner clusters together structures which would appear similar in an X-ray diffraction (XRD) pattern, where the hydrogen atoms are effectively invisible.

Many identical heavy-atom lattices can host very different hydrogen lattices.
For example, the orthorhombic prototypes H$_3$X-$Cmcm$, H$_4$X-$Cmcm$, H$_5$X-$Cmcm$, and H$_{11}$X-$Cmc2_1$ shown in FIG.~\ref{fig:struc1}a could all be indexed from an XRD pattern using a conventional $Cmcm$ unit cell with heavy atoms on the same 4a site.
These prototypes are stable with a range of heavy atoms, shown in Supplementary Table~S8, though it is useful to focus briefly on the alkali and alkaline earth metals.
H$_3$X-$Cmcm$ is stable at 100\,GPa for X=Na and contains no H$_2$ bonds.
In contrast, H$_4$X-$Cmcm$ is not stable for any alkali metals but is stable for the alkaline earth metals, X=Mg and Sr, and contains a mixture of H$_2$ bonds and atomic hydrogen.
The more hydrogen-rich prototypes, H$_5$X-$Cmcm$ and H$_{11}$X-$Cmc2_1$, are stable for K and Rb and contain a mixture of H$_2$ bonds, as well as symmetric or asymmetric H$_3$ units.

To give another example, we look at the hexagonal heavy-atom lattices shown in FIG.~\ref{fig:struc2}d.
The prototype H$_9$X-$Cccm$ (stable for X=K and Cs) has a heavy-atom lattice with hexagonal symmetry.
An XRD pattern of this prototype would show a primitive simple-hexagonal unit cell.
When hydrogen atoms are incorporated into the lattice in a calculation, the hexagonal symmetry is broken, resulting in an orthorhombic lattice.
Conversely, the prototype H$_{11}$X-$P\overline{6}m2$, which is stable at 100\,GPa for X=Cl and stable at 50\,GPa for X=Br and Cl, retains its symmetry by forming triangular H$_3$ trimers.

These examples illustrate that, despite their identical heavy-atom lattices, the hydrogen chemistry uncovered by structure prediction can vary significantly.
By framing structures as prototypes based on the heavy-atom lattice or by the lattice formed by explicitly placing hydrogen atoms, we can better understand the diversity of hydrides revealed by structure prediction.

\subsection{Supermolecular Crystals}

\begin{figure}[t!]
  \centering
  \includegraphics[scale=0.35]{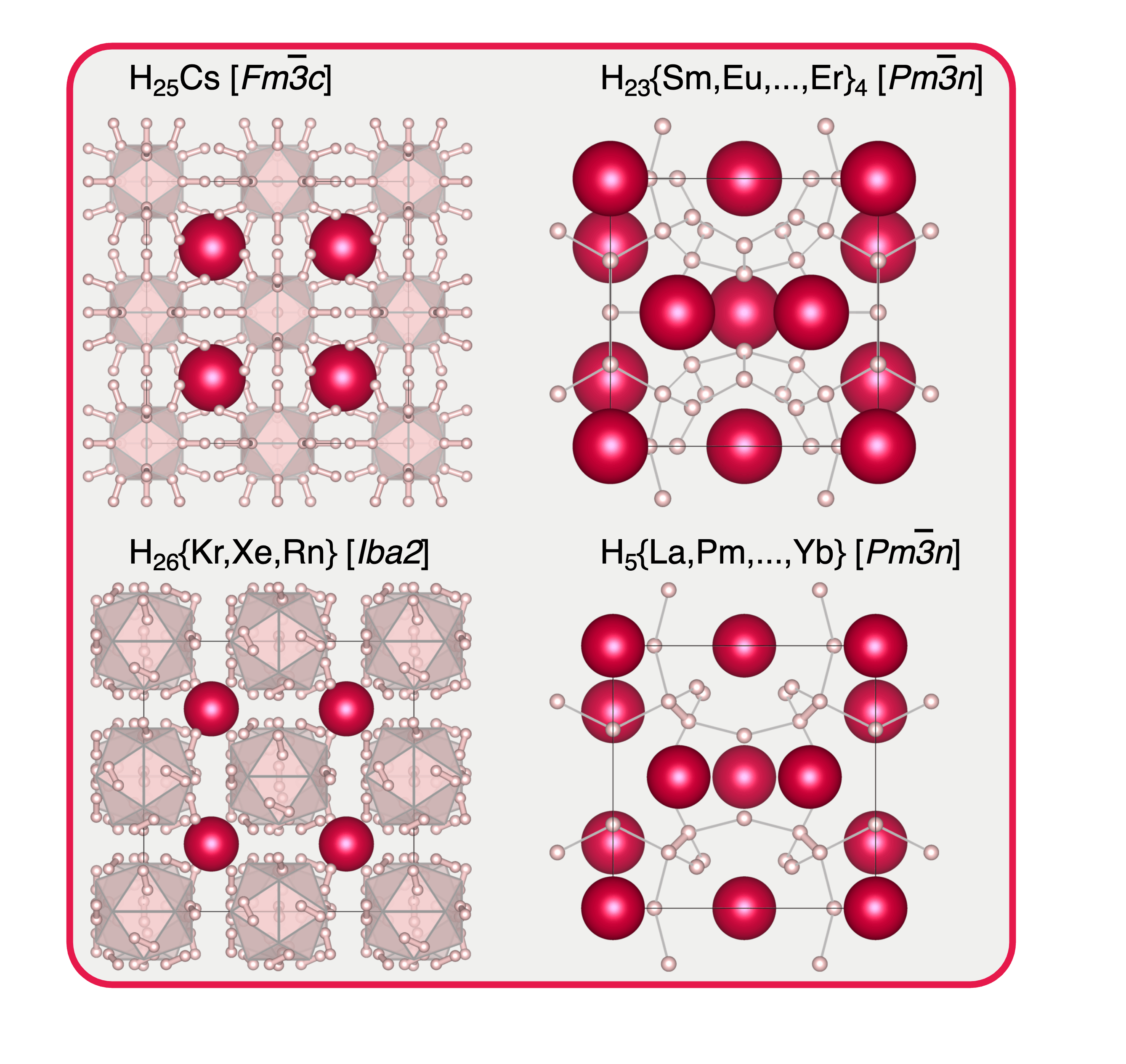}
  \caption{Structure prototypes with a simple cubic heavy-atom Bravais lattice.  The elements in braces represent elements for which the prototype is within 10\,meV/atom of the convex hull. The full lists are tabulated in the Supplementary Materials.}
  \label{fig:struc3}
\end{figure}

A remarkable result of these searches is the emergence of H$_{25}$X-$Fm\overline{3}c$ as a stable structure for X=Cs.
The structure contains a simple cubic heavy-atom lattice, with interstitial sites filled by an ionic supermolecule, H$^-$(H$_2$)$_{12}$, where each H$_2$ unit is oriented toward the central H$^-$, as shown in FIG.~\ref{fig:struc3}.
The structure could also be viewed as isostructural to CsCl, with the supermolecules occupying the Cl sites.
Species-swapping this structure type confirms that only Cs adopts this structure type, suggesting that Cs alone possesses the necessary valence chemistry and ionic radius to be consistent with the principles behind Pauling's radius-ratio rules and act as a counter-ion to H$^-$(H$_2$)$_{12}$.
Notably, this structure remains on the convex hull at least down to 50\,GPa and up to 250\,GPa, indicating significant stability across a wide pressure range.

The searches also reveal closely related structures for noble gas hydrides.
The prototype, H$_{26}$X-$Iba2$, is stable for X=Kr, Xe, and Rn at 100\,GPa (also shown in FIG.~\ref{fig:struc3}).
Although the structure still resembles CsCl, it forms a crystal of neutral, rather than ionic, icosahedral (H$_2$)$_{13}$ supermolecules.
This configuration is consistent with the experimentally observed hydride of hydrogen iodide by Howie and co-workers~\cite{Binns2018}.
The same group also noted that similar supermolecules could form in hydrides with inert elements such as Xe~\cite{Ackland2018}, though no ordered structure or thermodynamic explanation for their stability had been provided.
Unlike the ionic structure, these appear to be less stable across pressure ranges -- only between 15 and 50\,GPa.

These icosahedral motifs could not have appeared in any of the training data formed only of structures with 20 hydrogen atoms or fewer, but were still found in the searches which targeted systems with 80 hydrogen atoms or fewer.
A similar observation was made in \textsc{EDDP} searches for high-pressure boron, where the icosahedral motif appears in the searches despite not appearing in the training data~\cite{Pickard2022}.

\subsection{Substoichiometric Hydrides}

Typically, structure prediction methods are not able to describe non-stoichiometric compounds -- those with formulas X$_m$H$_n$ where $m$ and $n$ are not small integers.
With sufficiently large supercells, models can be constructed to approximate non-integer values of $m$ and $n$;
For example, XH$_{3}$ could be modelled by structure prediction using unit cells containing multiples of 4 atoms.
A substoichiometric version of this structure, XH$_{2.9}$ could be modelled by a unit cell containing multiples of 39 atoms to give X$_{10}$H$_{29}$.
While such methods cannot capture the full continuum of stoichiometries or disorder within these compounds, \textsc{EDDP}-assisted structure searches at least allow for structure prediction for formulas in which $m$ and $m$ are moderate integers ($10$--$100$).

The ability to form stable substoichiometric states allows switchable mirrors to be formed from ambient-pressure films of Y and La hydrides~\cite{Huiberts1996}.
These films undergo a metal-insulator transition when the composition is gradually tuned from H$_2$X to H$_3$X.
Such tunability is only possible if there is a low thermodynamic barrier to hydrogen addition and removal.

\begin{figure*}[t!]
  \centering
  \includegraphics[scale=0.35]{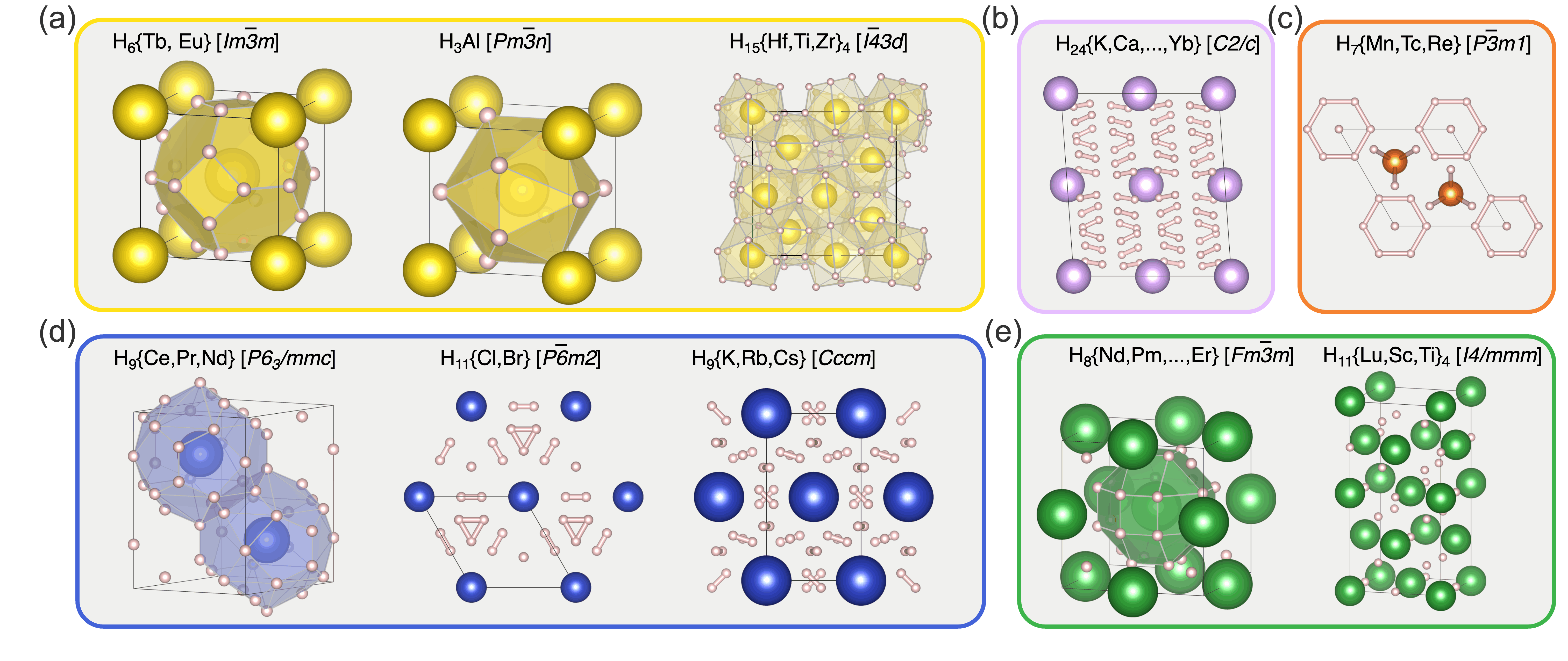}
  \caption{Structure prototypes grouped by their heavy-atom Bravais lattice. The elements in braces represent elements for which the prototype is within 10\,meV/atom of the convex hull. The full lists are tabulated in the Supplementary Materials. (a) Body-centred cubic. (b) Monoclinic. (c) Rhombohedral. (d) Hexagonal. (e) Face-centred cubic.}
  \label{fig:struc2}
\end{figure*}

In the high-pressure regime, the stoichiometric H$_3$X-$Fm\overline{3}m$ prototype -- common to Sc, Ti, Y, Er, Tm, Yb, and Lu at 100\,GPa, readily appears in the searches.
However, several pseudo-substoichiometric variants such as H$_{2.75}$X and H$_{2.86}$X also emerge.
For example, Ti stabilises in prototypes H$_{11}$X$_{4}$-$I\overline{4}2m$, H$_{11}$X$_{4}$-$I4/mmm$ (shown in FIG.~\ref{fig:struc2}e), H$_{20}$X$_{7}$-$C2$, and H$_{20}$X$_{7}$-$R3$, all of which are on or close to the convex hull.
Each of these has an fcc heavy-atom lattice, and can be understood as ordered substoichiometric variants of H$_3$X-$Fm\overline{3}m$.
Similar substoichiometric prototypes are stable for Sc and Lu, but less common for Y, Er, Tm and Yb.
Since \textsc{AIRSS} is a sampling technique, this suggests that at 100\,GPa, Sc and Lu are more tunable in composition than Y, Er, Tm and Yb.
The searches also recover known complex transition metal hydride geometries such as H$_{15}$X$_4$-$I\overline{4}3d$, shown in FIG.~\ref{fig:struc2}a.

Clathrates are another typical class of high-pressure hydrides, consisting of cage-like hydrogen atom configurations surrounding the heavy atoms.
They include prototypes such as H$_{24}$X$_4$-$Pm\overline{3}m$, H$_6$X-$Im\overline{3}m$, H$_9$X-$P6_3/mmc$ and H$_{10}$X-$Fm\overline{3}m$.

The H$_{23}$X$_4$-$Pm\overline{3}n$ prototype (FIG.~\ref{fig:struc3}) consists of heavy atoms forming a simple cubic A15 lattice and H atoms forming the Weaire–Phelan clathrate structure.
Other stable prototypes uncovered by the searches include H$_5$X-$Pm\overline{3}n$, H$_{11}$X$_{2}$-$R3c$, H$_{11}$X$_{2}$-$Cc$, H$_{19}$X$_{4}$-$P\overline{4}3n$, H$_{31}$X$_{8}$-$R3$.
Grouping the structures by the symmetry of their heavy atom lattices identifies them as ordered substoichiometric variants of H$_{24}$X$_4$-$Pm\overline{3}m$.
Just as in the cubic trihydrides, we note that the appearance of these structures depends on the species of the heavy atom, the total hydrogen content and external pressure.
For example, at 100\,GPa, H$_5$La-$Pm\overline{3}n$ and H$_{11}$Ca$_2$-$Cc$ are on the convex hull, but the hydrogen-saturated compositions of H$_{23}$La$_4$-$Pm\overline{3}n$ and H$_{23}$Ca$_4$-$Pm\overline{3}n$ are not.
This suggests that the A15 heavy-atom lattice supports a wide range of hydrogen contents which could be tuned.

In contrast, substoichiometric variants of the H$_6$X-$Im\overline{3}m$ prototype (FIG.~\ref{fig:struc2}a) with formula such as H$_{11}$X$_2$ do not appear, indicating that this prototype is less tunable.
The closest similar structure is H$_3$X-$Pm\overline{3}n$, the structure of the well-known aluminium hydride~\cite{Pickard2007, Goncharenko2008}.

At 100\,GPa, the prototype, H$_{10}$X-$Fm\overline{3}m$, does not appear as stable for any $X$.
The structure contains 8 hydrogen atoms forming a cage network and 2 hydrogen atoms in tetrahedrally coordinated interstitial sites.
Variants of H$_{10}$X-$Fm\overline{3}m$, such as H$_9$X-$F\overline{4}3m$ and H$_8$X-$Fm\overline{3}m$ do emerge, where one or both of the tetrahedral hydrogen sites are vacated.
Substoichiometric variants like H$_{35}$X$_4$-$Cm$, H$_{17}$X$_2$-$P\overline{4}m2$, H$_{33}$X$_4$-$P\overline{4}3m$ and so on also emerge from the searches.
Unlike the H$_{23}$X$_4$-$Pm\overline{3}n$ structures, however, the stoichiometric structures, H$_9$X-$F\overline{4}3m$ and H$_8$X-$Fm\overline{3}m$, are always the most stable.
For example, H$_8$Eu-$Fm\overline{3}m$ (FIG.~\ref{fig:struc2}e) is on the hull, but H$_{33}$Eu$_4$-$P\overline{4}3m$ and H$_{17}$Eu$_2$-$Ibam$ are increasingly further from the hull.
There do not appear to be any structures of this type with H$_x$X with $x<8$, suggesting that only the tetrahedral sites in H$_{10}$X-$Fm\overline{3}m$ can be removed.

Fewer substoichiometric variants of the hcp clathrate prototype, H$_9$X-$P6_3/mmc$, are observed.
However, the searches do show an ordered substoichiometric variant, H$_{26}$Ce$_3$-$P321$, which aligns with the experimental observation of H$_{9-\delta}$Ce ($\delta \le 0.85$) by Li \textit{et al.}~\cite{Li2019}.
The presence of H$_8$Pr-$P6_3mc$ and H$_{31}$Pr$_4$-$Cm$ on the convex hull could therefore imply the existence of a similar phase of H$_{9-\delta}$Pr.

A final example where substoichiometry may be important is found in structures which form from layered structural motifs such as the high-pressure hydrides of Fe studied by P\'epin \textit{et al.}~\cite{Pepin2017}; H$_3$Fe and H$_5$Fe.
A prototype found in these searches, H$_{13}$X$_3$-$I4/mmm$, which is stable for X=Fe, represents yet another layered structure with a stoichiometry not usually sampled in structure prediction.
Similarly, H$_{13}$X$_3$-$I\overline{4}m2$ as shown in FIG.~\ref{fig:struc1}b is stable for X=Cs and is formed of layers of the motifs seen in the hydrogen-poorer and hydrogen-richer structures.

\subsection{Extremely hydrogen-rich structures}

The \textsc{EDDP}-assisted structure search allows for structure prediction of large unit cells with many atoms, revealing extremely hydrogen-rich prototypes on the convex hull, such as H$_{22}$X-$Immm$, H$_{24}$X-$C2/c$ (shown in FIG~\ref{fig:struc2}b), H$_{32}$X-$C2/m$, H$_{37}$X-$R\overline{3}$ and H$_{42}$X-$P\overline{1}$.

For instance, H$_{22}$Sr-$C2m$, H$_{22}$Sr-$Immm$, and H$_{23}$Sr-$C2/c$ are stable or near the convex hull at 100\,GPa.
They all possess an $I4/mmm$ heavy-atom lattice and so would be nearly indistinguishable in XRD measurements.
Notably, these structures are consistent with the measurements of `pseudotetragonal' high-hydrides of Sr synthesised by Semenok \textit{et al.}~\cite{Semenok2022}, and are more stable than the $P1$ structural model proposed in that work.
Semenok \textit{et al.} also provide experimental evidence for H$_{21-23}$Ba with a similar heavy atom lattice and propose there may be other high-hydrogen content hydrides.
This observation is consistent with the appearance of the prototype H$_{23}$X-$C2/m$ in these searches which is stable for a range of heavy atoms, X=Sr, Ba, K, Rb, and La.

Additionally, the searches reveal a prototype structure of H$_{22}$(XH)-$Cc$ for X=Br and Cl, where the Br and Cl occupy the sites of a $I4_1/amd$ lattice (forming polarised XH molecules), while the H$_2$ molecules form a monoclinic lattice.
This structure type is shown in FIG~\ref{fig:struc1}b.

Other hydrogen-rich prototypes with monoclinic heavy-atom lattices include H$_{24}$X-$C2/c$ (X=Ca, Sr, Yb, and La) and H$_{32}$X-$P\overline{1}$ (X=Na, Ca, and Mg).
Even more hydrogen-rich are structures observed for Mg, Li, and Na, form extremely dilute solid solutions that may be stable or metastable such as H$_{37}$X-$R\overline{3}$ and H$_{42}$X-$P\overline{1}$.
With such high hydrogen contents, these may be better considered as incorporations of dopants into a pure hydrogen lattice.
This concept is analogous to the structure of (CH$_4$)$_3$(H$_2$)$_{25}$ observed by Ranieri \textit{et al.}, where CH$_4$ molecules form a commensurate rhombohedral lattice in a hcp lattice of molecular hydrogen~\cite{Ranieri2022}.

\subsection{Higher pressure hydrides}

Using the linear extrapolation of enthalpies, we can also identify structures stable at higher pressures.
This approach uncovers several previously unseen stable structures, such as H$_5$Pb-$C2/m$, H$_{23}$Pb-$C2/m$, H$_{22}$Na-$R\overline{3}m$, and H$_{24}$Rb-$I\overline{4}$ stable above 200\,GPa.

Similar to the Pb-H structures predicted by Chen \textit{et al.}~\cite{Chen2021}, the H$_5$Pb-C2/m prototype also exhibits polymerisation of hydrogen into H$_3$ chains.
The H$_{23}$Pb-$C2/m$ structure is identical to the prototype discussed in the previous section, emphasising the stability of this prototype in varying chemical and physical settings.
H$_{22}$Na-$R\overline{3}m$ appears to be a hcp lattice of H$_2$ molecules with Na atoms forming a rhombohedral lattice.
Lastly, H$_{24}$Rb-$I\overline{4}$ belongs to the same `pseudotetragonal' family of structures $I4/mmm$ seen by Semenok \textit{et al.}~\cite{Semenok2022} at lower pressures.

\section{Discussion}

The \textsc{EDDP}-assisted structure searches have uncovered previously unseen structures, contributing to a more comprehensive understanding of stable binary hydrides.
Crucially, this was possible with effectively zero prior knowledge of the systems with automatically generated machine-learning potentials constructed through random structure search.
This approach shows that models capable of generating extremely sensible structures and ranking them with a reasonable accuracy are sufficient to accelerate materials discovery.
Empowered by the high-throughput capabilities of CPU-optimised \textsc{EDDP}s and the ready availability of CPU resources, we can explore the chemical space in depth.
By doing so, we extend the diversity of hydride structural prototypes and provide another route for high-pressure hydride discovery~\cite{Ghahremanpour2018,Shipley2021,Saha2023}.

Many of the binary hydride systems explored here had previously been studied using conventional DFT-based structure prediction techniques.
However, structure prediction can easily miss structures due to computational constraints restricting the search to a narrow range of stoichiometries, or not exploring deep enough into the energy landscape.
For example, Frapper and co-workers used an evolutionary algorithm to search for all H$_m$Cl$_n$ with $m+n<30$~\cite{Zeng2017}.
While H$_{11}$Cl-$P\overline{6}m2$ (predicted here) contains only 12 atoms, and so would have eventually been found by their method, the \textsc{EDDP}-accelerated searches enable faster exhaustive explorations of each composition, uncovering chemical diversity that might otherwise be overlooked.
In addition to deeper exploration within each composition, the \textsc{EDDP} approach broadens the scope of exploration across compositions.
Previous studies, such as Zarifi~\textit{et al.}'s search for H$_n$Fe ($n=5-8$)~\cite{Zarifi2018}, missed H$_{13}$Fe$_3$-$I4/mmm$, while Chen \textit{et al.}'s search for H$_n$Pb ($n=2,4,6,8$)~\cite{Chen2021}, missed H$_5$Pb-$C2/m$ and H$_{23}$Pb-$C2/m$.

Our approach, both in terms of data generation and choice of model architecture, is designed for rapid exploration.
The dataset contains small unit cells which are fast to compute using DFT and trivially parallelisable over many CPU cores.
The \textsc{EDDP}s in this work take around 12 hours on 320 CPU cores to generate the data and train the ensemble of potentials.
The choice to train only on energies and with fixed feature vectors means they can be precomputed and stored in memory, significantly lowering the cost of training.
This also enables full-batch and second-order training schemes such as Levenberg-Marquardt.
As a rough guide, to fit an \textsc{EDDP} on a pre-made binary hydride dataset ($\sim$15,000 structures) would take around 30 minutes on an M1 MacBook Pro using 8 CPU cores, including the time to train 90 individual potentials and generate an ensemble using non-negative least-squares.

CPU-based software offers many practical advantages in the design and deployment of machine learning potentials.
It enables the development of code with complex control flows—including branching and loops—and supports both single and double precision without requiring specialised hardware.
This flexibility in precision is critical for scientific computing, particularly in optimisation problems.
Notably, an \textsc{EDDP} relaxation in double precision on just 4 CPU cores (M1 MacBook Pro) is 2–4 times faster than a single-precision MatterSim calculation on a high-end GPU (Nvidia A6000).

With fast training and inference of potentials, we can more easily search broadly in composition space.
By doing so, we can begin to tackle the challenge of predicting ordered, off-stoichiometry compounds that may be critical for understanding synthesis routes, and material properties.
We can also target hydrogen-rich compositions which may provide useful materials for hydrogen storage.
However, these searches represent an approximation of the energy landscape, neglecting effects from finite-temperature, quantum ionic motion, spin-orbit coupling, and strongly-correlated electrons.
The dataset accompanying this work will allow for more detailed studies on these materials.

Revisiting all of the binary hydrides with \textsc{EDDP}s has revealed many previously unexplored structures, suggesting that our understanding of these systems remains incomplete.
Therefore, it would be worth revisiting other previously studied systems in high-pressure hydrides and beyond with \textsc{EDDP}-accelerated structure prediction.

{\it Supplementary Material} Benchmarks of MLIPs for random structure search and tabulated structure prototypes are included in the supplementary material.

{\it Data availability} The data that support the findings of this article are openly available~\cite{conway_2025_14720015}.

{\it Acknowledgements.} We thank Ma\'elie Causs\'e and Kazuto Akagi for helpful discussions. Parts of this work used the ARCHER2 UK National Supercomputing Service (https://www.archer2.ac.uk)~\cite{Beckett2024} accessed through EPSRC RAP Open Access.


%

\onecolumngrid

\clearpage

\begin{center}
  \textbf{\large Supplementary information for `Accelerating Crystal Structure Prediction Using Data-Derived Potentials: High-Pressure Binary Hydrides'}
\end{center}
\vspace{1cm}

\section{Data Availability}
The full dataset, including training, searching, and final DFT data can be found here:

\href{https://zenodo.org/records/14720015}{https://zenodo.org/records/14720015}


\clearpage

\section{Comparison of MLIPs for Random Structure Search}

We performed benchmark AIRSS calculations on H$_{10}$La and $H_{25}$Cs with several open-source pretrained models; MatterSim-v1.0.0-1M~\cite{Yang2024}, ORB-V2~\cite{Neumann2024}, Mace-MP-0 and Mace-MPA-0~\cite{Batatia2024}.
The searches use the same input parameters to generate structures with two formula units of H$_{10}$La and H$_{25}$Cs which are optimised using the potentials at 100\,GPa.
The final relaxed structures are recalculated in \textsc{castep} using single-point energy calculations.
For the MatterSim, Orb, Mace and JAX\_EDDP searches the structures are optimised using the FIRE optimisation scheme~\cite{Bitzek2006} implemented in ASE~\cite{HjorthLarsen2017}.
For the EDDP searches, the structures are optimised using a two-point steepest descent method~\cite{Barzilai1988} implemented in the \texttt{repose} code~\cite{Pickard2022}.

For each of the calculators, we show the optimisation of a shaken reference structure on both CPU and GPU hardware. All CPU calculations were performed on an M1 MacBook Pro with OMP parallelism over 4 cores. All GPU calculations were performed on an NVIDIA RTX A6000. All data from the \texttt{repose} code is computed with double precision, the data from JAX\_EDDP is computed with mixed single and double precision, whereas all other codes are single precision.
\texttt{repose} can be compiled in single precision which is faster, but given the speed of double precision and since all modern CPUs support double precision (in contrast to most GPUs) we typically use double precision to ensure stable geometry optimisation.
\clearpage

\subsection{H$_{10}$La}

\begin{figure}[htpb]
  \centering
  \includegraphics[width=1\textwidth]{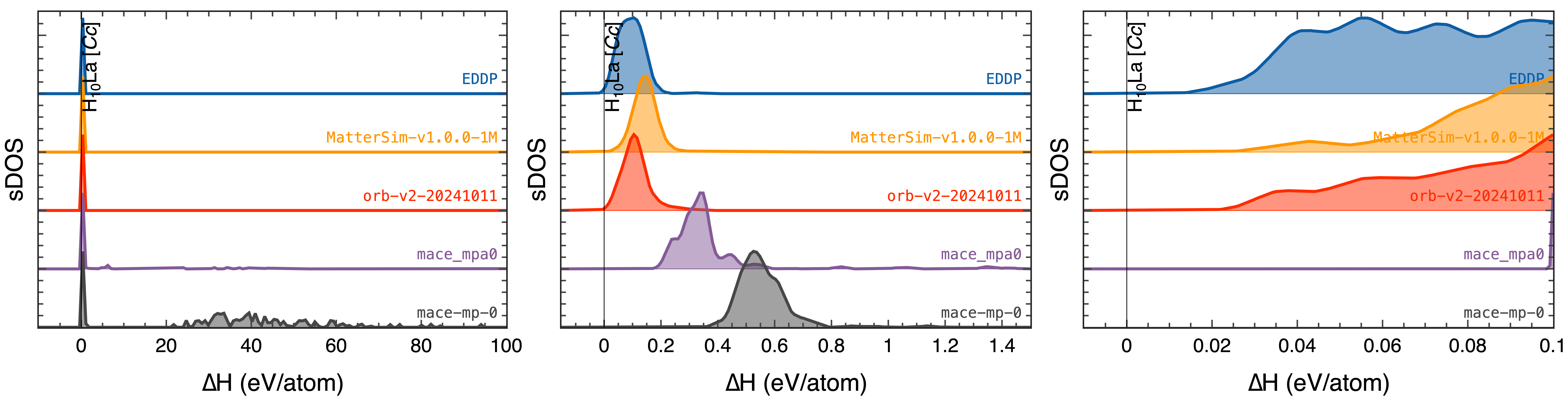}
  \caption{Structural Densities of States (sDOS) for AIRSS searches for 2 formula units of H$_{10}$La at 100\,GPa with various potentials. For visibility, the distributions are plotted on different energy scales and scaled to be visible in each plot. The total number of structures in each search was 512.}
  \label{fig:La-SDOS}
\end{figure}

Figure~\ref{fig:La-SDOS} shows the distribution of structures generated, relative to the lowest enthalpy structure.
For each potential, we performed AIRSS searches with the input shown in Listing~\ref{lst:La-AIRSS}.

Of the potentials used, only EDDP, MatterSim, and Orb models were able to reliably generate sensible candidate structures.
They each have a peak which is within 200\,meV/atom of the reference structure, however the peak of the EDDP sDOS is broader, indicating a better distribution of lower enthalpy structures.
Both MACE models struggle with this example, quickly running into instabilities and generating structures with many atomic close contacts, often relaxing into unphysical configurations with overlapping atoms, resulting in structures tens of eV/atom higher in enthalpy.
For the MACE-MP-0 model, the vast majority of structures are of this type.
They are less common with the MACE\_MPA0 model.

\begin{figure}[htpb]
  \centering
  \includegraphics[height=8cm]{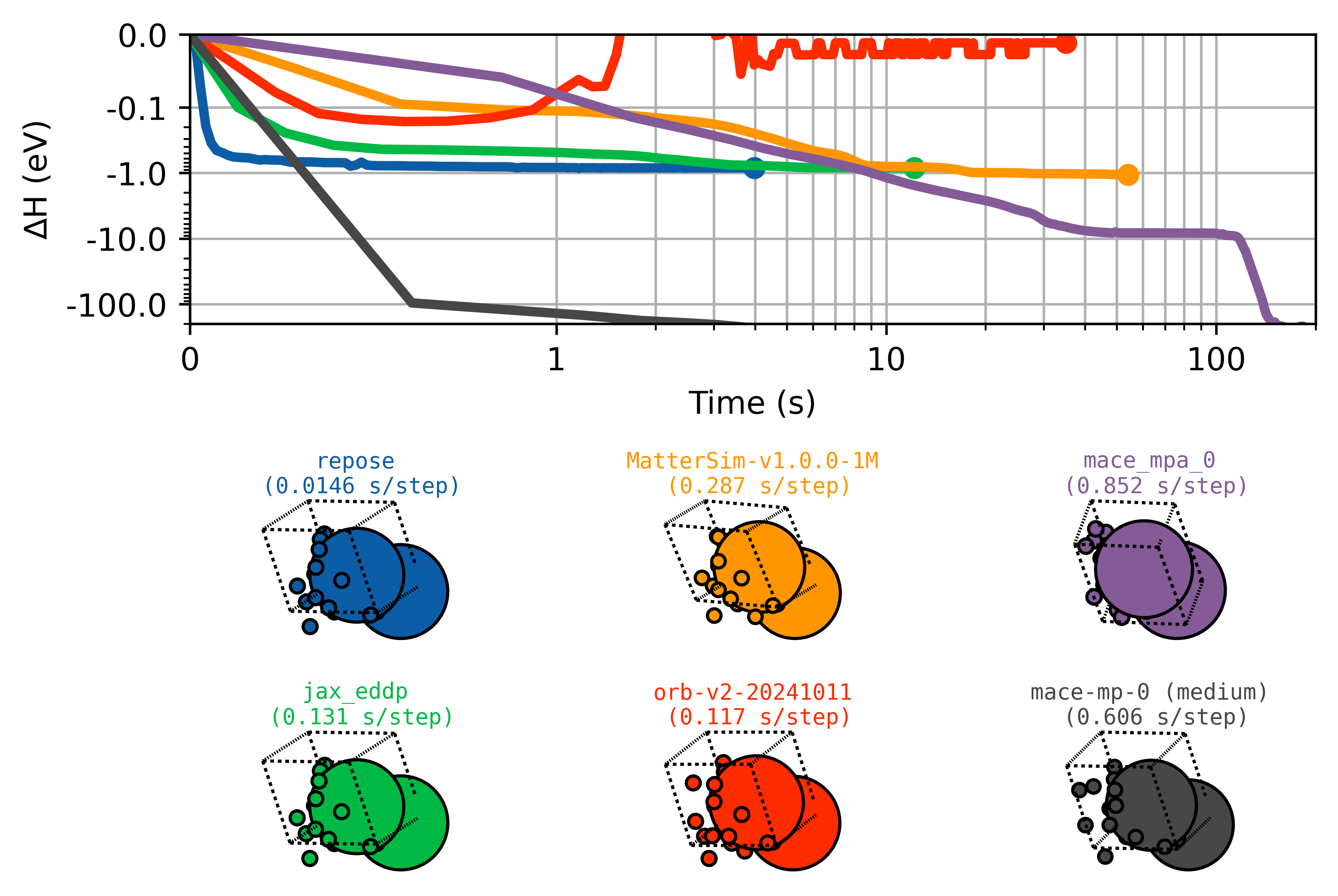}
  \caption{Geometry optimisations of shaken H$_{10}$La structures on \textbf{CPU}.}
  \label{fig:La-opt-cpu}
\end{figure}

On CPU (FIG~\ref{fig:La-opt-cpu}), the fastest universal potential is the orb model (0.1s/step), followed by MatterSim (0.3s/step).
Due to the non-conservative implementation of the Orb model, the optimisation is not stable. Despite this, it still generates sensible structures, successfully avoiding holes in the potential energy surface but failing to relax to the minimum.
It could be interpreted as an annealing stage.
The collapse of structures optimised using the MACE models is also shown in FIG~\ref{fig:La-opt-cpu}.
The models appear slower than usual because the later stages of the optimisation involve large numbers of atoms in the neighbour shells.

\begin{figure}[htpb]
  \centering
  \includegraphics[height=8cm]{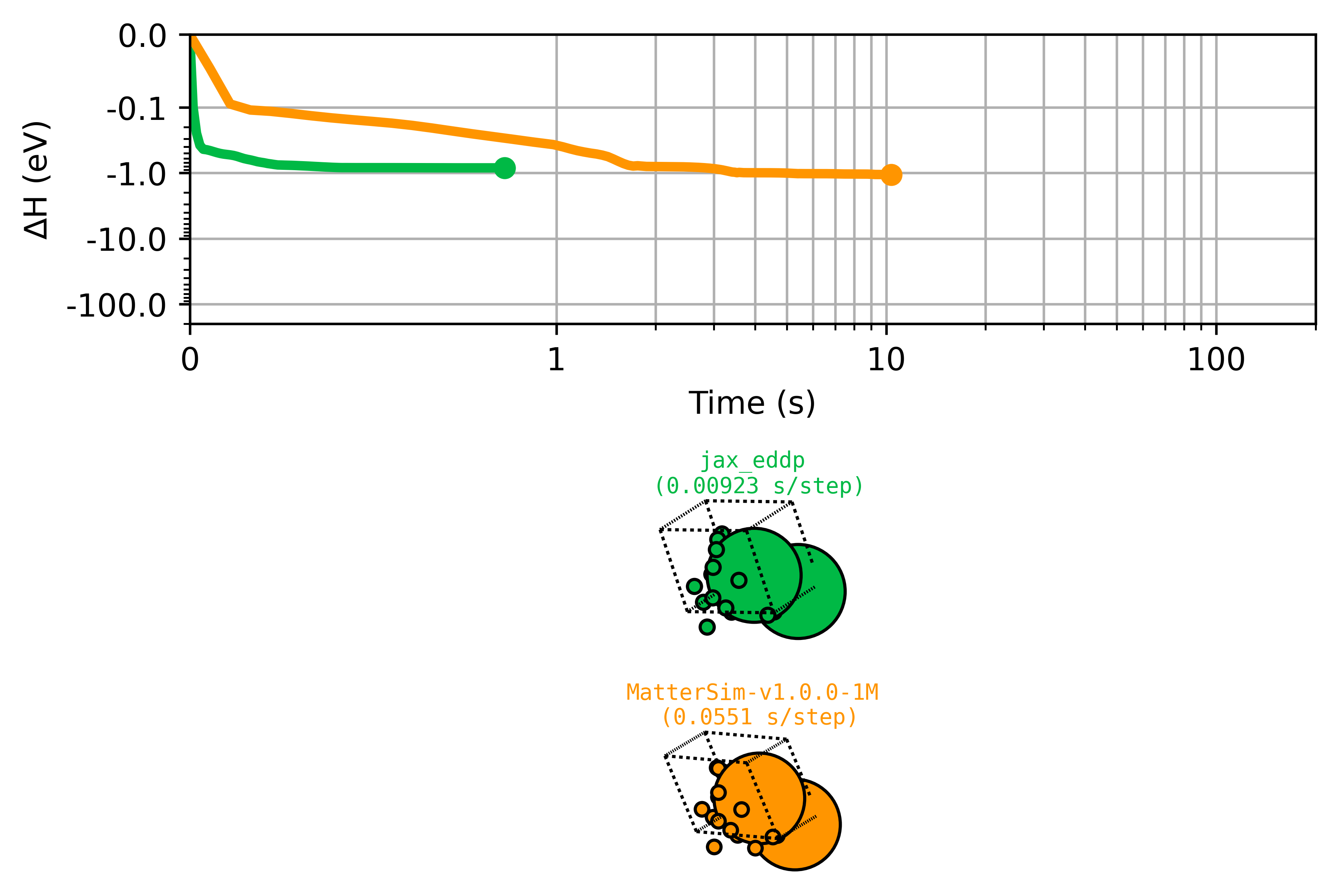}
  \caption{Geometry optimisations of shaken H$_{10}$La structures on \textbf{GPU}.}
  \label{fig:La-opt-gpu}
\end{figure}

On GPU (FIG~\ref{fig:La-opt-gpu}), we compare our JAX implementation of the EDDP framework with MatterSim-v1.
Given the much smaller computational load in the EDDP, it is approximately 10 times faster.
In production, it would be beneficial to batch structural optimisations to leverage the GPU parallelism.

It is likely that the reference structure was present in the training data of the EDDP. Similar structures are present in the Alexandria dataset and so are likely included in the training of the MatterSim, Orb, and Mace-MPA-0 models, but not the MACE-MP-0 model.

\begin{lstlisting}[float=tpb,language=Bash,caption={AIRSS input for H$_{10}$La structure generation.}, label={lst:La-AIRSS}]
%BLOCK POSITIONS_ABS
La      0.00000 0.00000 0.00000 # 1-R3 % NUM=1
H       0.00000 0.00000 0.00000 # 2-R3 % NUM=10
%ENDBLOCK POSITIONS_ABS
#FOCUS=2
#NFORM=2
#SYMMOPS=2-4
#SLACK=0.5
#OVERLAP=0.3
#COMPACT
#CELLADAPT
#TARGVOL=7.06
#MINSEP=1.0 La-La=3.66 La-H=2.12 H-H=0.87
KPOINTS_MP_SPACING 0.07
SYMMETRY_GENERATE
SNAP_TO_SYMMETRY
\end{lstlisting}

\clearpage

\clearpage

\subsection{H$_{25}$Cs}

\begin{figure}[htpb]
  \centering
  \includegraphics[width=\textwidth]{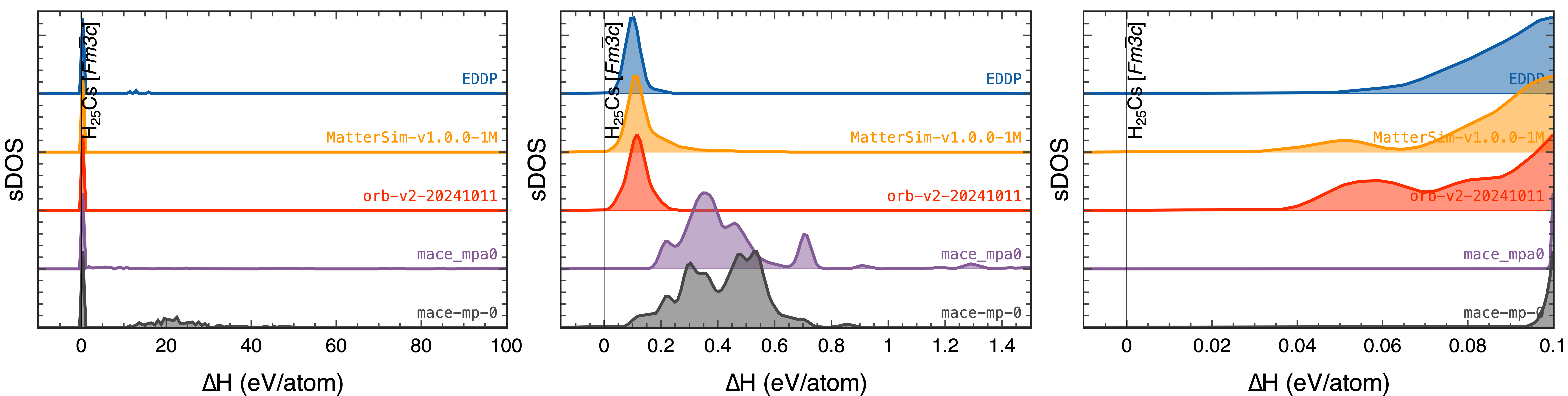}
  \caption{Structural Densities of States (sDOS) for AIRSS searches for 2 formula units of H$_{25}$Cs at 100\,GPa with various potentials. For visibility, the distributions are plotted on different energy scales and scaled to be visible in each plot. The total number of structures in each search was 512. All enthalpies were then recalculated using single-point DFT calculations for comparison and plotted relative to the lowest enthalpy structure.}
  \label{fig:Cs-SDOS}
\end{figure}

\begin{figure}[htpb]
  \centering
  \includegraphics[height=8cm]{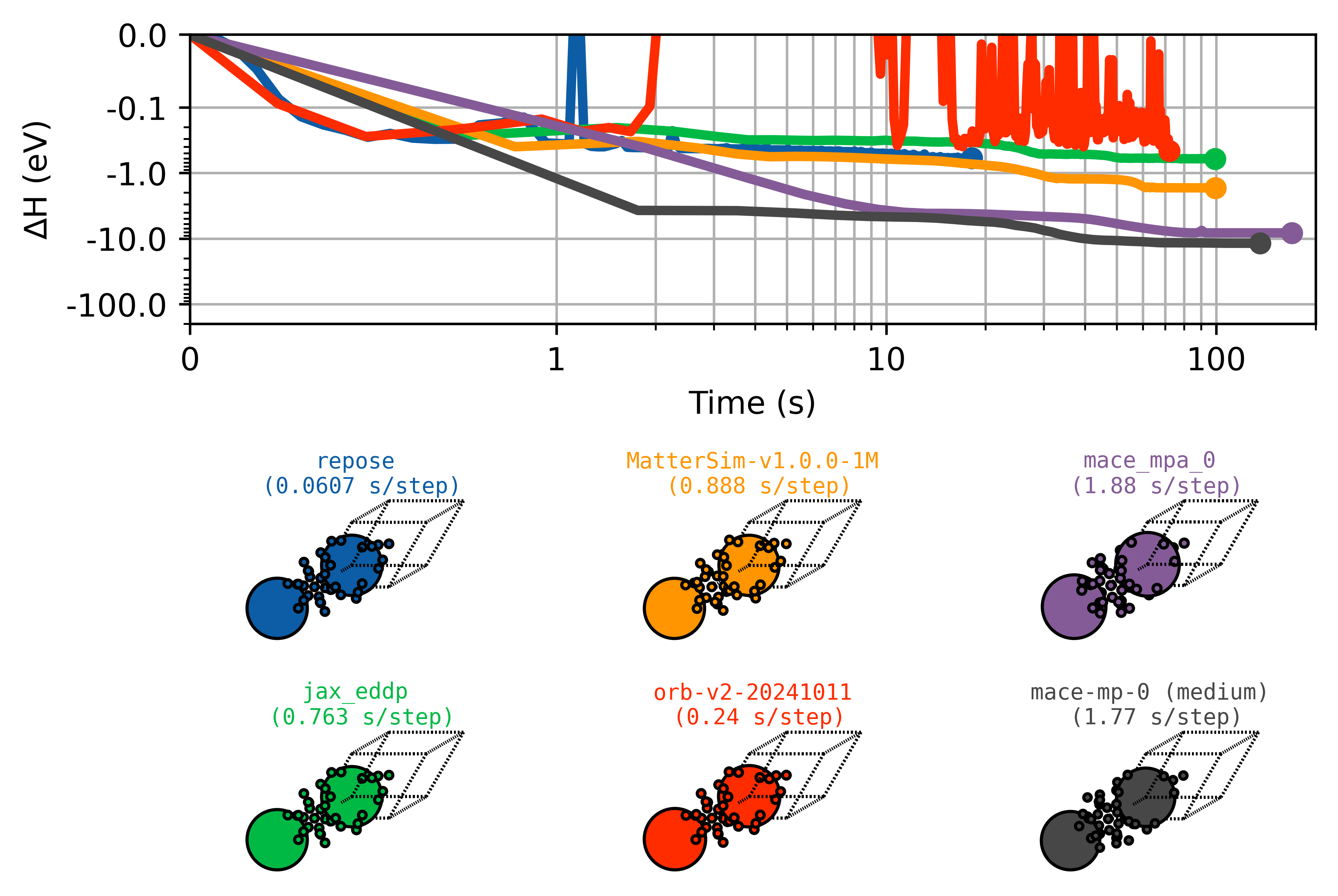}
  \caption{Geometry optimisations of shaken H$_{25}$Cs structures on \textbf{CPU}.}
  \label{fig:Cs-opt-cpu}
\end{figure}

\begin{figure}[htpb]
  \centering
  \includegraphics[height=8cm]{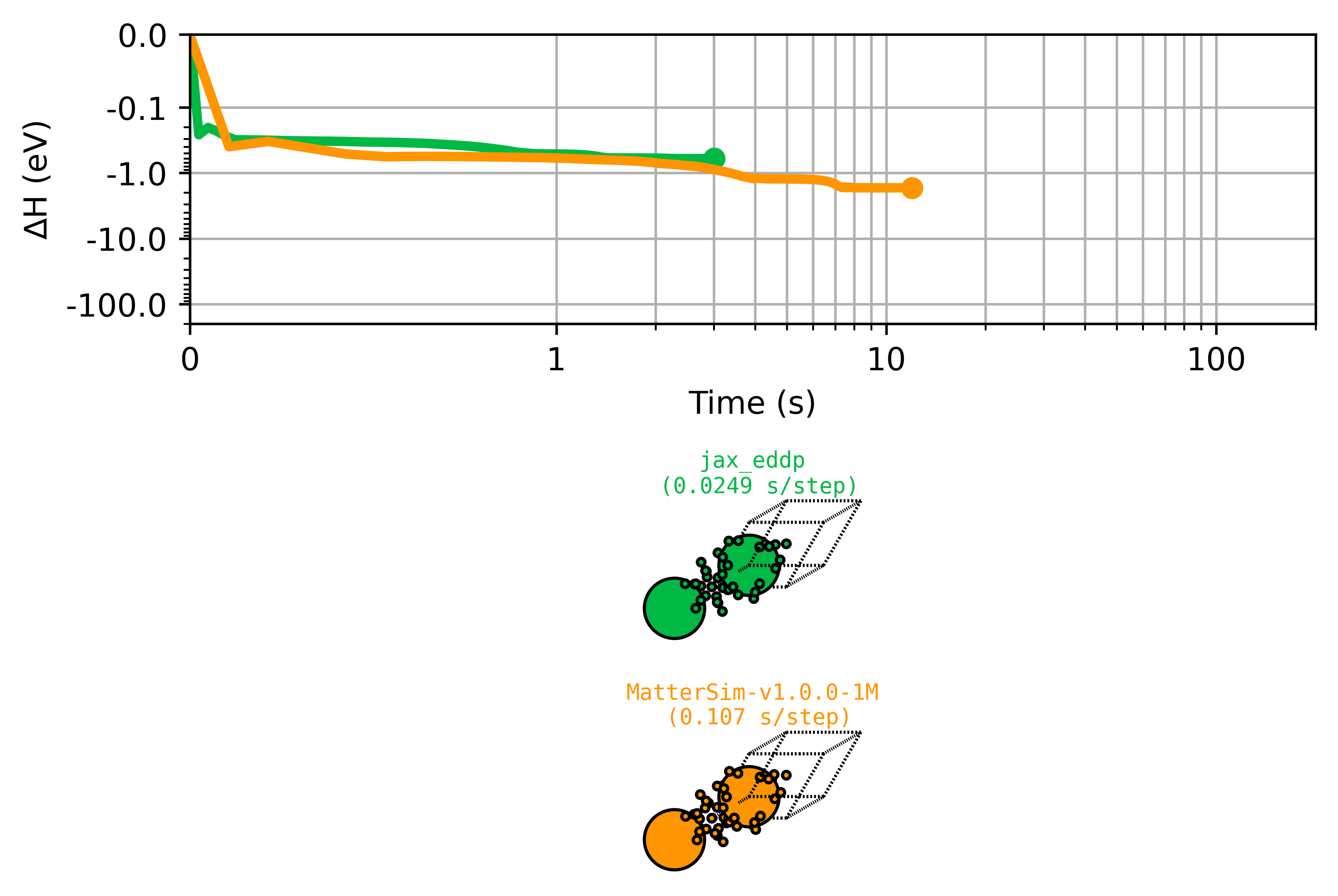}
  \caption{Geometry optimisations of shaken H$_{25}$Cs structures on \textbf{GPU}.}
  \label{fig:Cs-opt-gpu}
\end{figure}

Figure~\ref{fig:Cs-SDOS} shows the distribution of structures generated, relative to the lowest enthalpy structure.
For each potential, we performed AIRSS searches with the input shown in Listing~\ref{lst:Cs-AIRSS}.
FIG~\ref{fig:Cs-opt-cpu} and FIG~\ref{fig:Cs-opt-gpu} show CPU and GPU optimisation benchmarks, respectively.

The search results and timing benchmarks are consistent with the H$_{10}$La example.
The reference structure was not present in the EDDP training data and it is unlikely that the reference structure was present in the training data of any of the other models.

\begin{lstlisting}[float=htpb,language=Bash,caption={AIRSS input for H$_{25}$Cs structure generation.}, label={lst:Cs-AIRSS}]
%BLOCK POSITIONS_ABS
H       6.64214   5.44919   4.68684 # 1-R3 % NUM=25
Cs      7.46554   4.30074   1.22530 # 2-R3 % NUM=1
%ENDBLOCK POSITIONS_ABS
#SYMMOPS=2-48
#NFORM=2
#SLACK=0.25
#OVERLAP=0.1
#COMPACT
#CELLADAPT
#MINSEP=1.0 H-H=0.75 H-Cs=2.36 Cs-Cs=4.00
#TARGVOL=5.78
KPOINTS_MP_SPACING 0.07
SYMMETRY_GENERATE
SNAP_TO_SYMMETRY
\end{lstlisting}

\clearpage

\clearpage

\section{Tables of Structure Prototypes}

Structure prototypes are represented by the anonymized formula, H$_{m}$X$_{n}$, where X represents the heavy atom.
The prototypes are grouped by the Bravais lattice of the crystal, ignoring the hydrogen atoms.
For each prototype, the elements, X, for which they are within 15~meV/atom of the convex hull at 100~GPa are listed. Structures not on the convex hull are listed with their distance in meV/atom in parenthesis.

\clearpage

\begin{table*}[h]
  \centering
  \caption{Simple Cubic}
\begin{tabular}{lllp{0.5\linewidth}l}
    \hline\hline
    Formula & Spacegroup & n.\ form & X ($\Delta H$ [meV/atom]) & Heavy atom Spacegroup \\
    HX                                 & $Pm\overline{3}m$                  & 2                                  & K, Na                                                            & $Pm\overline{3}m$                  \\
    HX                                 & $Pm\overline{3}m$                  & 1                                  & Nd, Pr, Sm (2), Sr (12)                                          & $Pm\overline{3}m$                  \\
    H$_{8}$X$_{3}$                     & $P2/m$                             & 1                                  & Rh (8)                                                           & $Pm\overline{3}m$                  \\
    H$_{8}$X$_{3}$                     & $Pm\overline{3}m$                  & 1                                  & Ru (9), Fe (14)                                                  & $Pm\overline{3}m$                  \\
    H$_{14}$X$_{5}$                    & $P2/m$                             & 1                                  & Co (9)                                                           & $Pm\overline{3}m$                  \\
    H$_{17}$X$_{6}$                    & $P2/m$                             & 1                                  & Rh, Co (2), Ir (4)                                               & $Pm\overline{3}m$                  \\
    H$_{3}$X                           & $Pm\overline{3}m$                  & 1                                  & Fe (5), Co (6), Ru (12)                                          & $Pm\overline{3}m$                  \\
\hline \hline
  \end{tabular}
\end{table*}


\begin{table*}[t]
  \centering
  \caption{FCC}
  \begin{tabular}{lllp{0.5\linewidth}l}
    \hline\hline
    Formula & Spacegroup & n.\ form & X ($\Delta H$ [meV/atom]) & Heavy atom Spacegroup \\

    HX$_{3}$                           & $C2/c$                             & 2                                  & Pd, Yb (3), Er (12)                                              & $Fm\overline{3}m$                  \\
    H$_{2}$X$_{5}$                     & $C2/m$                             & 1                                  & Pd                                                               & $Fm\overline{3}m$                  \\
    HX$_{2}$                           & $R\overline{3}m$                   & 1                                  & Ag, Pt, Os (7), Pd (9), Ni (9)                                   & $Fm\overline{3}m$                  \\
    HX$_{2}$                           & $I4_1/amd$                         & 2                                  & Er, Yb, Pd (7), Tm (13), Lu (13)                                 & $Fm\overline{3}m$                  \\
    HX                                 & $Immm$                             & 1                                  & Ta (8), Nb (14)                                                  & $Fm\overline{3}m$                  \\
    HX                                 & $Fm\overline{3}m$                  & 1                                  & Co, Tm, Sc, Ru, Rh, Pd, Y, Os, Lu, Li, La, Fe, Cu, Ni, Yb, Er, Ti (6), Pt (8) & $Fm\overline{3}m$                  \\
    H$_{2}$X                           & $Fm\overline{3}m$                  & 2                                  & V                                                                & $Fm\overline{3}m$                  \\
\hline \hline
  \end{tabular}
\end{table*}


\begin{table*}[t]
  \centering
  \caption{BCC}
  \begin{tabular}{lllp{0.5\linewidth}l}
    \hline\hline
    Formula & Spacegroup & n.\ form & X ($\Delta H$ [meV/atom]) & Heavy atom Spacegroup \\

    H$_{3}$X                           & $I\overline{4}3d$                  & 8                                  & Ta, Nb (3)                                                       & $I\overline{4}3d$                  \\
    H$_{15}$X$_{4}$                    & $I\overline{4}3d$                  & 2                                  & Hf, Ti, Zr                                                       & $I\overline{4}3d$                  \\
    & & & & \\
    H$_{2}$X                           & $I4/mmm$                           & 1                                  & Hf, Zr                                                           & $Im\overline{3}m$                  \\
    H$_{3}$X                           & $Pm\overline{3}n$                  & 2                                  & Al                                                               & $Im\overline{3}m$                  \\
    H$_{6}$X                           & $Im\overline{3}m$                  & 1                                  & Tb (8), Eu (12)                                                  & $Im\overline{3}m$                  \\
    & & & & \\

    \hline \hline
  \end{tabular}
\end{table*}


\begin{table*}[t]
  \centering
  \caption{Hexagonal}
  \begin{tabular}{lllp{0.5\linewidth}l}
    \hline\hline
    Formula & Spacegroup & n.\ form & X ($\Delta H$ [meV/atom]) & Heavy atom Spacegroup \\

    HX$_{3}$                           & $P6_3/m$                           & 2                                  & Ge (5)                                                           & $P6_3/m$                           \\
    & & & & \\
    H$_{4}$X$_{7}$                     & $C2/m$                             & 1                                  & Br, Cl                                                           & $P6/mmm$                           \\
    H$_{2}$X                           & $P\overline{3}m1$                  & 1                                  & Be                                                               & $P6/mmm$                           \\
    H$_{2}$X                           & $C2/m$                             & 1                                  & La (2)                                                           & $P6/mmm$                           \\
    H$_{2}$X                           & $P6/mmm$                           & 1                                  & Ba, Sr, La (4)                                                   & $P6/mmm$                           \\
    H$_{3}$X                           & $Cmmm$                             & 2                                  & Cs (12)                                                          & $P6/mmm$                           \\
\hline \hline
  \end{tabular}
\end{table*}


\begin{table*}[t]
  \centering
  \caption{Trigonal}
  \begin{tabular}{lllp{0.5\linewidth}l}
    \hline\hline
    Formula & Spacegroup & n.\ form & X ($\Delta H$ [meV/atom]) & Heavy atom Spacegroup \\

    HX$_{2}$                           & $C2/m$                             & 2                                  & Pm                                                               & $P\overline{3}m1$                  \\
    HX$_{2}$                           & $C2/m$                             & 1                                  & Dy, Gd, Ho, Ta, Tb, Er (7)                                       & $P\overline{3}m1$                  \\
    H$_{3}$X                           & $Cm$                               & 2                                  & Pb (6)                                                           & $P\overline{3}m1$                  \\
    H$_{7}$X                           & $P\overline{3}m1$                  & 2                                  & Tc, Re (12)                                                      & $P\overline{3}m1$                  \\
    & & & & \\
    HX                                 & $R\overline{3}m$                   & 1                                  & Nb, Ta                                                           & $R\overline{3}m$                   \\
    H$_{7}$X$_{3}$                     & $P2_1/m$                           & 2                                  & Nb, Ta, V                                                        & $R\overline{3}m$                   \\
\hline \hline
  \end{tabular}
\end{table*}


\begin{table*}[t]
  \centering
  \caption{Simple Tetragonal}
  \begin{tabular}{lllp{0.5\linewidth}l}
    \hline\hline
    Formula & Spacegroup & n.\ form & X ($\Delta H$ [meV/atom]) & Heavy atom Spacegroup \\

    HX$_{2}$                           & $P4/mmm$                           & 1                                  & Sm, Lu (5), Pd (8)                                               & $P4/mmm$                           \\
    H$_{6}$X                           & $P4/mmm$                           & 1                                  & Ba (8)                                                           & $P4/mmm$                           \\
    H$_{7}$X                           & $P4/nmm$                           & 2                                  & Cs                                                               & $P4/mmm$                           \\
    H$_{9}$X                           & $Pm$                               & 1                                  & Mn (13)                                                          & $P4/mmm$                           \\
    & & & & \\
    H$_{2}$X                           & $P4/mbm$                           & 4                                  & Li (9)                                                           & $P4/mbm$                           \\
    & & & & \\
\hline \hline
  \end{tabular}
\end{table*}


\begin{table*}[t]
  \centering
  \caption{Body Centred Tetragonal}
  \begin{tabular}{lllp{0.5\linewidth}l}
    \hline\hline
    Formula & Spacegroup & n.\ form & X ($\Delta H$ [meV/atom]) & Heavy atom Spacegroup \\

    H$_{7}$X$_{3}$                     & $I4cm$                             & 2                                  & Pm (13)                                                          & $I4cm$                             \\
    & & & & \\
    H$_{13}$X$_{3}$                    & $I\overline{4}m2$                  & 1                                  & Cs                                                               & $I\overline{4}m2$                  \\
    & & & & \\
    HX$_{2}$                           & $I4/mmm$                           & 1                                  & Nd                                                               & $I4/mmm$                           \\
    HX                                 & $I4/mmm$                           & 1                                  & Hf, Ti, Zr (4), Ta (8)                                           & $I4/mmm$                           \\
    H$_{2}$X                           & $I4/mmm$                           & 2                                  & Co, Mn, Ni, Fe (2)                                               & $I4/mmm$                           \\
\hline \hline
  \end{tabular}
\end{table*}


\begin{table*}[t]
  \centering
  \caption{Primitive Orthorhombic}
  \begin{tabular}{lllp{0.5\linewidth}l}
    \hline\hline
    Formula & Spacegroup & n.\ form & X ($\Delta H$ [meV/atom]) & Heavy atom Spacegroup \\

    H$_{3}$X                           & $P2_12_12_1$                       & 4                                  & Cl, Br                                                           & $P2_12_12_1$                       \\
    & & & & \\
    H$_{6}$X                           & $Pmn2_1$                           & 2                                  & Kr (12)                                                          & $Pmmn$                             \\
    H$_{12}$X                          & $P2_1/m$                           & 2                                  & Ba                                                               & $Pmmn$                             \\
    & & & & \\
    H$_{2}$X                           & $P2_12_12_1$                       & 4                                  & Ar (11), Kr (13)                                                 & $Pnma$                             \\
    H$_{2}$X                           & $Pnma$                             & 4                                  & Cr, Dy, Mg, Mo, Nb, V, Ta (1), Pm (1), Kr (7), Ar (11), Xe (14)  & $Pnma$                             \\
    & & & & \\

    \hline \hline
  \end{tabular}
\end{table*}


\begin{table*}[t]
  \centering
  \caption{Base Centred Orthorhombic}
  \begin{tabular}{lllp{0.5\linewidth}l}
    \hline\hline
    Formula & Spacegroup & n.\ form & X ($\Delta H$ [meV/atom]) & Heavy atom Spacegroup \\

    H$_{2}$X                           & $Amm2$                             & 4                                  & Cs (7)                                                           & $Amm2$                             \\
    & & & & \\
    HX                                 & $Cmcm$                             & 2                                  & Cl, Cs, Dy, Er, Gd, Ho, Rb, Sm, Tb, Eu (12)                      & $Cmcm$                             \\
    H$_{2}$X                           & $Cmc2_1$                           & 2                                  & Ar (15)                                                          & $Cmcm$                             \\
    H$_{2}$X                           & $Cmcm$                             & 2                                  & At, I, Ar (13), Kr (14)                                          & $Cmcm$                             \\
    H$_{2}$X                           & $P2_1/c$                           & 4                                  & Rh                                                               & $Cmcm$                             \\
    H$_{5}$X$_{2}$                     & $Cmcm$                             & 2                                  & Co (14)                                                          & $Cmcm$                             \\
\hline \hline
  \end{tabular}
\end{table*}


\begin{table*}[t]
  \centering
  \caption{Face Centred Orthorhombic}
  \begin{tabular}{lllp{0.5\linewidth}l}
    \hline\hline
    Formula & Spacegroup & n.\ form & X ($\Delta H$ [meV/atom]) & Heavy atom Spacegroup \\

    H$_{2}$X                           & $Fddd$                             & 2                                  & Tb (14)                                                          & $Fddd$                             \\
    & & & & \\

    \hline \hline
  \end{tabular}
\end{table*}


\begin{table*}[t]
  \centering
  \caption{Body Centred Orthorhombic}
  \begin{tabular}{lllp{0.5\linewidth}l}
    \hline\hline
    Formula & Spacegroup & n.\ form & X ($\Delta H$ [meV/atom]) & Heavy atom Spacegroup \\

    H$_{11}$X                          & $Imm2$                             & 1                                  & K (2)                                                            & $Immm$                             \\
    & & & & \\
    H$_{2}$X                           & $Ibam$                             & 4                                  & Br                                                               & $Ibam$                             \\
    & & & & \\
    H$_{2}$X                           & $Imma$                             & 2                                  & Tb, Eu (6), Dy (8)                                               & $Imma$                             \\
    H$_{3}$X                           & $C2/m$                             & 2                                  & Rb                                                               & $Imma$                             \\
    H$_{4}$X                           & $C2/c$                             & 2                                  & Ar                                                               & $Imma$                             \\
    H$_{4}$X                           & $Imma$                             & 2                                  & Ar (7)                                                           & $Imma$                             \\
    & & & & \\

    \hline \hline
  \end{tabular}
\end{table*}


\begin{table*}[t]
  \centering
  \caption{Primitive Monoclinic}
  \begin{tabular}{lllp{0.5\linewidth}l}
    \hline\hline
    Formula & Spacegroup & n.\ form & X ($\Delta H$ [meV/atom]) & Heavy atom Spacegroup \\

    H$_{3}$X                           & $P1$                               & 8                                  & S (8)                                                            & $Pm$                               \\
    & & & & \\
    H$_{2}$X                           & $P2_1/m$                           & 2                                  & Eu, Gd, Tb (14)                                                  & $P2_1/m$                           \\
    H$_{5}$X$_{2}$                     & $P2_1/m$                           & 2                                  & Rh                                                               & $P2_1/m$                           \\
    H$_{7}$X                           & $P21$                              & 2                                  & Cl (1), Br (6)                                                   & $P2_1/m$                           \\
    & & & & \\

    \hline \hline
  \end{tabular}
\end{table*}


\begin{table*}[t]
  \centering
  \caption{Base Centred Monoclinic}
  \begin{tabular}{lllp{0.5\linewidth}l}
    \hline\hline
    Formula & Spacegroup & n.\ form & X ($\Delta H$ [meV/atom]) & Heavy atom Spacegroup \\

    H$_{11}$X$_{3}$                    & $P1$                               & 2                                  & Br (3)                                                           & $Cm$                               \\
    & & & & \\
    H$_{3}$X$_{5}$                     & $C2/m$                             & 1                                  & S                                                                & $C2/m$                             \\
    H$_{8}$X$_{3}$                     & $P\overline{1}$                    & 1                                  & Ir                                                               & $C2/m$                             \\
    H$_{14}$X$_{5}$                    & $P1$                               & 1                                  & Ir                                                               & $C2/m$                             \\
    H$_{10}$X$_{3}$                    & $C2/m$                             & 1                                  & Gd, Eu (4), Ho (6)                                               & $C2/m$                             \\
    H$_{7}$X$_{2}$                     & $P1$                               & 3                                  & Ru                                                               & $C2/m$                             \\
\hline \hline
  \end{tabular}
\end{table*}


\begin{table*}[t]
  \centering
  \caption{Triclinic}
  \begin{tabular}{lllp{0.5\linewidth}l}
    \hline\hline
    Formula & Spacegroup & n.\ form & X ($\Delta H$ [meV/atom]) & Heavy atom Spacegroup \\

    H$_{11}$X$_{2}$                    & $P1$                               & 4                                  & Ba (8)                                                           & $P1$                               \\
    & & & & \\
    H$_{9}$X$_{4}$                     & $P\overline{1}$                    & 1                                  & Eu (10)                                                          & $P\overline{1}$                    \\
    H$_{7}$X$_{2}$                     & $P1$                               & 2                                  & Ru (6)                                                           & $P\overline{1}$                    \\
    H$_{17}$X                          & $P\overline{1}$                    & 2                                  & Ar, Kr, Li (8), Na (11)                                          & $P\overline{1}$                    \\
    H$_{42}$X                          & $P\overline{1}$                    & 2                                  & Mg (8), Na (9), Yb (13), Li (14)                                 & $P\overline{1}$                    \\

    \hline \hline
  \end{tabular}
\end{table*}
\clearpage

\section{Summary of Searches and Potentials}

The summaries for each binary system in the data repository contain the following:
\begin{itemize}
  \item Convex hull for training, searching, single-point and final DFT optimised structures at 100\,GPa.
  \item List of structures within 15~meV of the convex hull at 100\,GPa.
  \item Parity plot of DFT energies and predictions in the test set.
  \item Delta plot of DFT energies and prediction errors in the test set. Red line = RMSE, Blue line = MAE. Dashed lines are cumulative averages.
  \item Testing, training and validation errors.
\end{itemize}
\clearpage

\flushleft{
\subsection{Ag-H}}
\subsubsection*{Searching}
\centering
\includegraphics[width=0.4\textwidth]{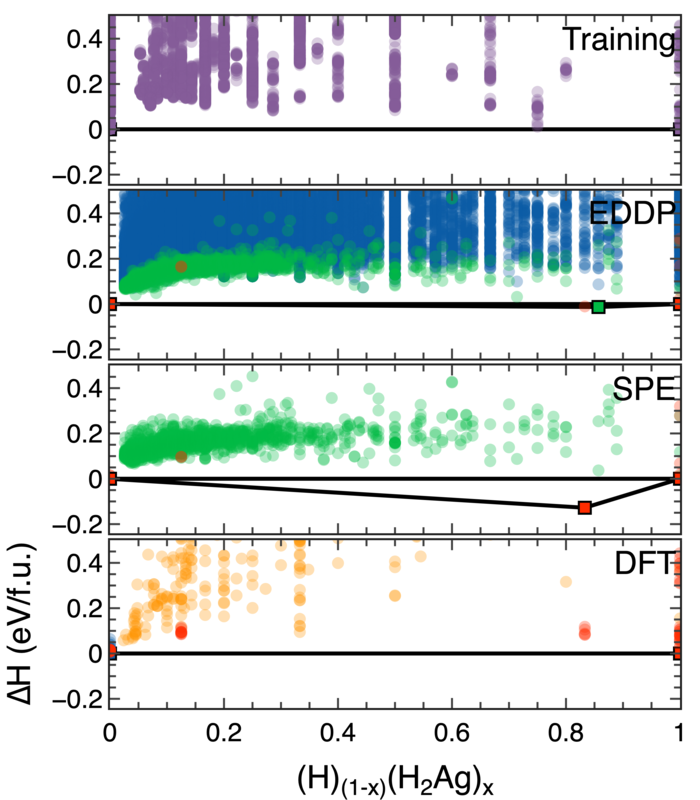}
\footnotesize


\flushleft{
\subsubsection*{\textsc{EDDP}}}
\centering
\includegraphics[width=0.3\textwidth]{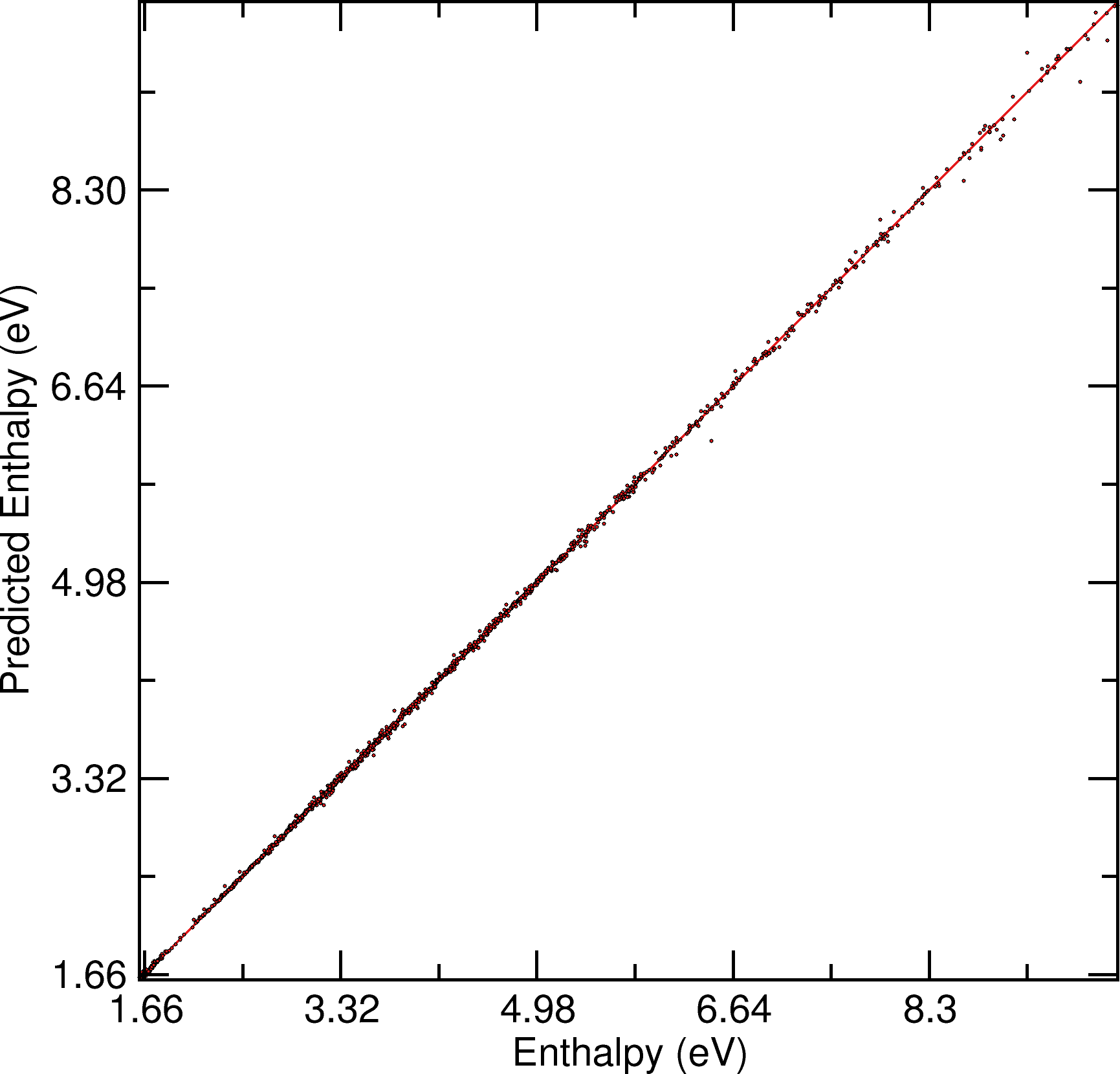}
\includegraphics[width=0.3\textwidth]{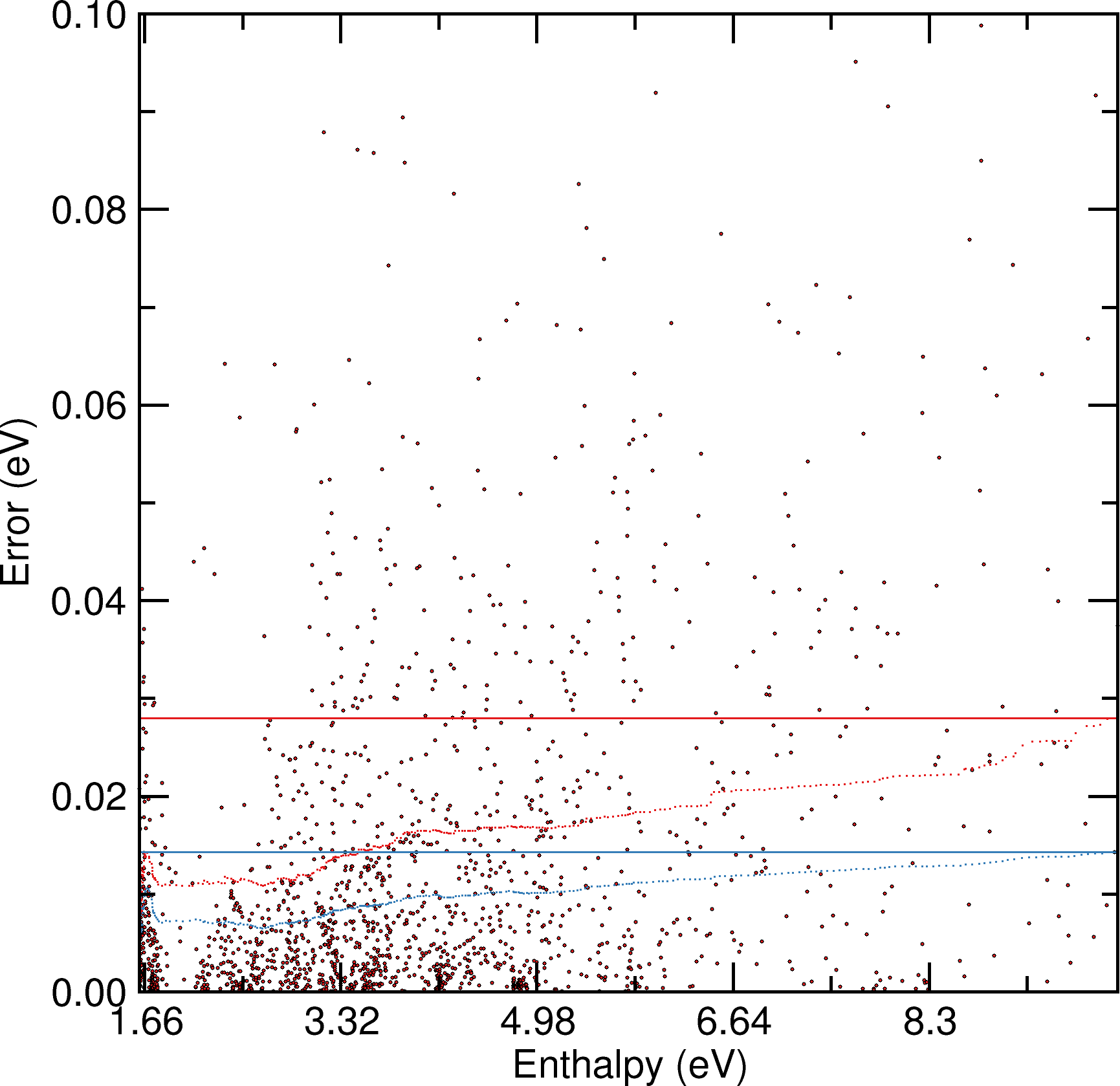}
\centering\begin{verbatim}
training    RMSE/MAE:  13.47  8.45   meV  Spearman  :  0.99990
validation  RMSE/MAE:  23.49  13.74  meV  Spearman  :  0.99985
testing     RMSE/MAE:  27.95  14.28  meV  Spearman  :  0.99986
\end{verbatim}
\clearpage

\flushleft{
\subsection{Al-H}}
\subsubsection*{Searching}
\centering
\includegraphics[width=0.4\textwidth]{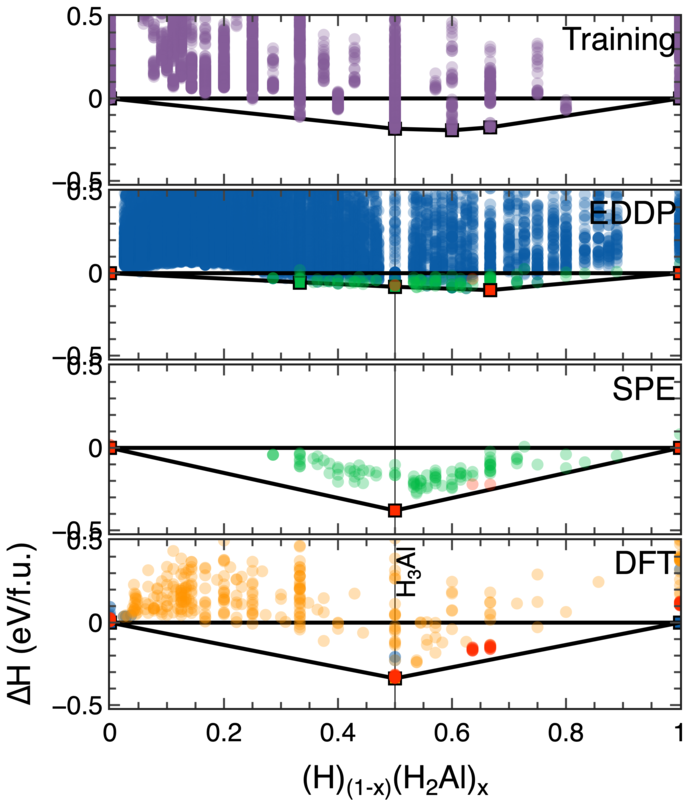}
\footnotesize


\flushleft{
\subsubsection*{\textsc{EDDP}}}
\centering
\includegraphics[width=0.3\textwidth]{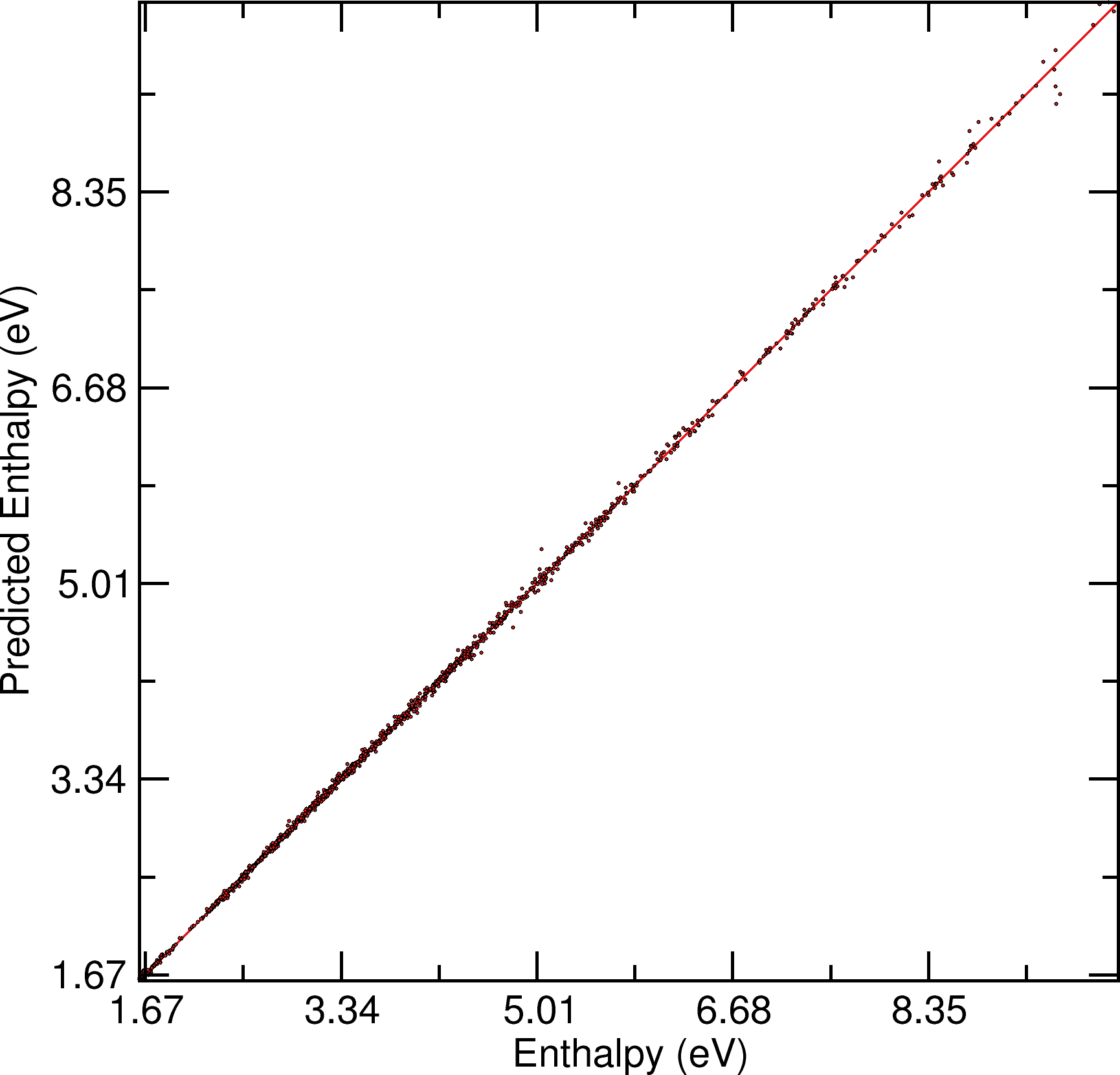}
\includegraphics[width=0.3\textwidth]{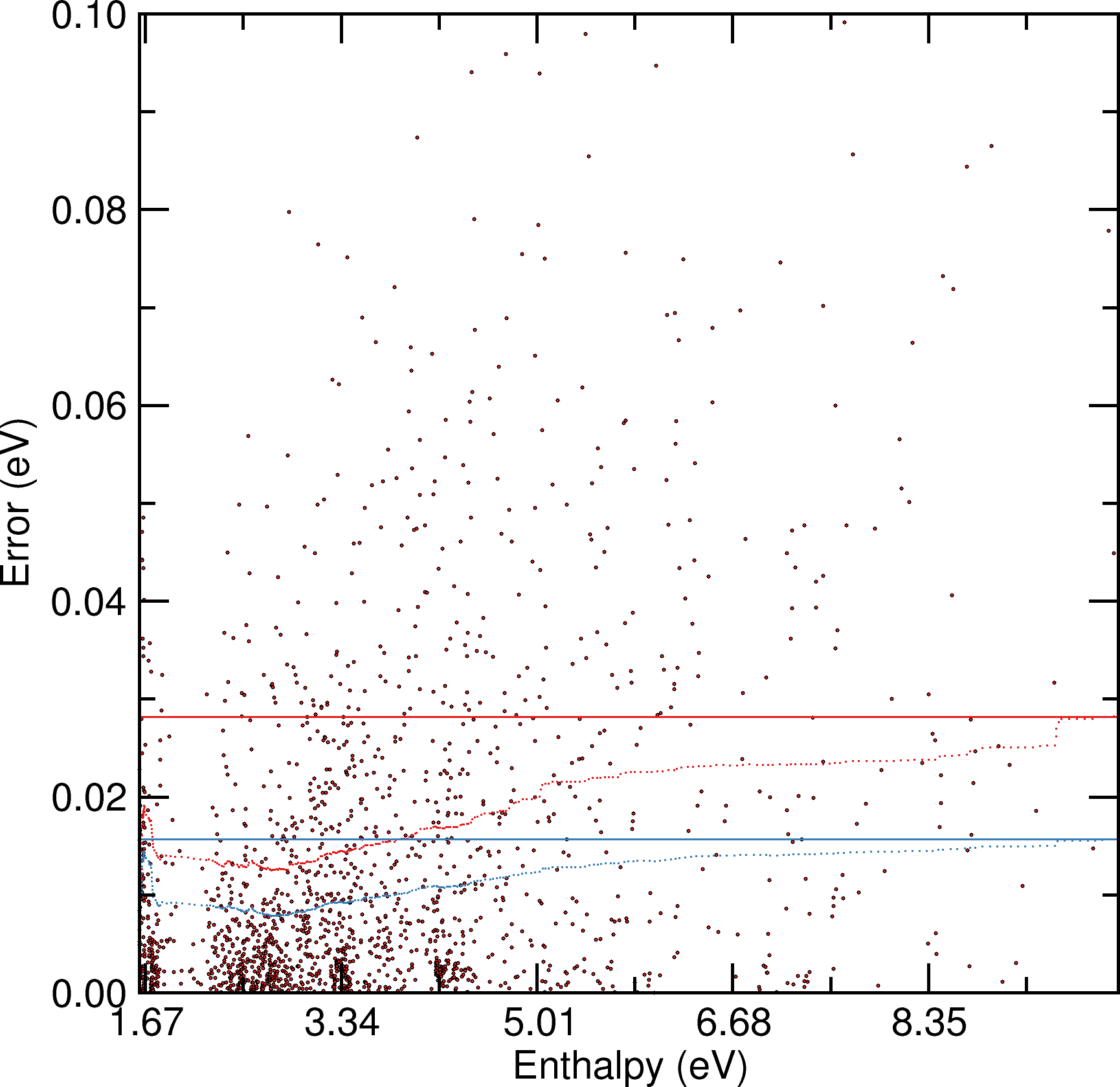}
\centering\begin{verbatim}
training    RMSE/MAE:  16.88  10.44  meV  Spearman  :  0.99987
validation  RMSE/MAE:  24.94  15.44  meV  Spearman  :  0.99980
testing     RMSE/MAE:  28.20  15.68  meV  Spearman  :  0.99982
\end{verbatim}
\clearpage

\flushleft{
\subsection{Ar-H}}
\subsubsection*{Searching}
\centering
\includegraphics[width=0.4\textwidth]{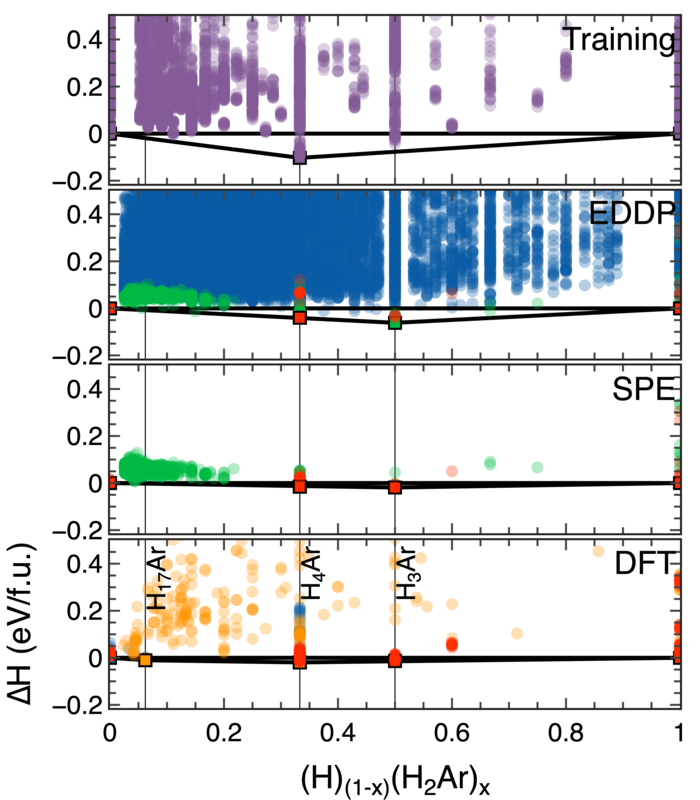}
\footnotesize


\flushleft{
\subsubsection*{\textsc{EDDP}}}
\centering
\includegraphics[width=0.3\textwidth]{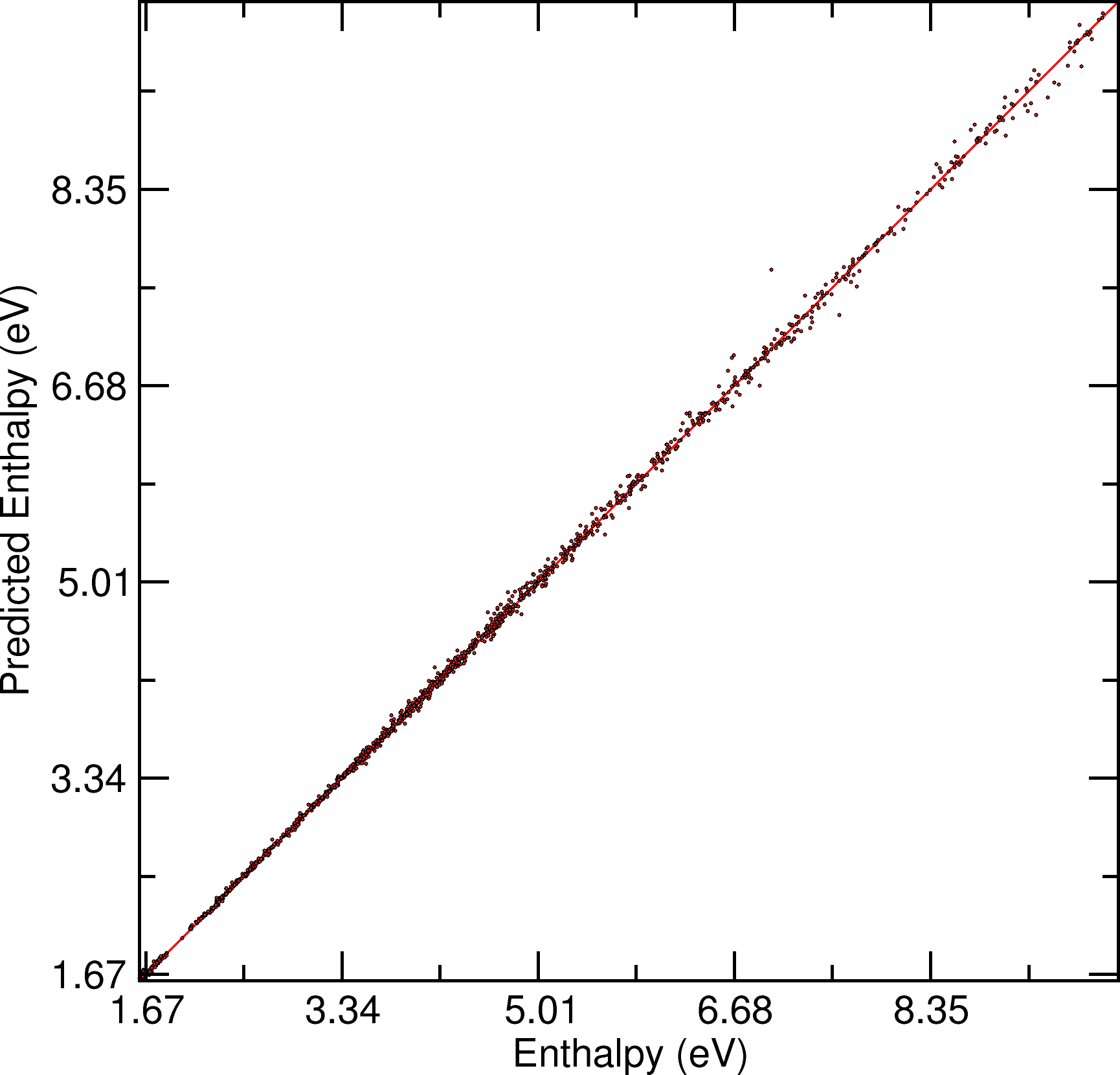}
\includegraphics[width=0.3\textwidth]{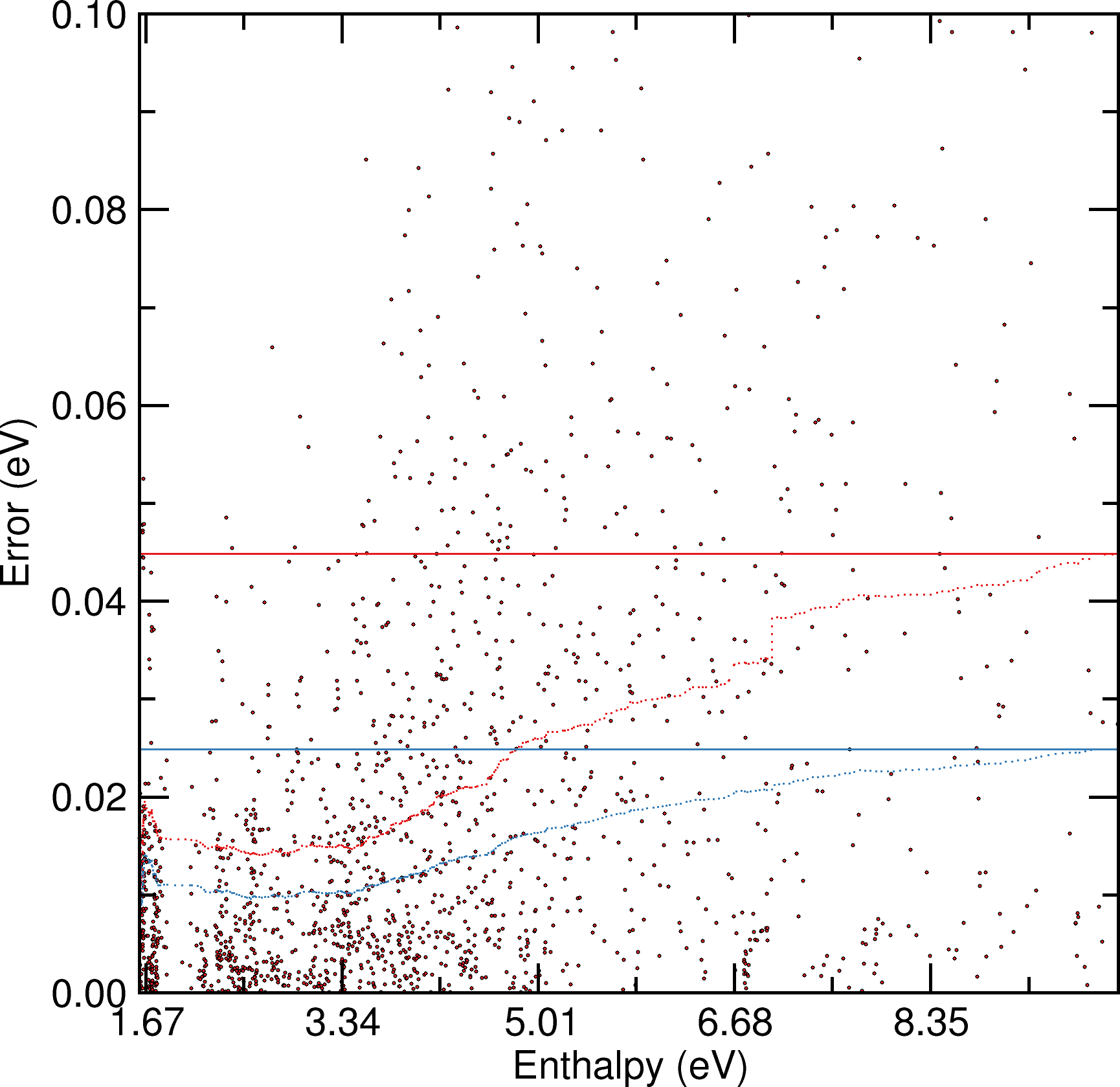}
\centering\begin{verbatim}
training    RMSE/MAE:  27.30  16.16  meV  Spearman  :  0.99985
validation  RMSE/MAE:  38.51  22.68  meV  Spearman  :  0.99979
testing     RMSE/MAE:  44.81  24.88  meV  Spearman  :  0.99977
\end{verbatim}
\clearpage

\flushleft{
\subsection{As-H}}
\subsubsection*{Searching}
\centering
\includegraphics[width=0.4\textwidth]{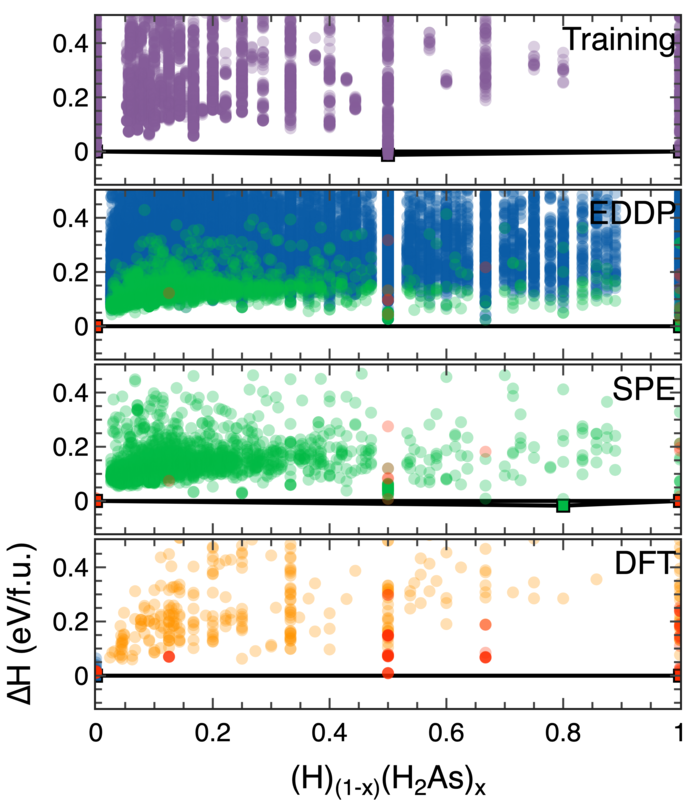}
\footnotesize


\flushleft{
\subsubsection*{\textsc{EDDP}}}
\centering
\includegraphics[width=0.3\textwidth]{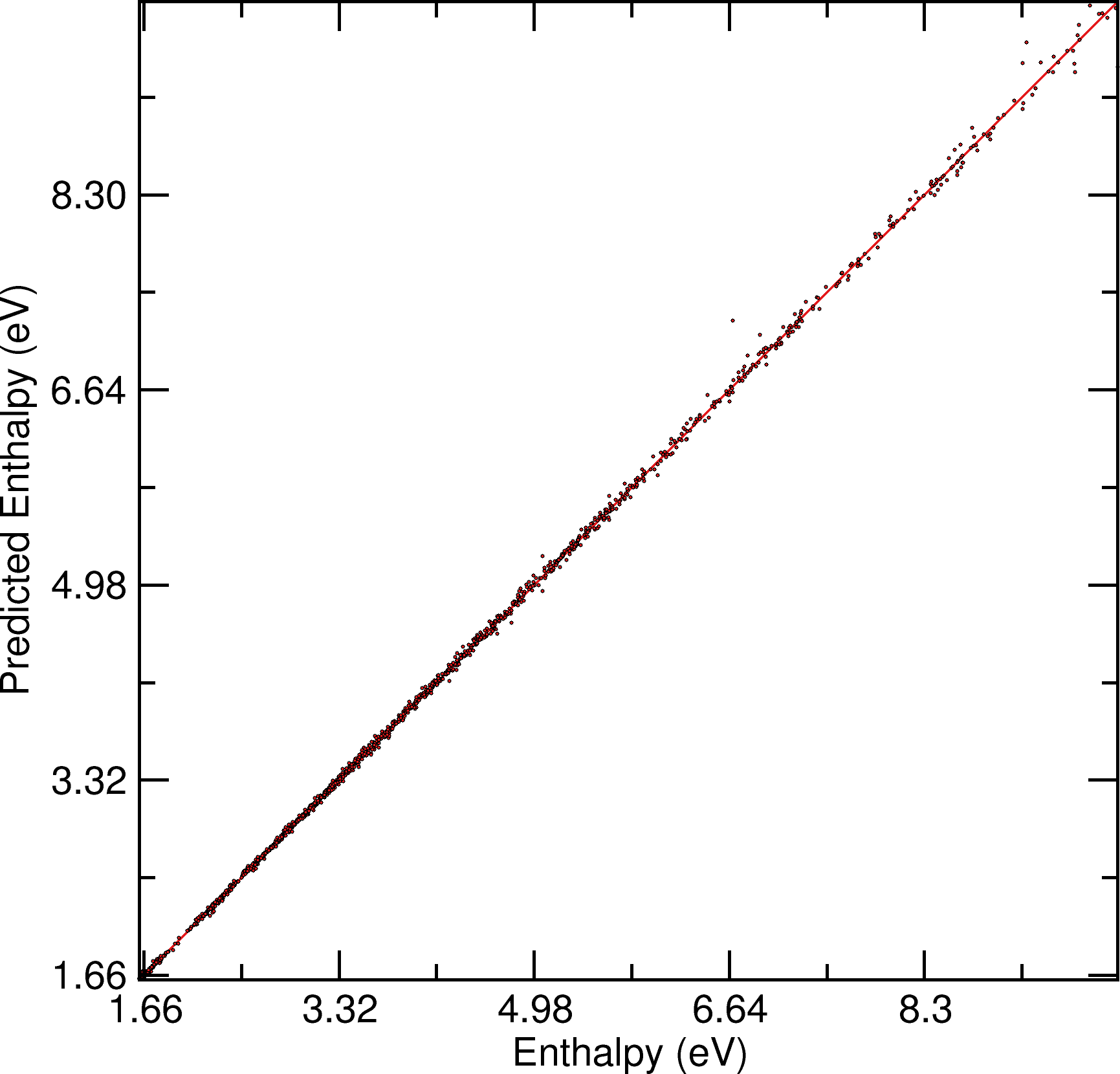}
\includegraphics[width=0.3\textwidth]{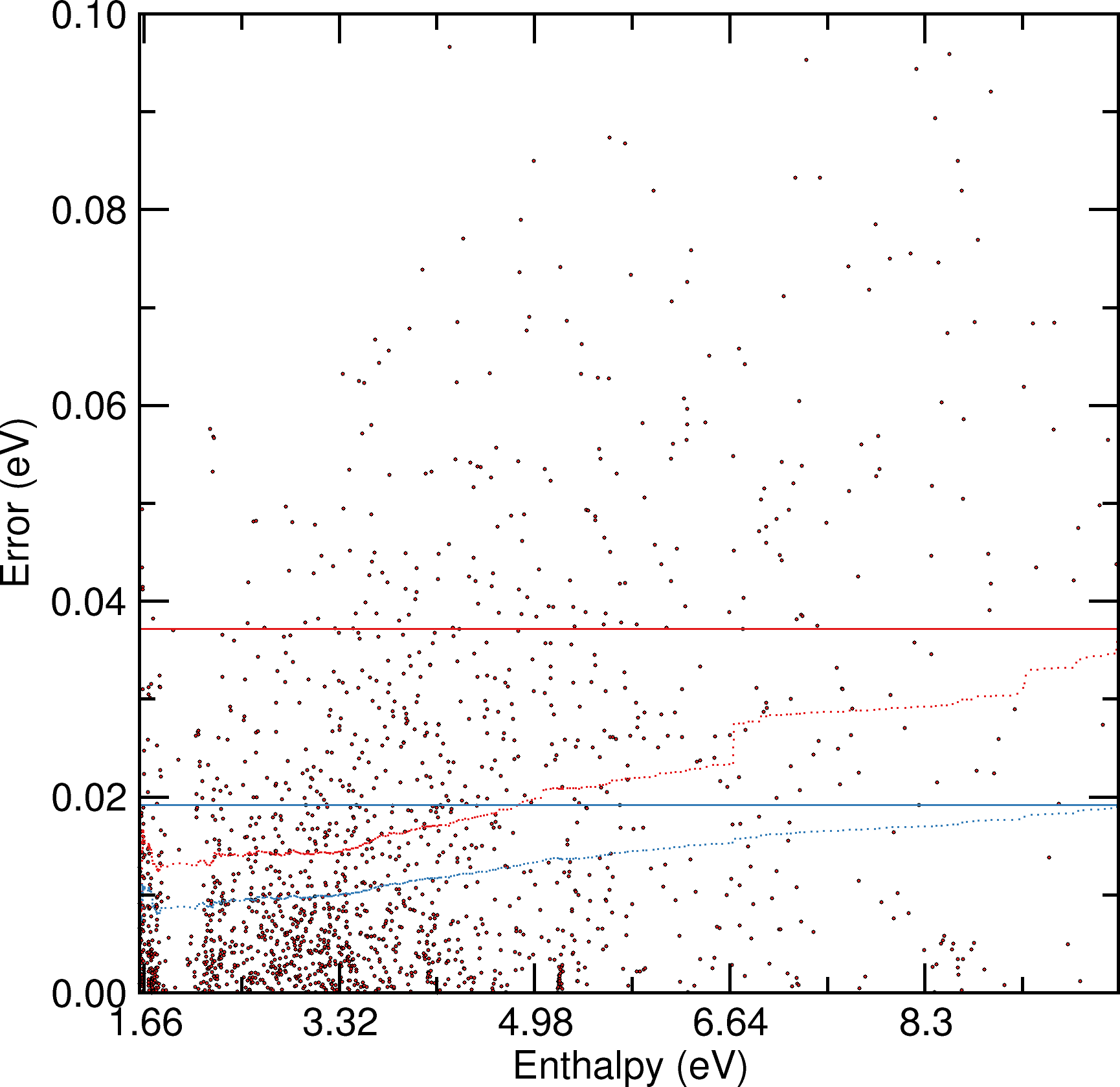}
\centering\begin{verbatim}
training    RMSE/MAE:  18.72  12.00  meV  Spearman  :  0.99987
validation  RMSE/MAE:  26.90  16.72  meV  Spearman  :  0.99979
testing     RMSE/MAE:  37.18  19.18  meV  Spearman  :  0.99986
\end{verbatim}
\clearpage

\flushleft{
\subsection{Au-H}}
\subsubsection*{Searching}
\centering
\includegraphics[width=0.4\textwidth]{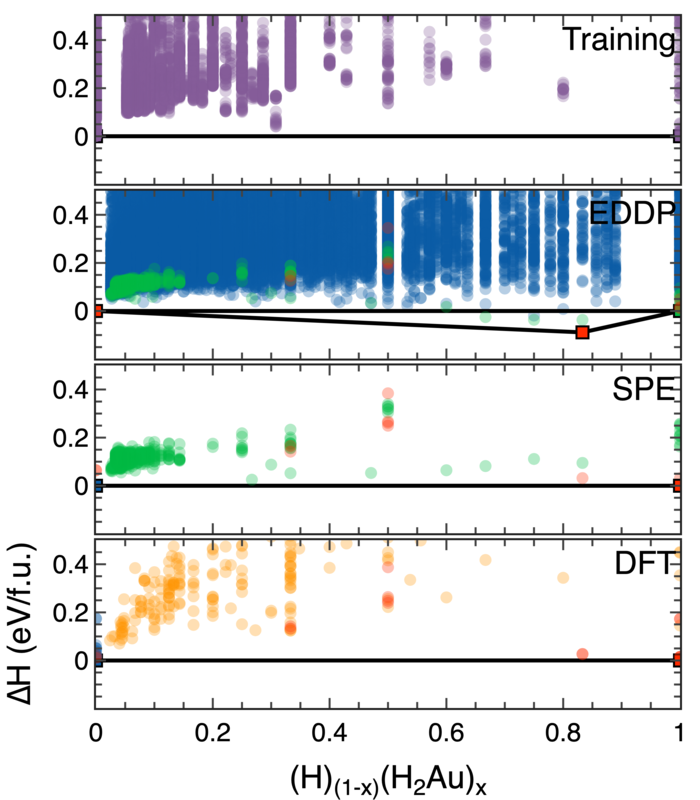}
\footnotesize


\flushleft{
\subsubsection*{\textsc{EDDP}}}
\centering
\includegraphics[width=0.3\textwidth]{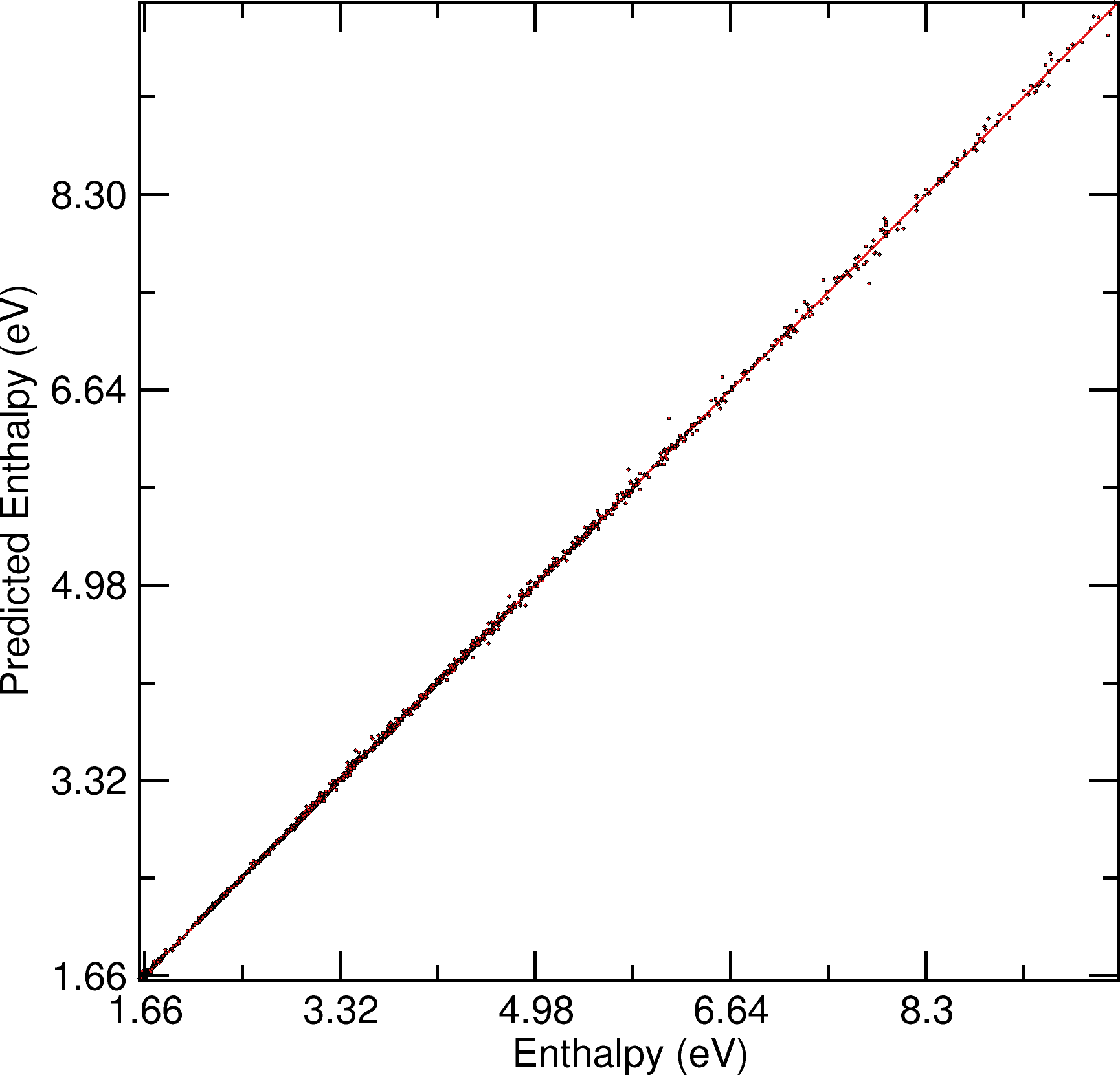}
\includegraphics[width=0.3\textwidth]{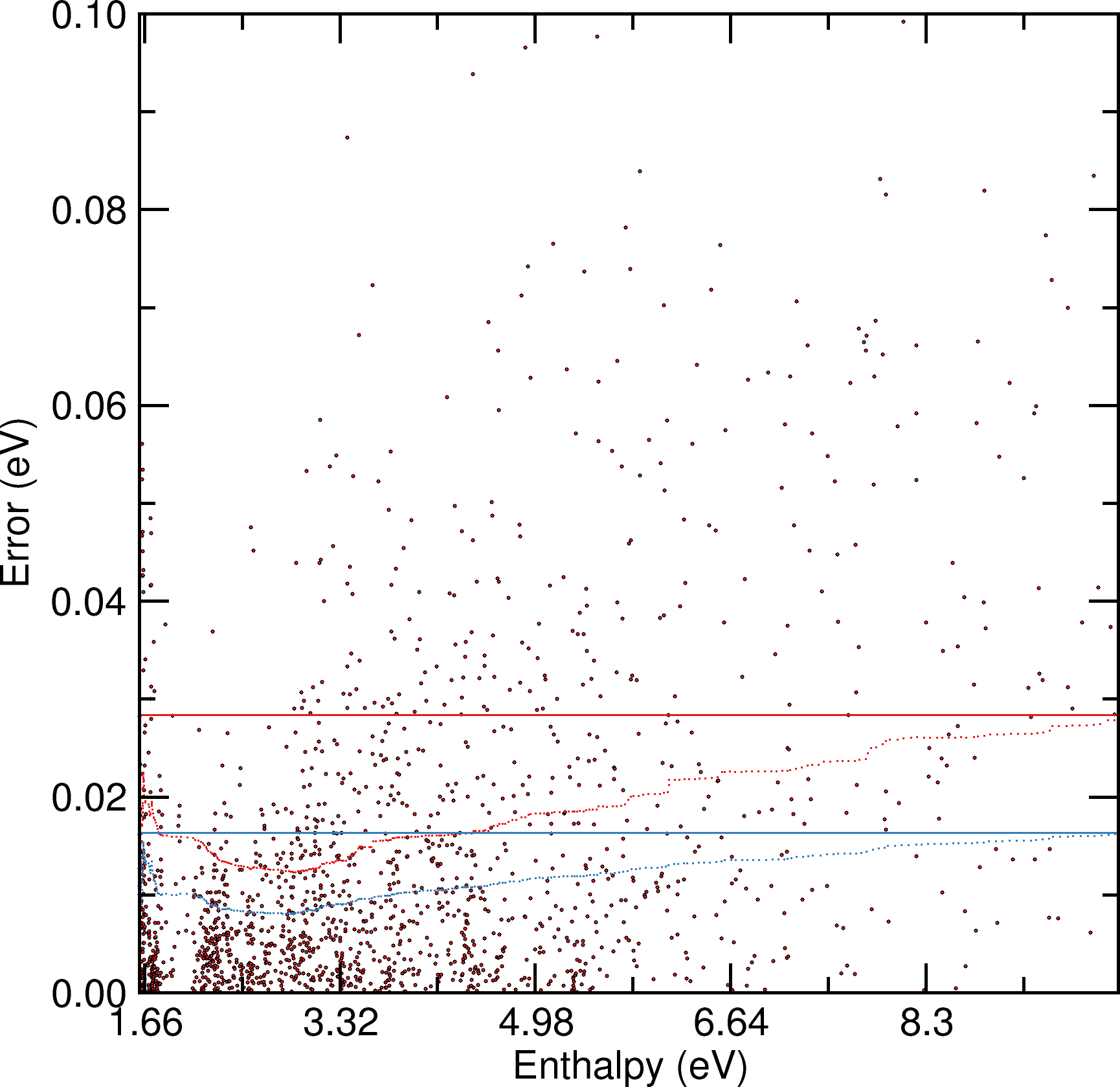}
\centering\begin{verbatim}
training    RMSE/MAE:  15.19  9.94   meV  Spearman  :  0.99989
validation  RMSE/MAE:  24.96  15.76  meV  Spearman  :  0.99983
testing     RMSE/MAE:  28.39  16.32  meV  Spearman  :  0.99980
\end{verbatim}
\clearpage

\flushleft{
\subsection{Ba-H}}
\subsubsection*{Searching}
\centering
\includegraphics[width=0.4\textwidth]{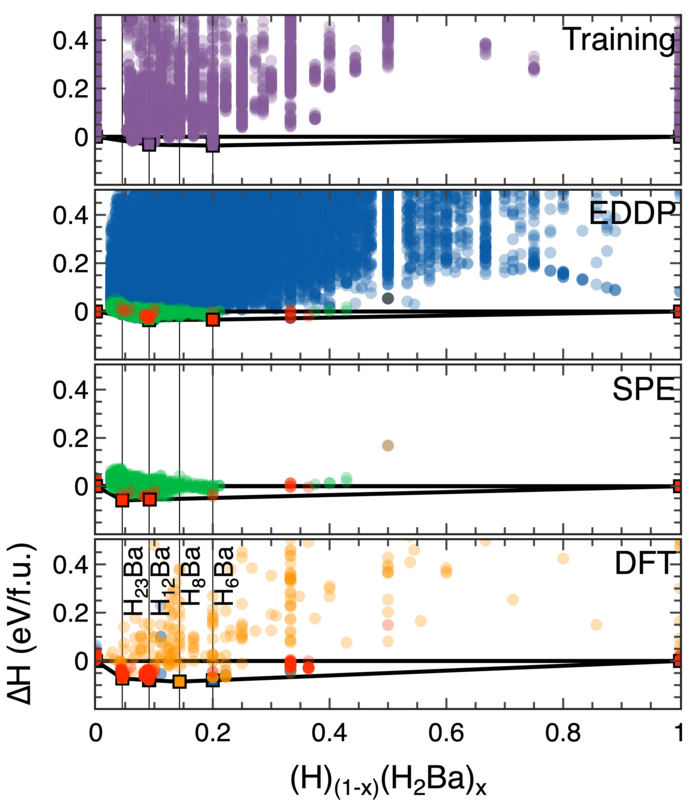}
\footnotesize


\flushleft{
\subsubsection*{\textsc{EDDP}}}
\centering
\includegraphics[width=0.3\textwidth]{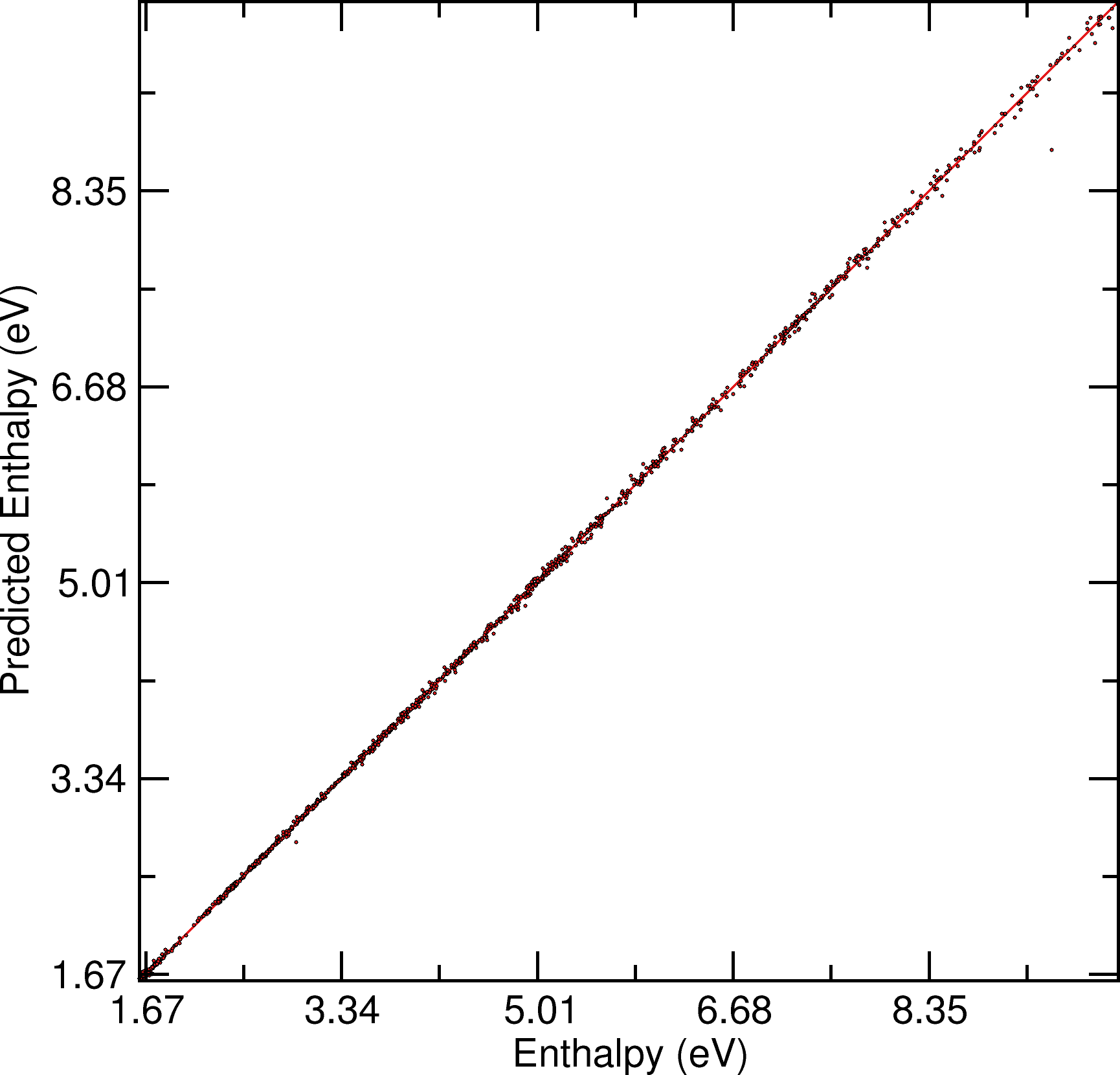}
\includegraphics[width=0.3\textwidth]{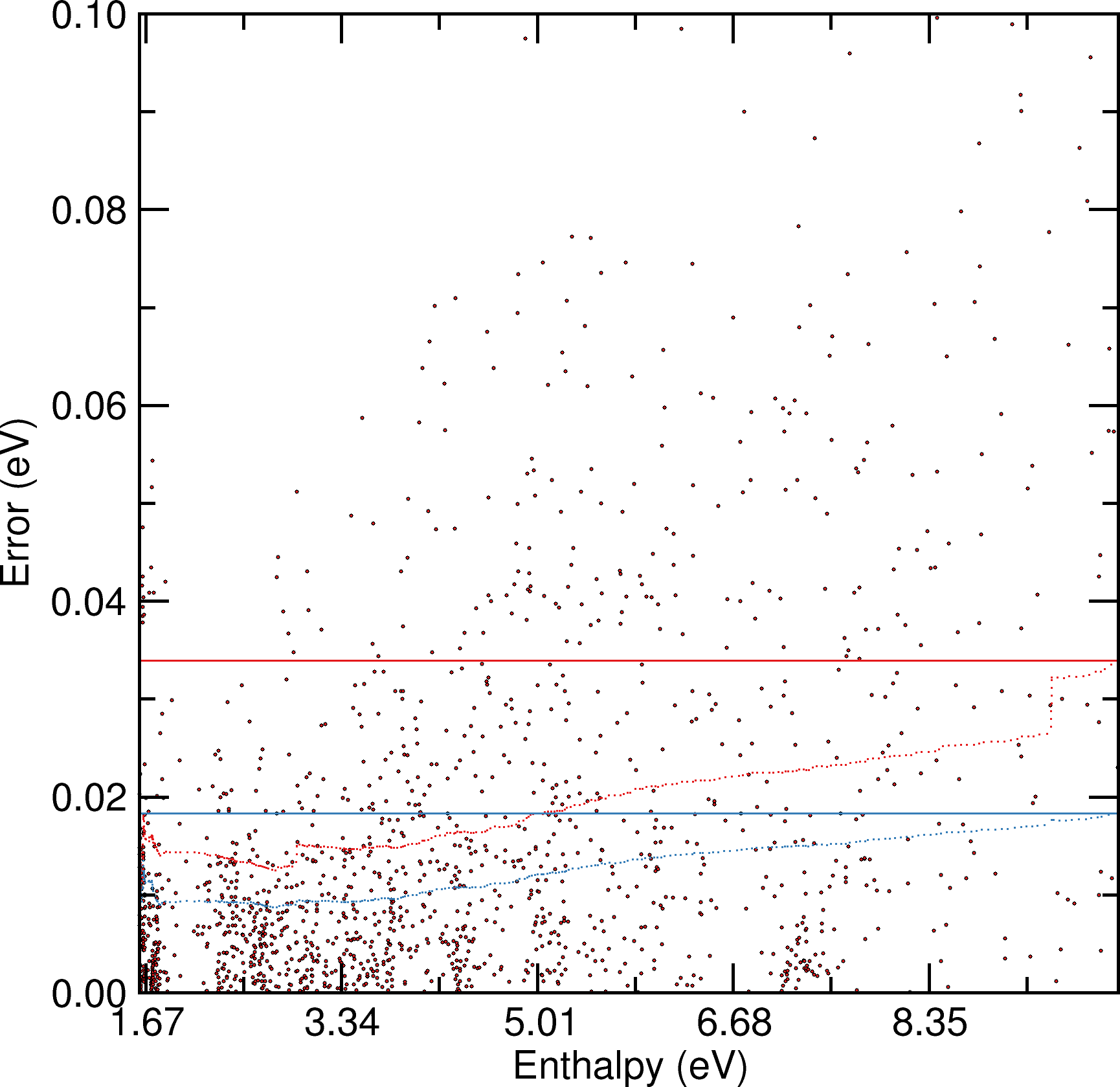}
\centering\begin{verbatim}
training    RMSE/MAE:  19.55  12.47  meV  Spearman  :  0.99987
validation  RMSE/MAE:  28.93  18.28  meV  Spearman  :  0.99980
testing     RMSE/MAE:  33.90  18.35  meV  Spearman  :  0.99983
\end{verbatim}
\clearpage

\flushleft{
\subsection{Be-H}}
\subsubsection*{Searching}
\centering
\includegraphics[width=0.4\textwidth]{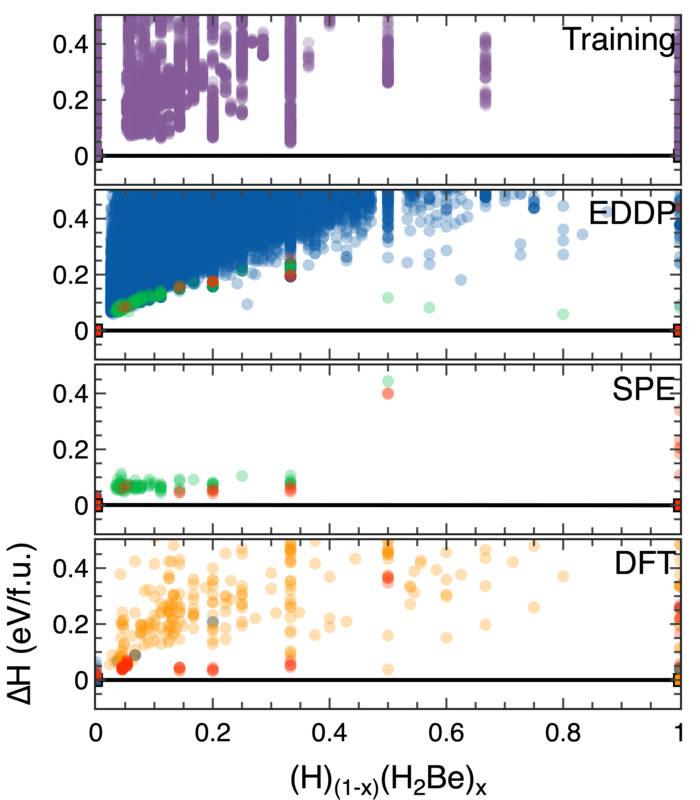}
\footnotesize


\flushleft{
\subsubsection*{\textsc{EDDP}}}
\centering
\includegraphics[width=0.3\textwidth]{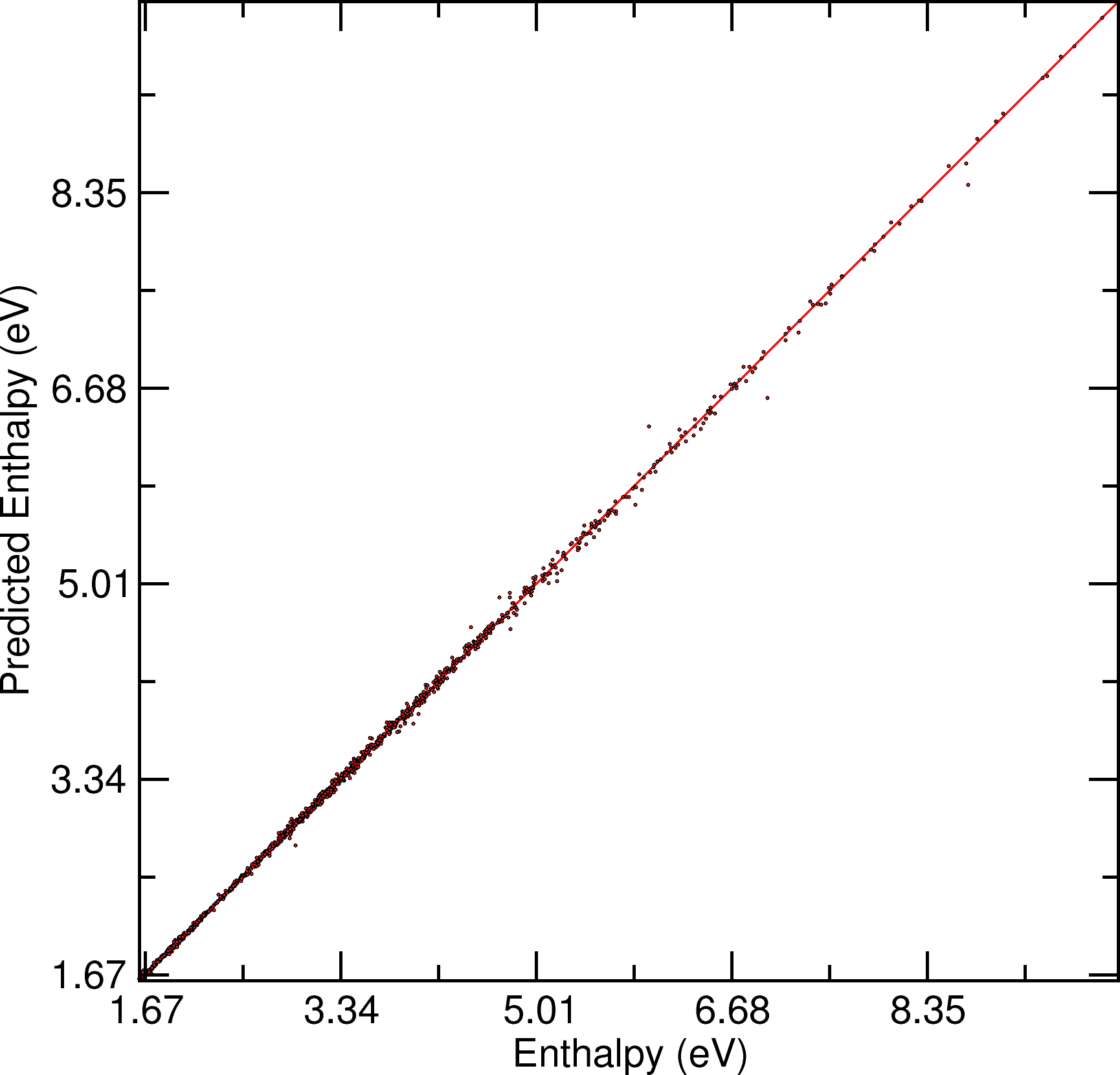}
\includegraphics[width=0.3\textwidth]{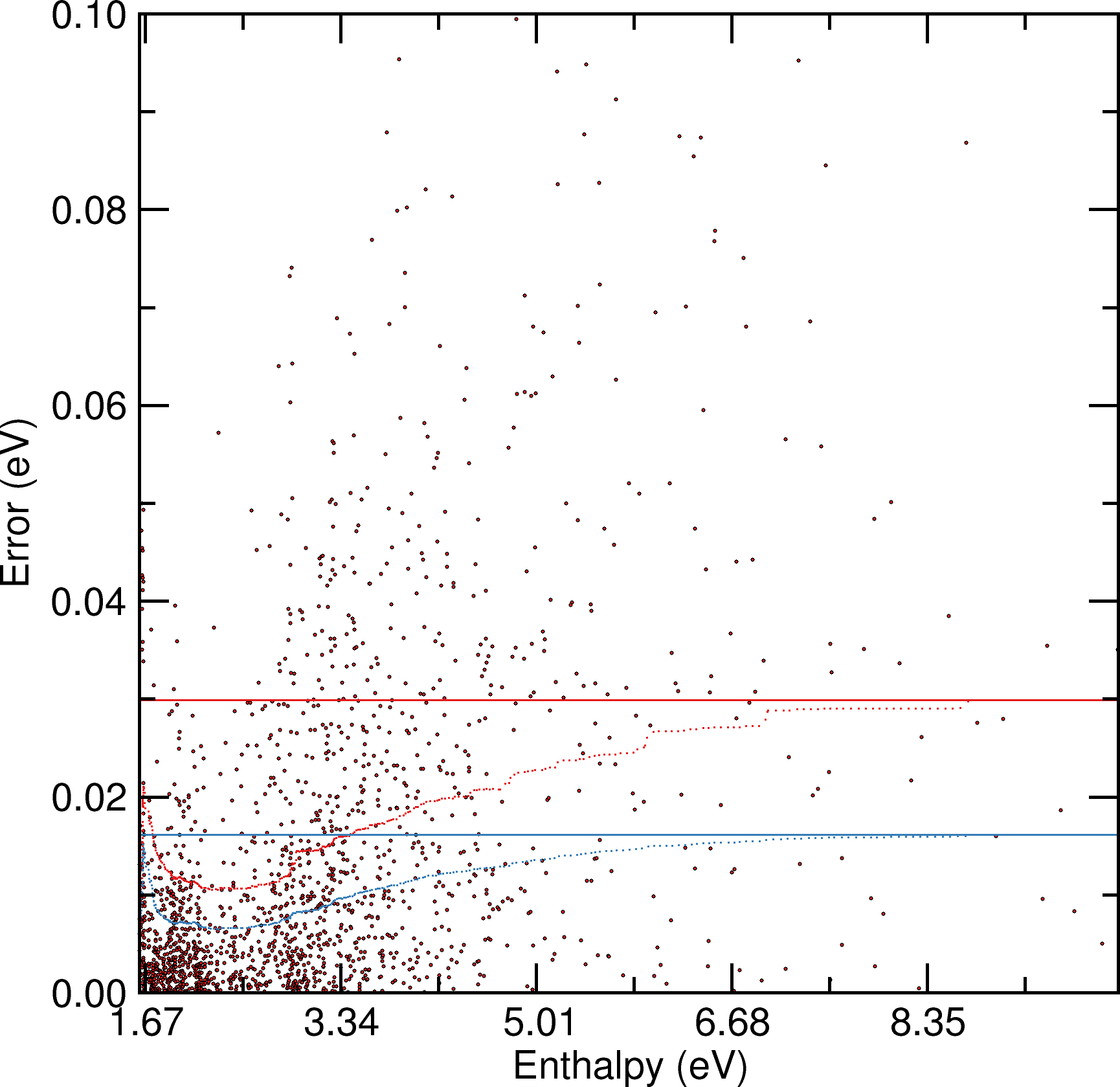}
\centering\begin{verbatim}
training    RMSE/MAE:  16.87  10.11  meV  Spearman  :  0.99986
validation  RMSE/MAE:  23.47  14.09  meV  Spearman  :  0.99978
testing     RMSE/MAE:  29.88  16.15  meV  Spearman  :  0.99974
\end{verbatim}
\clearpage

\flushleft{
\subsection{Bi-H}}
\subsubsection*{Searching}
\centering
\includegraphics[width=0.4\textwidth]{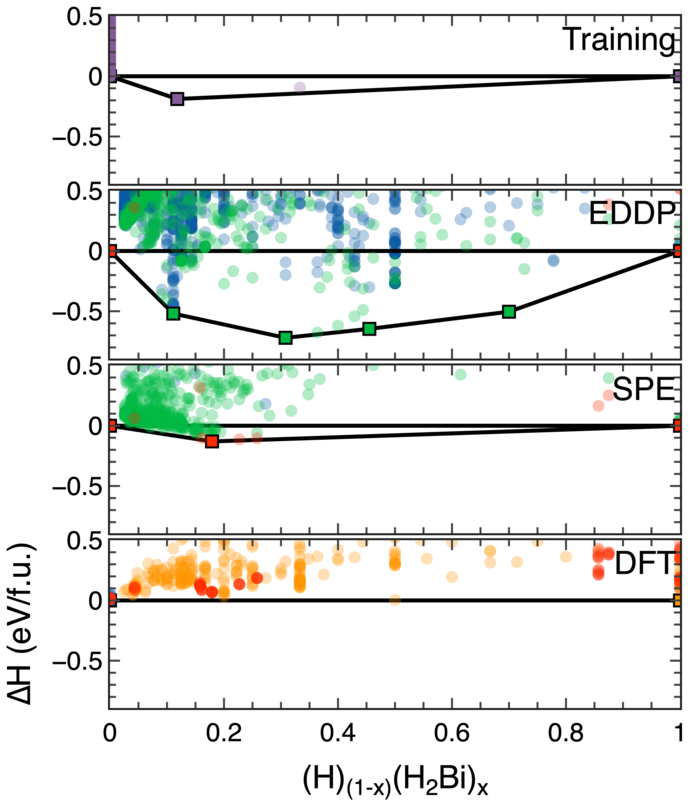}
\footnotesize


\flushleft{
\subsubsection*{\textsc{EDDP}}}
\centering
\includegraphics[width=0.3\textwidth]{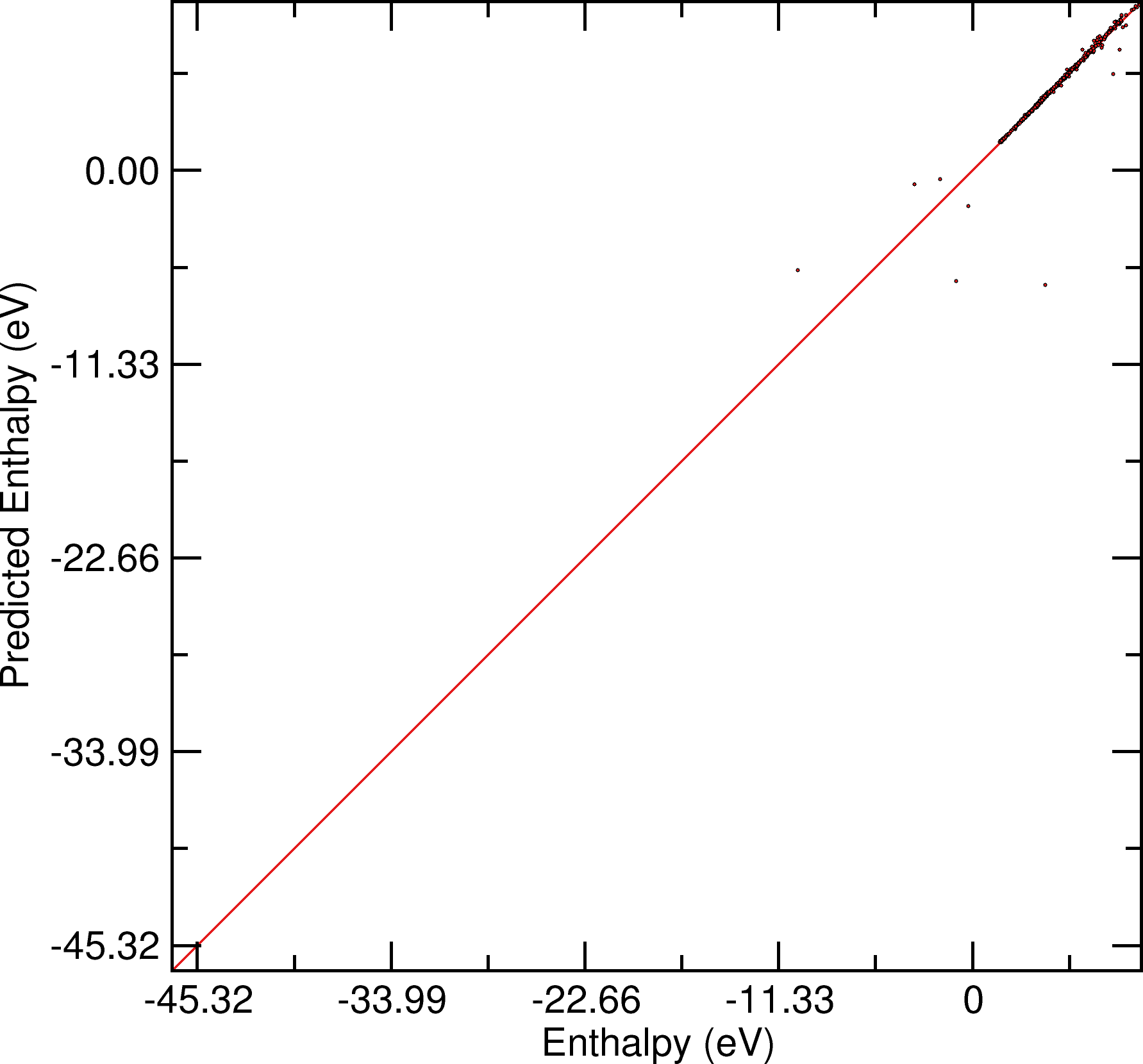}
\includegraphics[width=0.3\textwidth]{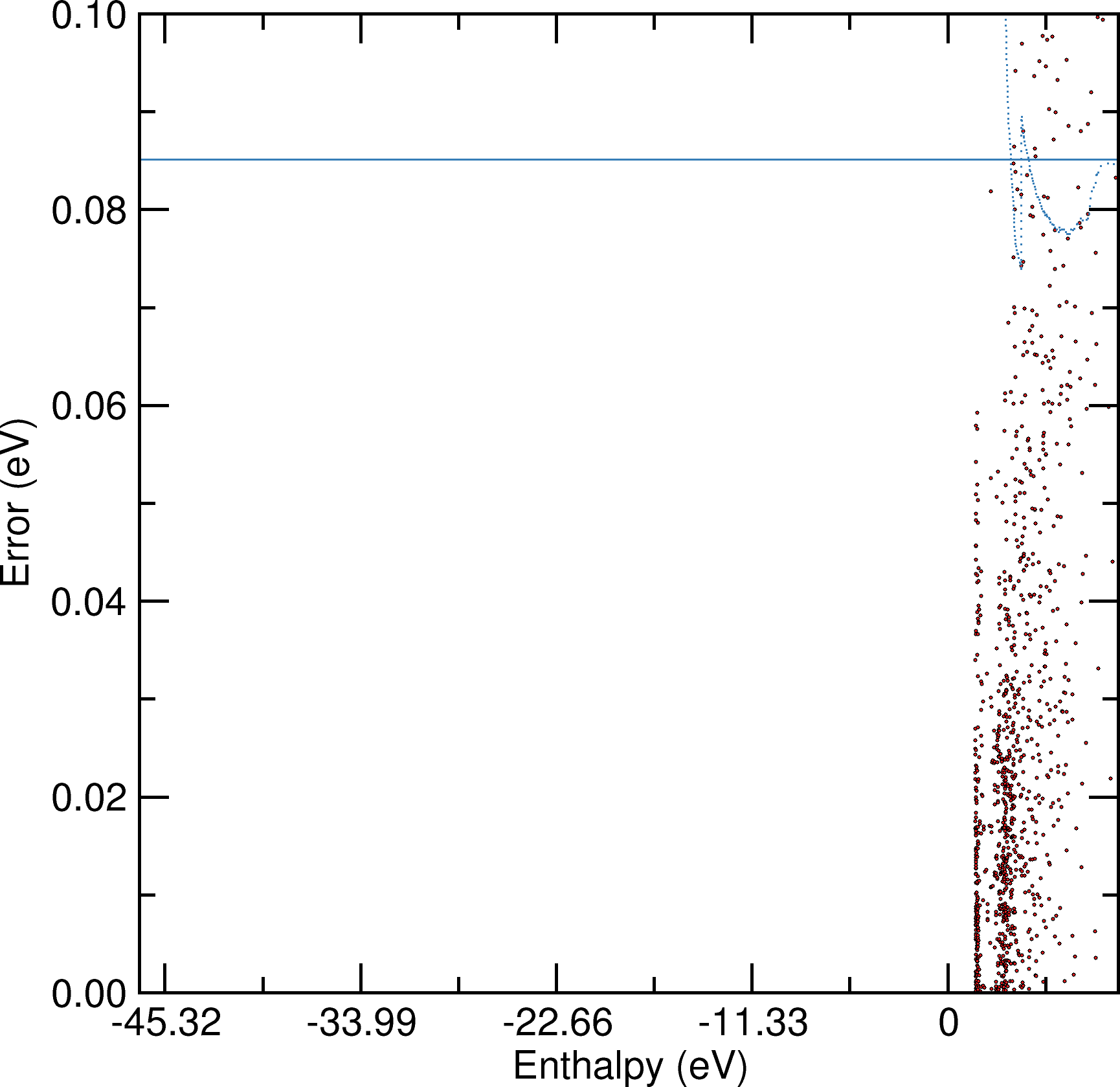}
\centering\begin{verbatim}
training    RMSE/MAE:  1305.11  79.96  meV  Spearman  :  0.99847
validation  RMSE/MAE:  641.59   92.39  meV  Spearman  :  0.99210
testing     RMSE/MAE:  777.92   85.11  meV  Spearman  :  0.99685
\end{verbatim}
\clearpage

\flushleft{
\subsection{Br-H}}
\subsubsection*{Searching}
\centering
\includegraphics[width=0.4\textwidth]{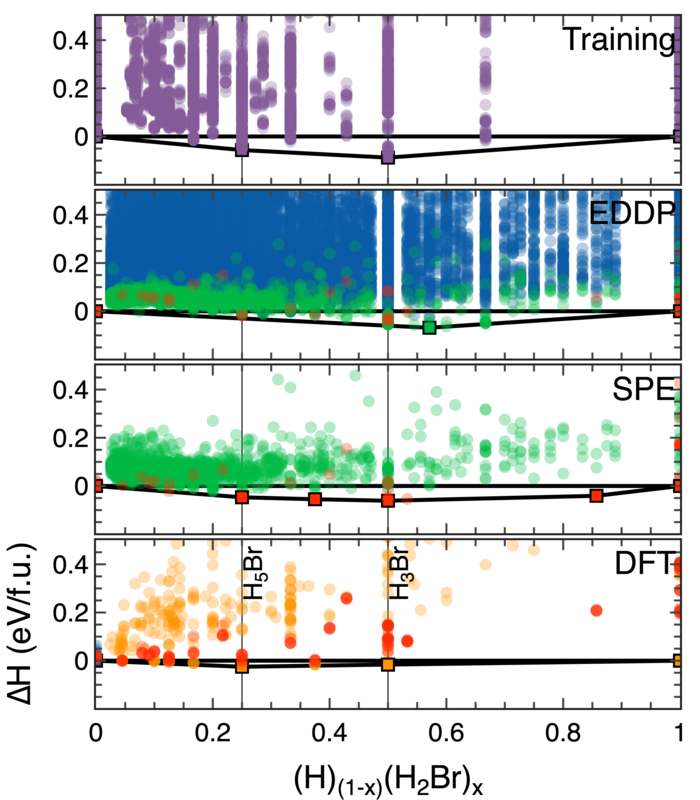}
\footnotesize


\flushleft{
\subsubsection*{\textsc{EDDP}}}
\centering
\includegraphics[width=0.3\textwidth]{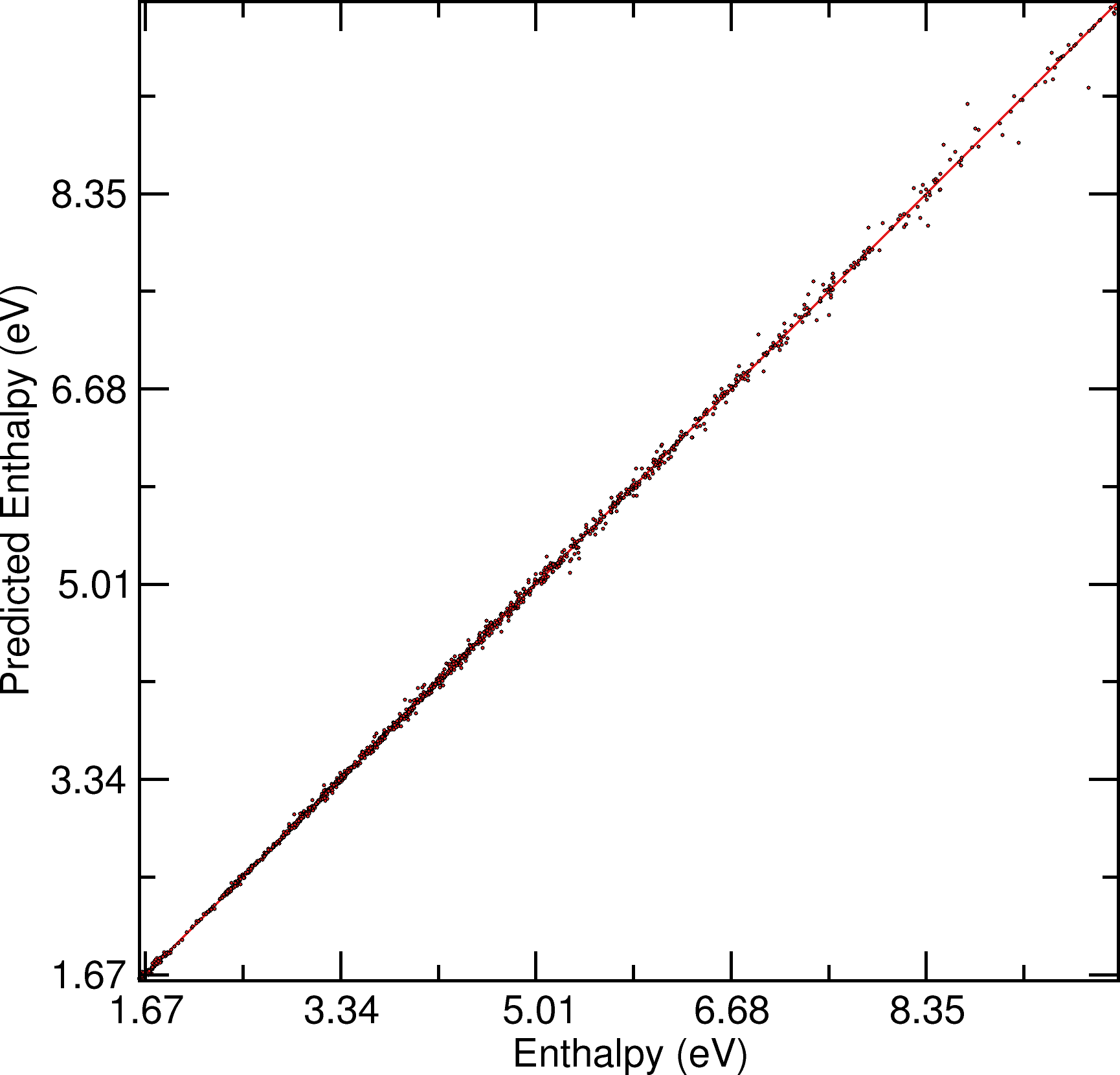}
\includegraphics[width=0.3\textwidth]{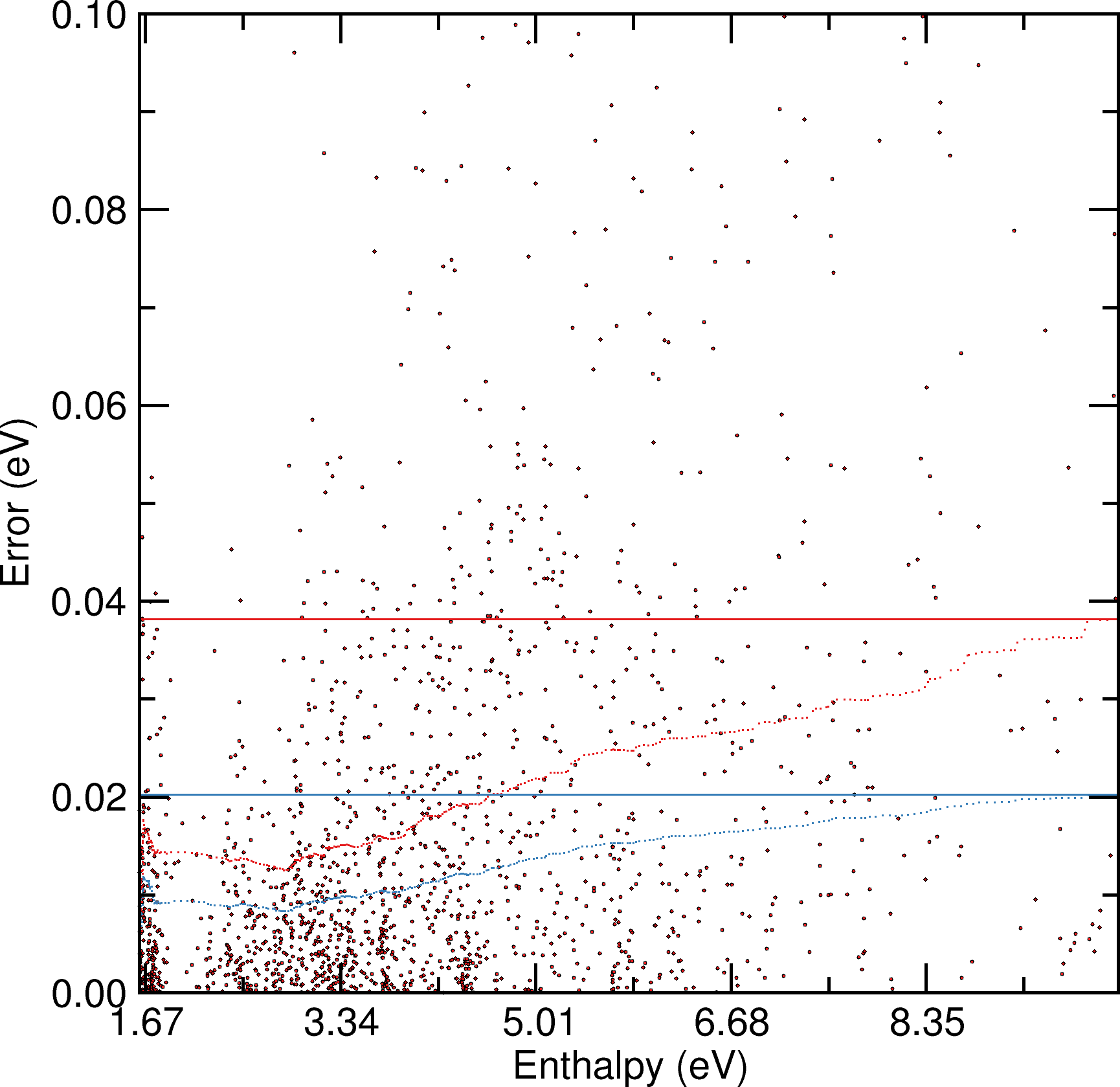}
\centering\begin{verbatim}
training    RMSE/MAE:  21.65  13.00  meV  Spearman  :  0.99986
validation  RMSE/MAE:  31.65  19.22  meV  Spearman  :  0.99978
testing     RMSE/MAE:  38.16  20.21  meV  Spearman  :  0.99982
\end{verbatim}
\clearpage

\flushleft{
\subsection{Ca-H}}
\subsubsection*{Searching}
\centering
\includegraphics[width=0.4\textwidth]{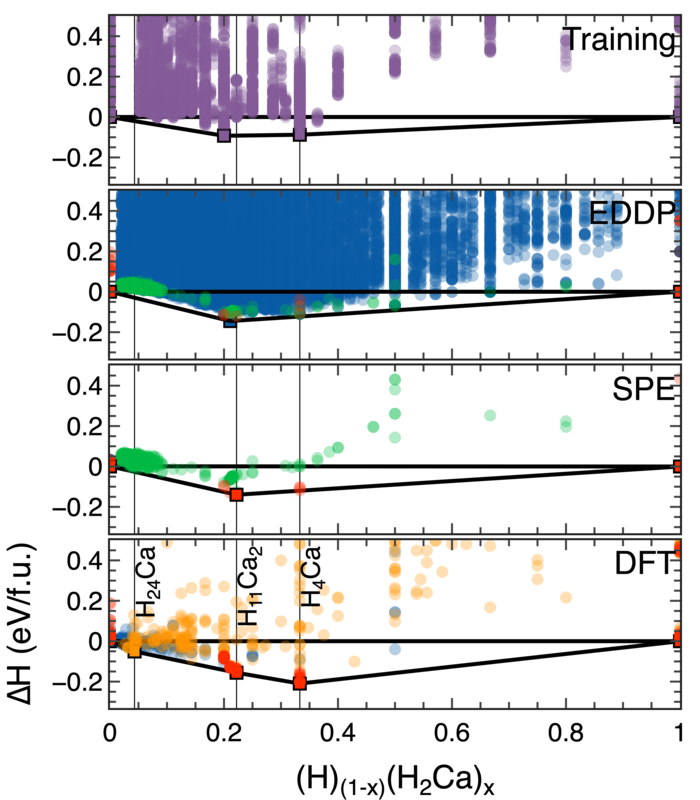}
\footnotesize


\flushleft{
\subsubsection*{\textsc{EDDP}}}
\centering
\includegraphics[width=0.3\textwidth]{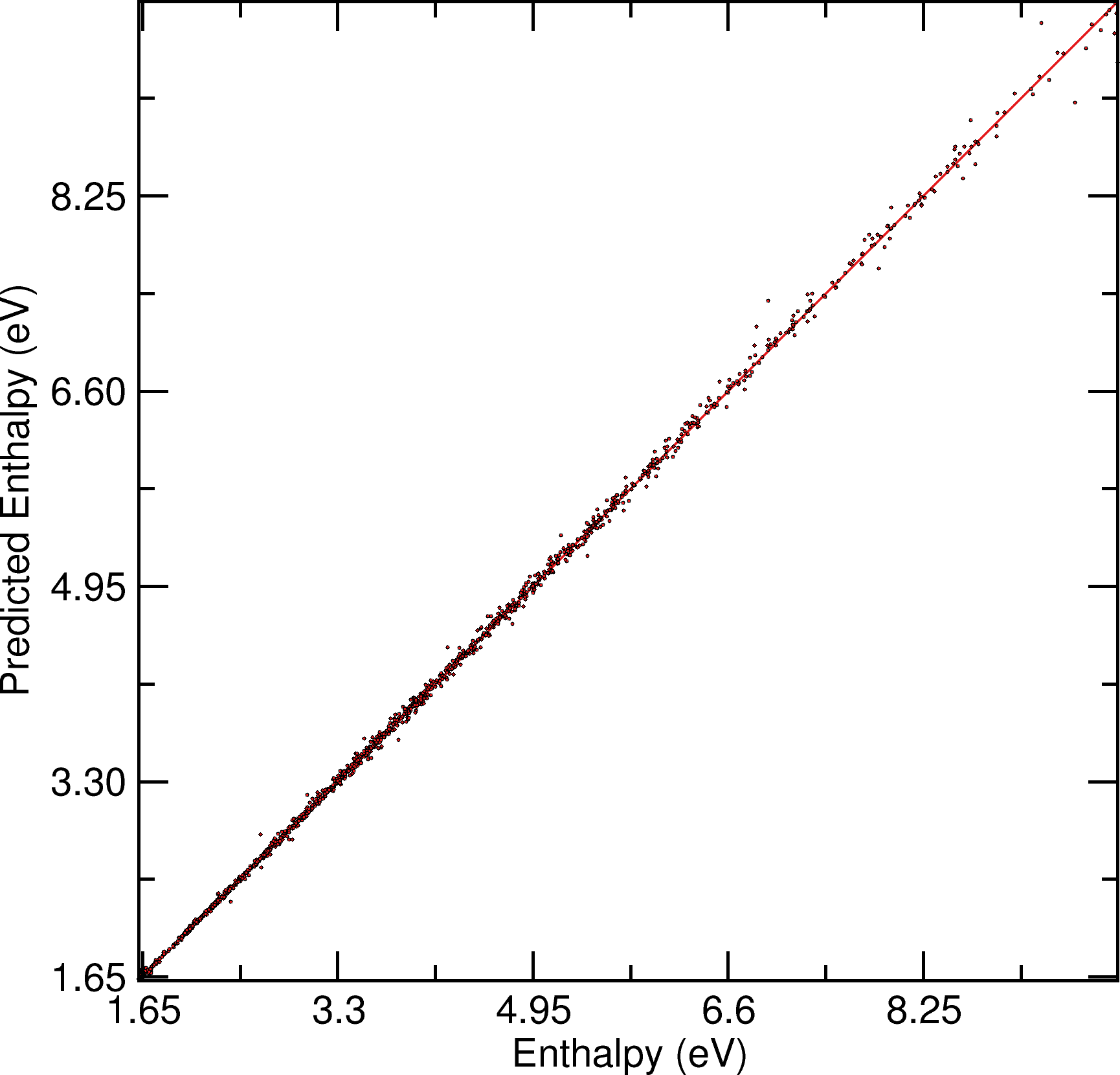}
\includegraphics[width=0.3\textwidth]{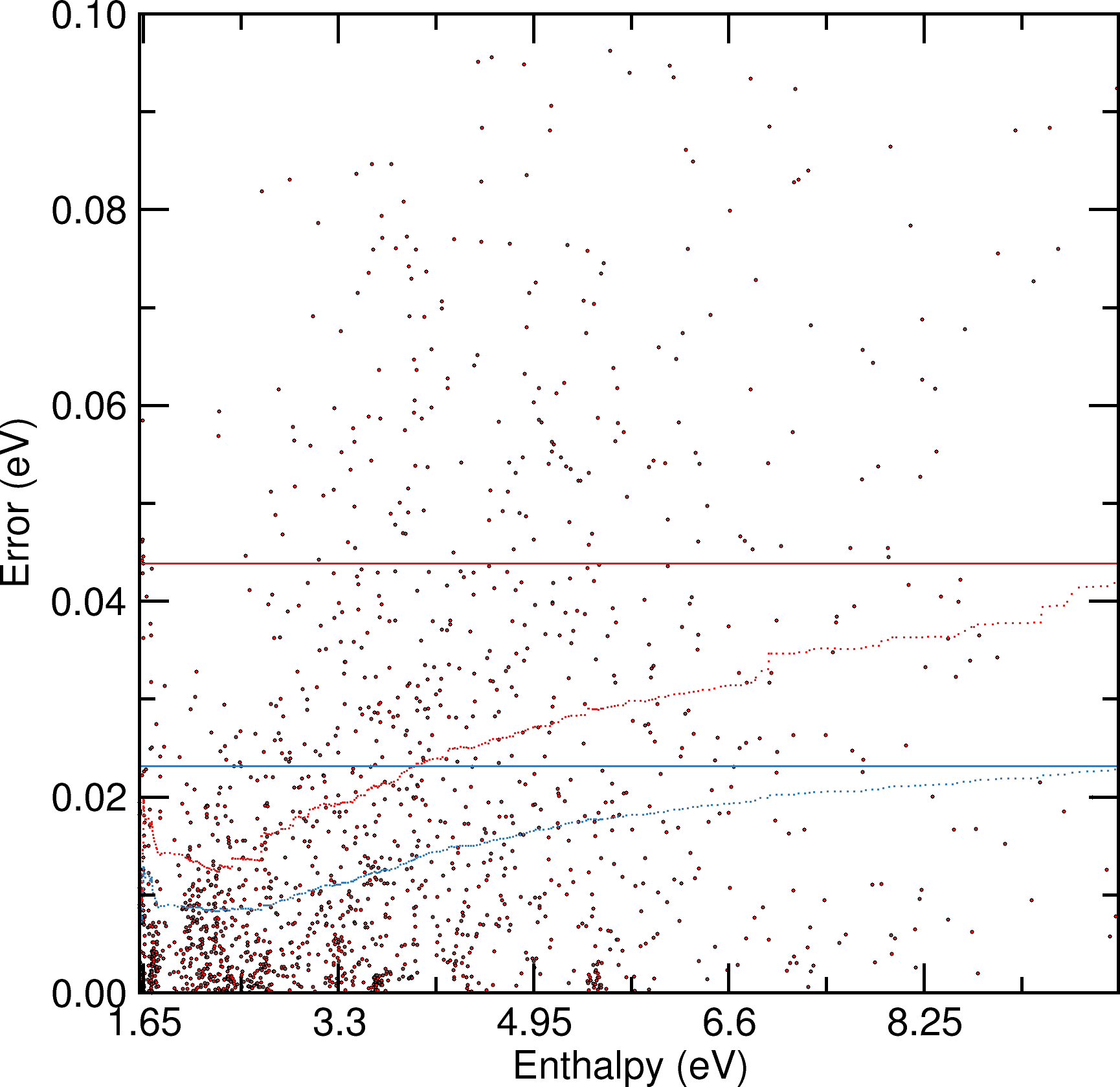}
\centering\begin{verbatim}
training    RMSE/MAE:  22.69  13.99  meV  Spearman  :  0.99987
validation  RMSE/MAE:  37.17  21.96  meV  Spearman  :  0.99982
testing     RMSE/MAE:  43.86  23.14  meV  Spearman  :  0.99977
\end{verbatim}
\clearpage

\flushleft{
\subsection{Cd-H}}
\subsubsection*{Searching}
\centering
\includegraphics[width=0.4\textwidth]{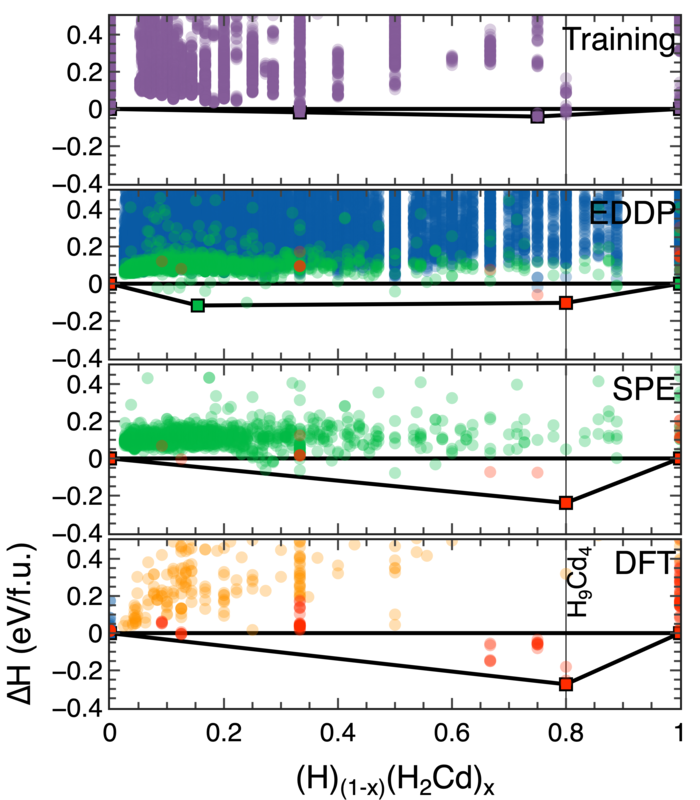}
\footnotesize


\flushleft{
\subsubsection*{\textsc{EDDP}}}
\centering
\includegraphics[width=0.3\textwidth]{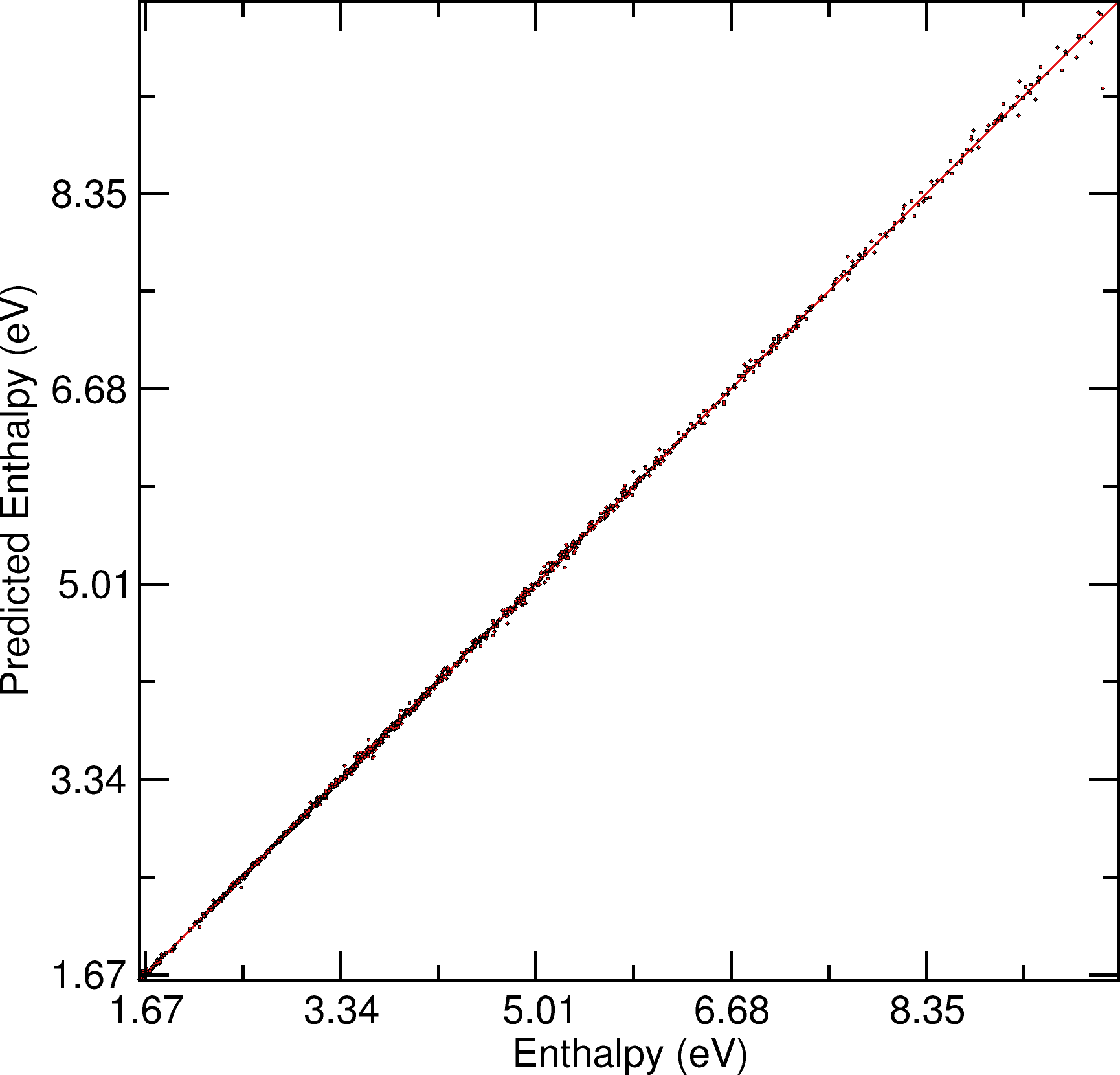}
\includegraphics[width=0.3\textwidth]{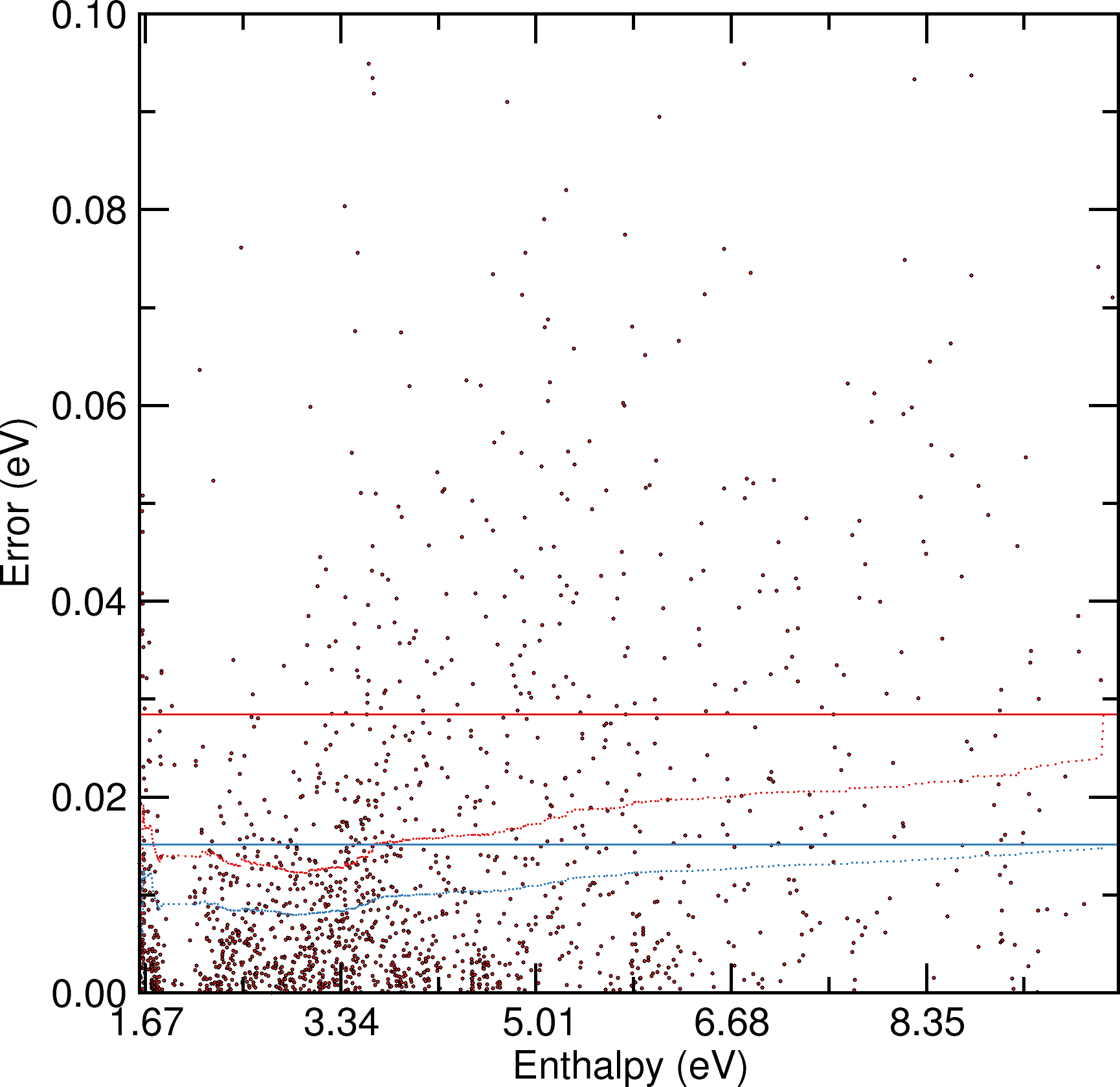}
\centering\begin{verbatim}
training    RMSE/MAE:  14.70  9.47   meV  Spearman  :  0.99990
validation  RMSE/MAE:  21.33  13.54  meV  Spearman  :  0.99984
testing     RMSE/MAE:  28.45  15.14  meV  Spearman  :  0.99986
\end{verbatim}
\clearpage

\flushleft{
\subsection{Ce-H}}
\subsubsection*{Searching}
\centering
\includegraphics[width=0.4\textwidth]{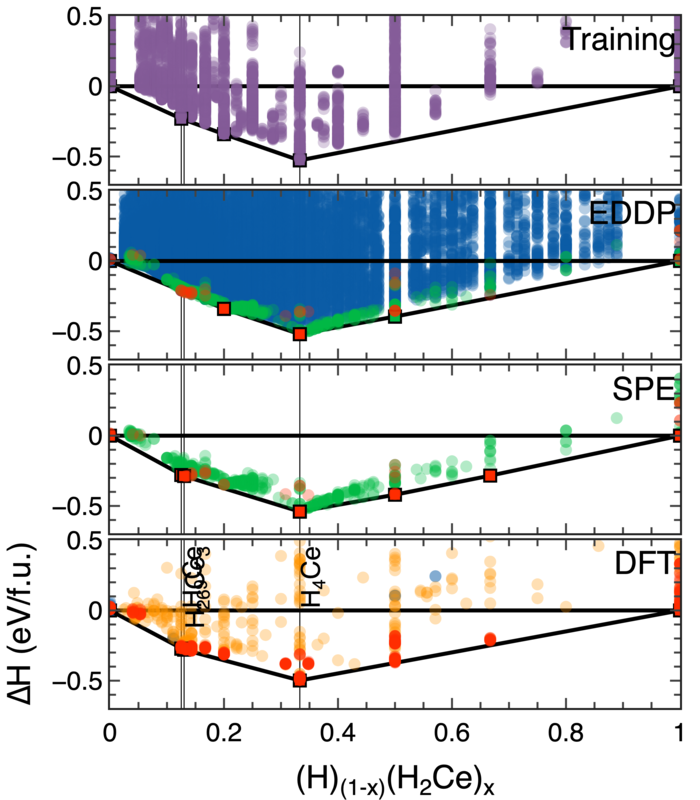}
\footnotesize


\flushleft{
\subsubsection*{\textsc{EDDP}}}
\centering
\includegraphics[width=0.3\textwidth]{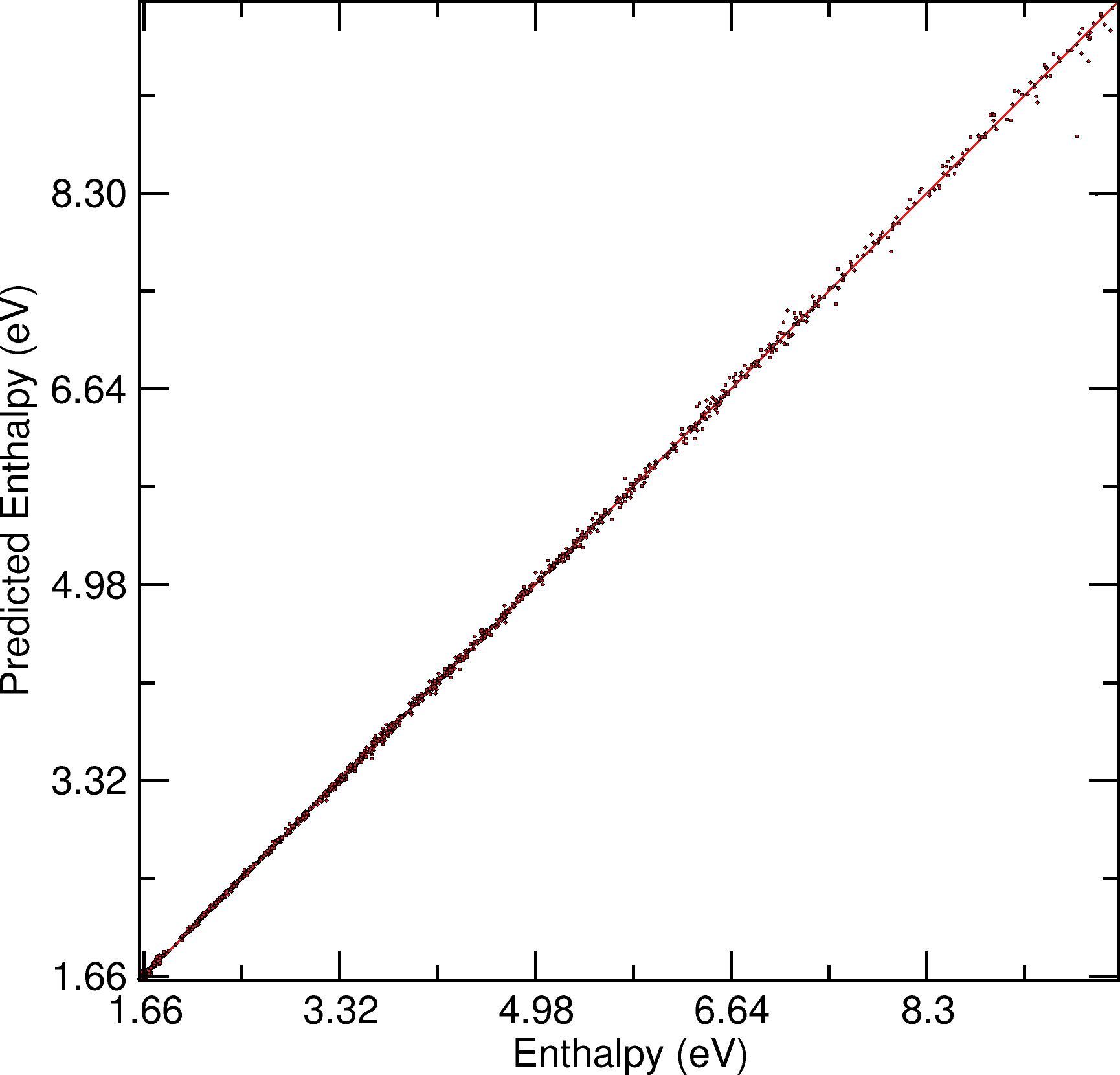}
\includegraphics[width=0.3\textwidth]{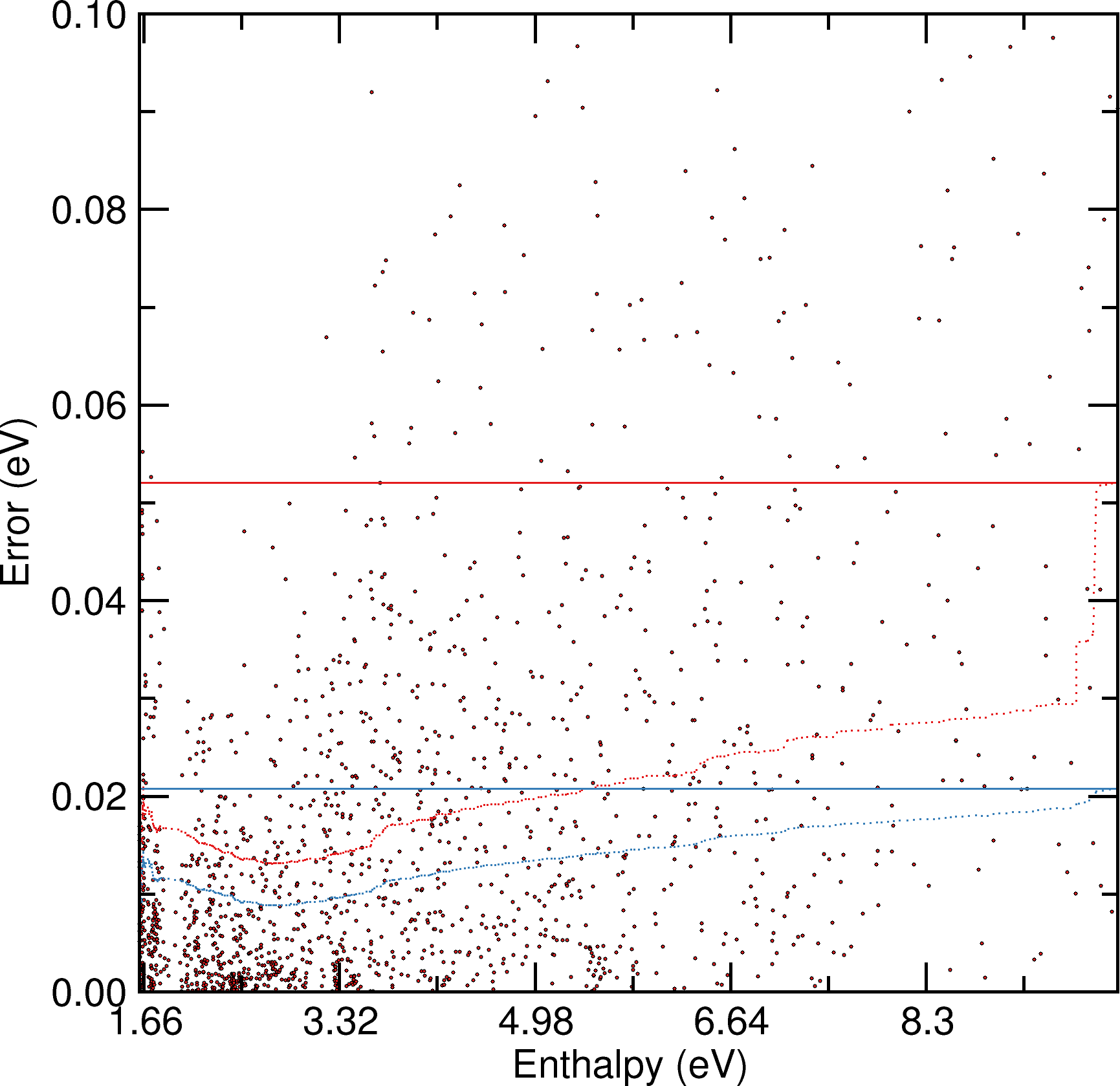}
\centering\begin{verbatim}
training    RMSE/MAE:  20.86  13.54  meV  Spearman  :  0.99986
validation  RMSE/MAE:  27.97  18.45  meV  Spearman  :  0.99984
testing     RMSE/MAE:  52.03  20.76  meV  Spearman  :  0.99981
\end{verbatim}
\clearpage

\flushleft{
\subsection{Cl-H}}
\subsubsection*{Searching}
\centering
\includegraphics[width=0.4\textwidth]{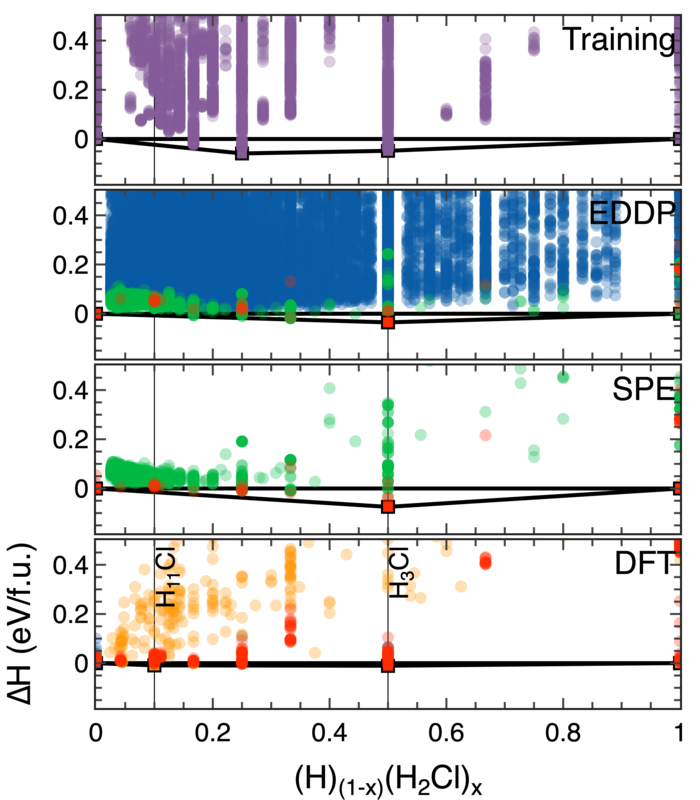}
\footnotesize


\flushleft{
\subsubsection*{\textsc{EDDP}}}
\centering
\includegraphics[width=0.3\textwidth]{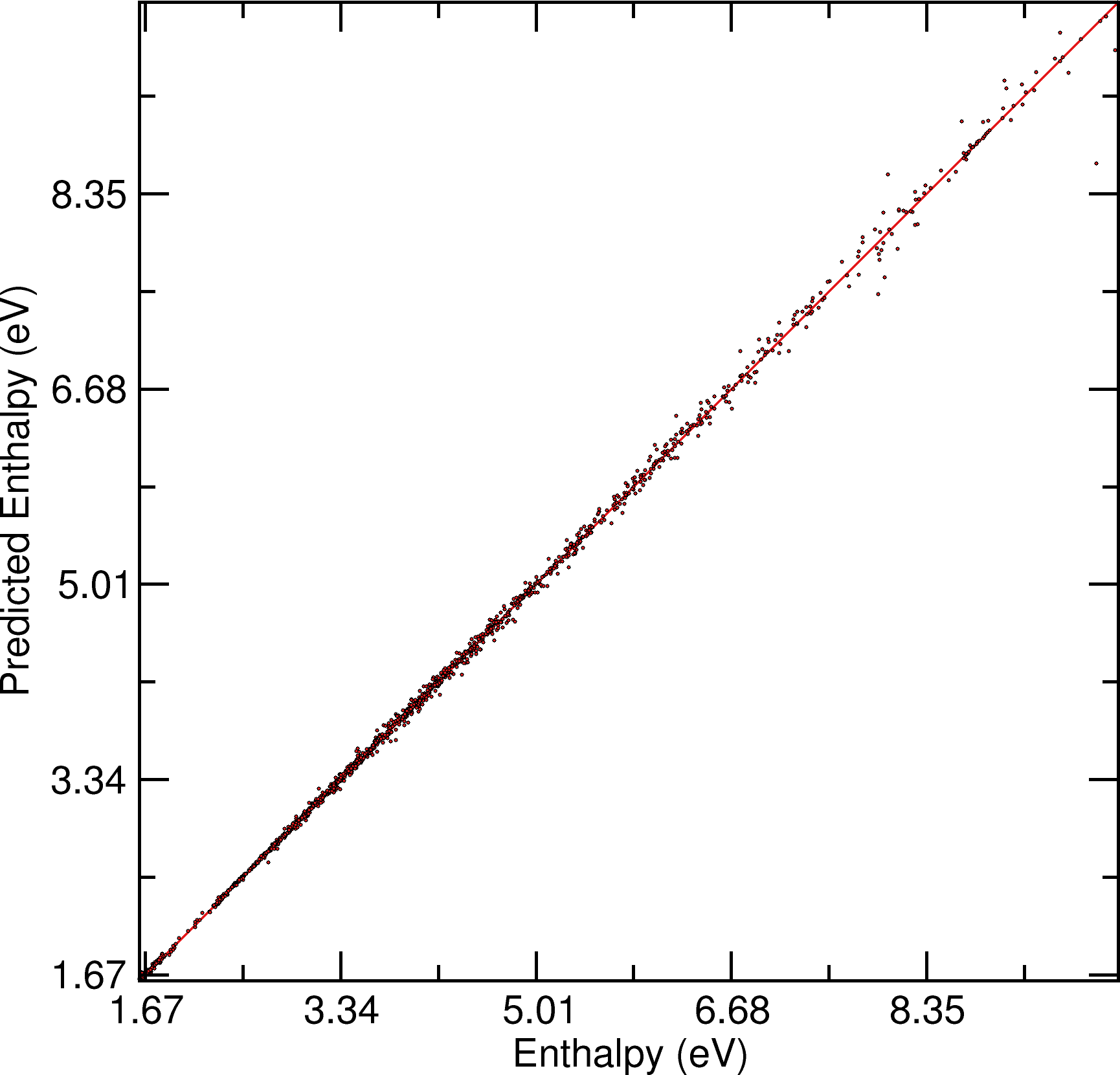}
\includegraphics[width=0.3\textwidth]{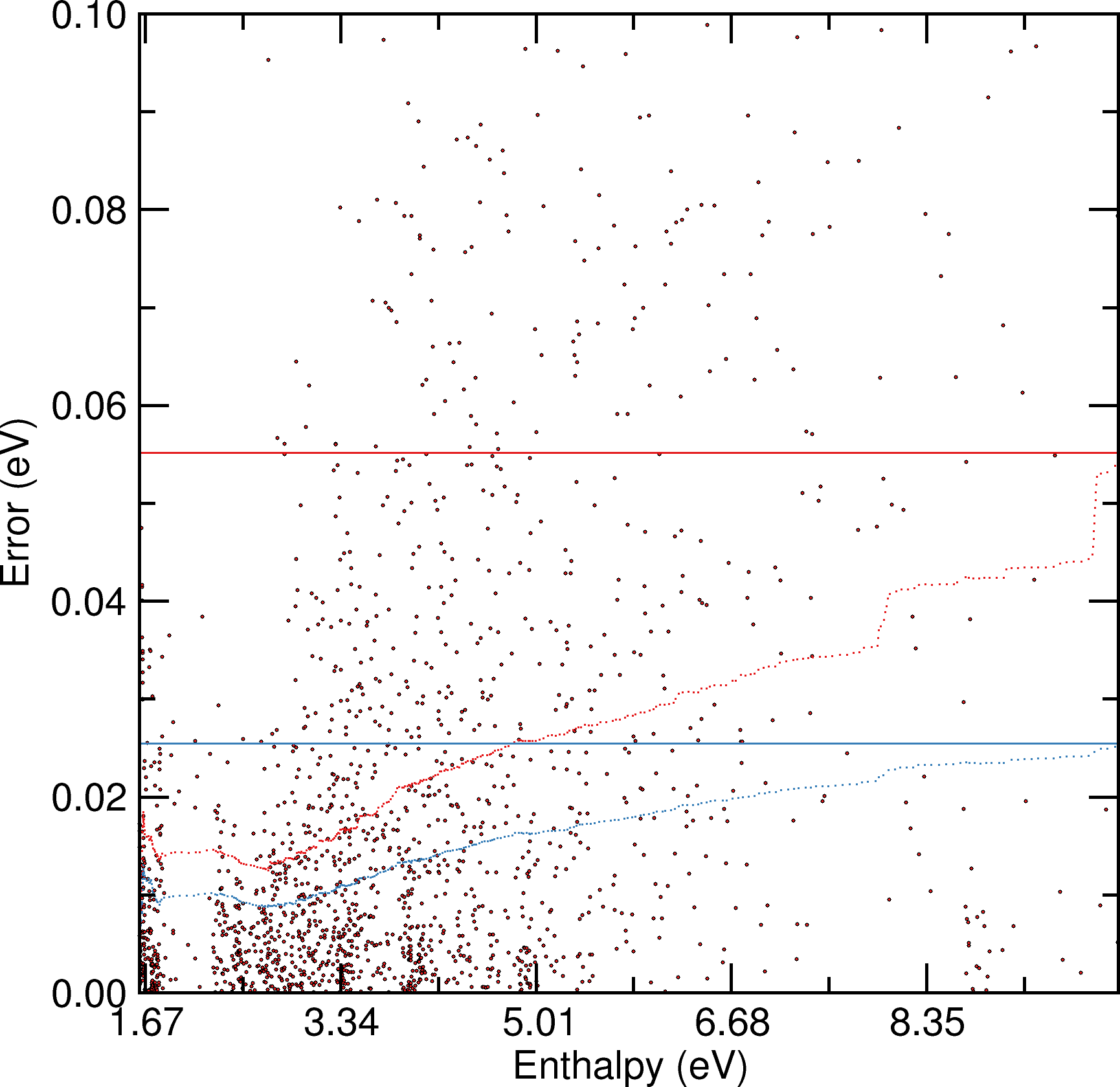}
\centering\begin{verbatim}
training    RMSE/MAE:  23.01  13.84  meV  Spearman  :  0.99983
validation  RMSE/MAE:  40.32  22.21  meV  Spearman  :  0.99971
testing     RMSE/MAE:  55.15  25.45  meV  Spearman  :  0.99973
\end{verbatim}
\clearpage

\flushleft{
\subsection{Co-H}}
\subsubsection*{Searching}
\centering
\includegraphics[width=0.4\textwidth]{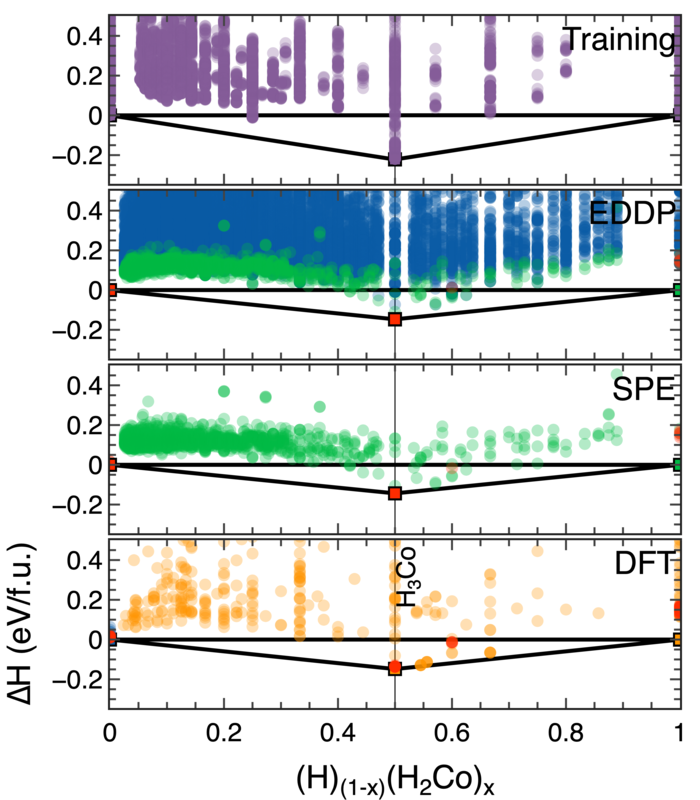}
\footnotesize


\flushleft{
\subsubsection*{\textsc{EDDP}}}
\centering
\includegraphics[width=0.3\textwidth]{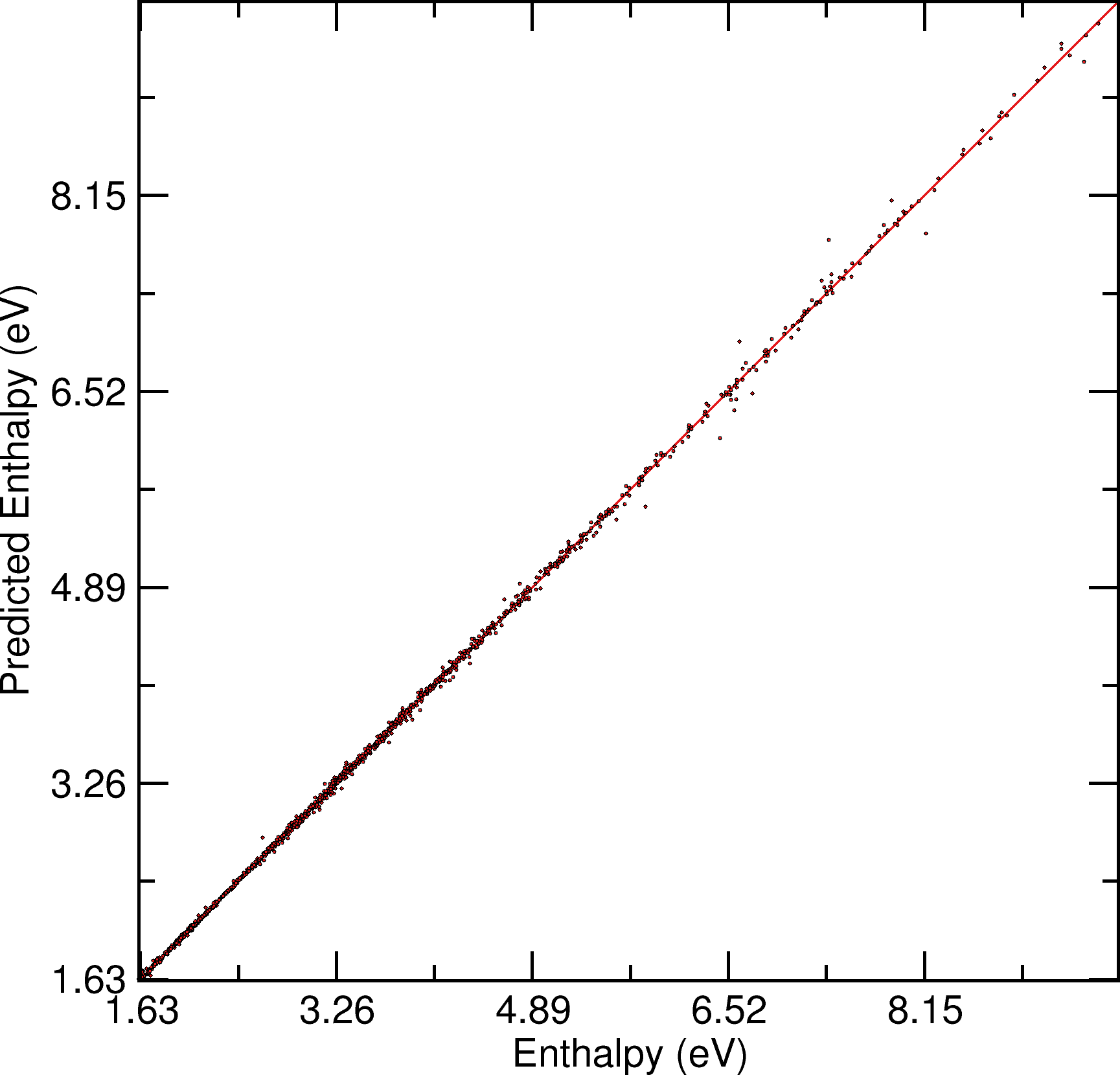}
\includegraphics[width=0.3\textwidth]{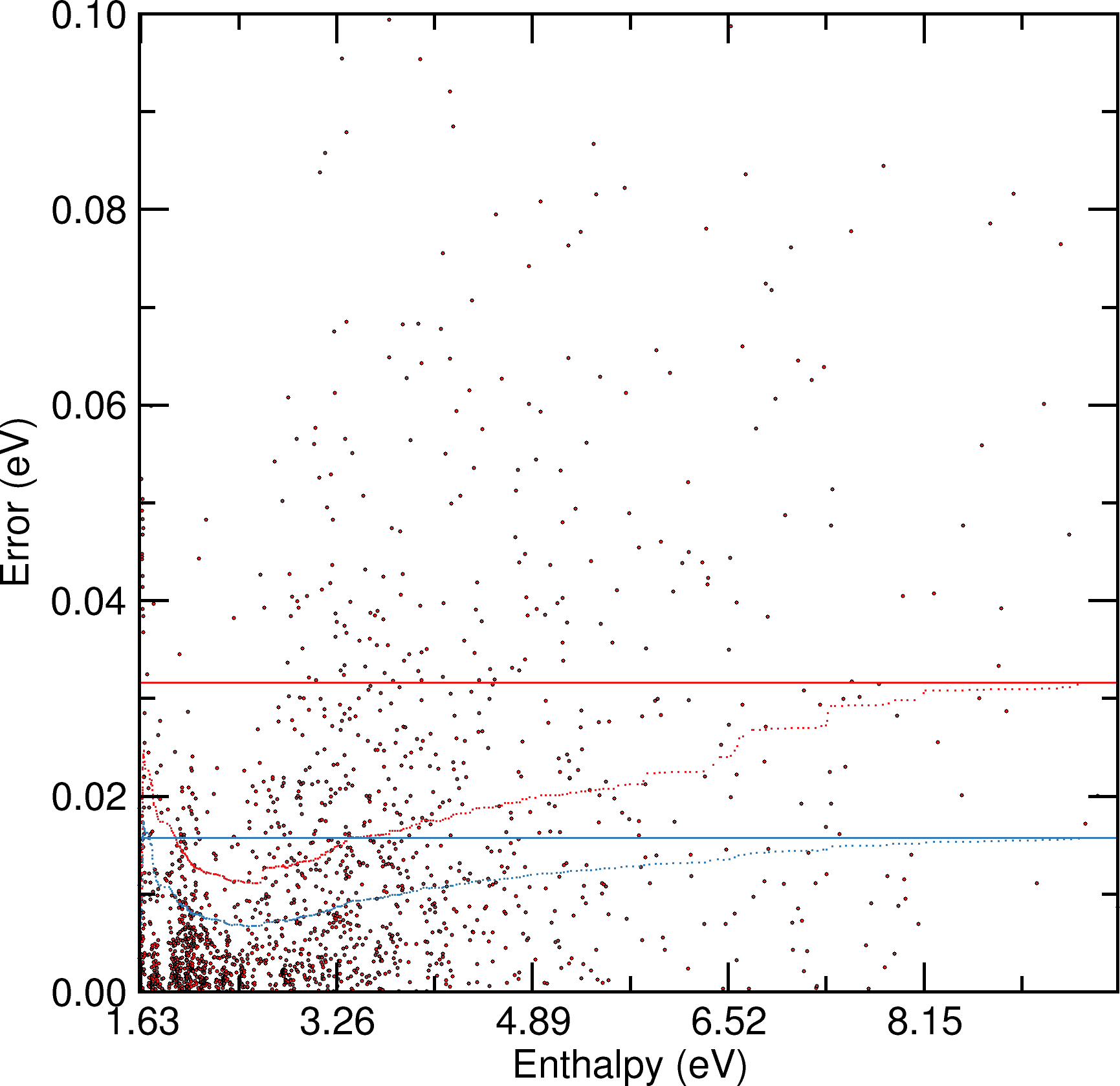}
\centering\begin{verbatim}
training    RMSE/MAE:  16.81  10.43  meV  Spearman  :  0.99984
validation  RMSE/MAE:  22.49  14.29  meV  Spearman  :  0.99977
testing     RMSE/MAE:  31.59  15.74  meV  Spearman  :  0.99981
\end{verbatim}
\clearpage

\flushleft{
\subsection{Cr-H}}
\subsubsection*{Searching}
\centering
\includegraphics[width=0.4\textwidth]{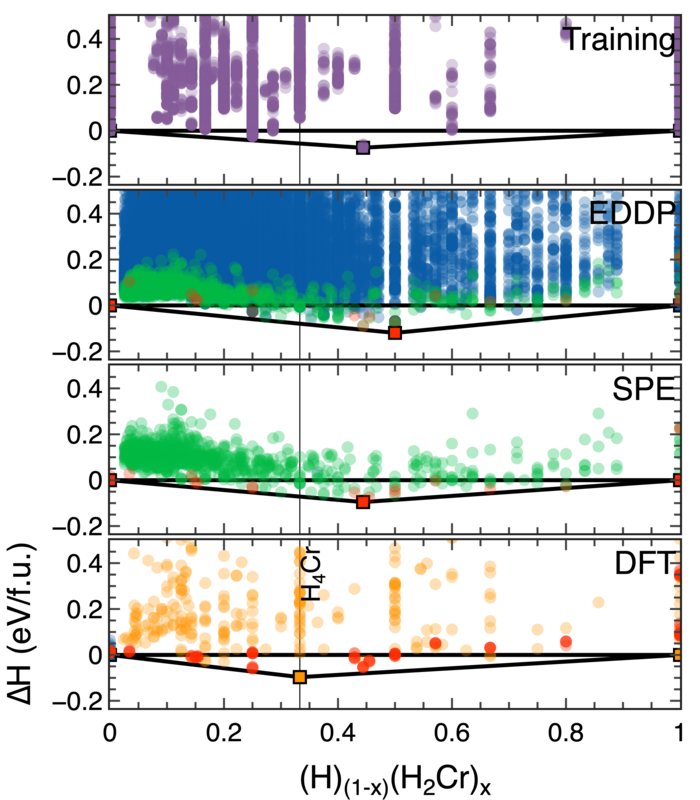}
\footnotesize


\flushleft{
\subsubsection*{\textsc{EDDP}}}
\centering
\includegraphics[width=0.3\textwidth]{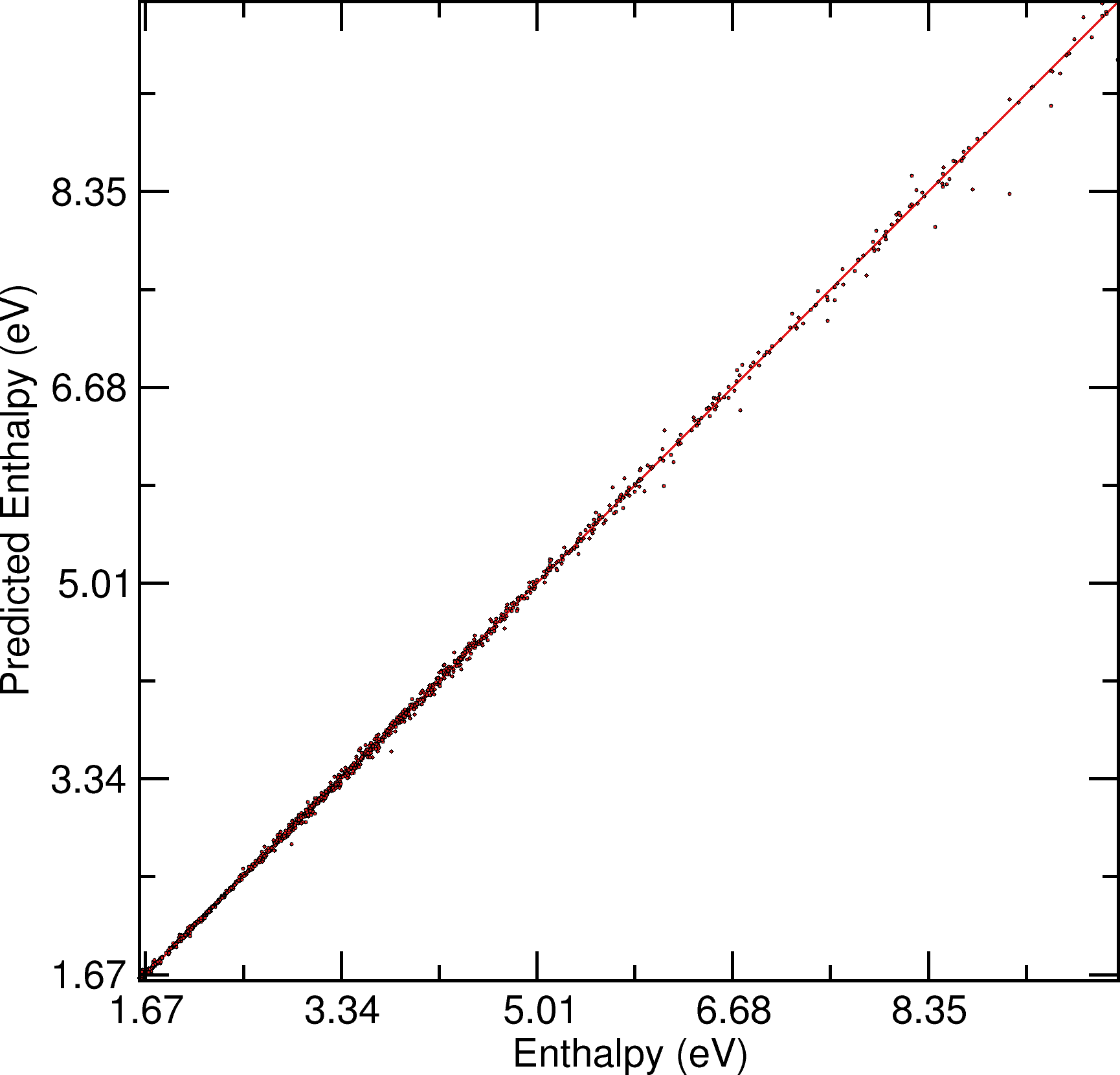}
\includegraphics[width=0.3\textwidth]{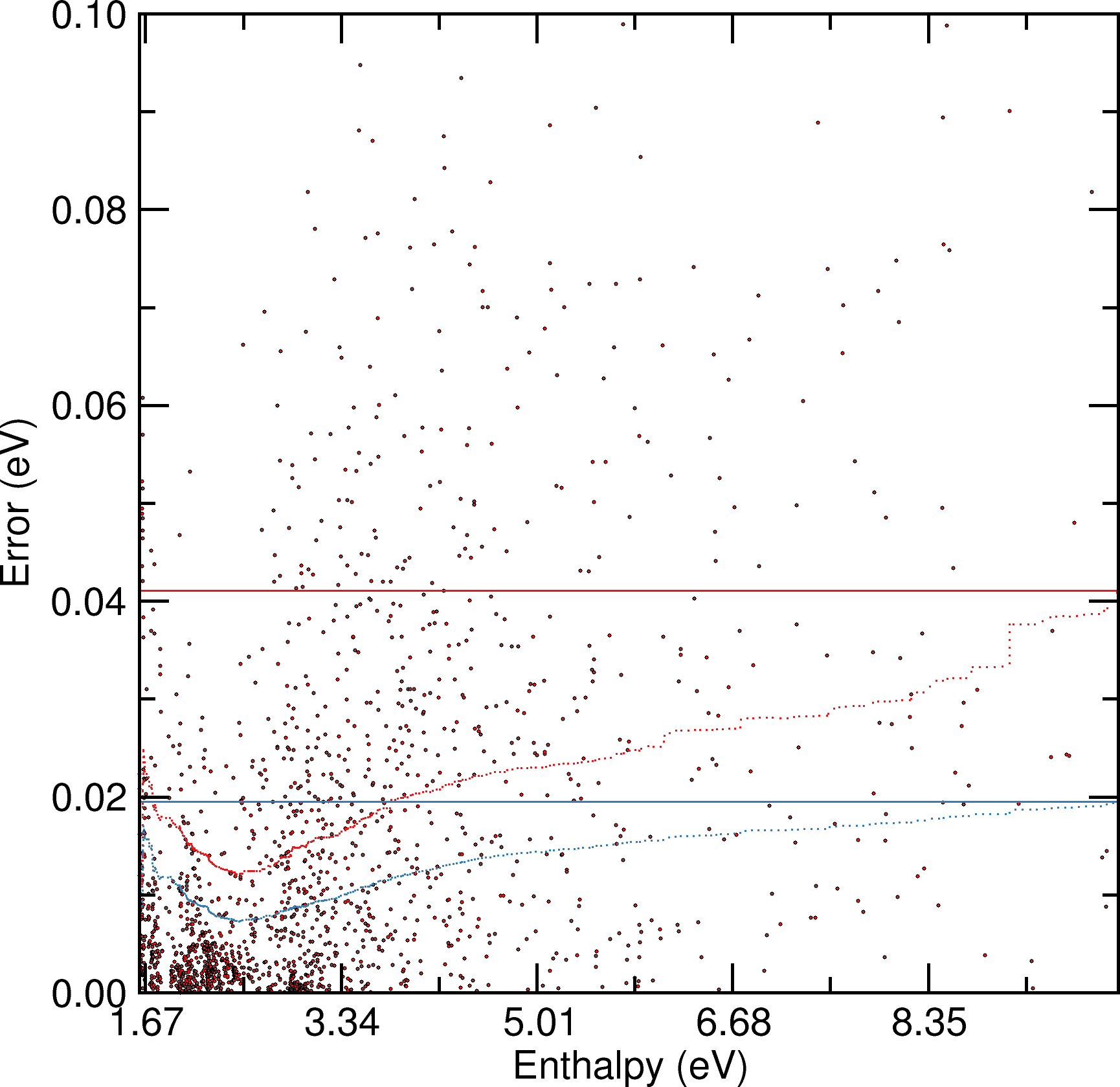}
\centering\begin{verbatim}
training    RMSE/MAE:  20.08  12.66  meV  Spearman  :  0.99982
validation  RMSE/MAE:  31.55  19.04  meV  Spearman  :  0.99977
testing     RMSE/MAE:  41.10  19.54  meV  Spearman  :  0.99976
\end{verbatim}
\clearpage

\flushleft{
\subsection{Cs-H}}
\subsubsection*{Searching}
\centering
\includegraphics[width=0.4\textwidth]{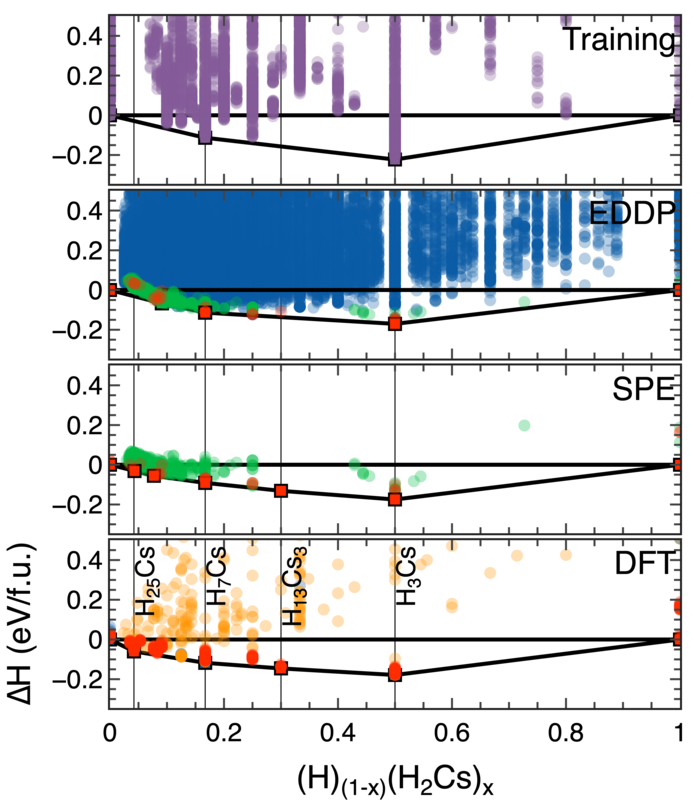}
\footnotesize


\flushleft{
\subsubsection*{\textsc{EDDP}}}
\centering
\includegraphics[width=0.3\textwidth]{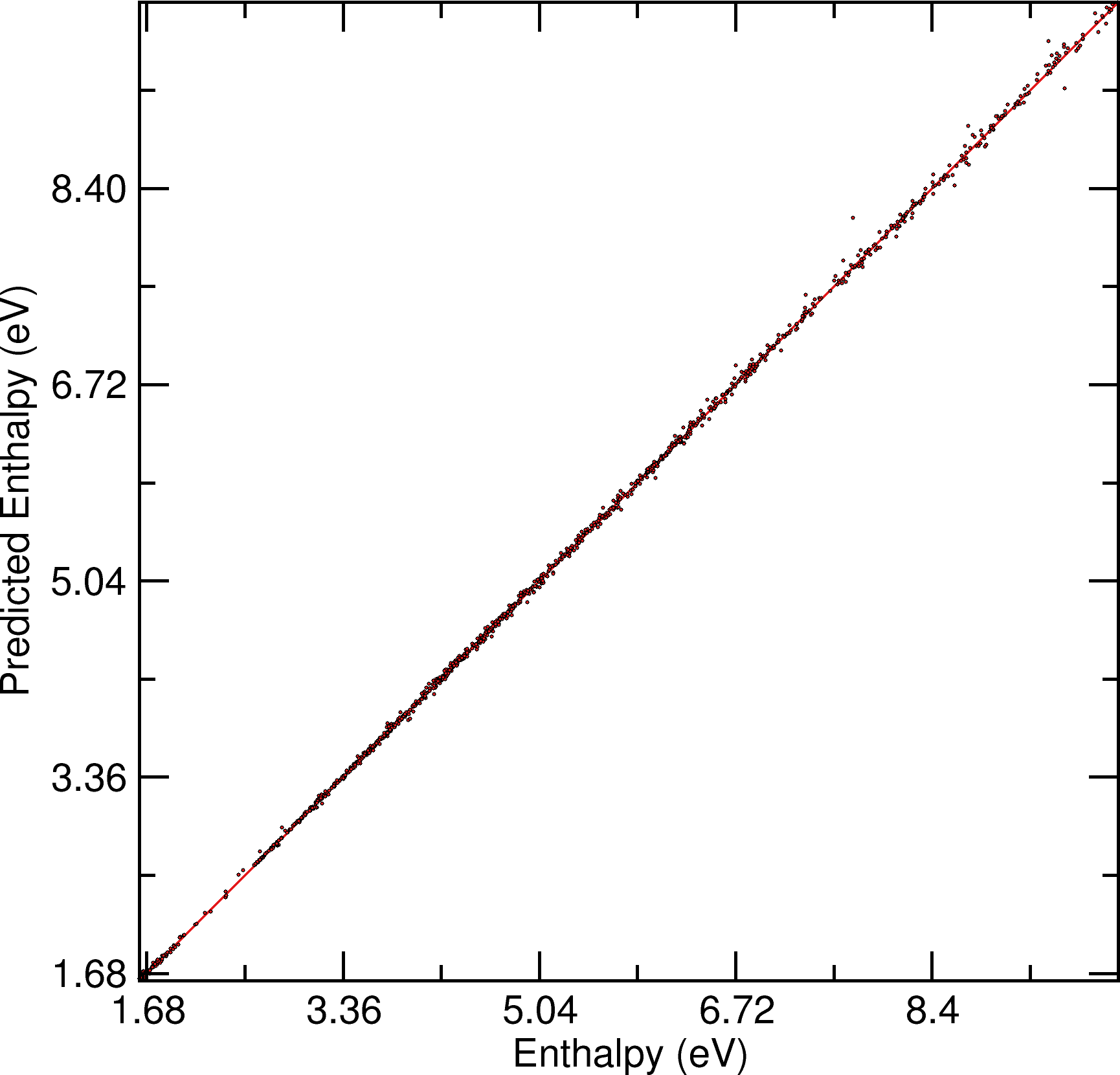}
\includegraphics[width=0.3\textwidth]{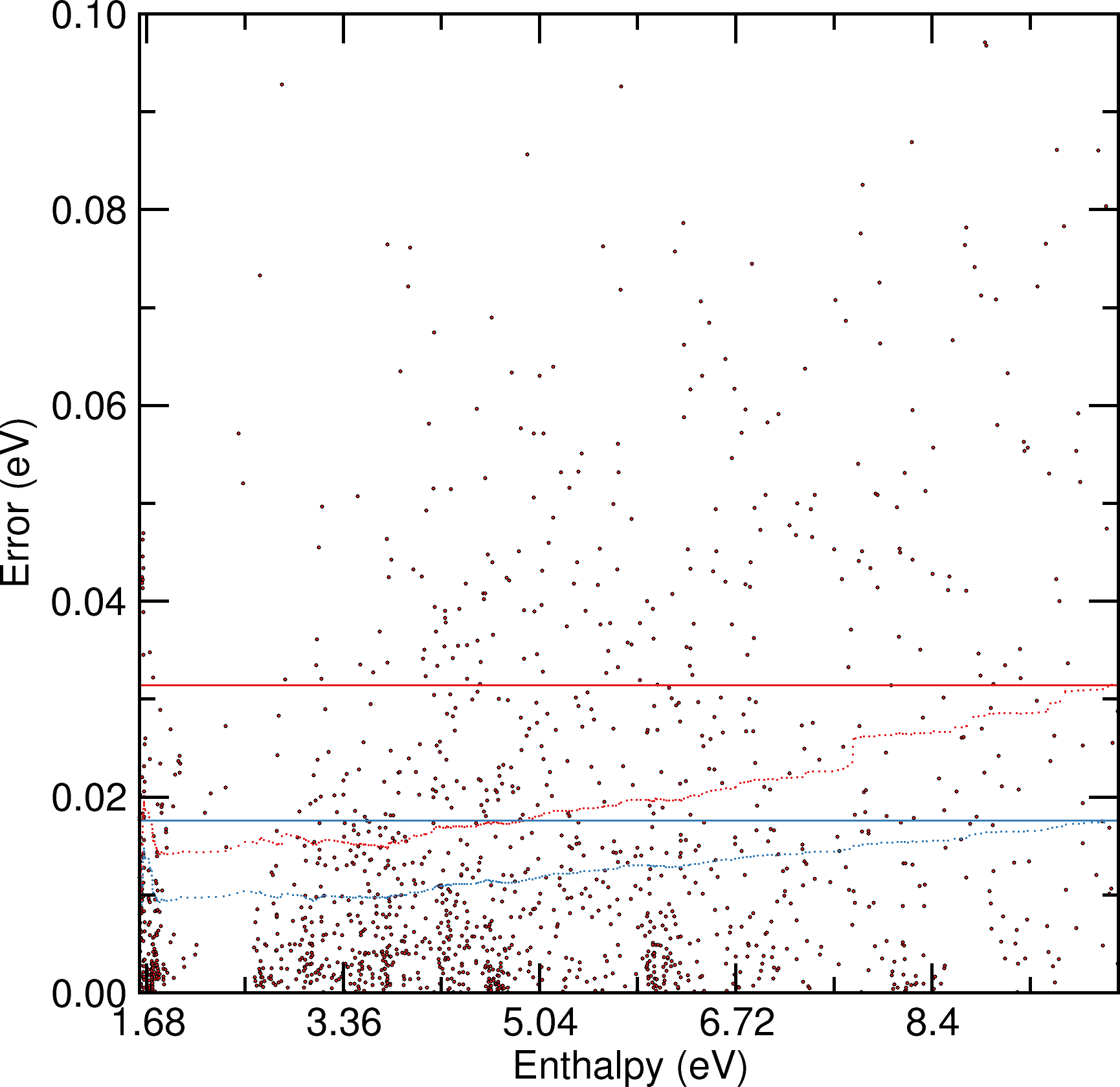}
\centering\begin{verbatim}
training    RMSE/MAE:  18.29  11.14  meV  Spearman  :  0.99987
validation  RMSE/MAE:  28.14  17.24  meV  Spearman  :  0.99982
testing     RMSE/MAE:  31.44  17.59  meV  Spearman  :  0.99983
\end{verbatim}
\clearpage

\flushleft{
\subsection{Cu-H}}
\subsubsection*{Searching}
\centering
\includegraphics[width=0.4\textwidth]{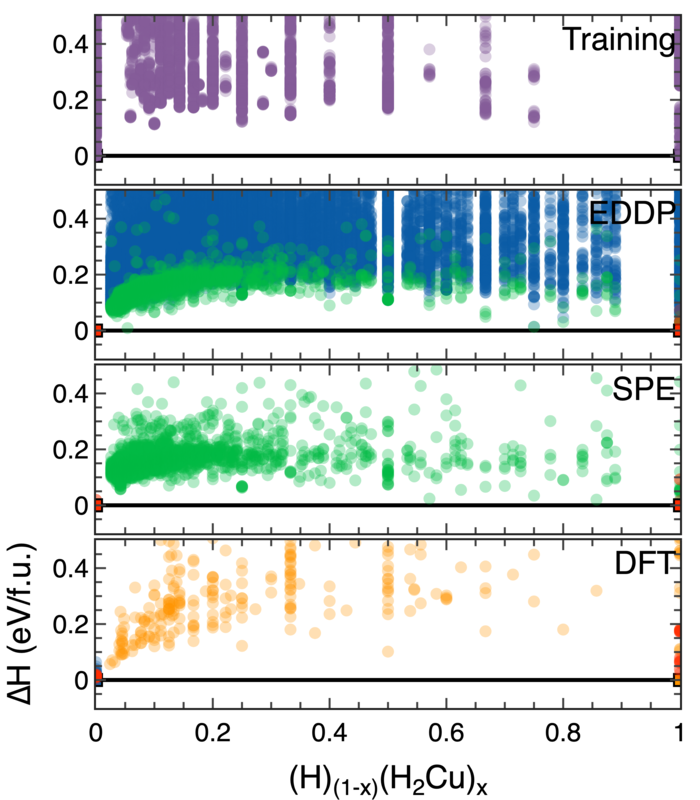}
\footnotesize


\flushleft{
\subsubsection*{\textsc{EDDP}}}
\centering
\includegraphics[width=0.3\textwidth]{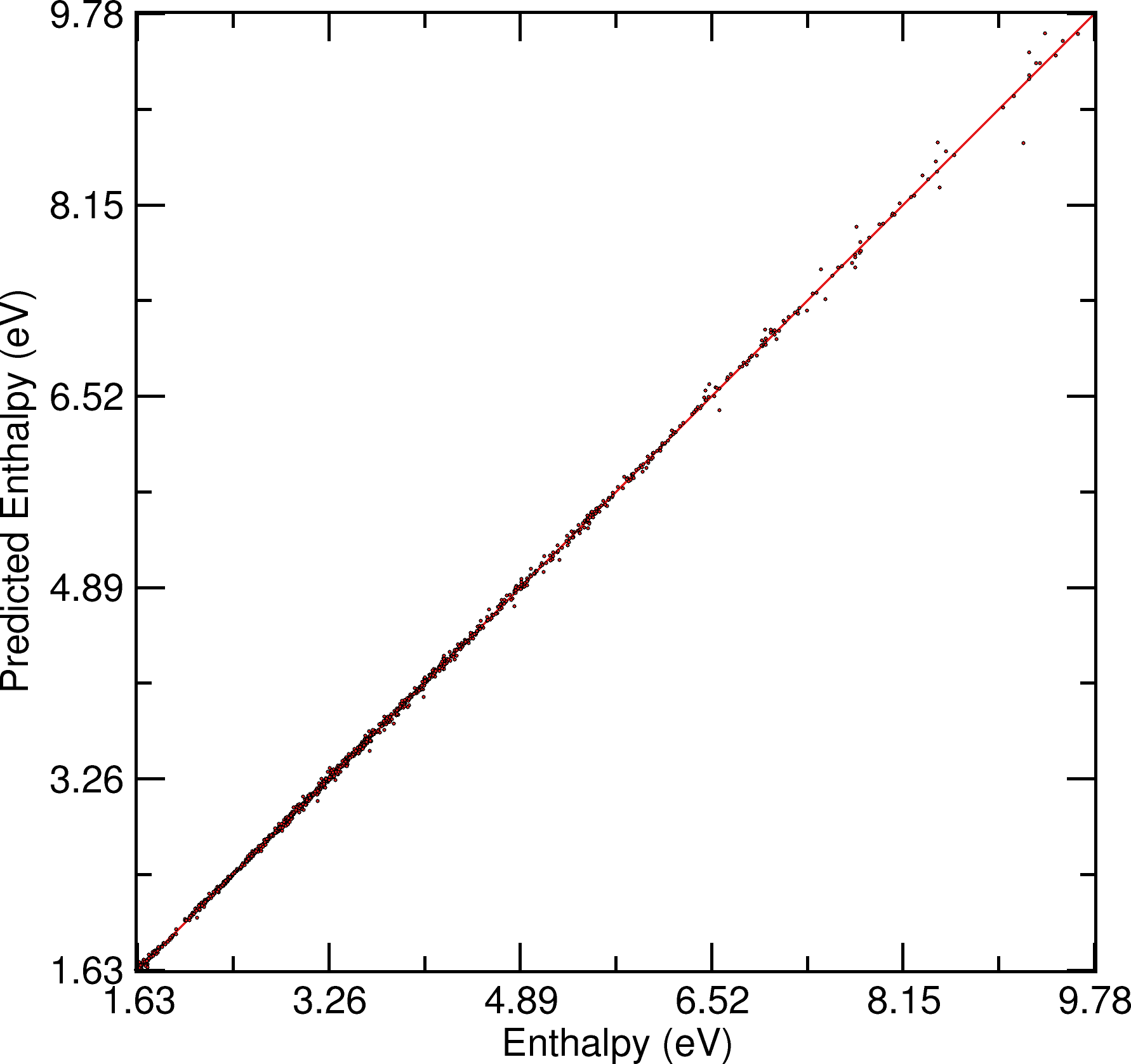}
\includegraphics[width=0.3\textwidth]{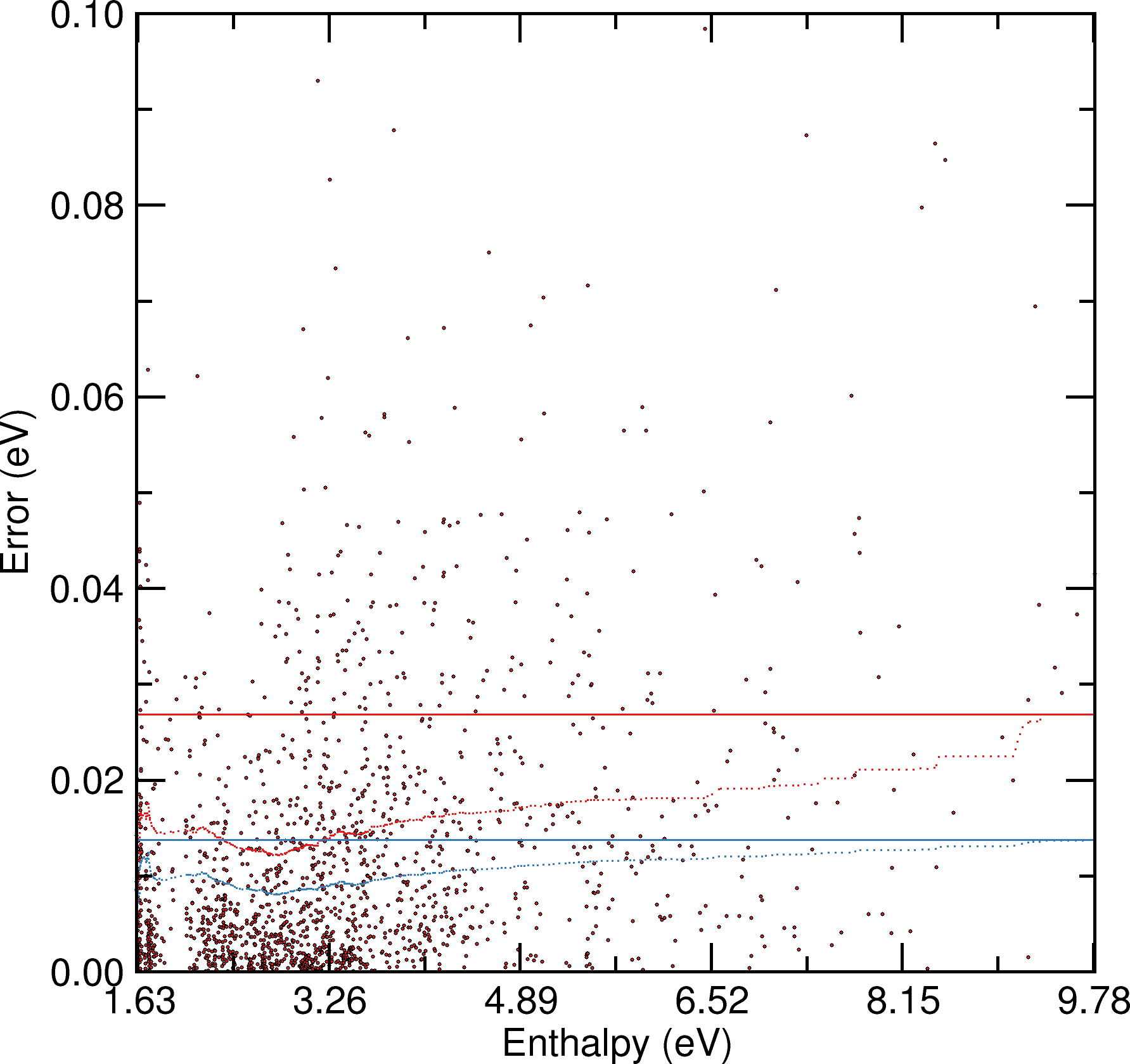}
\centering\begin{verbatim}
training    RMSE/MAE:  16.59  9.54   meV  Spearman  :  0.99986
validation  RMSE/MAE:  20.99  13.03  meV  Spearman  :  0.99980
testing     RMSE/MAE:  26.86  13.73  meV  Spearman  :  0.99980
\end{verbatim}
\clearpage

\flushleft{
\subsection{Dy-H}}
\subsubsection*{Searching}
\centering
\includegraphics[width=0.4\textwidth]{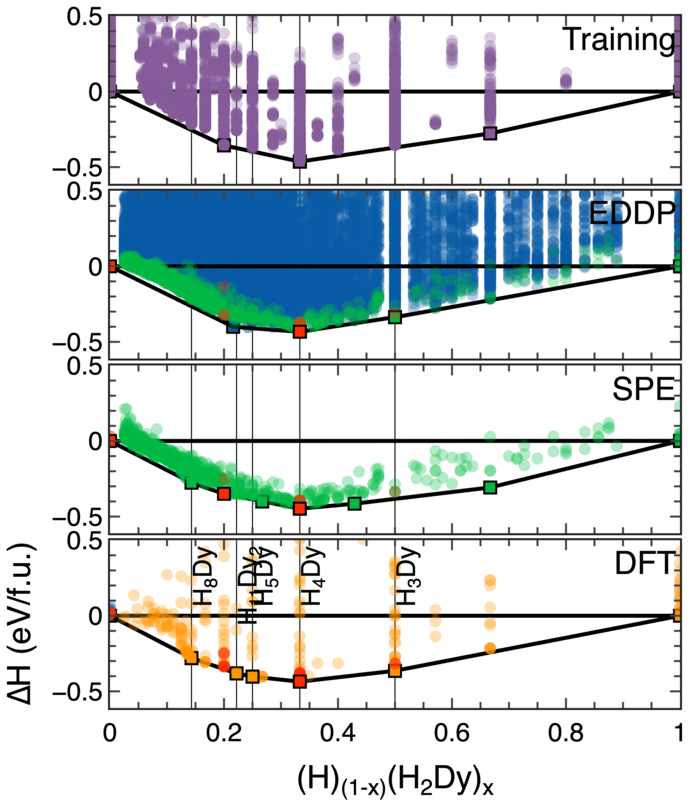}
\footnotesize


\flushleft{
\subsubsection*{\textsc{EDDP}}}
\centering
\includegraphics[width=0.3\textwidth]{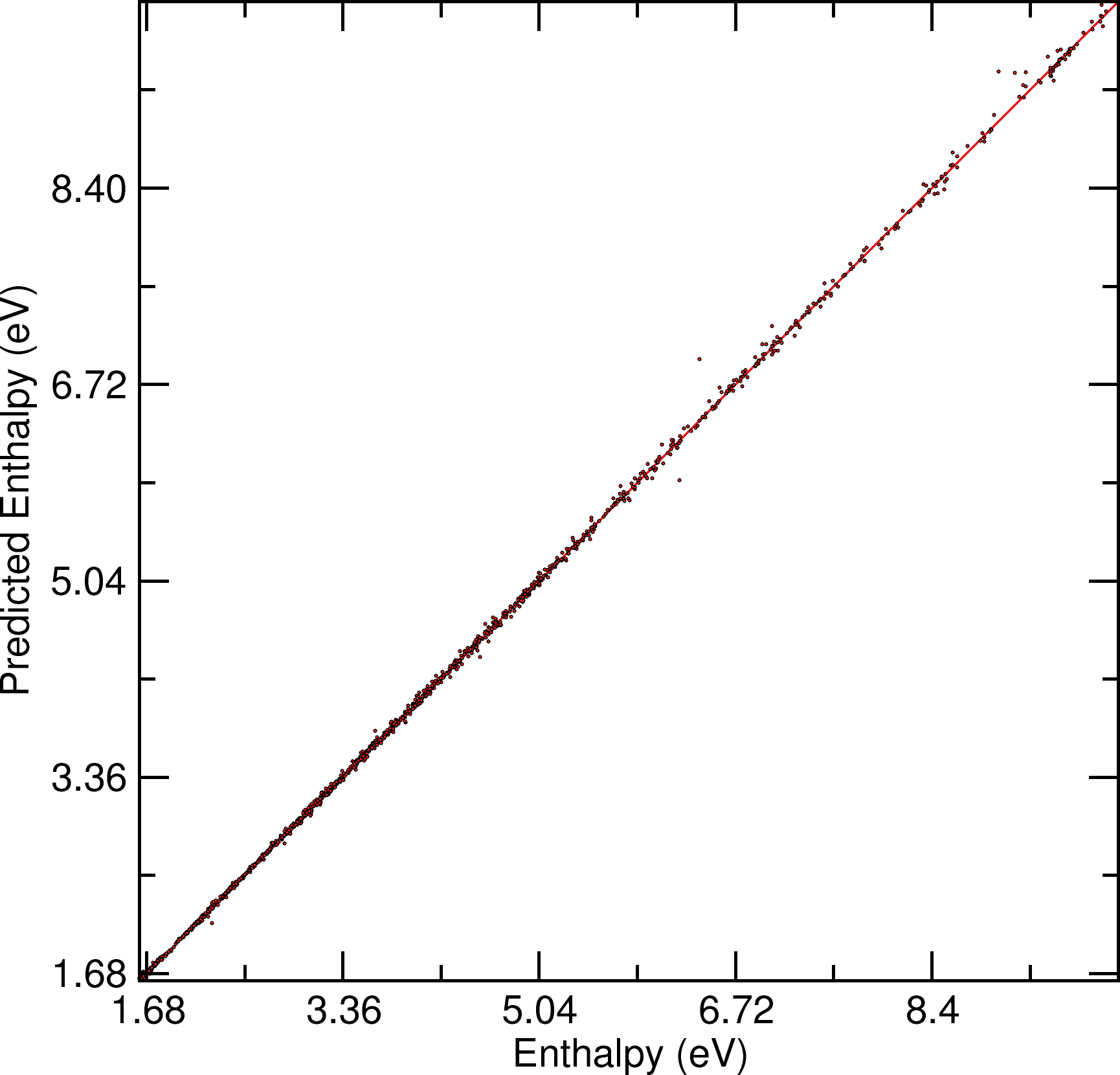}
\includegraphics[width=0.3\textwidth]{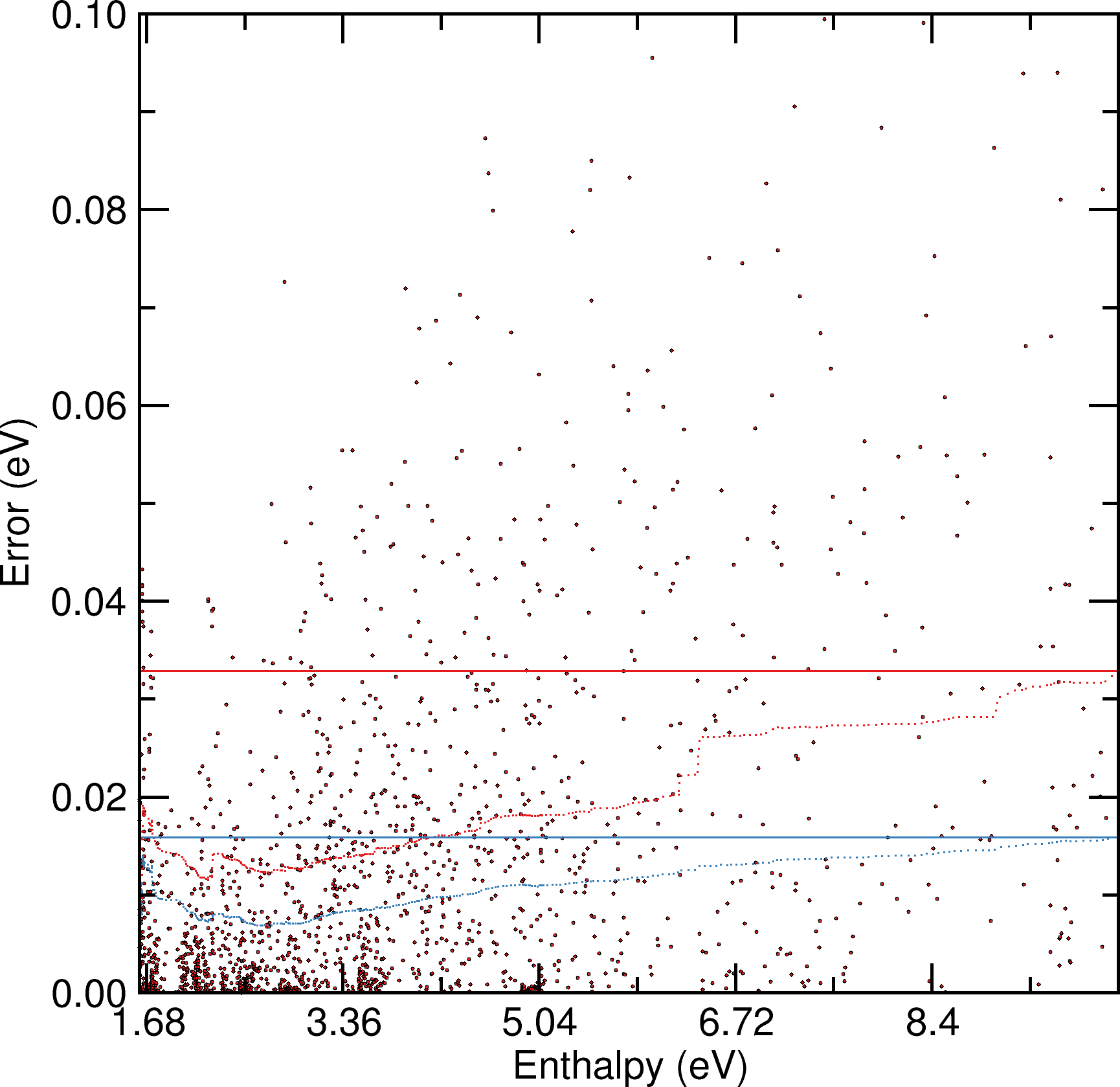}
\centering\begin{verbatim}
training    RMSE/MAE:  15.20  9.65   meV  Spearman  :  0.99990
validation  RMSE/MAE:  22.60  14.63  meV  Spearman  :  0.99984
testing     RMSE/MAE:  32.86  15.89  meV  Spearman  :  0.99986
\end{verbatim}
\clearpage

\flushleft{
\subsection{Er-H}}
\subsubsection*{Searching}
\centering
\includegraphics[width=0.4\textwidth]{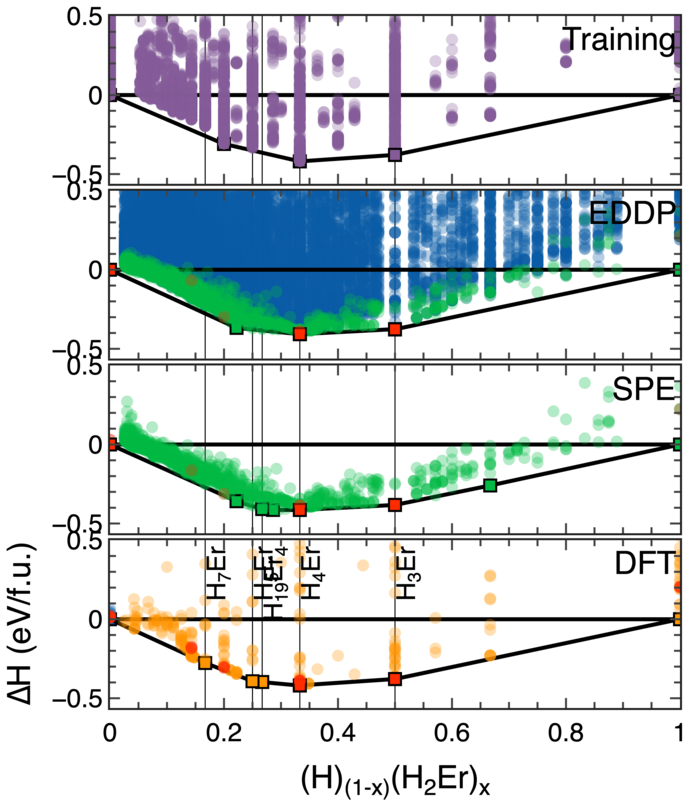}
\footnotesize


\flushleft{
\subsubsection*{\textsc{EDDP}}}
\centering
\includegraphics[width=0.3\textwidth]{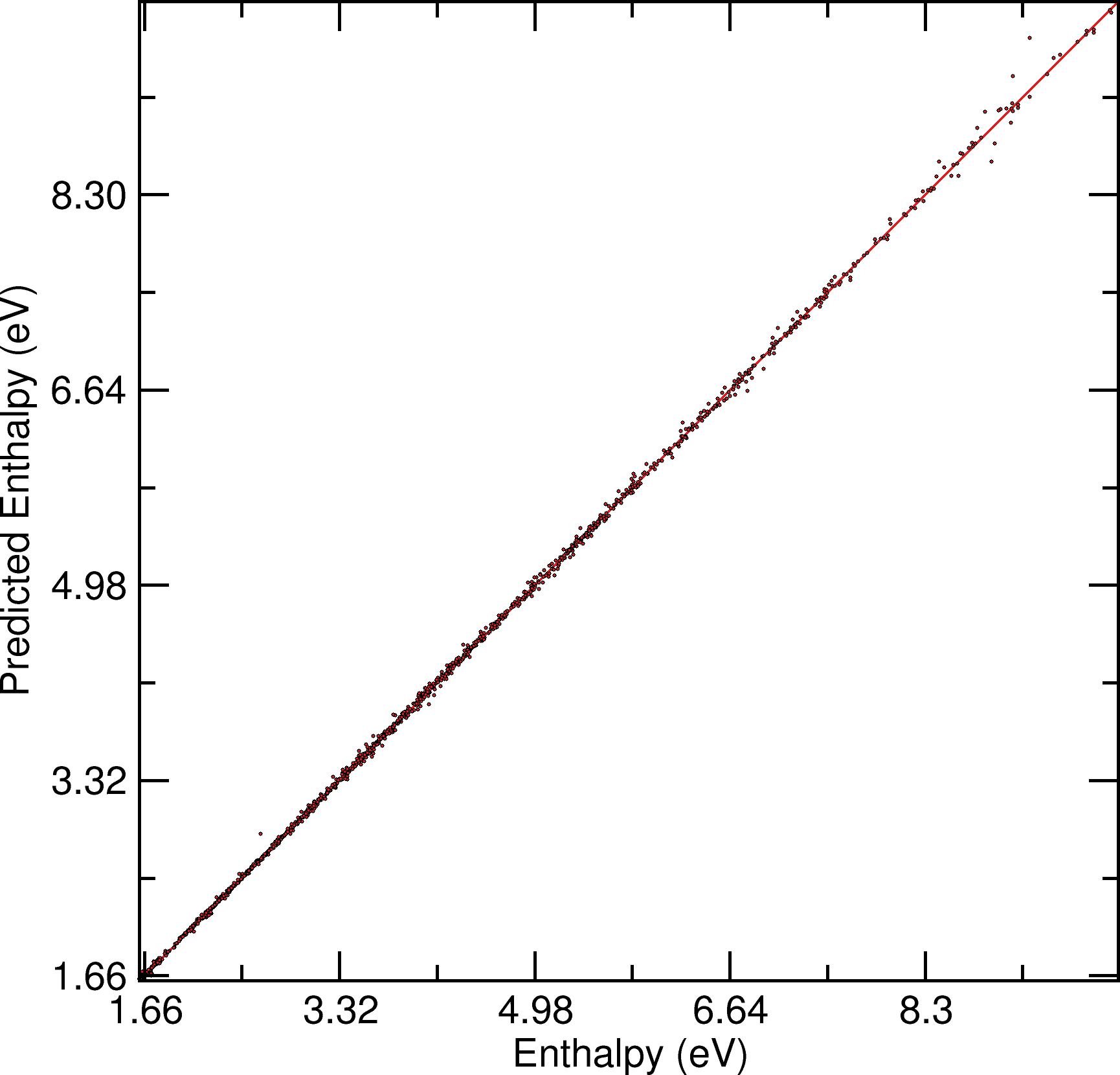}
\includegraphics[width=0.3\textwidth]{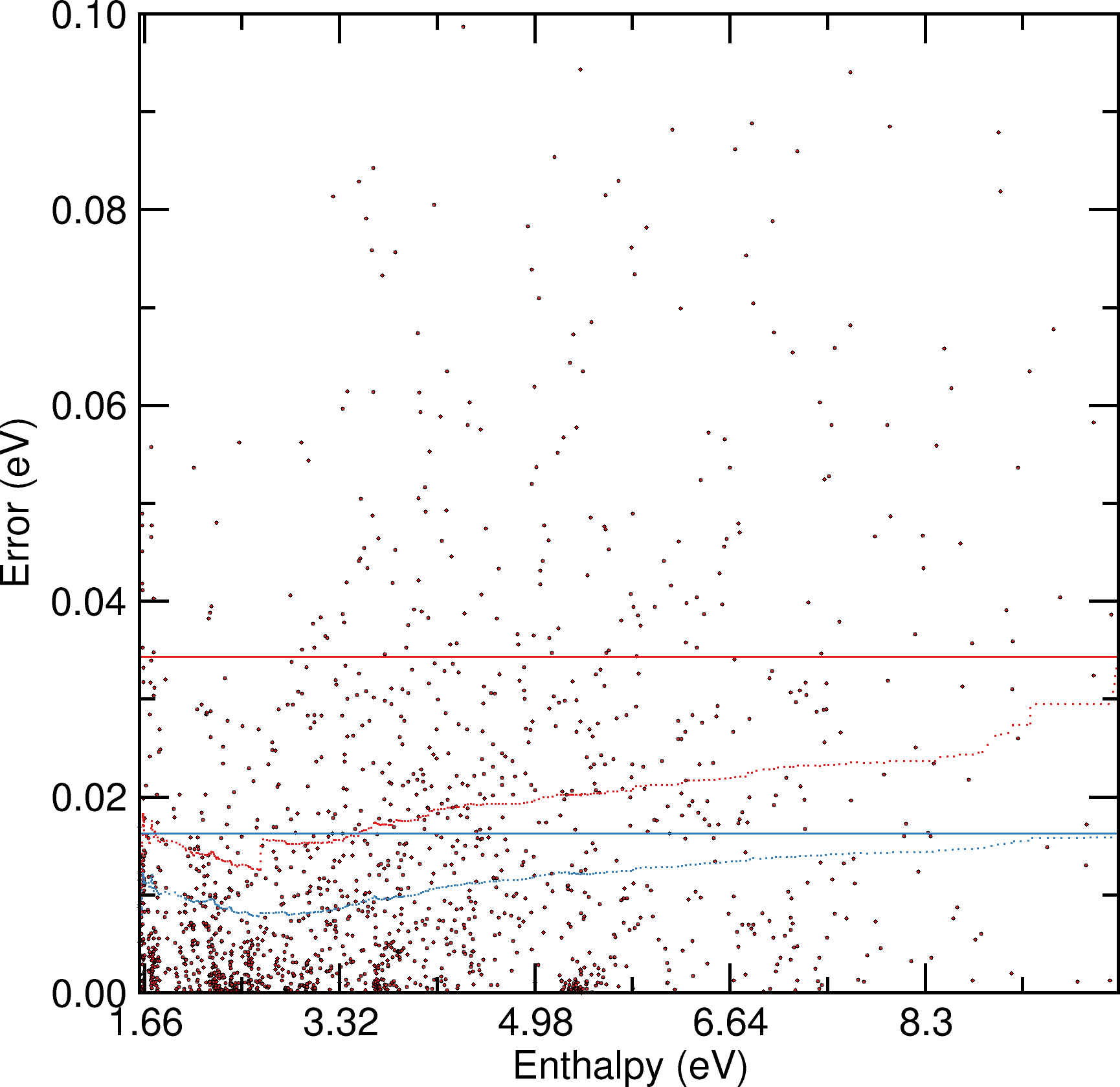}
\centering\begin{verbatim}
training    RMSE/MAE:  16.13  10.10  meV  Spearman  :  0.99989
validation  RMSE/MAE:  25.03  15.44  meV  Spearman  :  0.99987
testing     RMSE/MAE:  34.34  16.28  meV  Spearman  :  0.99986
\end{verbatim}
\clearpage

\flushleft{
\subsection{Eu-H}}
\subsubsection*{Searching}
\centering
\includegraphics[width=0.4\textwidth]{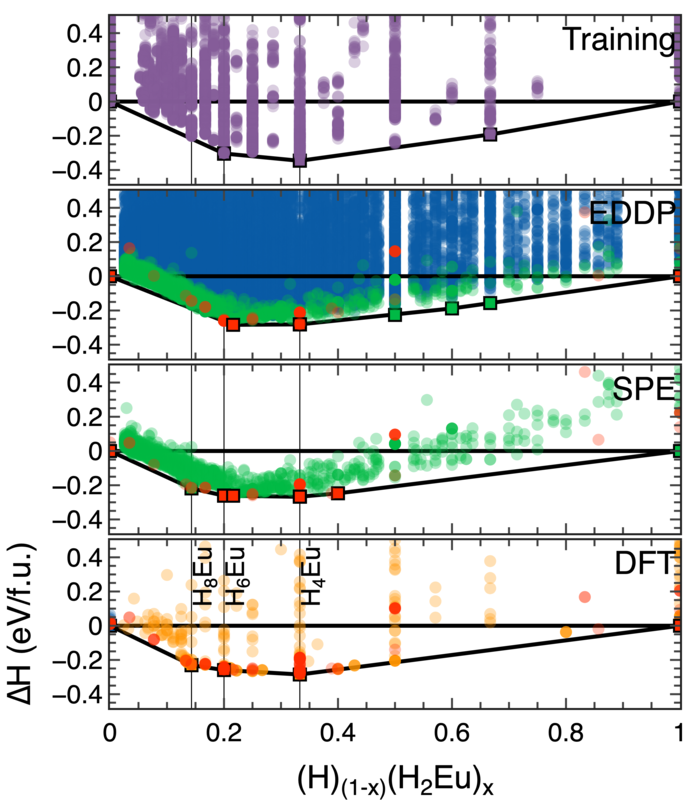}
\footnotesize


\flushleft{
\subsubsection*{\textsc{EDDP}}}
\centering
\includegraphics[width=0.3\textwidth]{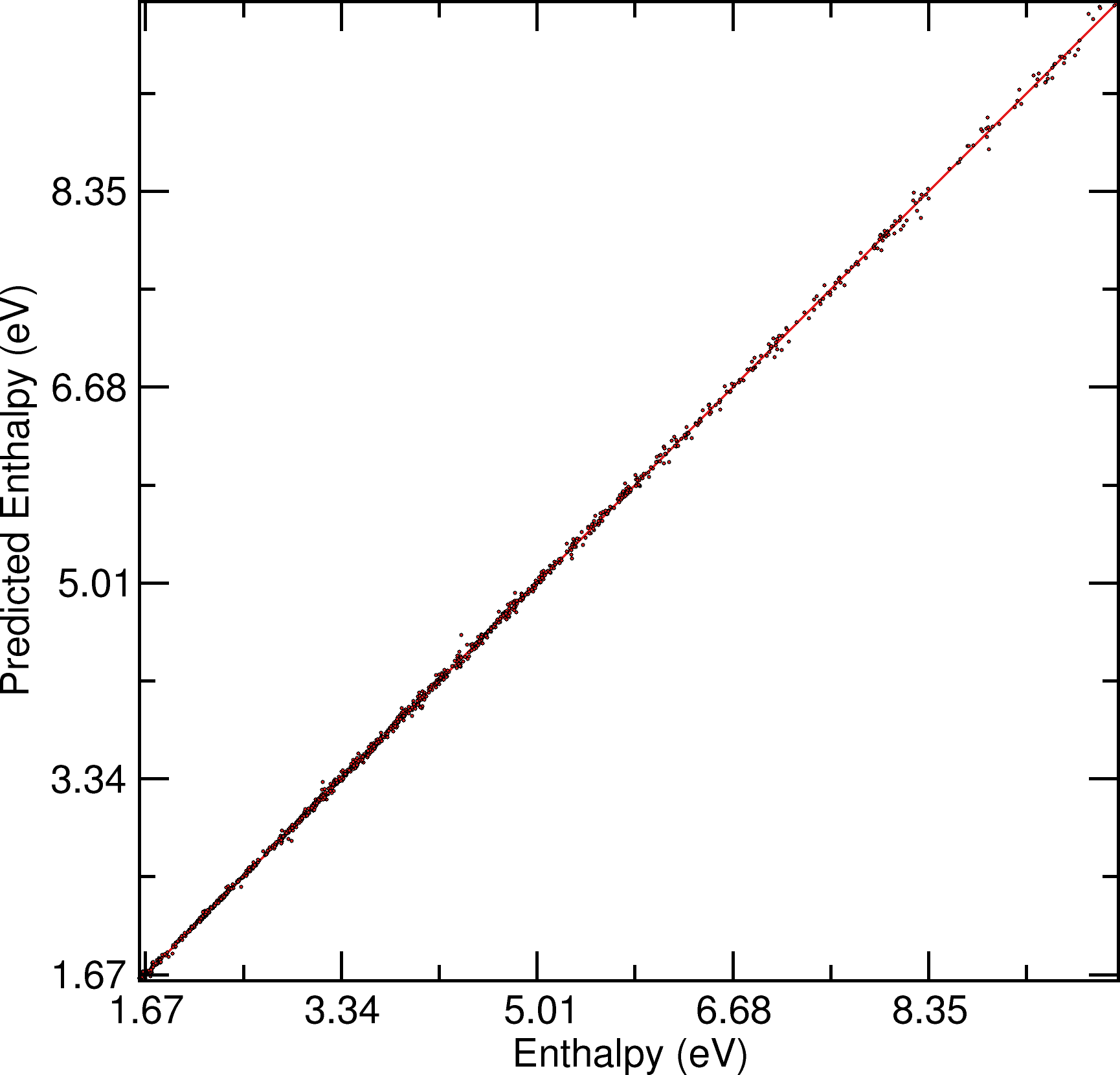}
\includegraphics[width=0.3\textwidth]{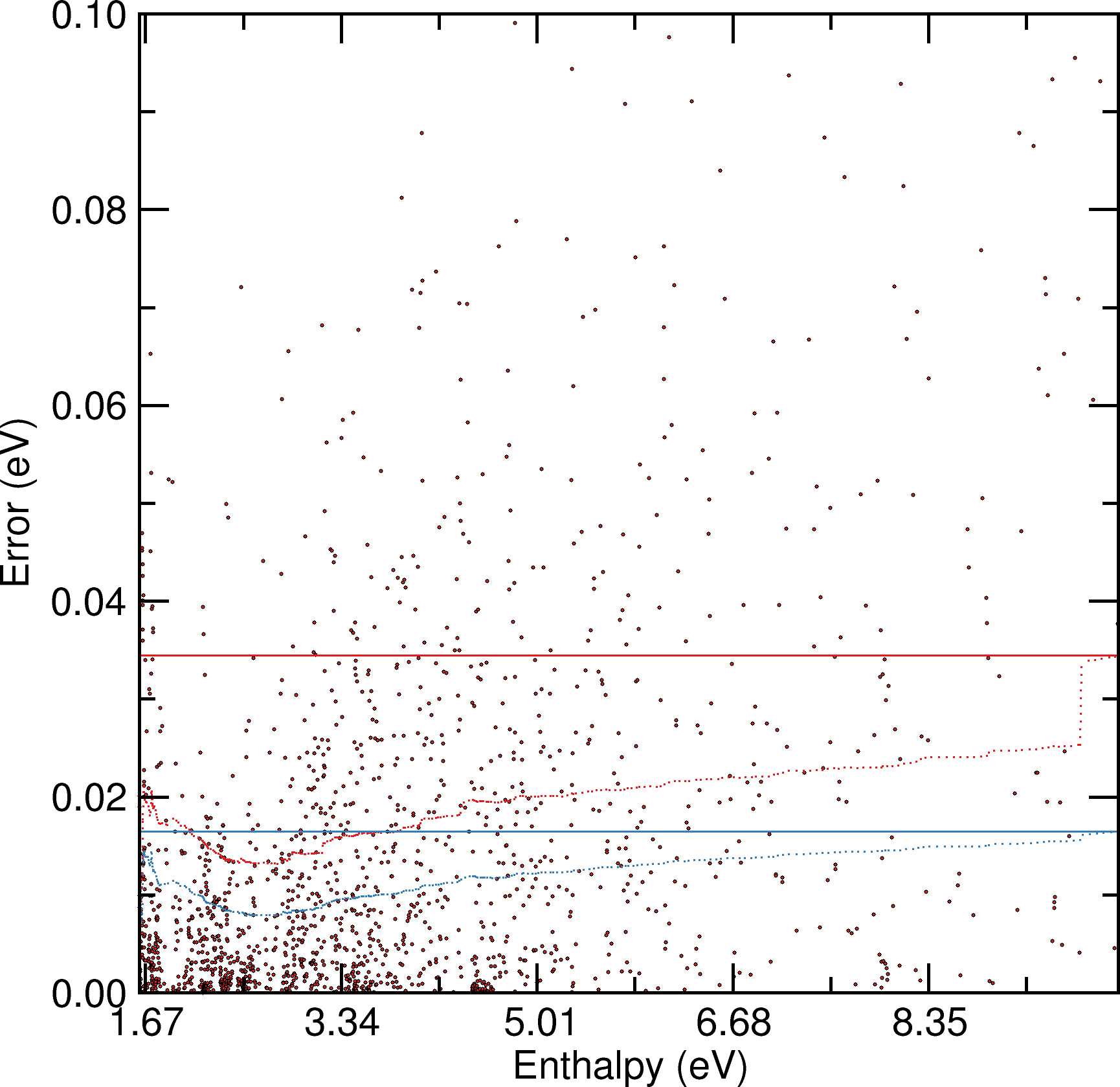}
\centering\begin{verbatim}
training    RMSE/MAE:  17.11  10.94  meV  Spearman  :  0.99987
validation  RMSE/MAE:  25.26  15.89  meV  Spearman  :  0.99986
testing     RMSE/MAE:  34.48  16.50  meV  Spearman  :  0.99984
\end{verbatim}
\clearpage

\flushleft{
\subsection{Fe-H}}
\subsubsection*{Searching}
\centering
\includegraphics[width=0.4\textwidth]{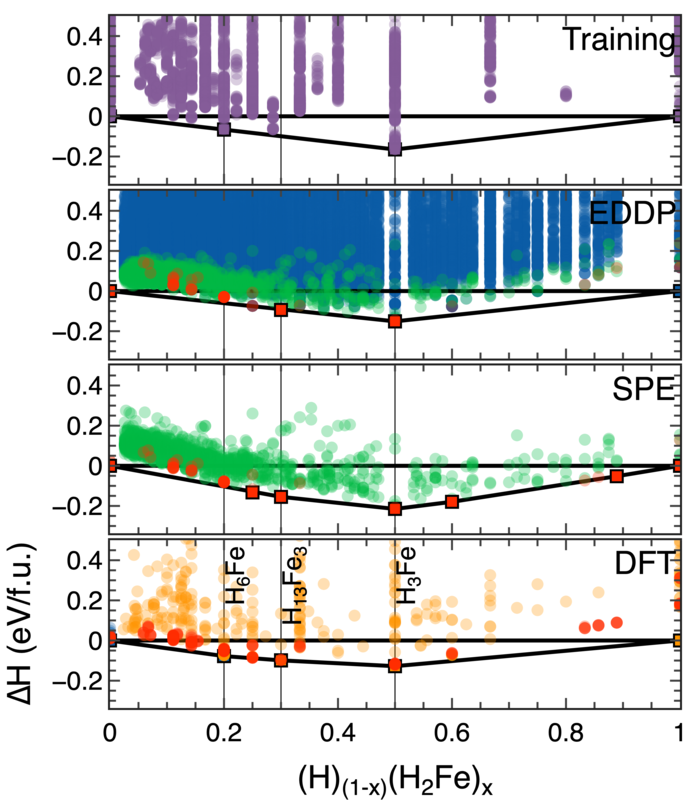}
\footnotesize


\flushleft{
\subsubsection*{\textsc{EDDP}}}
\centering
\includegraphics[width=0.3\textwidth]{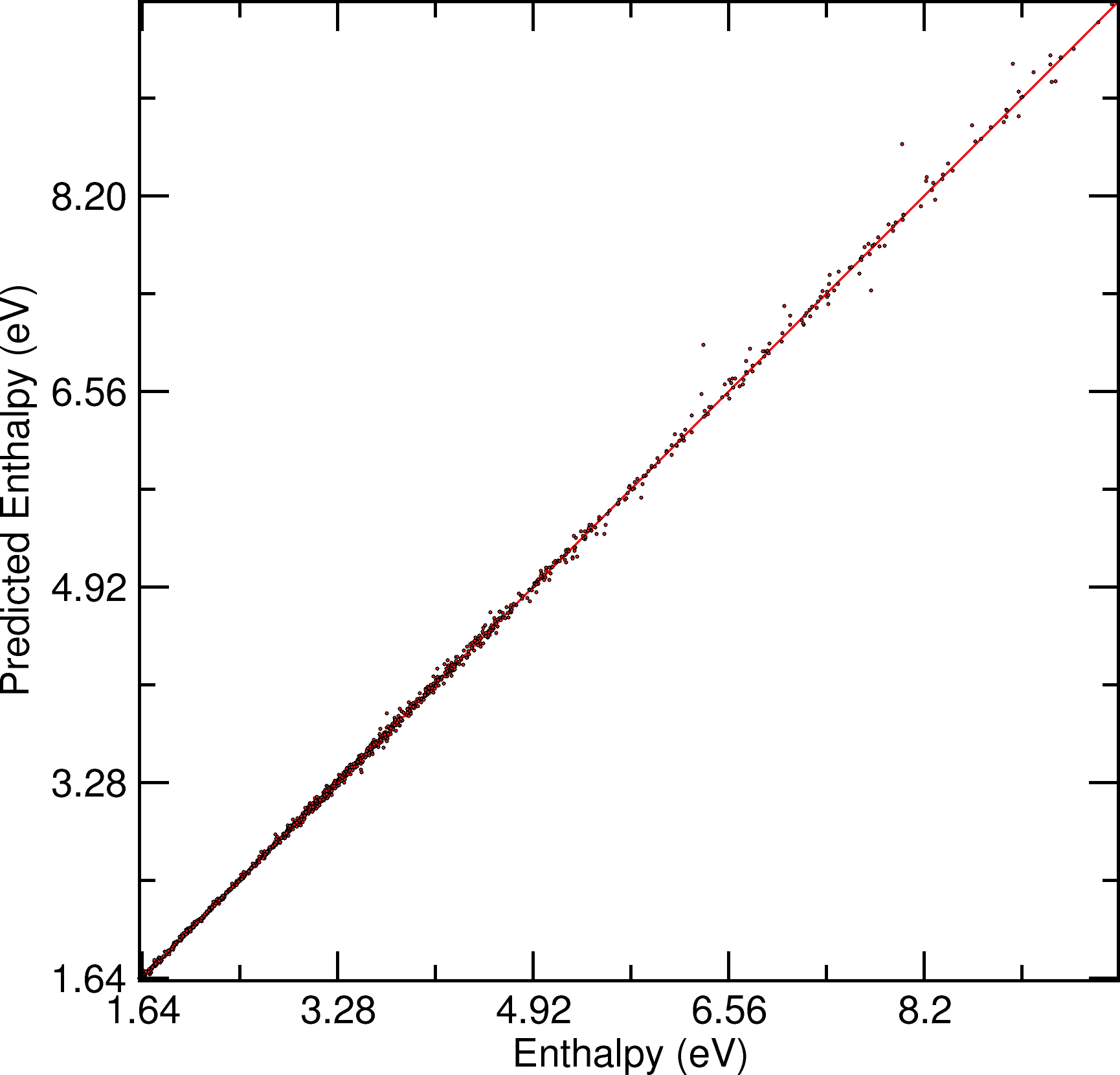}
\includegraphics[width=0.3\textwidth]{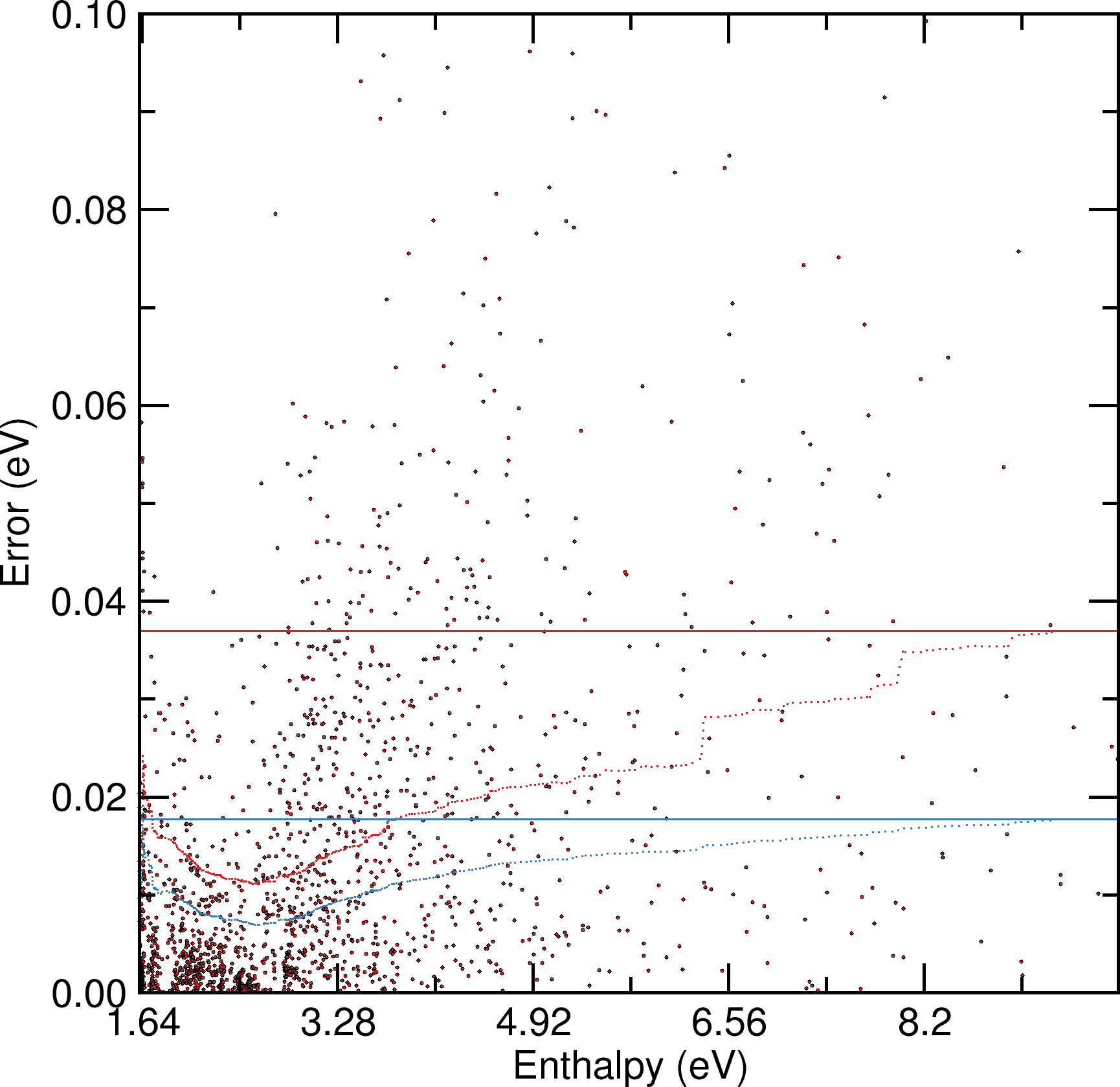}
\centering\begin{verbatim}
training    RMSE/MAE:  20.50  11.96  meV  Spearman  :  0.99985
validation  RMSE/MAE:  25.23  15.55  meV  Spearman  :  0.99983
testing     RMSE/MAE:  36.95  17.73  meV  Spearman  :  0.99977
\end{verbatim}
\clearpage

\flushleft{
\subsection{Ga-H}}
\subsubsection*{Searching}
\centering
\includegraphics[width=0.4\textwidth]{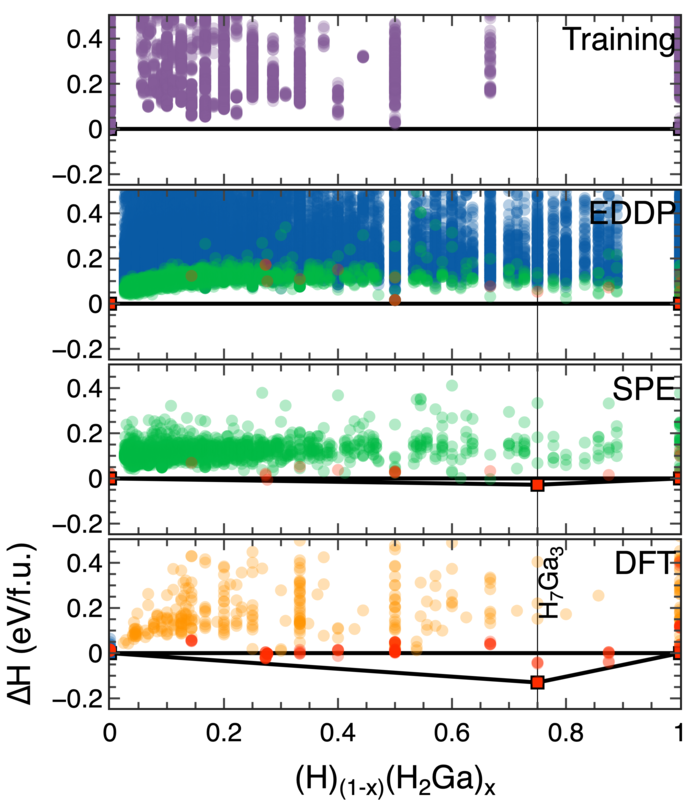}
\footnotesize


\flushleft{
\subsubsection*{\textsc{EDDP}}}
\centering
\includegraphics[width=0.3\textwidth]{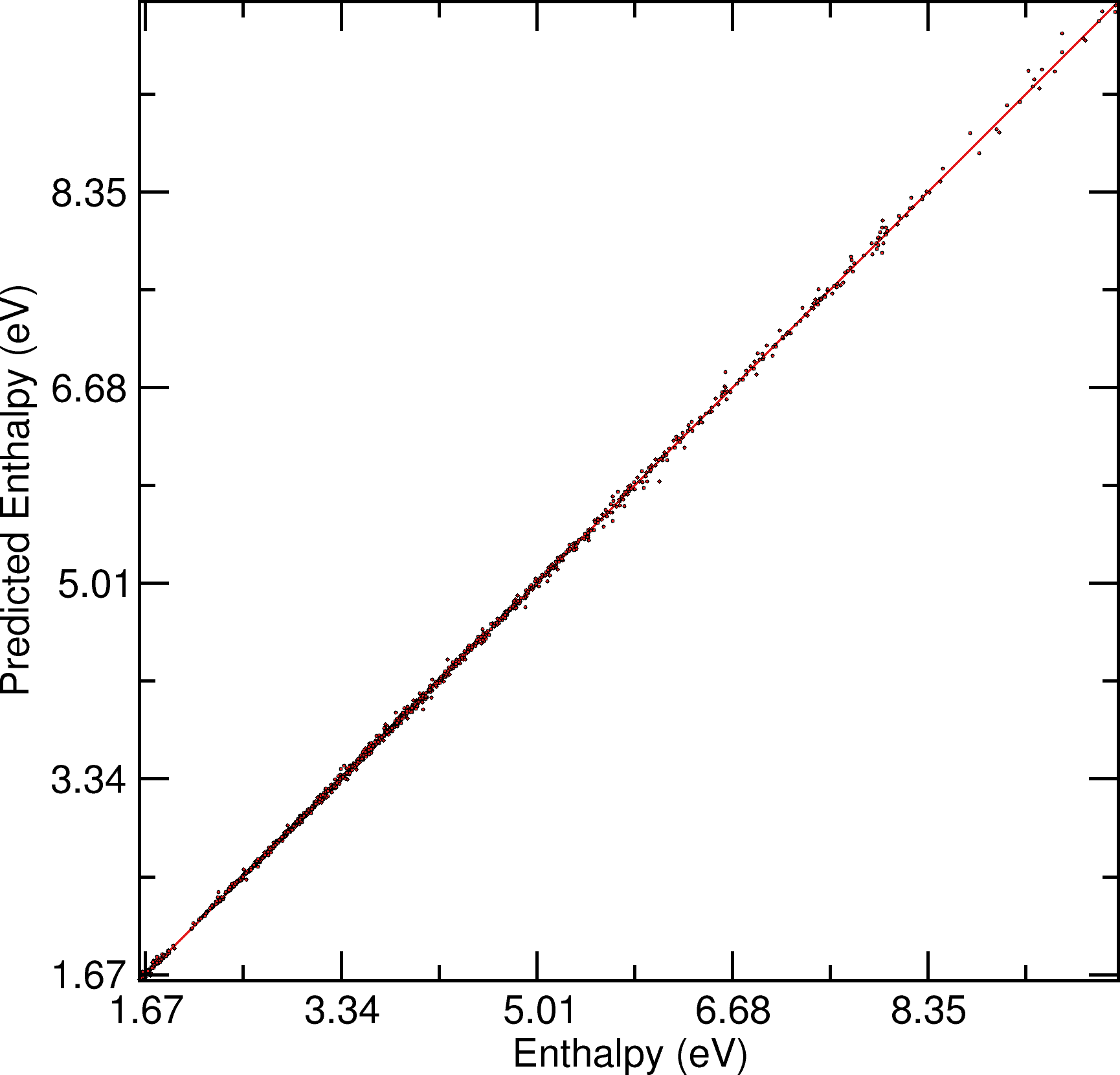}
\includegraphics[width=0.3\textwidth]{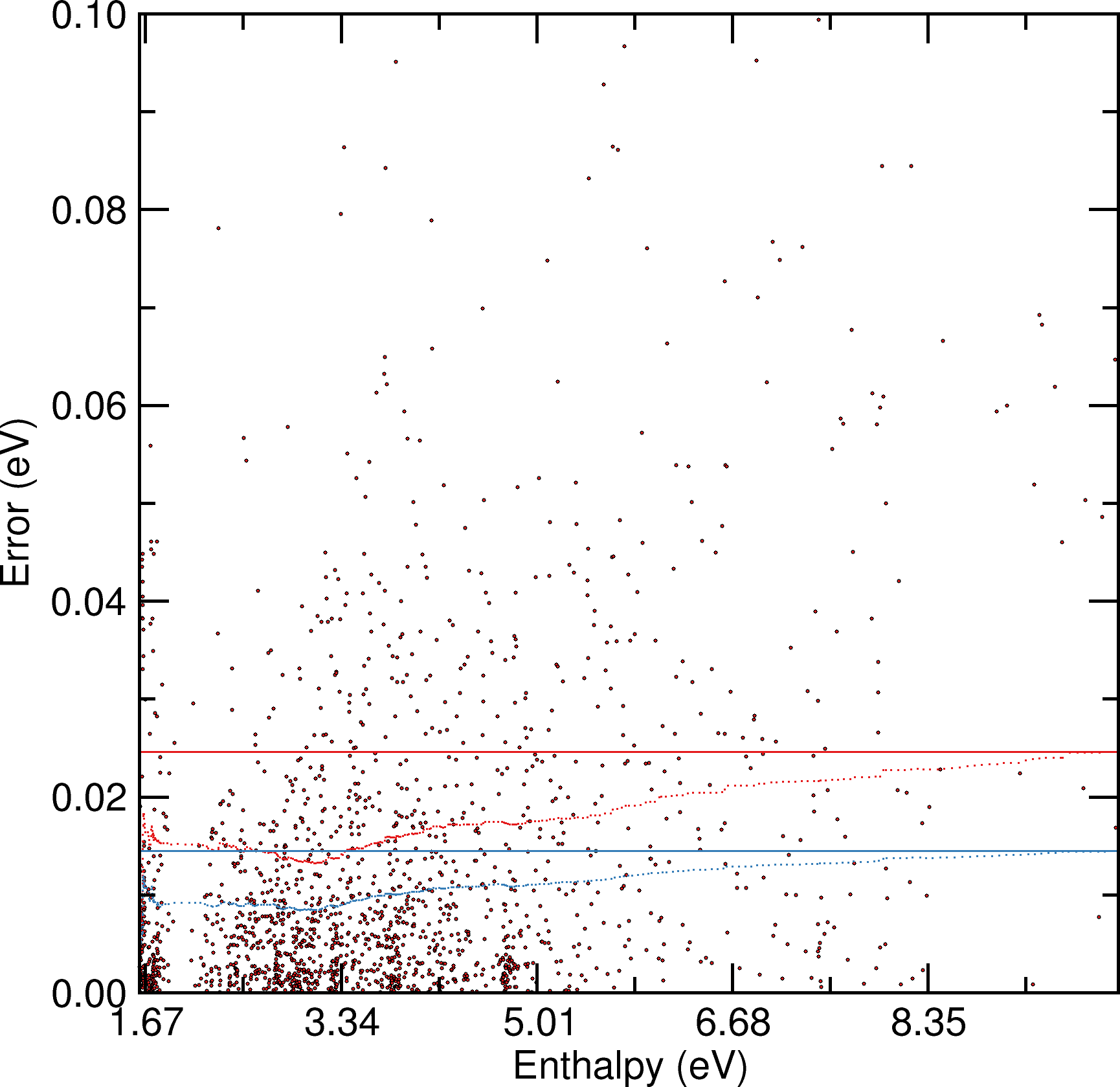}
\centering\begin{verbatim}
training    RMSE/MAE:  15.40  9.56   meV  Spearman  :  0.99988
validation  RMSE/MAE:  22.87  13.97  meV  Spearman  :  0.99982
testing     RMSE/MAE:  24.61  14.47  meV  Spearman  :  0.99984
\end{verbatim}
\clearpage

\flushleft{
\subsection{Gd-H}}
\subsubsection*{Searching}
\centering
\includegraphics[width=0.4\textwidth]{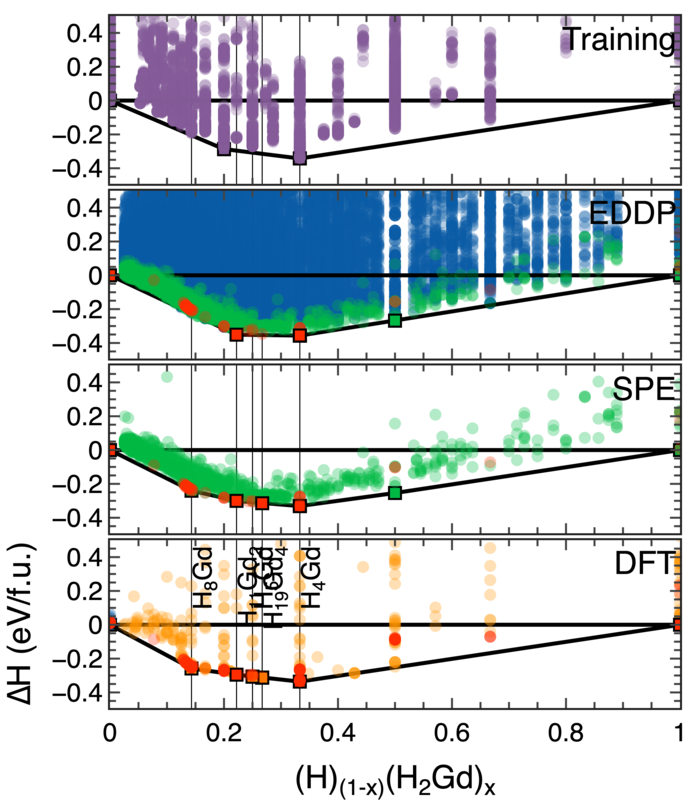}
\footnotesize


\flushleft{
\subsubsection*{\textsc{EDDP}}}
\centering
\includegraphics[width=0.3\textwidth]{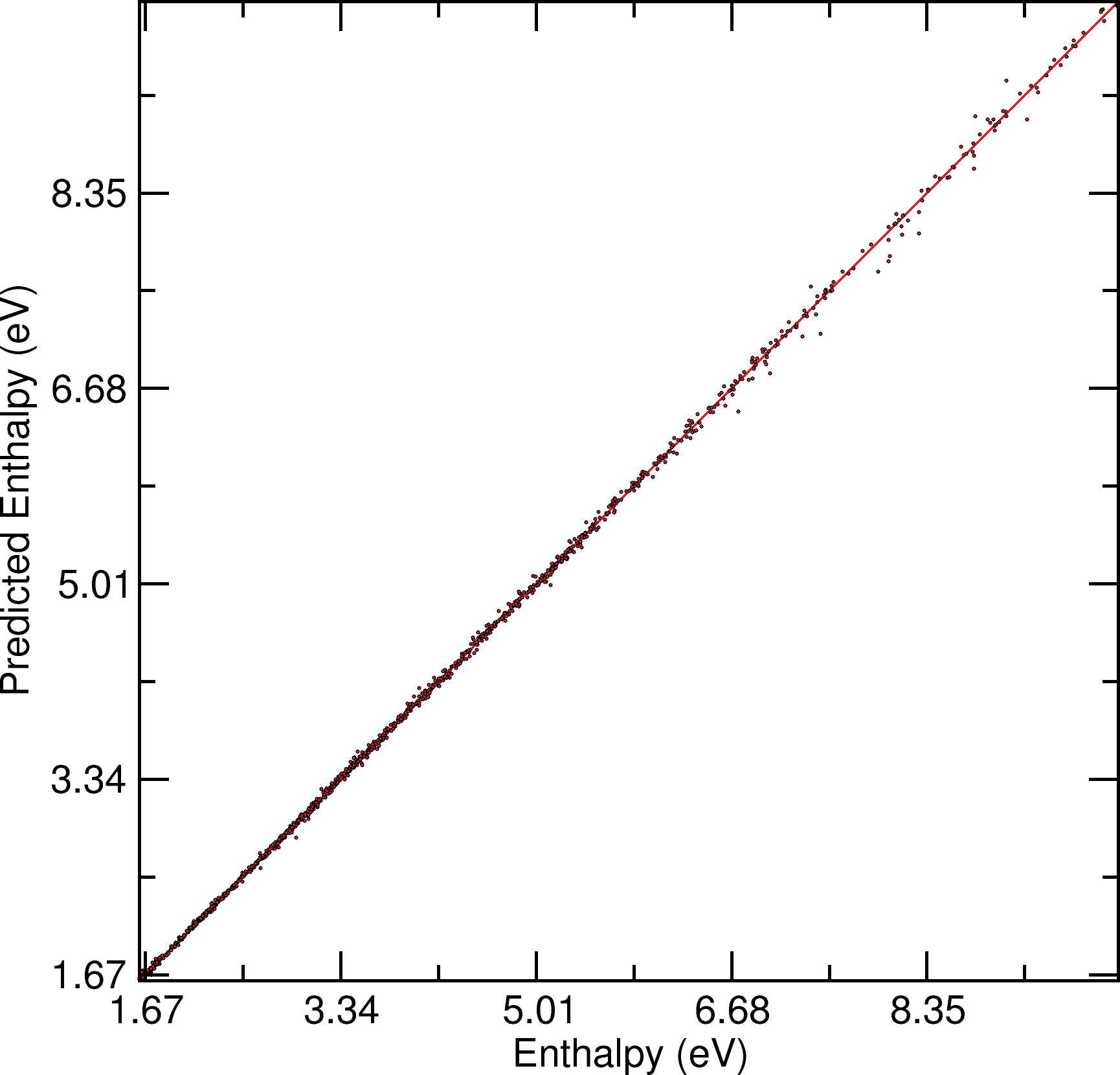}
\includegraphics[width=0.3\textwidth]{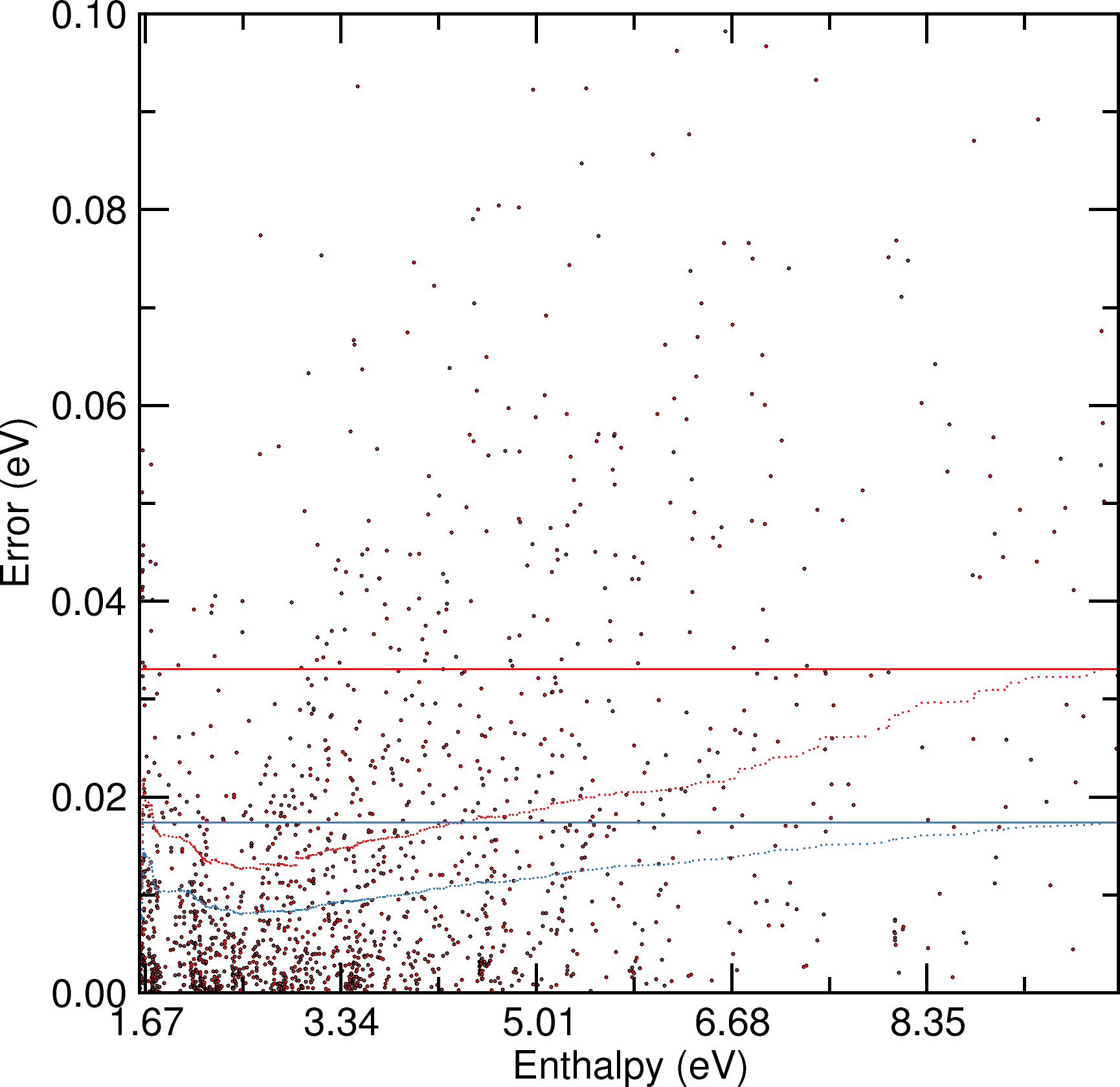}
\centering\begin{verbatim}
training    RMSE/MAE:  17.13  11.02  meV  Spearman  :  0.99987
validation  RMSE/MAE:  25.67  15.69  meV  Spearman  :  0.99983
testing     RMSE/MAE:  33.04  17.38  meV  Spearman  :  0.99982
\end{verbatim}
\clearpage

\flushleft{
\subsection{Ge-H}}
\subsubsection*{Searching}
\centering
\includegraphics[width=0.4\textwidth]{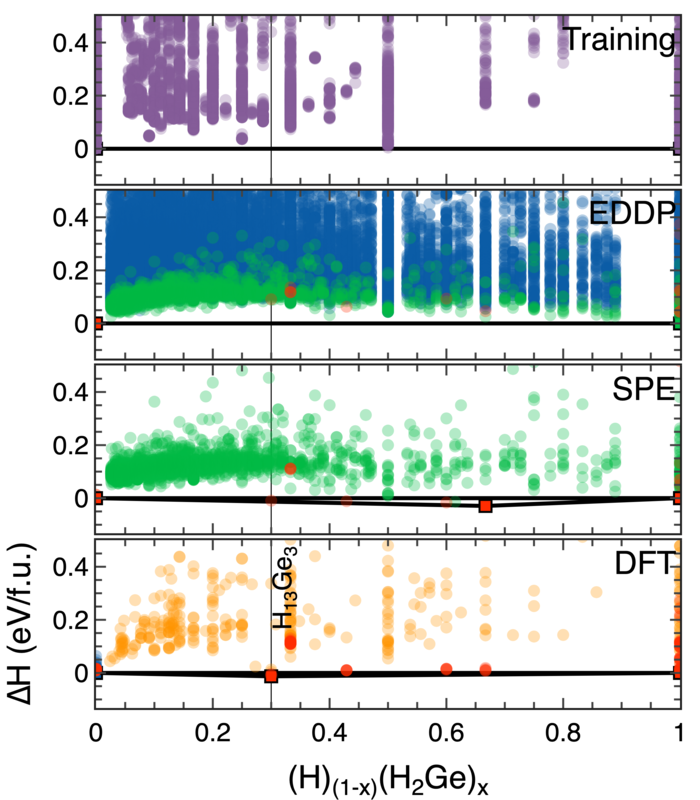}
\footnotesize


\flushleft{
\subsubsection*{\textsc{EDDP}}}
\centering
\includegraphics[width=0.3\textwidth]{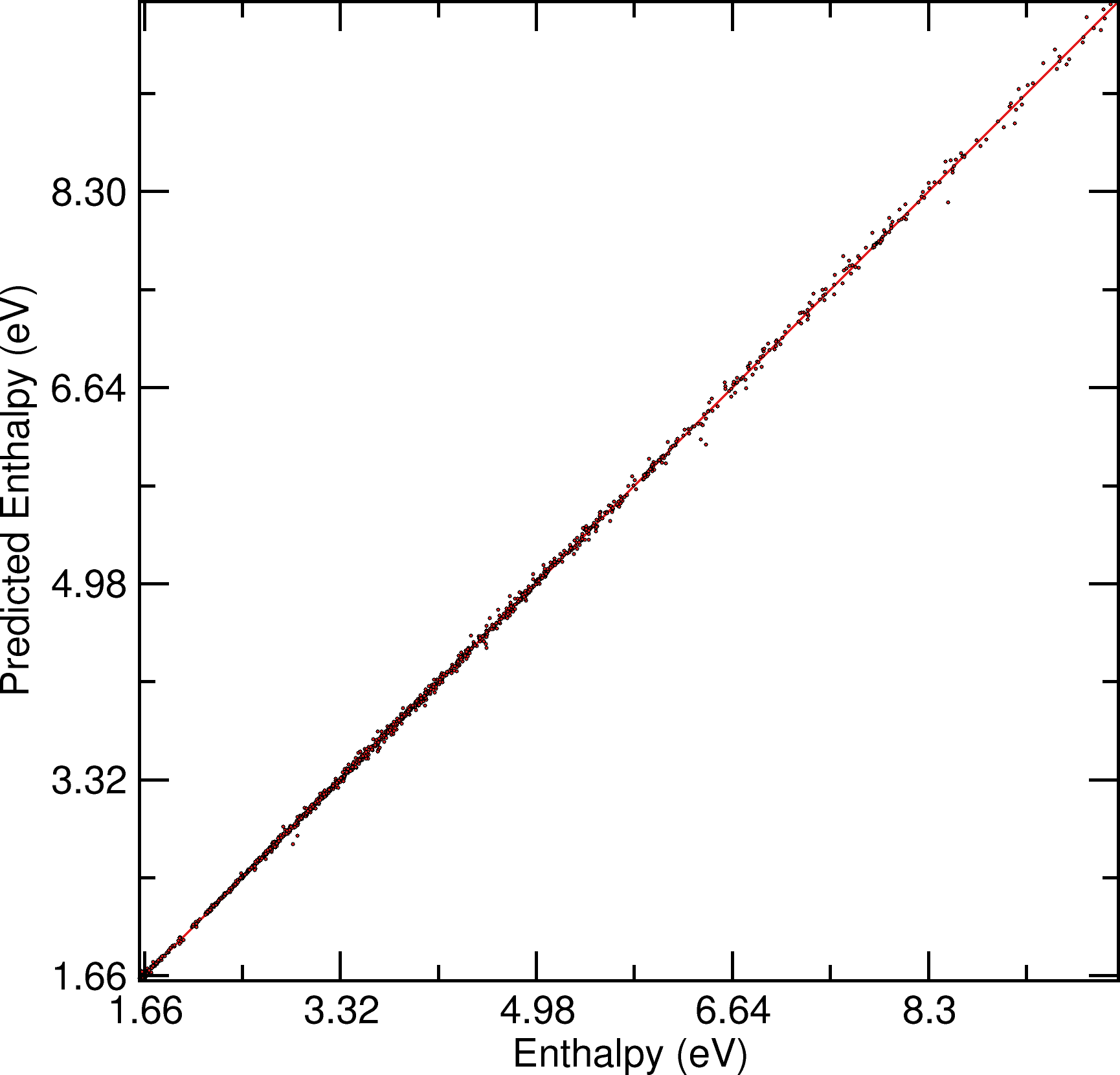}
\includegraphics[width=0.3\textwidth]{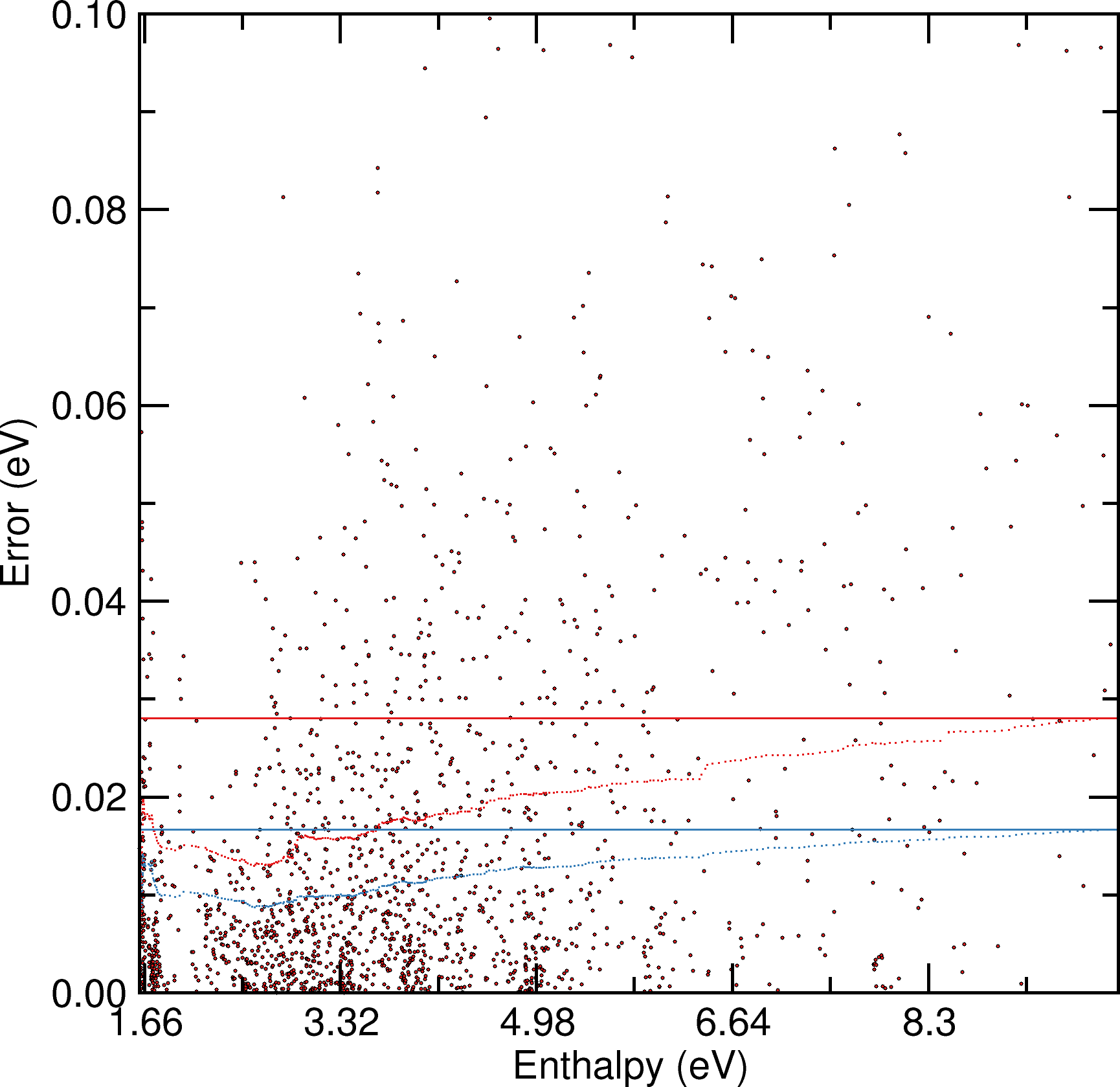}
\centering\begin{verbatim}
training    RMSE/MAE:  17.34  10.95  meV  Spearman  :  0.99987
validation  RMSE/MAE:  28.09  16.96  meV  Spearman  :  0.99981
testing     RMSE/MAE:  28.07  16.68  meV  Spearman  :  0.99981
\end{verbatim}
\clearpage

\flushleft{
\subsection{He-H}}
\subsubsection*{Searching}
\centering
\includegraphics[width=0.4\textwidth]{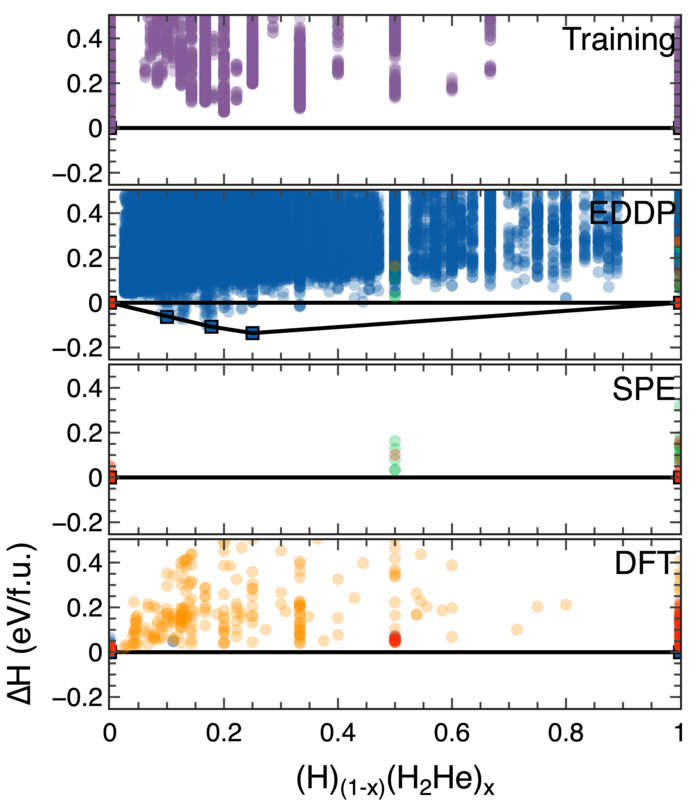}
\footnotesize


\flushleft{
\subsubsection*{\textsc{EDDP}}}
\centering
\includegraphics[width=0.3\textwidth]{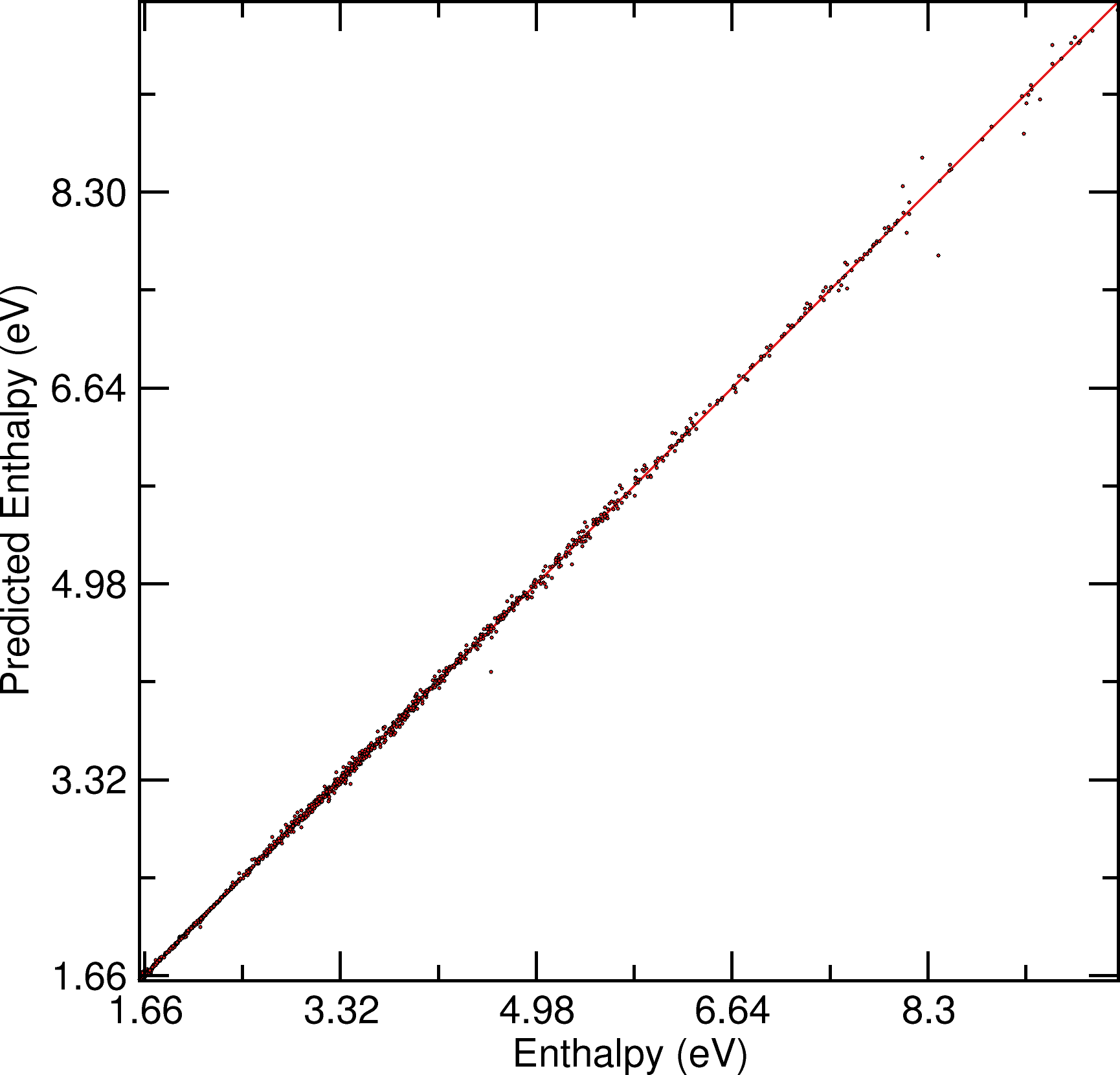}
\includegraphics[width=0.3\textwidth]{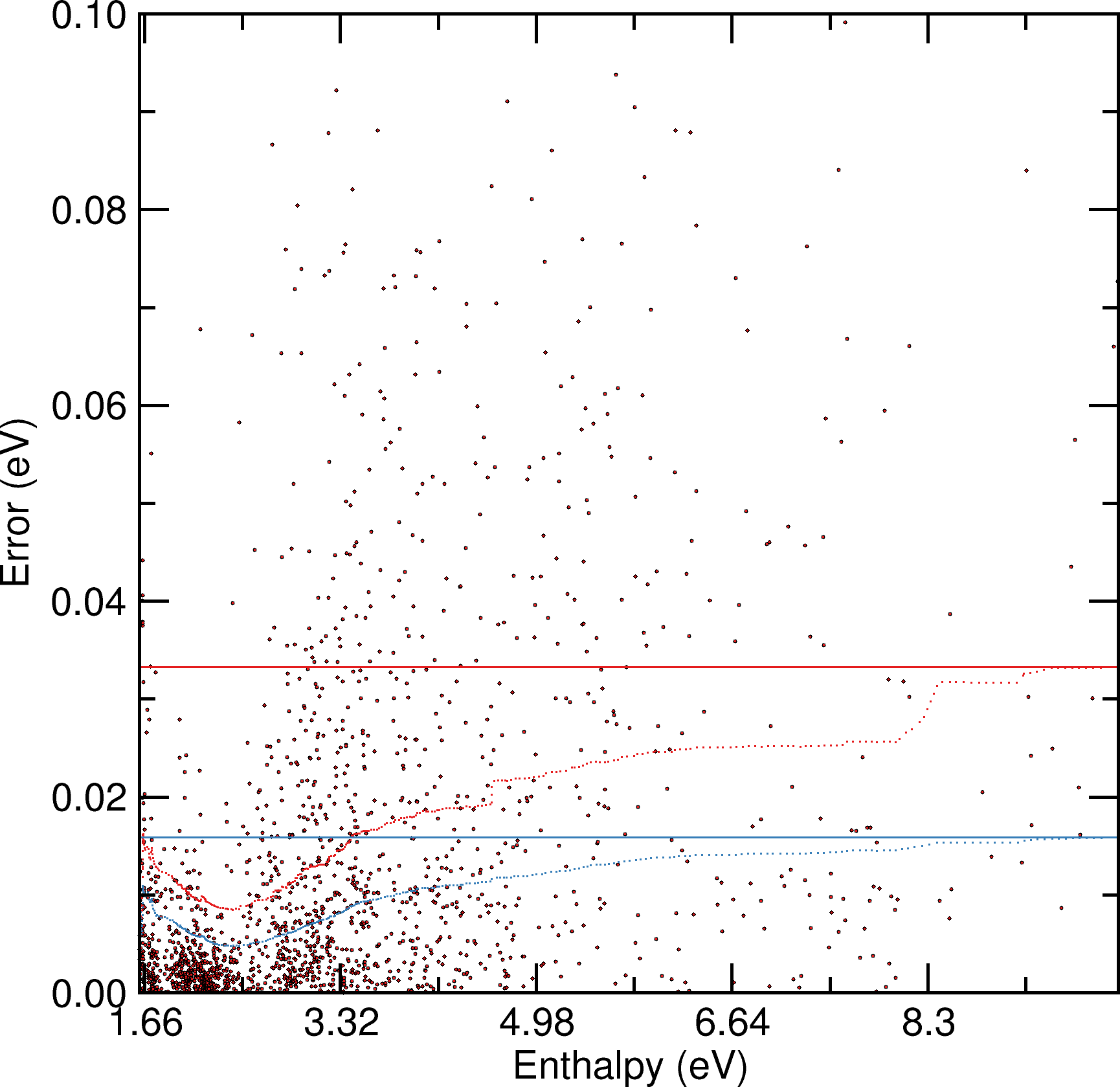}
\centering\begin{verbatim}
training    RMSE/MAE:  16.66  9.46   meV  Spearman  :  0.99988
validation  RMSE/MAE:  24.83  14.58  meV  Spearman  :  0.99979
testing     RMSE/MAE:  33.27  15.88  meV  Spearman  :  0.99982
\end{verbatim}
\clearpage

\flushleft{
\subsection{Hf-H}}
\subsubsection*{Searching}
\centering
\includegraphics[width=0.4\textwidth]{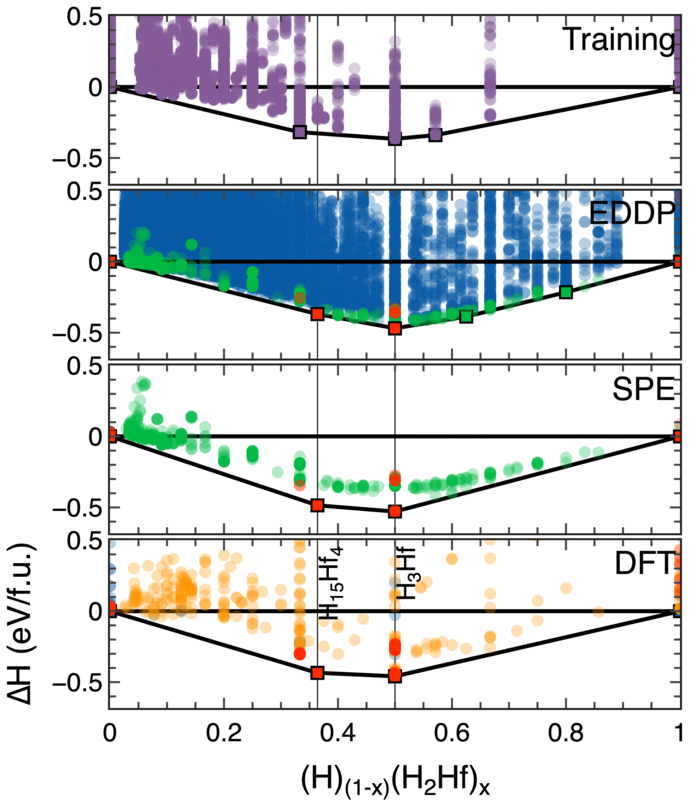}
\footnotesize


\flushleft{
\subsubsection*{\textsc{EDDP}}}
\centering
\includegraphics[width=0.3\textwidth]{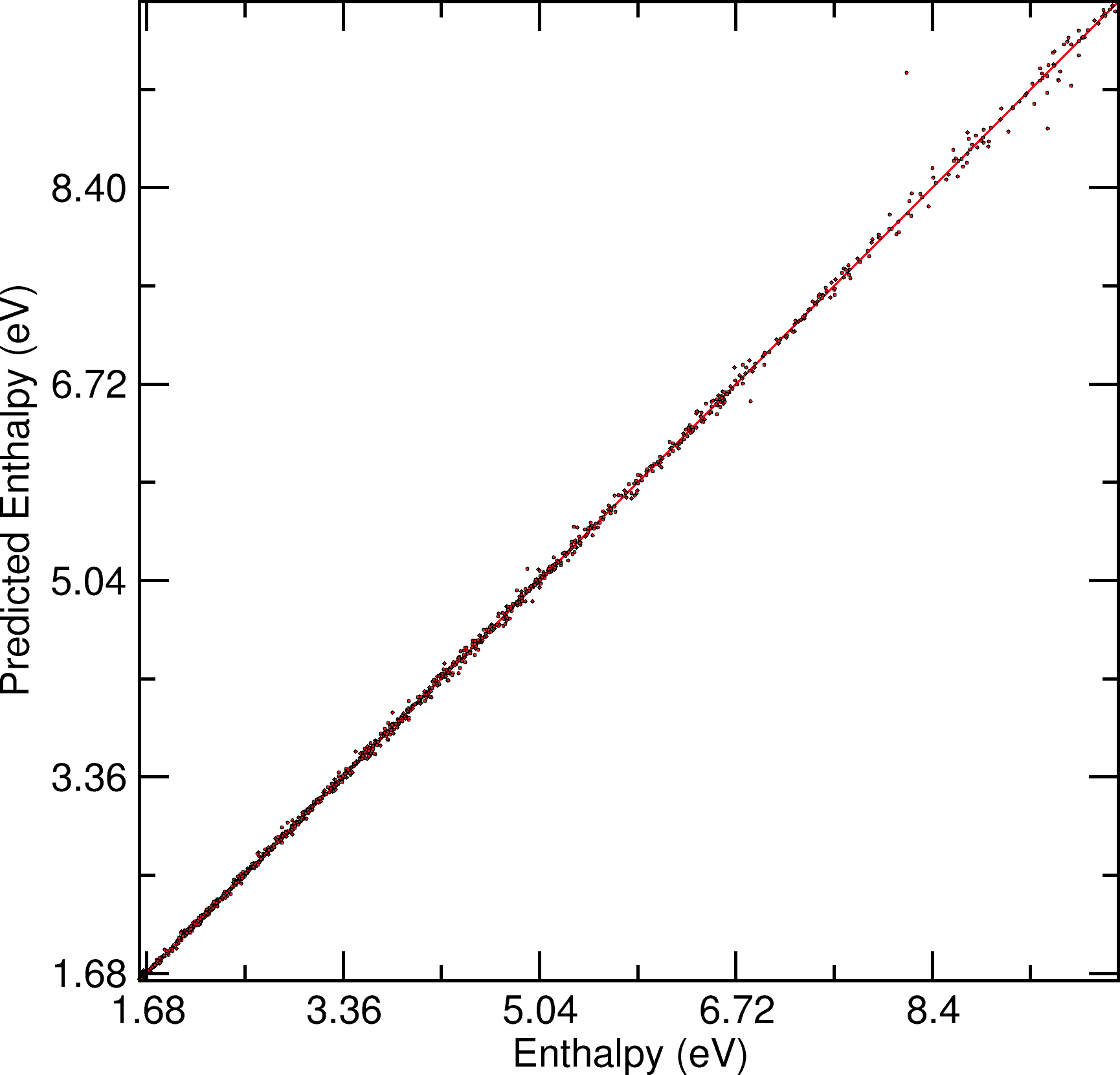}
\includegraphics[width=0.3\textwidth]{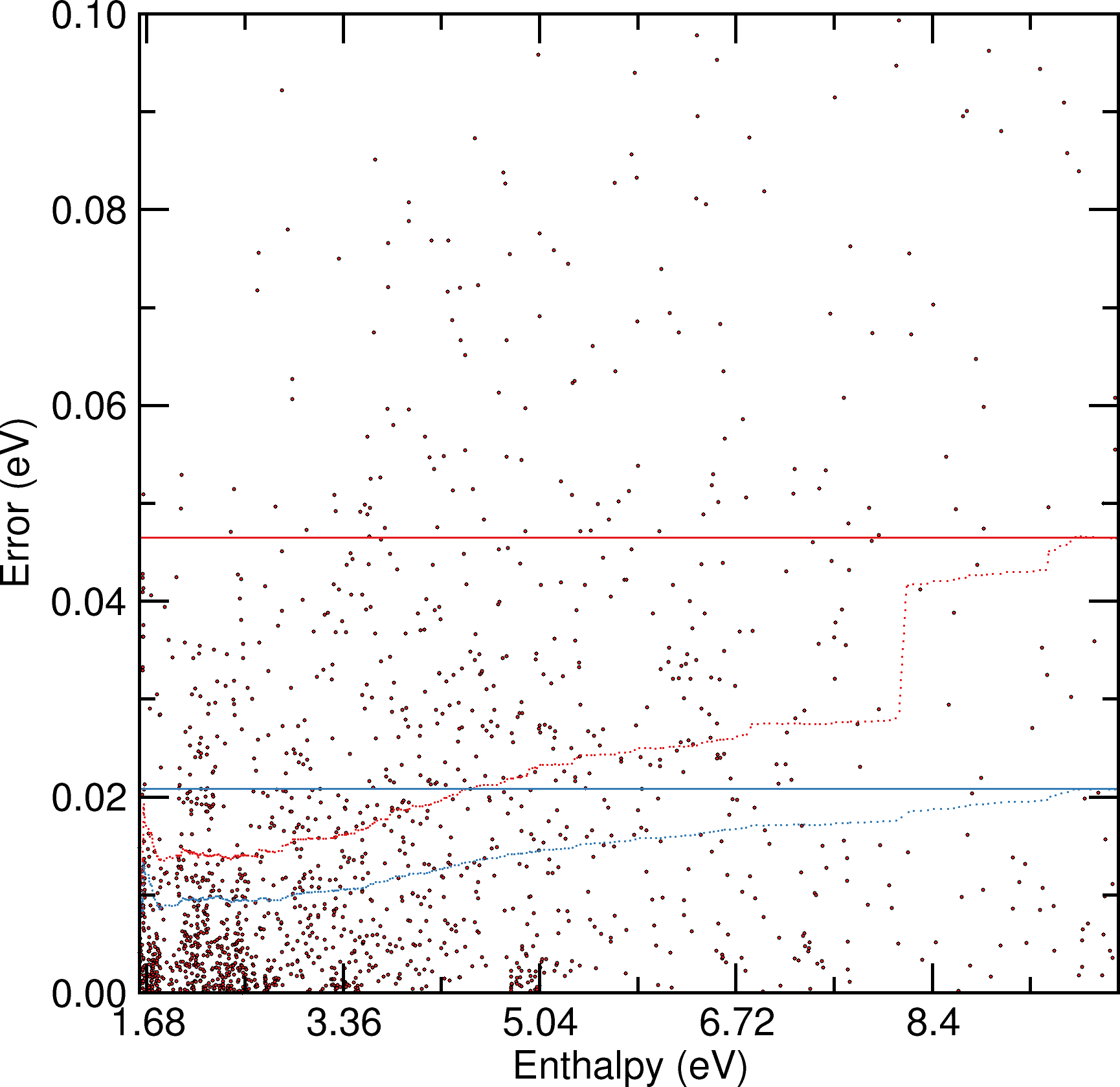}
\centering\begin{verbatim}
training    RMSE/MAE:  19.51  12.44  meV  Spearman  :  0.99986
validation  RMSE/MAE:  29.22  18.85  meV  Spearman  :  0.99975
testing     RMSE/MAE:  46.49  20.86  meV  Spearman  :  0.99982
\end{verbatim}
\clearpage

\flushleft{
\subsection{Hg-H}}
\subsubsection*{Searching}
\centering
\includegraphics[width=0.4\textwidth]{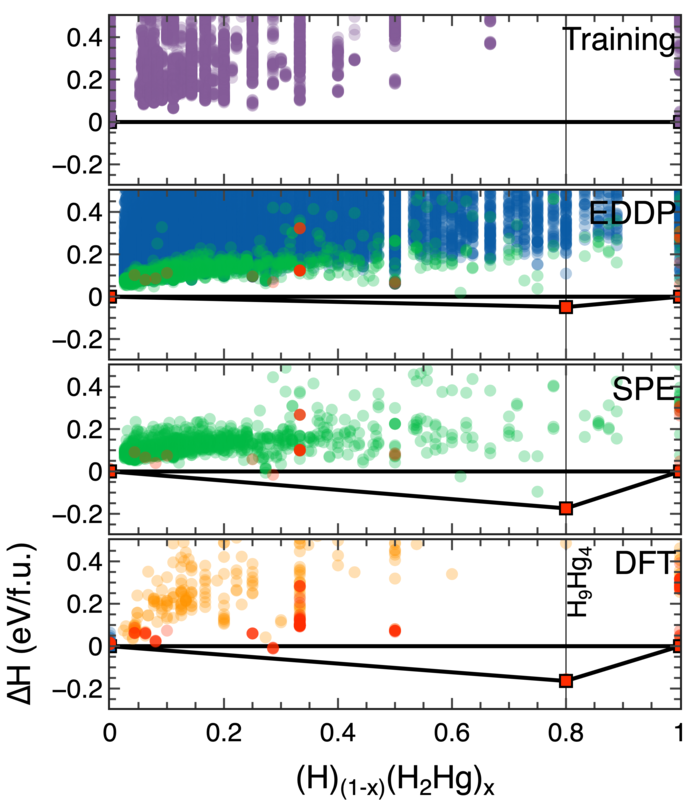}
\footnotesize


\flushleft{
\subsubsection*{\textsc{EDDP}}}
\centering
\includegraphics[width=0.3\textwidth]{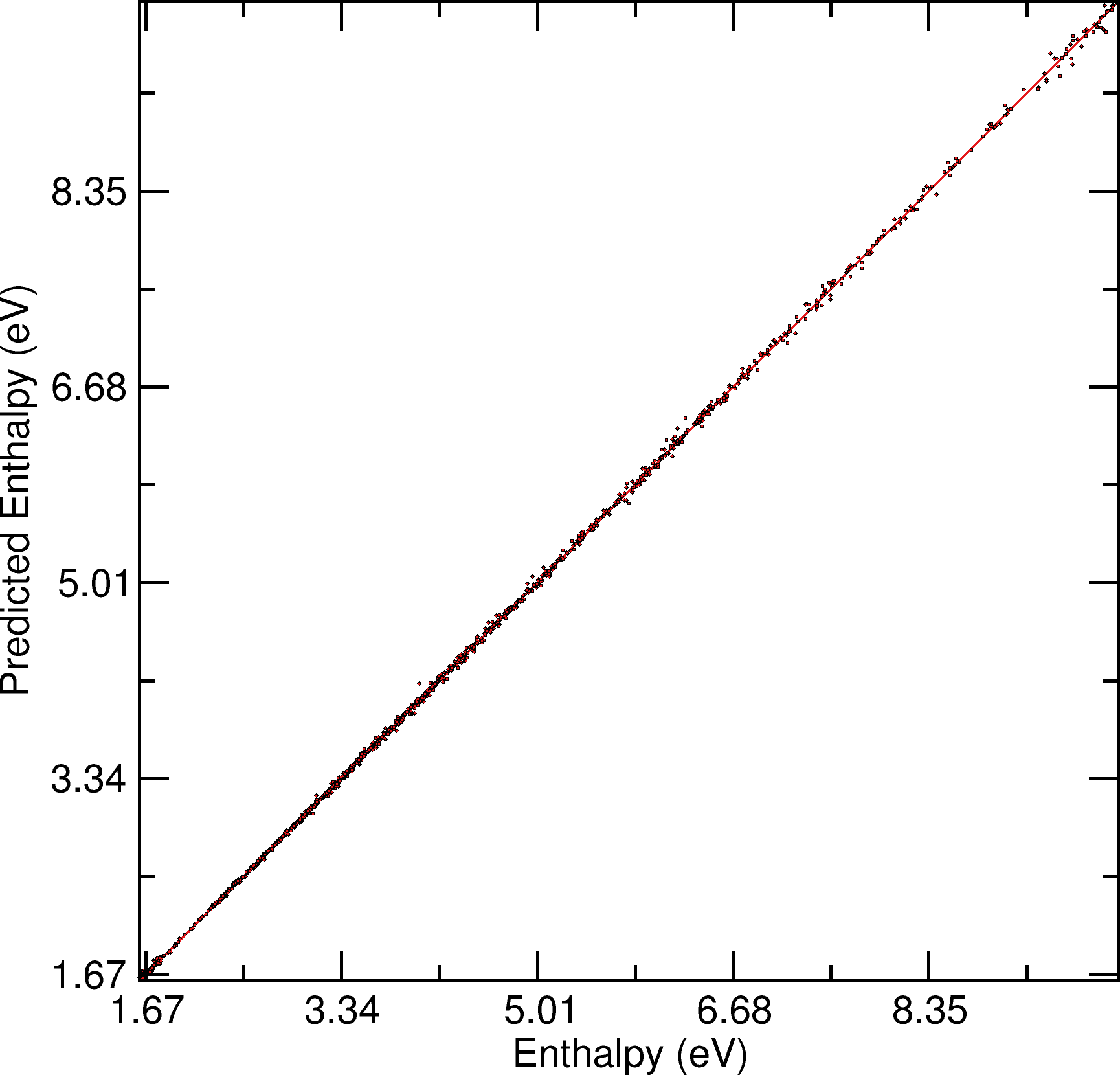}
\includegraphics[width=0.3\textwidth]{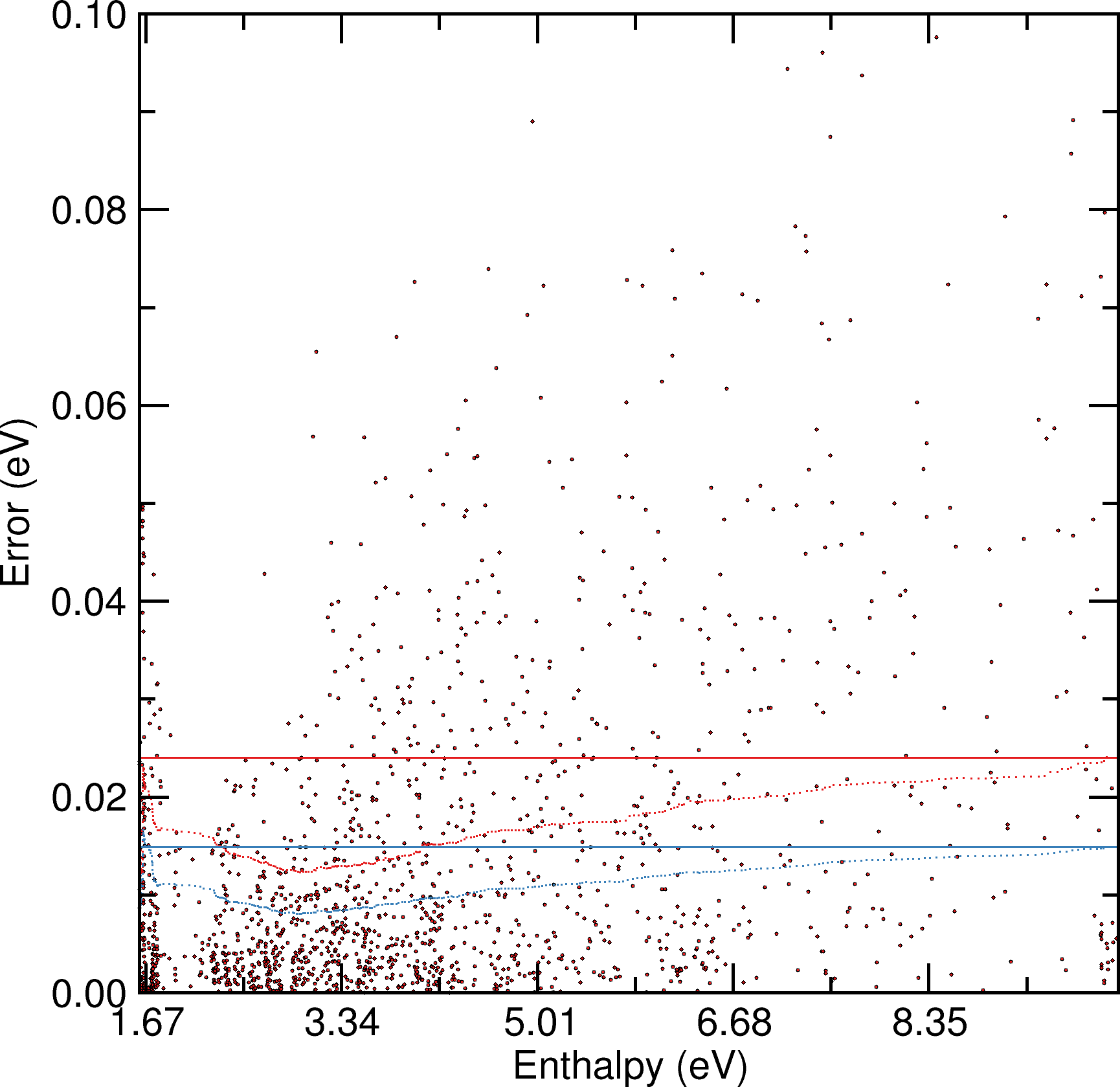}
\centering\begin{verbatim}
training    RMSE/MAE:  15.89  10.16  meV  Spearman  :  0.99989
validation  RMSE/MAE:  22.90  14.97  meV  Spearman  :  0.99984
testing     RMSE/MAE:  24.01  14.85  meV  Spearman  :  0.99982
\end{verbatim}
\clearpage

\flushleft{
\subsection{Ho-H}}
\subsubsection*{Searching}
\centering
\includegraphics[width=0.4\textwidth]{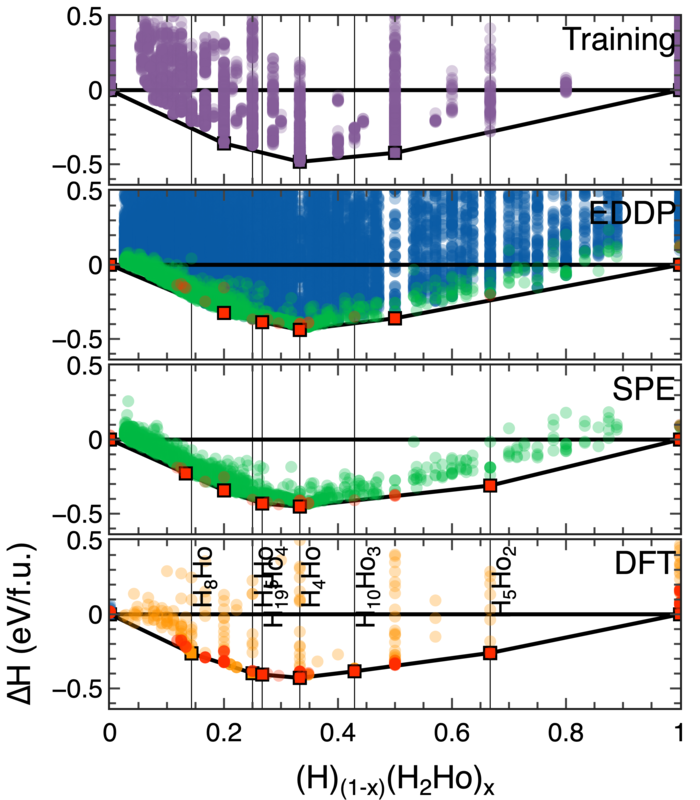}
\footnotesize


\flushleft{
\subsubsection*{\textsc{EDDP}}}
\centering
\includegraphics[width=0.3\textwidth]{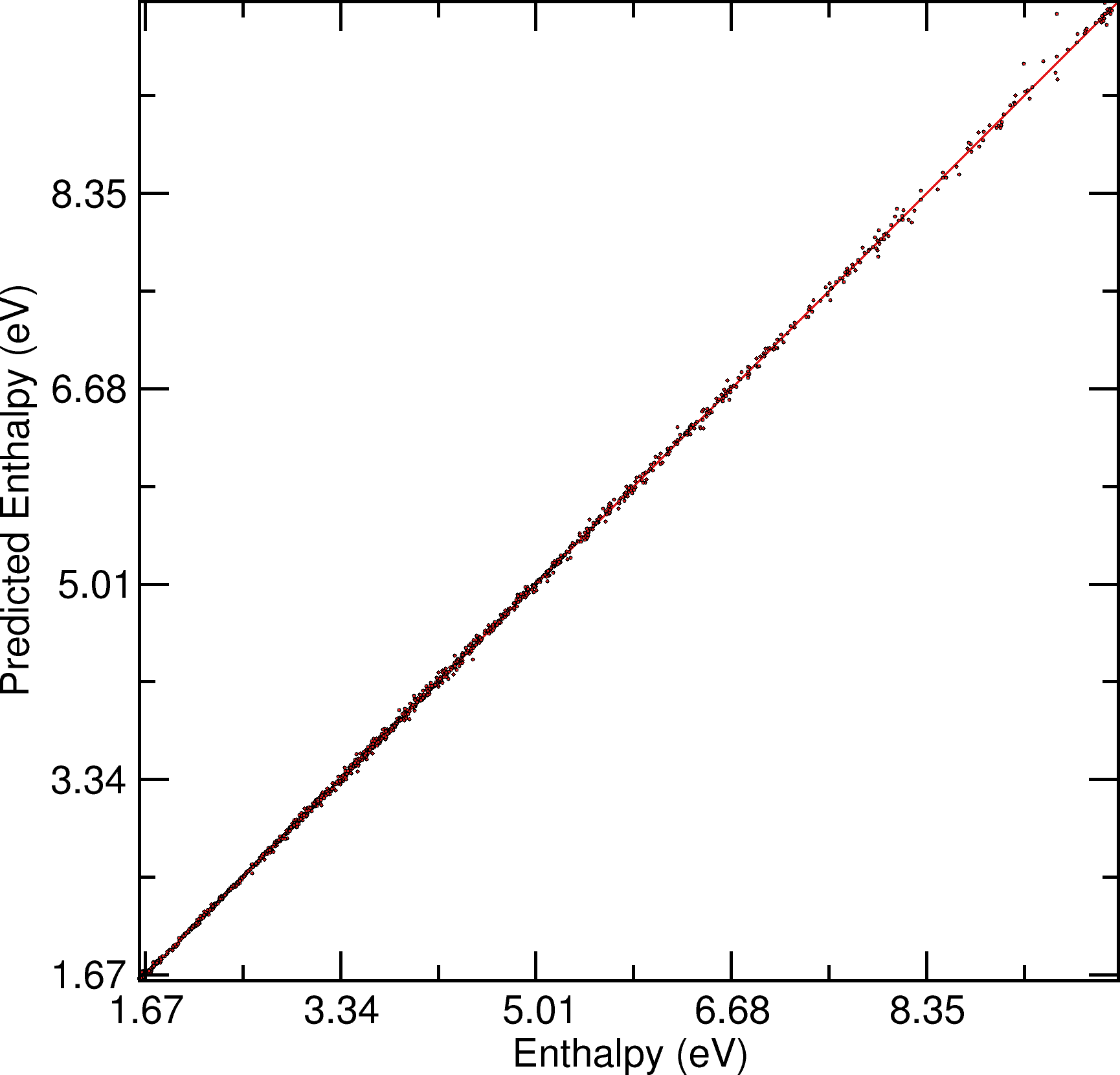}
\includegraphics[width=0.3\textwidth]{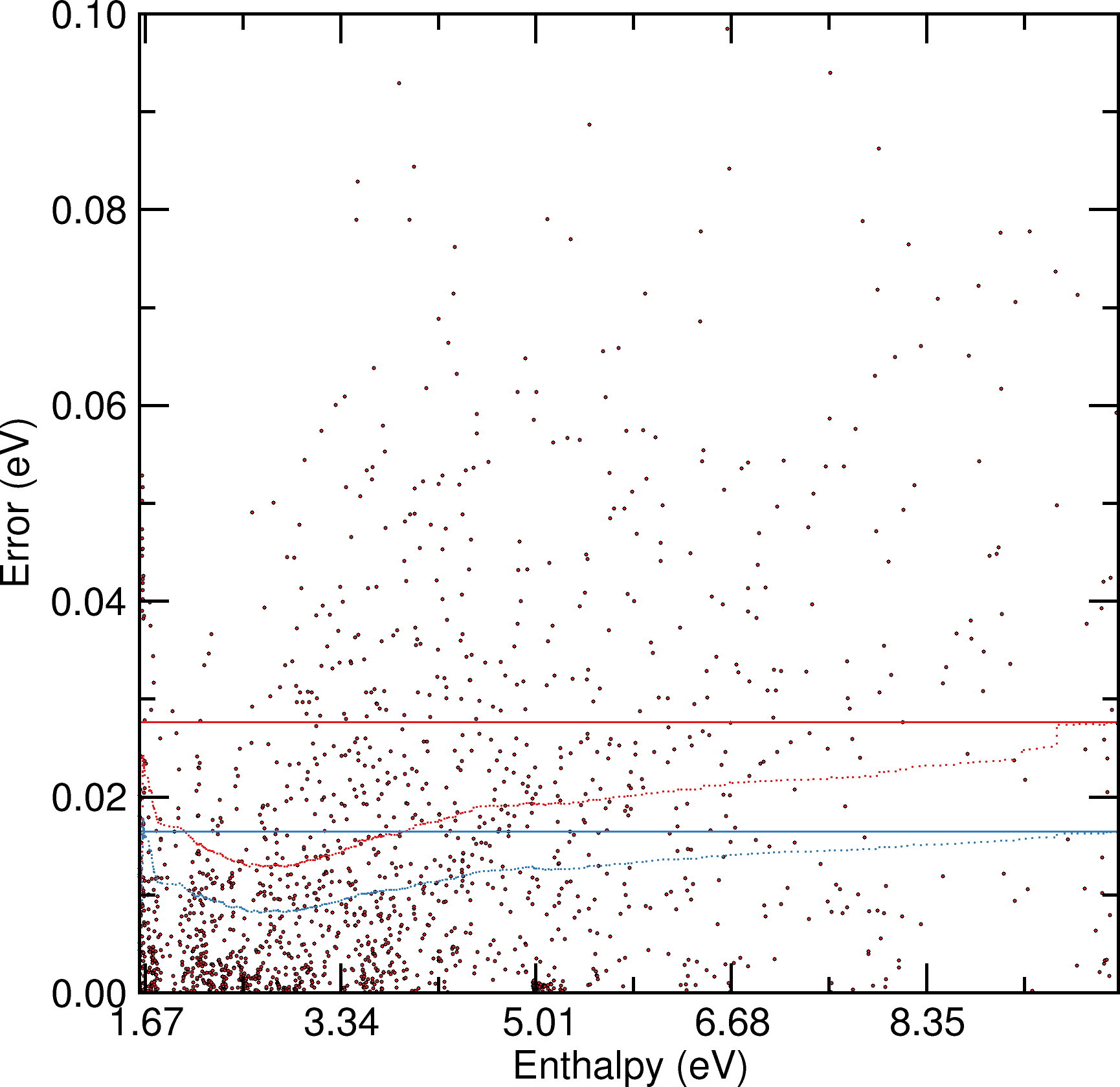}
\centering\begin{verbatim}
training    RMSE/MAE:  16.07  10.23  meV  Spearman  :  0.99989
validation  RMSE/MAE:  23.28  14.79  meV  Spearman  :  0.99986
testing     RMSE/MAE:  27.63  16.44  meV  Spearman  :  0.99984
\end{verbatim}
\clearpage

\flushleft{
\subsection{I-H}}
\subsubsection*{Searching}
\centering
\includegraphics[width=0.4\textwidth]{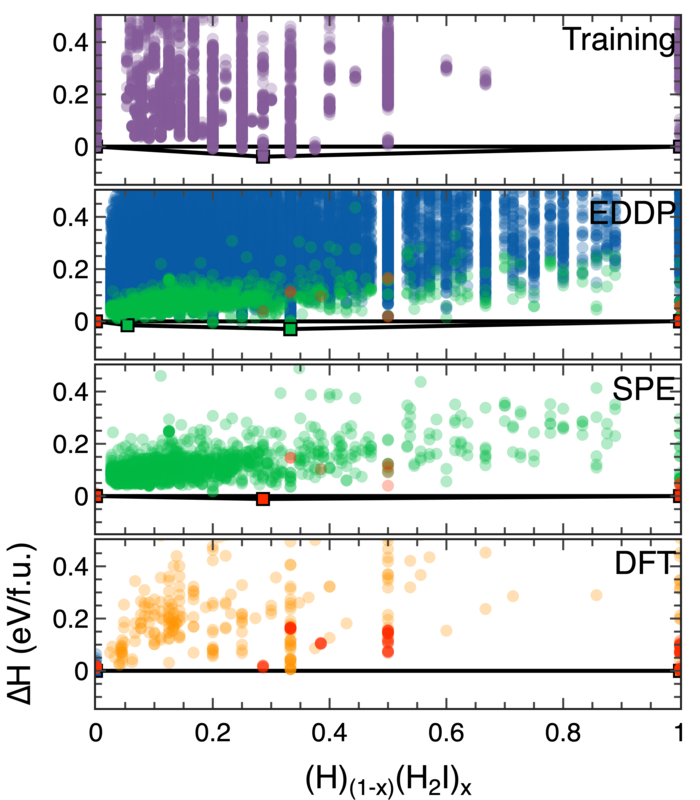}
\footnotesize


\flushleft{
\subsubsection*{\textsc{EDDP}}}
\centering
\includegraphics[width=0.3\textwidth]{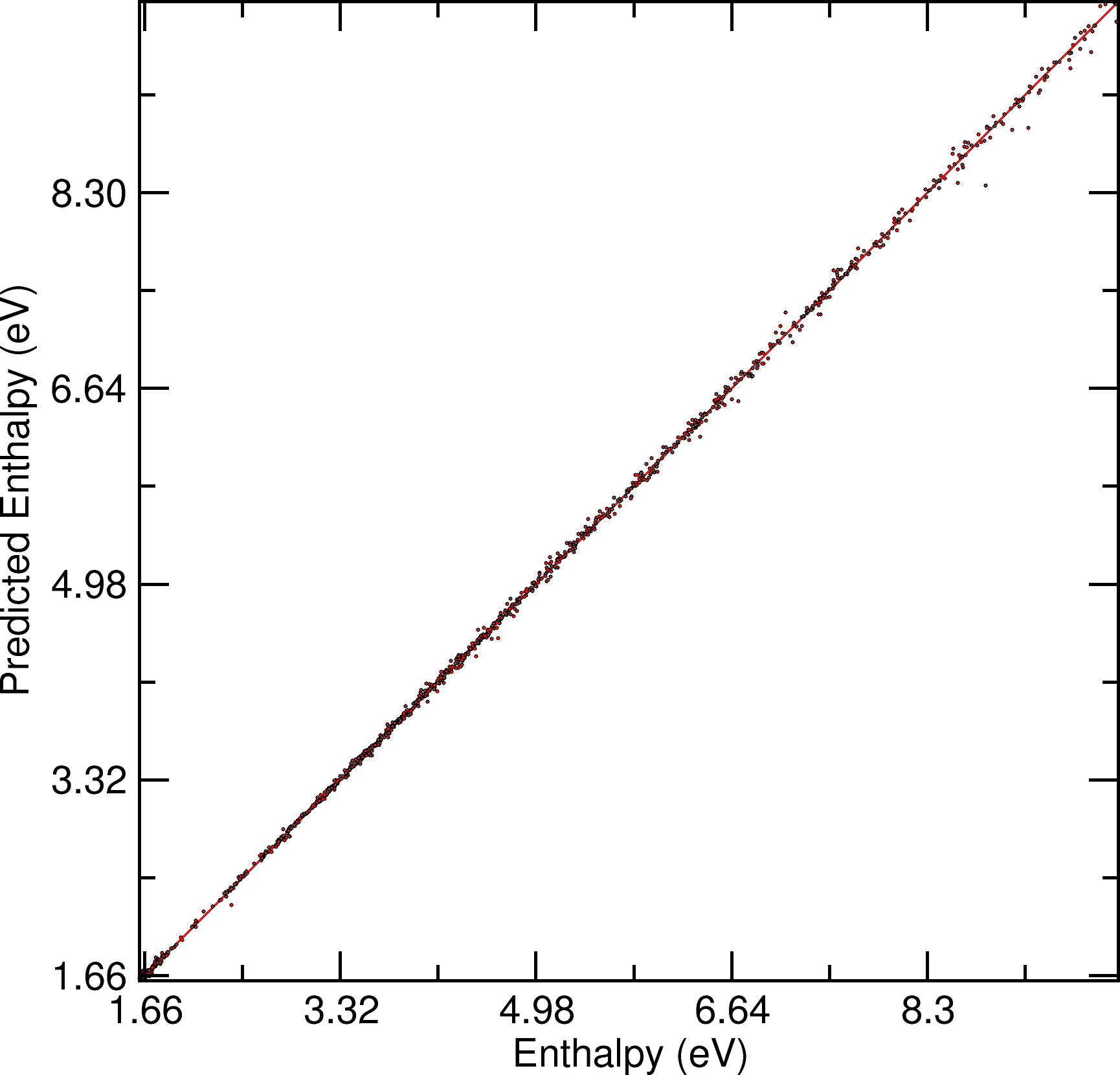}
\includegraphics[width=0.3\textwidth]{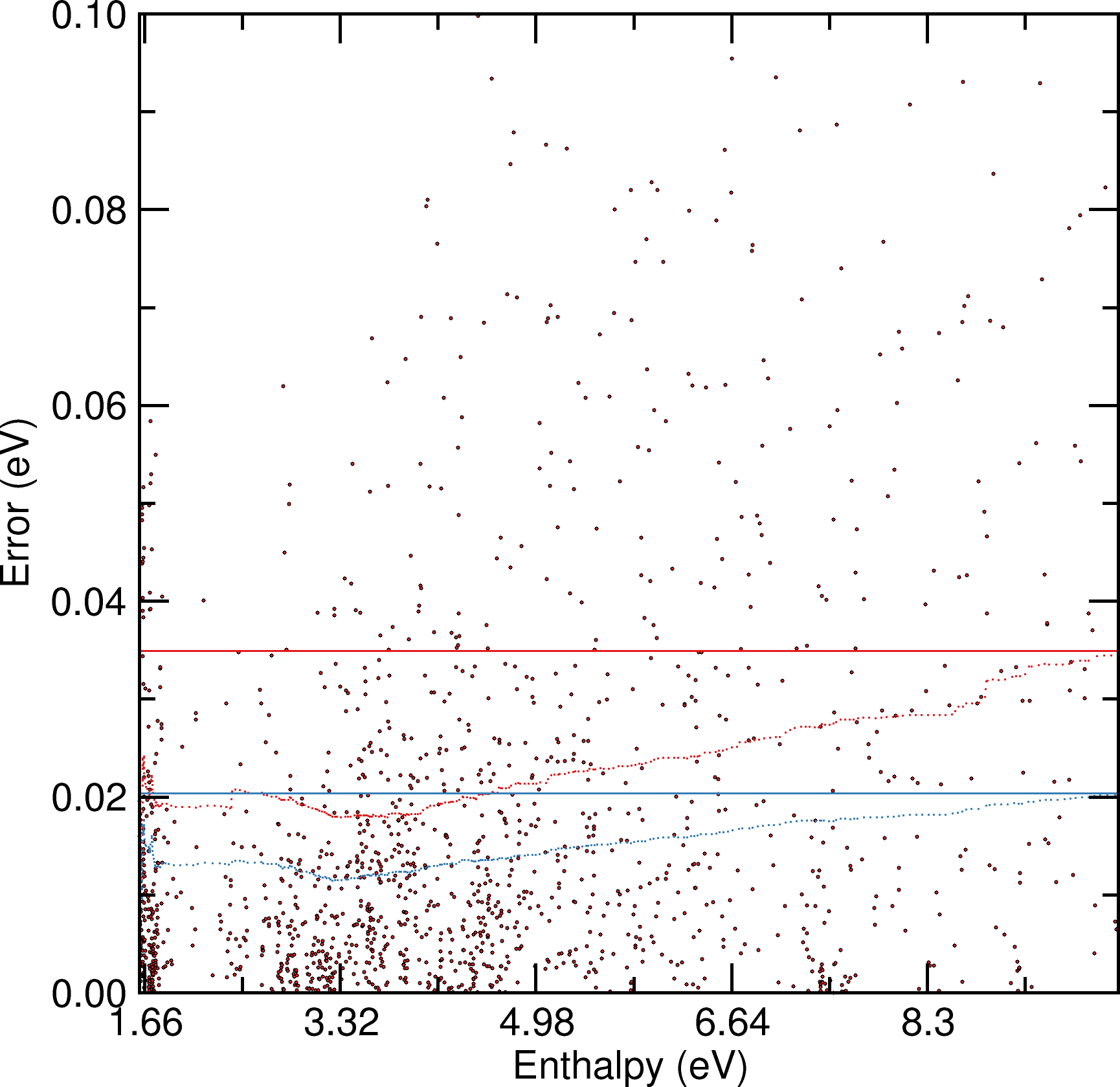}
\centering\begin{verbatim}
training    RMSE/MAE:  161.26  17.89  meV  Spearman  :  0.99971
validation  RMSE/MAE:  30.81   19.17  meV  Spearman  :  0.99977
testing     RMSE/MAE:  34.90   20.40  meV  Spearman  :  0.99969
\end{verbatim}
\clearpage

\flushleft{
\subsection{Ir-H}}
\subsubsection*{Searching}
\centering
\includegraphics[width=0.4\textwidth]{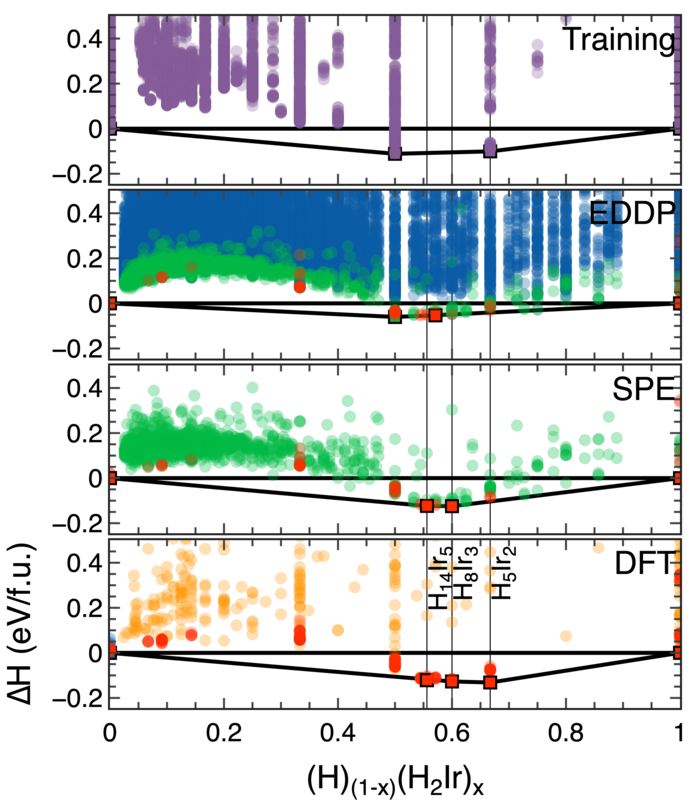}
\footnotesize


\flushleft{
\subsubsection*{\textsc{EDDP}}}
\centering
\includegraphics[width=0.3\textwidth]{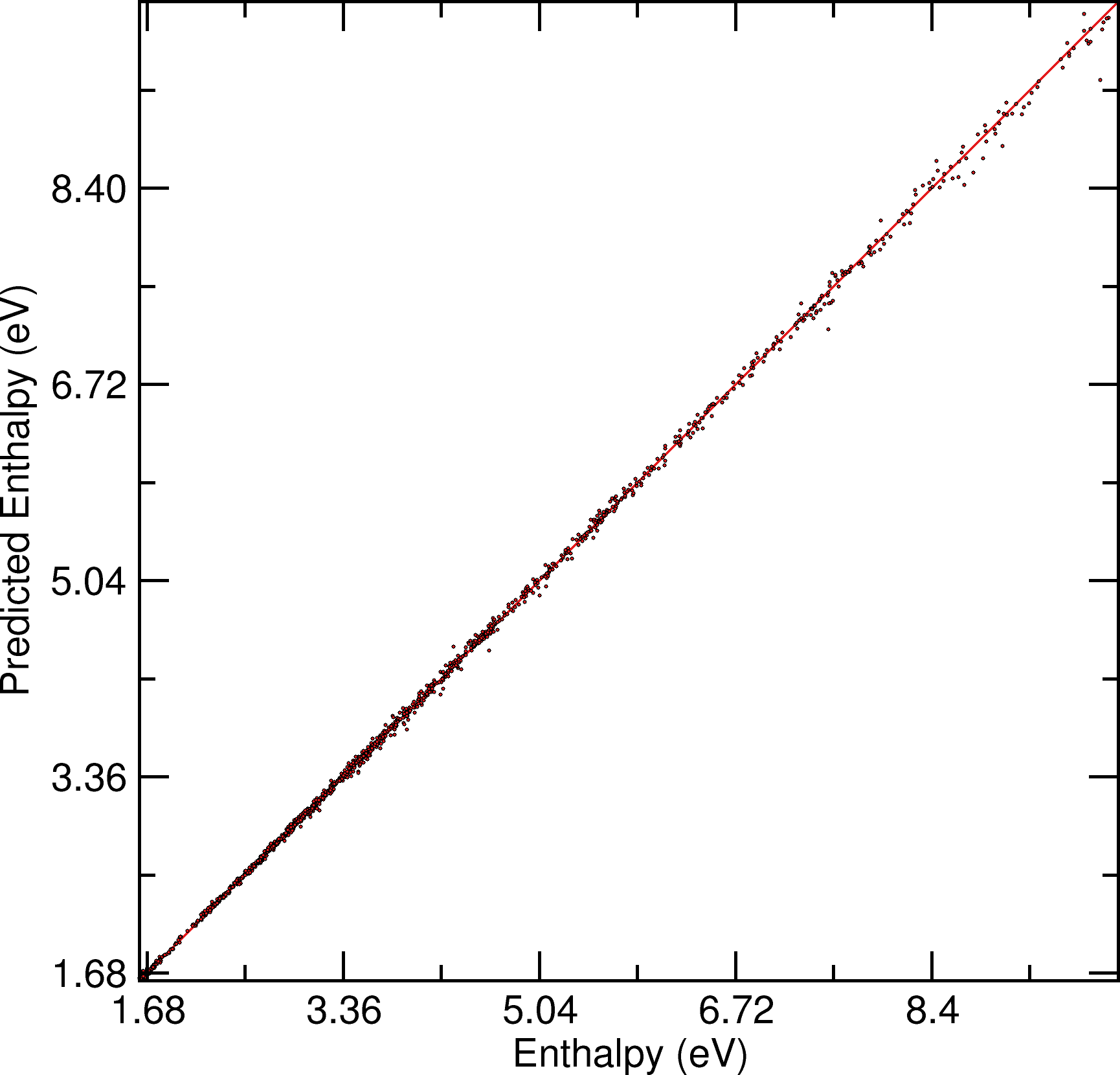}
\includegraphics[width=0.3\textwidth]{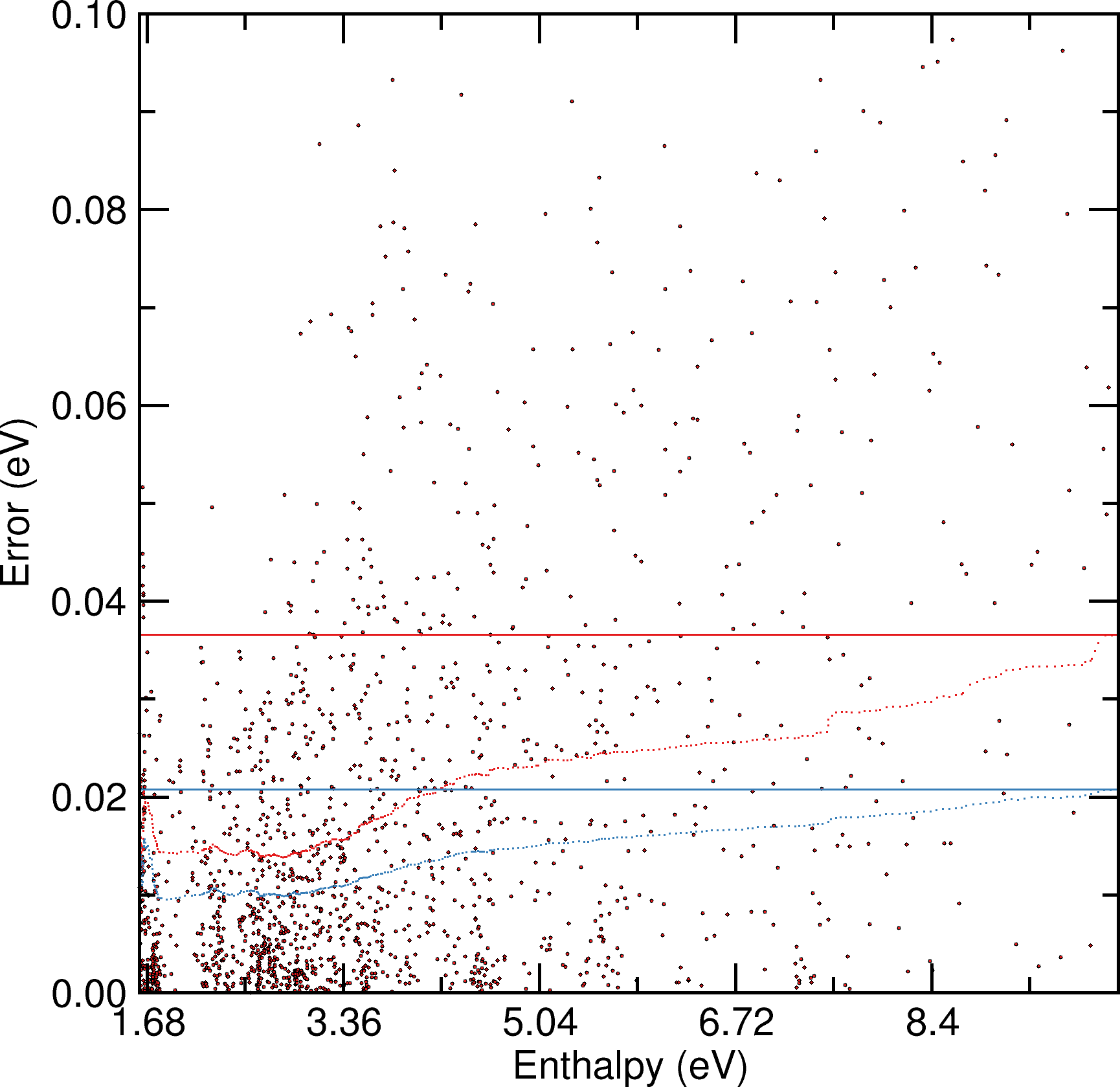}
\centering\begin{verbatim}
training    RMSE/MAE:  19.94  12.62  meV  Spearman  :  0.99984
validation  RMSE/MAE:  29.17  18.44  meV  Spearman  :  0.99979
testing     RMSE/MAE:  36.60  20.77  meV  Spearman  :  0.99978
\end{verbatim}
\clearpage

\flushleft{
\subsection{K-H}}
\subsubsection*{Searching}
\centering
\includegraphics[width=0.4\textwidth]{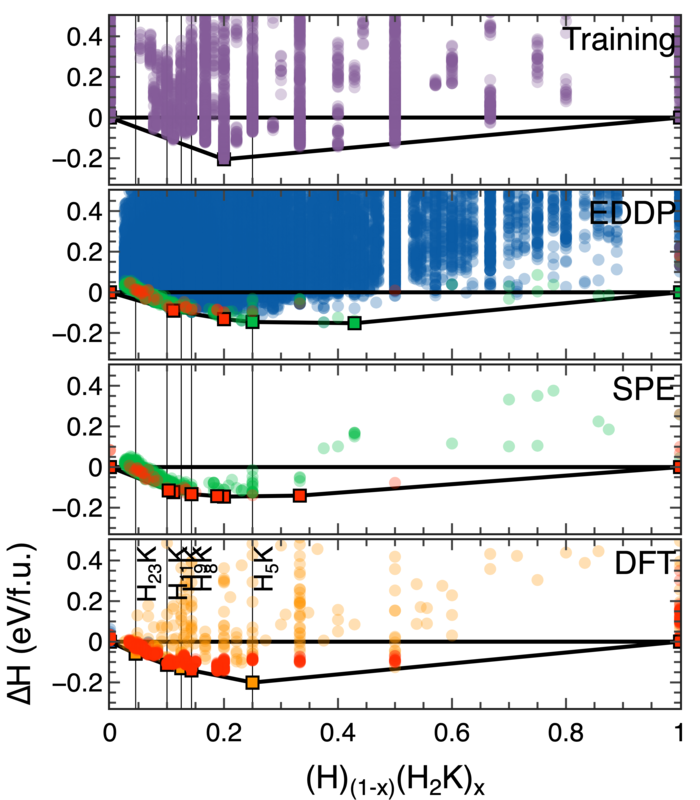}
\footnotesize


\flushleft{
\subsubsection*{\textsc{EDDP}}}
\centering
\includegraphics[width=0.3\textwidth]{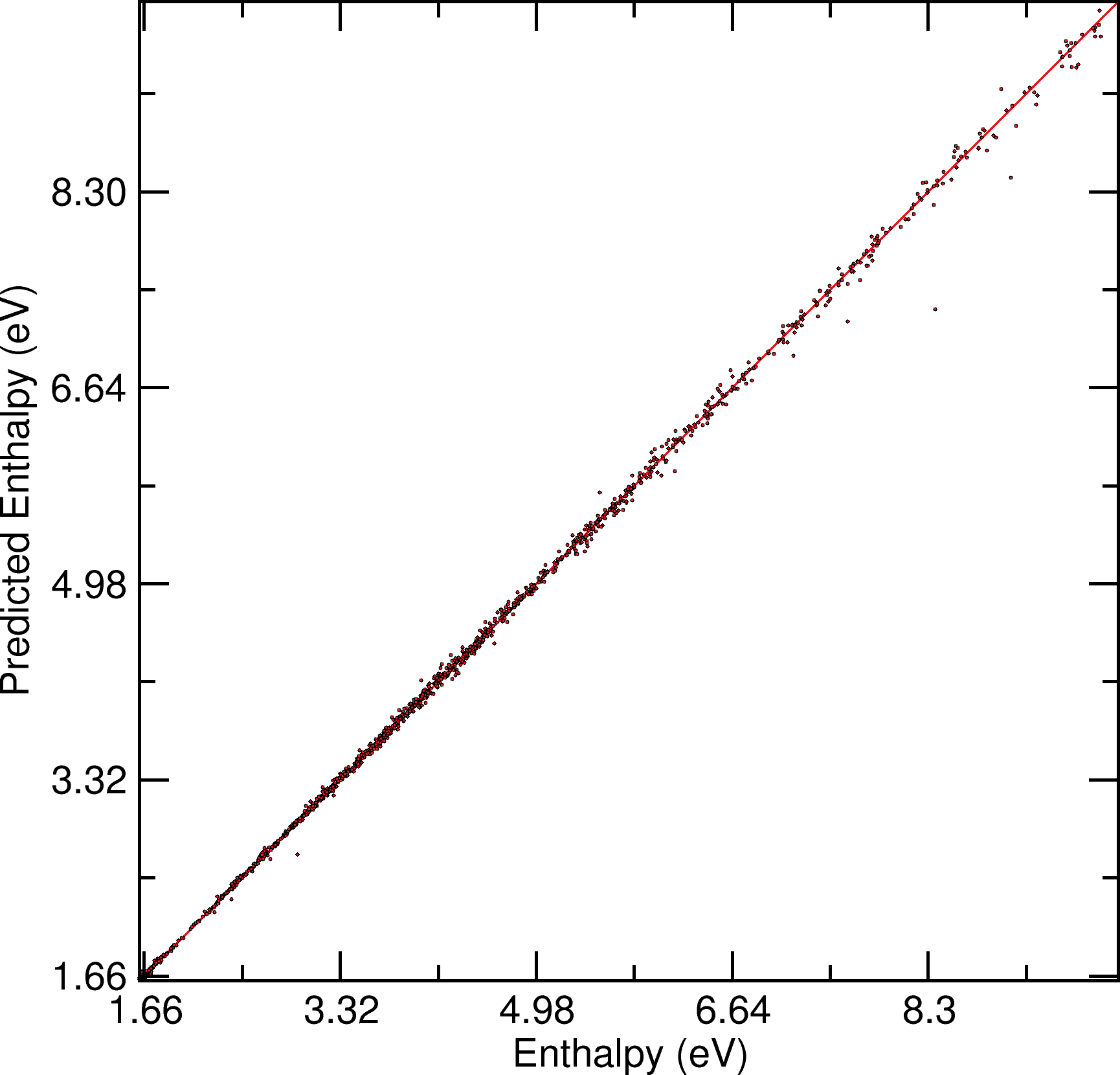}
\includegraphics[width=0.3\textwidth]{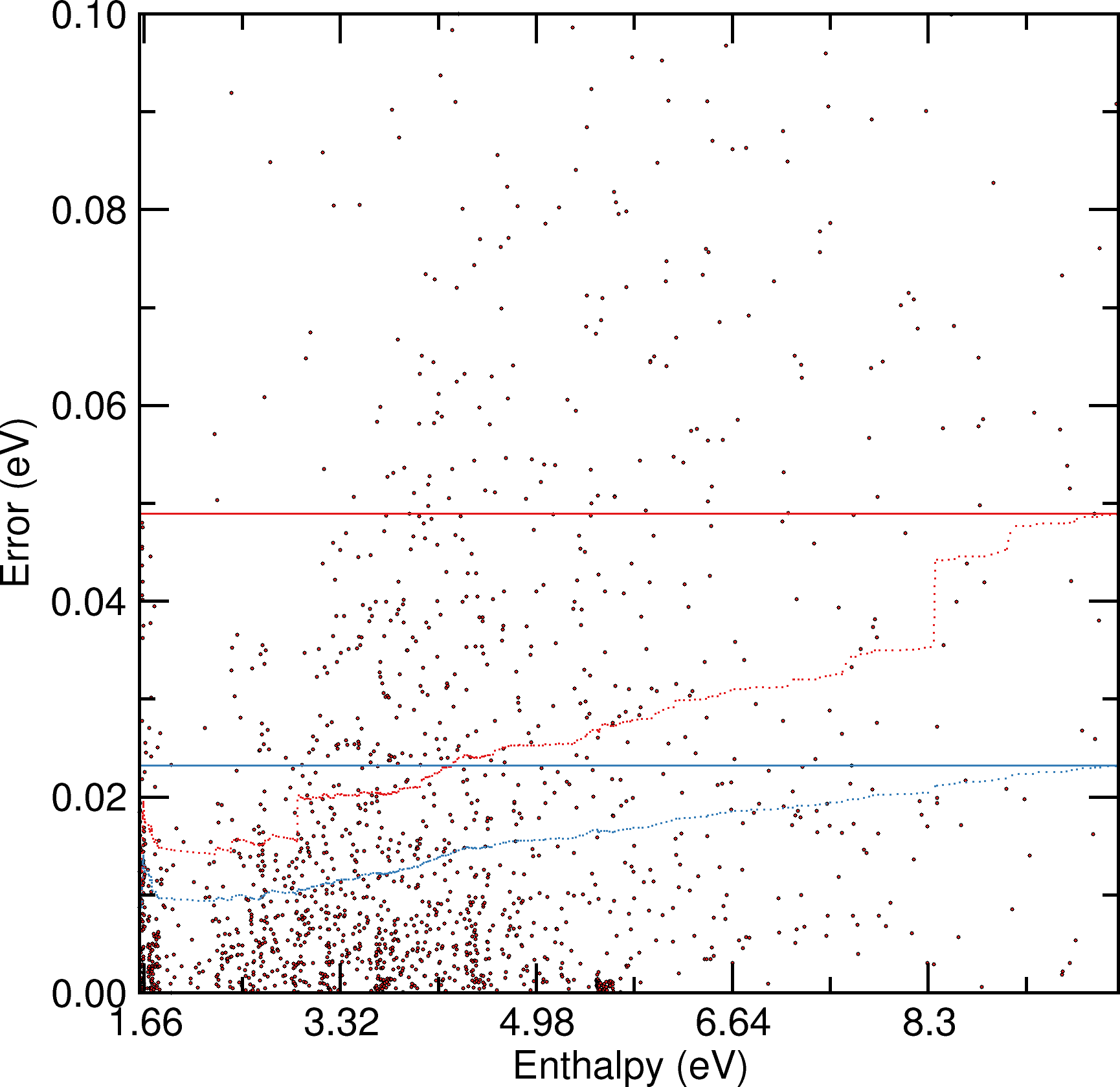}
\centering\begin{verbatim}
training    RMSE/MAE:  25.21  15.12  meV  Spearman  :  0.99984
validation  RMSE/MAE:  35.78  21.91  meV  Spearman  :  0.99976
testing     RMSE/MAE:  48.93  23.23  meV  Spearman  :  0.99972
\end{verbatim}
\clearpage

\flushleft{
\subsection{Kr-H}}
\subsubsection*{Searching}
\centering
\includegraphics[width=0.4\textwidth]{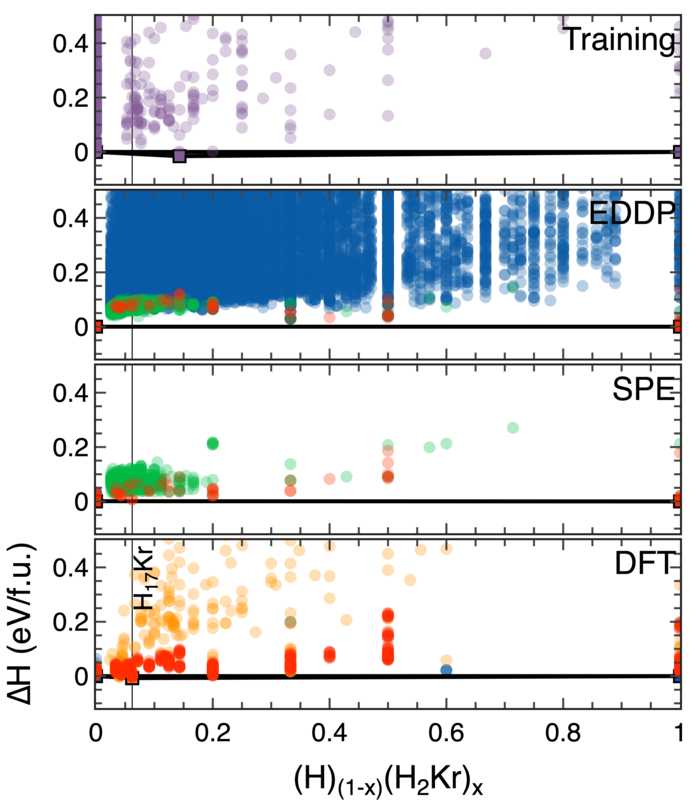}
\footnotesize


\flushleft{
\subsubsection*{\textsc{EDDP}}}
\centering
\includegraphics[width=0.3\textwidth]{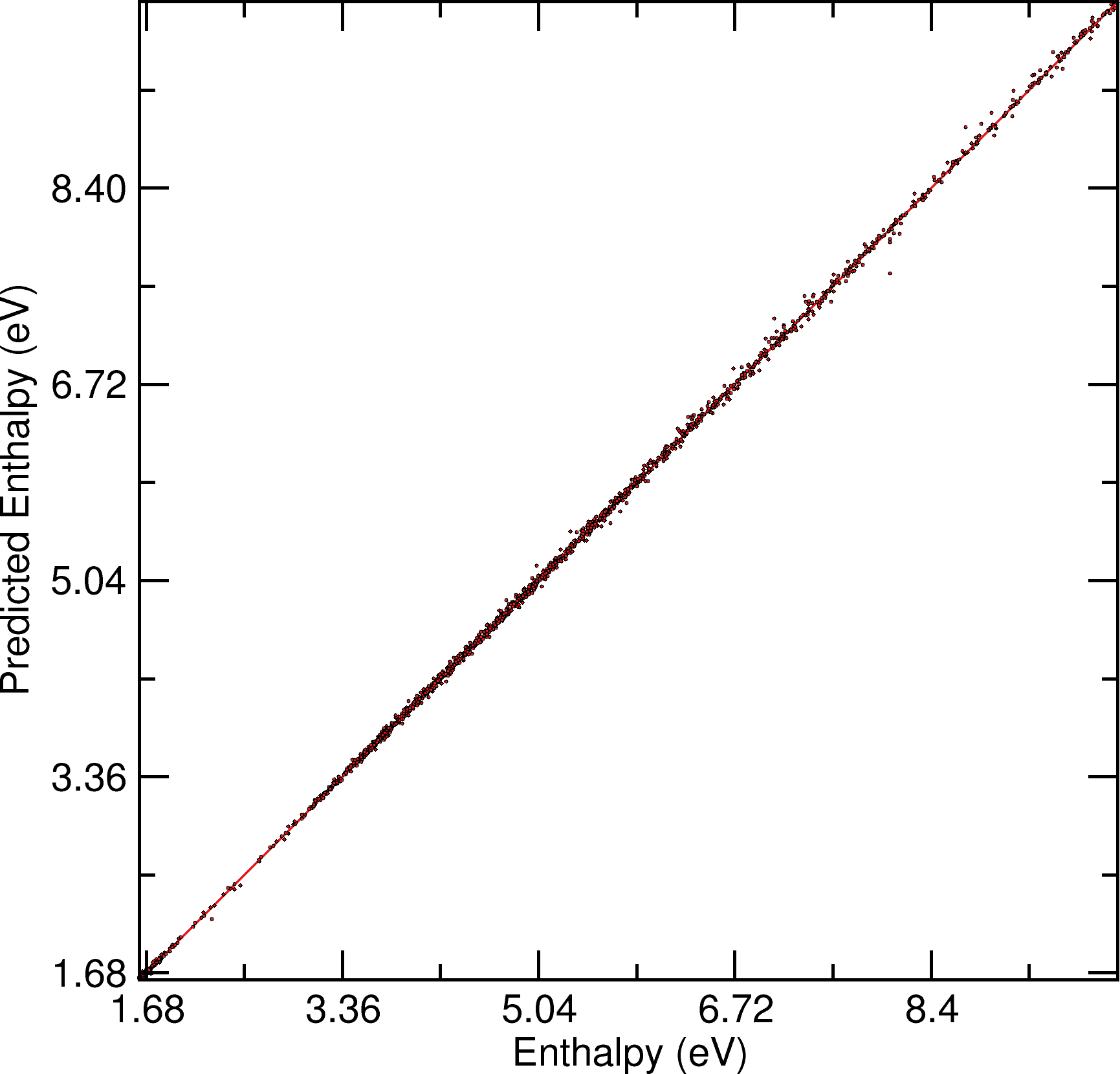}
\includegraphics[width=0.3\textwidth]{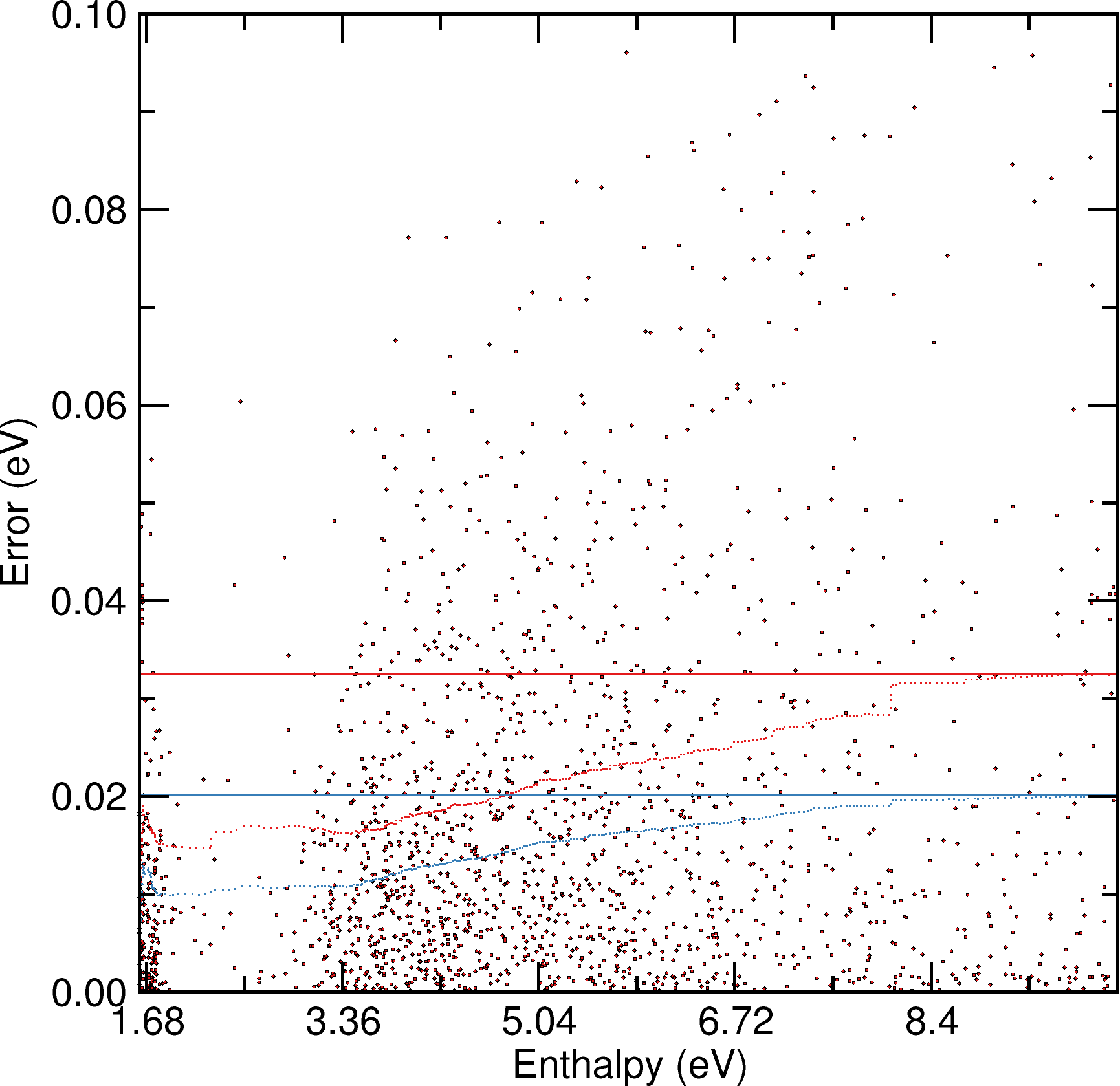}
\centering\begin{verbatim}
training    RMSE/MAE:  23.36  14.85  meV  Spearman  :  0.99988
validation  RMSE/MAE:  28.86  18.73  meV  Spearman  :  0.99984
testing     RMSE/MAE:  32.50  20.10  meV  Spearman  :  0.99984
\end{verbatim}
\clearpage

\flushleft{
\subsection{La-H}}
\subsubsection*{Searching}
\centering
\includegraphics[width=0.4\textwidth]{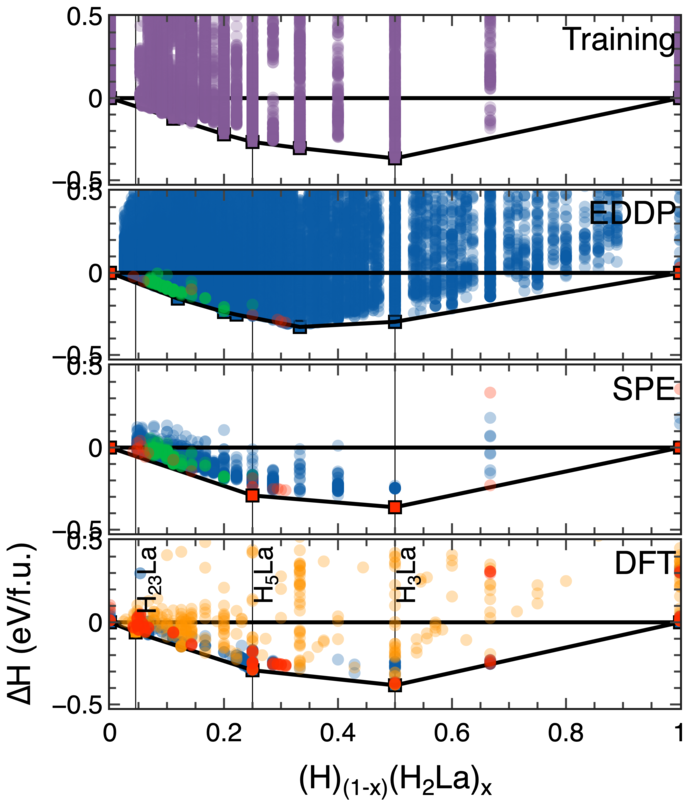}
\footnotesize


\flushleft{
\subsubsection*{\textsc{EDDP}}}
\centering
\includegraphics[width=0.3\textwidth]{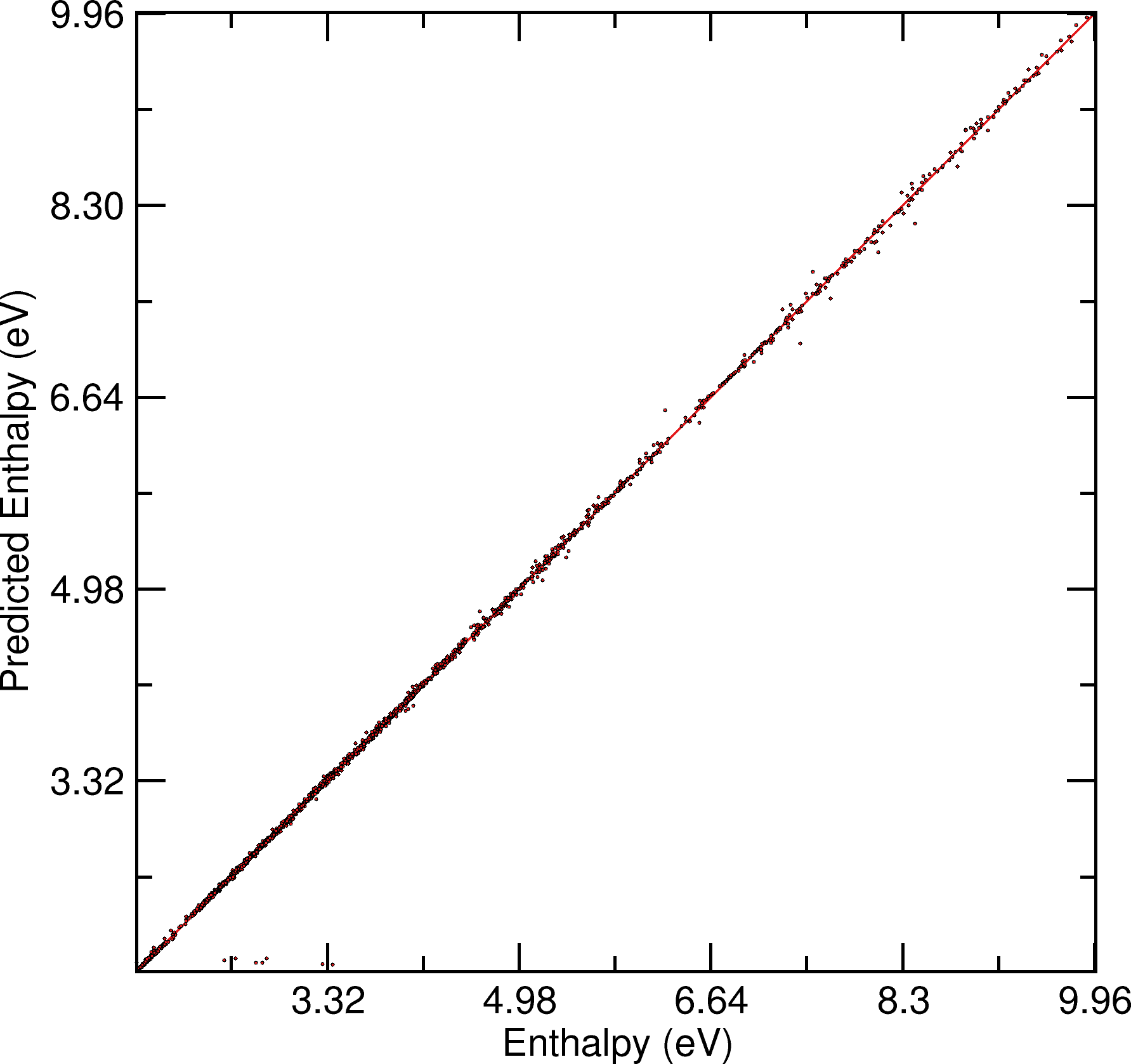}
\includegraphics[width=0.3\textwidth]{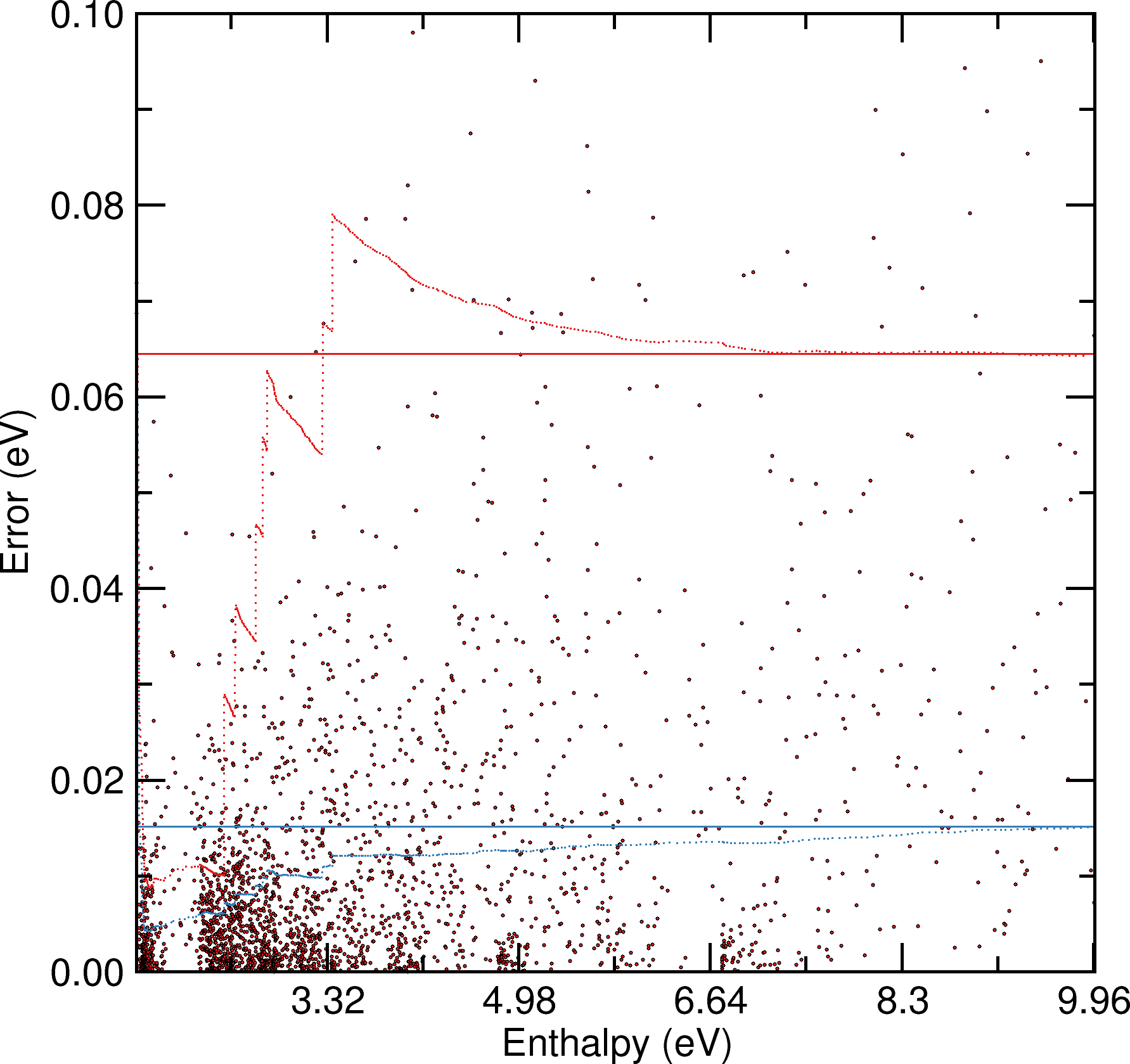}
\centering\begin{verbatim}
training    RMSE/MAE:  43.28  9.80   meV  Spearman  :  0.99884
validation  RMSE/MAE:  34.17  12.95  meV  Spearman  :  0.99955
testing     RMSE/MAE:  64.47  15.13  meV  Spearman  :  0.99757
\end{verbatim}
\clearpage

\flushleft{
\subsection{Li-H}}
\subsubsection*{Searching}
\centering
\includegraphics[width=0.4\textwidth]{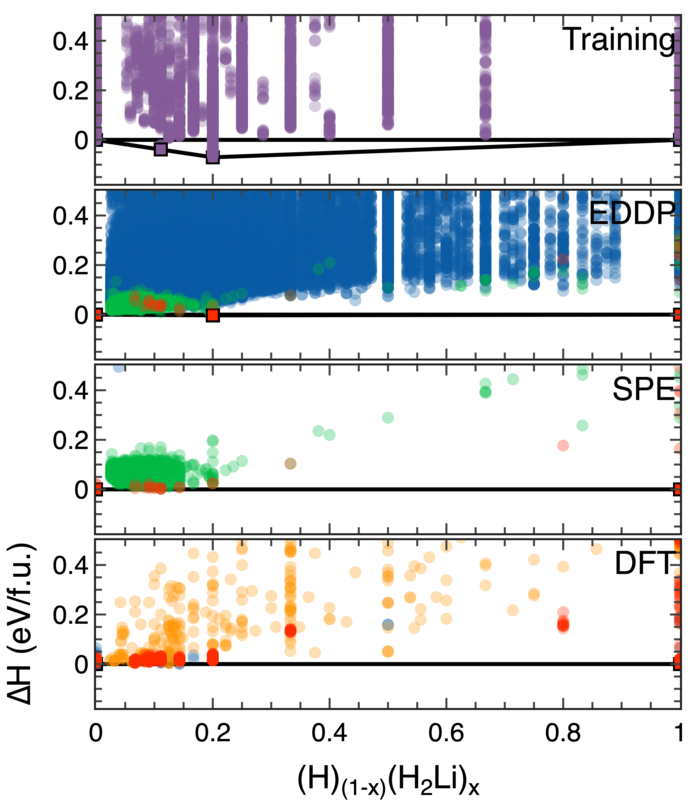}
\footnotesize


\flushleft{
\subsubsection*{\textsc{EDDP}}}
\centering
\includegraphics[width=0.3\textwidth]{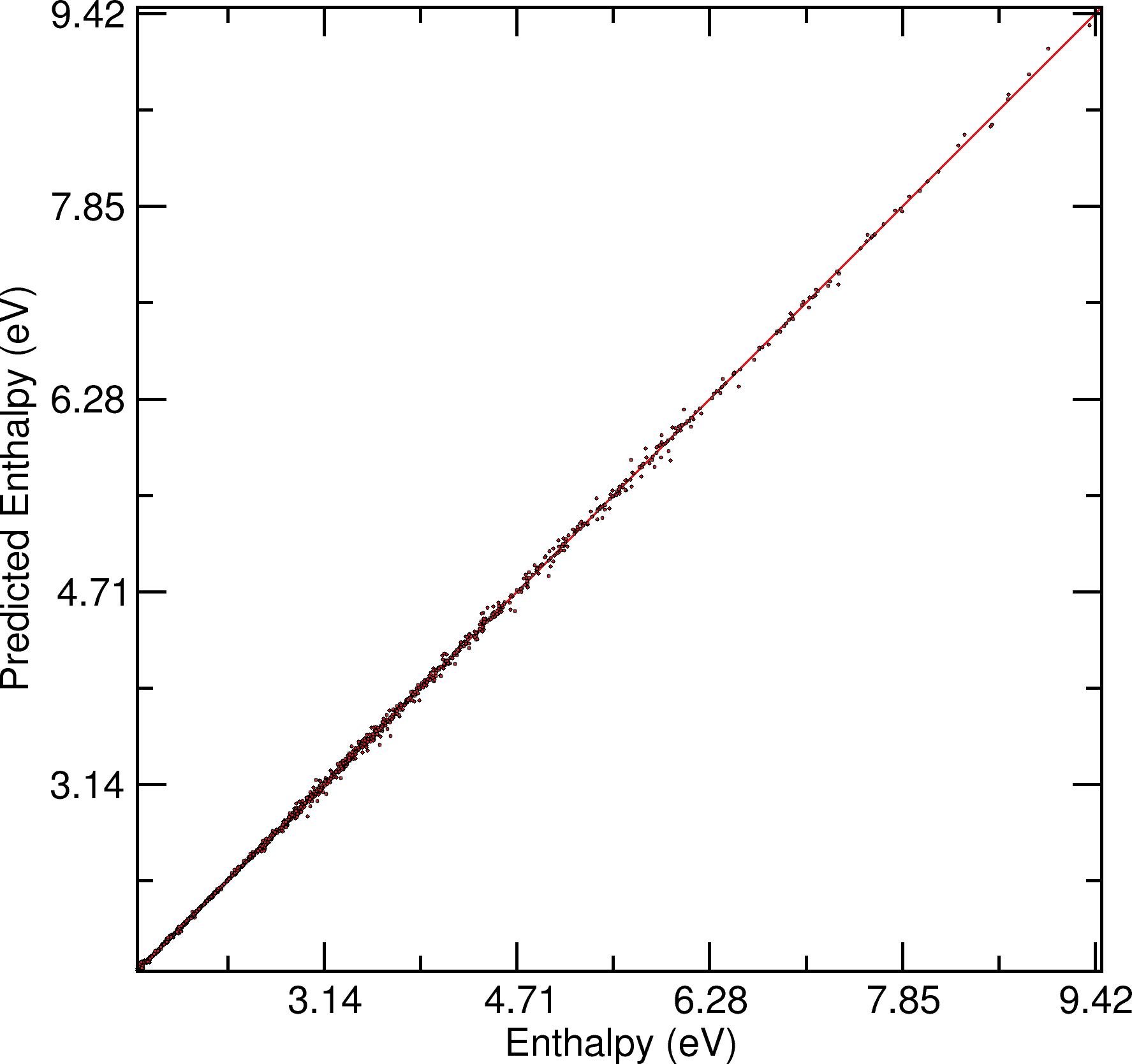}
\includegraphics[width=0.3\textwidth]{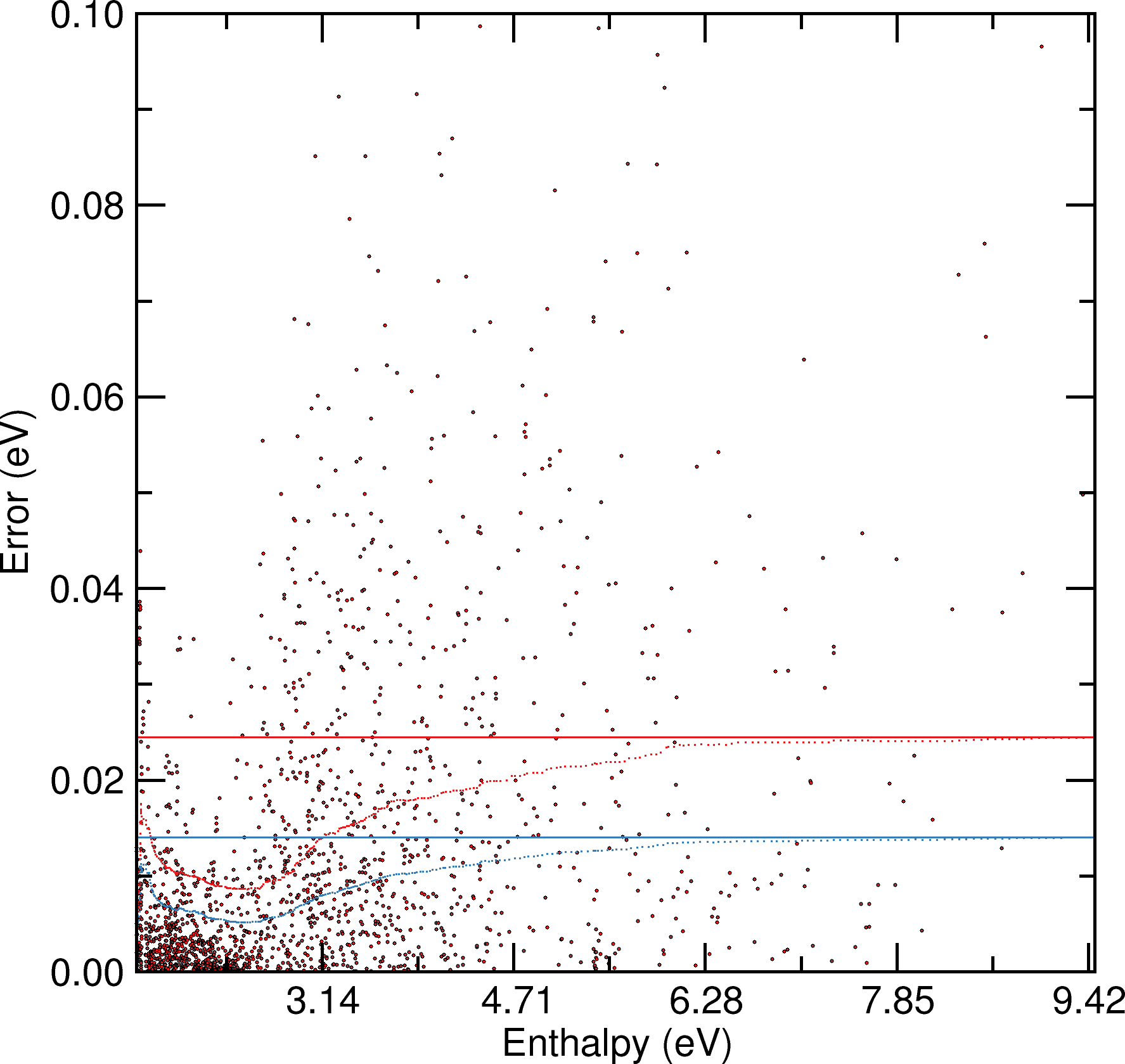}
\centering\begin{verbatim}
training    RMSE/MAE:  16.15  9.39   meV  Spearman  :  0.99988
validation  RMSE/MAE:  23.08  14.15  meV  Spearman  :  0.99979
testing     RMSE/MAE:  24.44  14.02  meV  Spearman  :  0.99982
\end{verbatim}
\clearpage

\flushleft{
\subsection{Lu-H}}
\subsubsection*{Searching}
\centering
\includegraphics[width=0.4\textwidth]{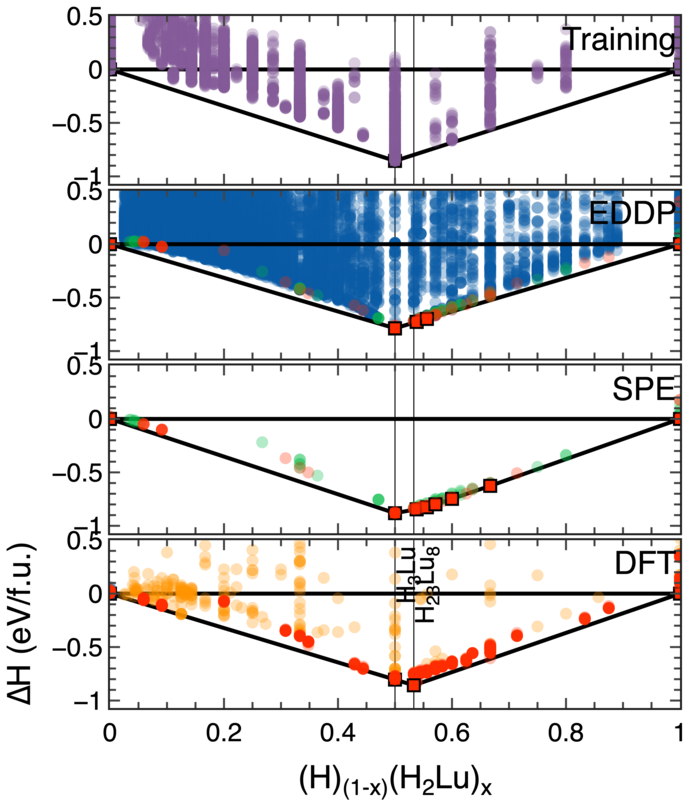}
\footnotesize


\flushleft{
\subsubsection*{\textsc{EDDP}}}
\centering
\includegraphics[width=0.3\textwidth]{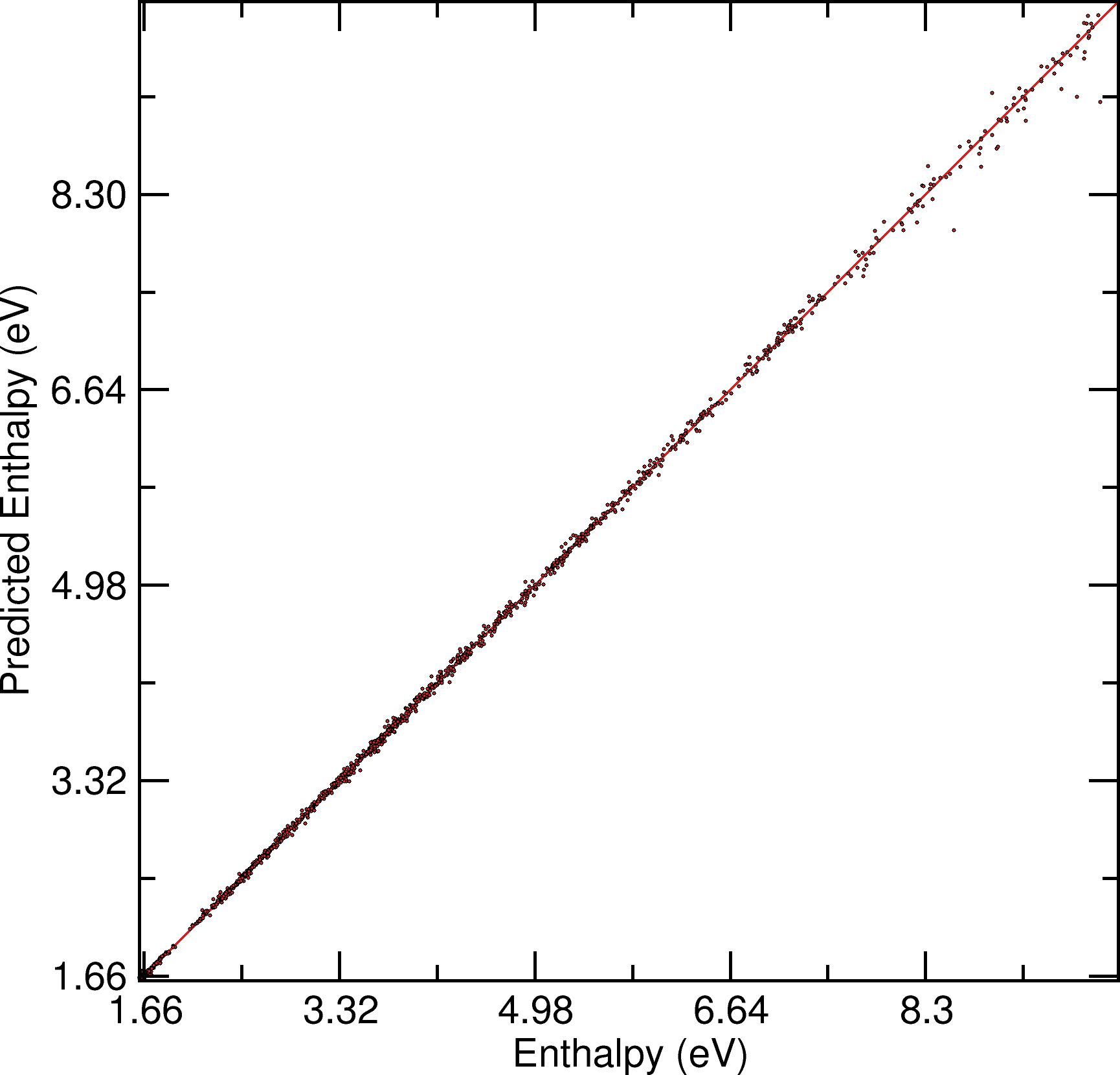}
\includegraphics[width=0.3\textwidth]{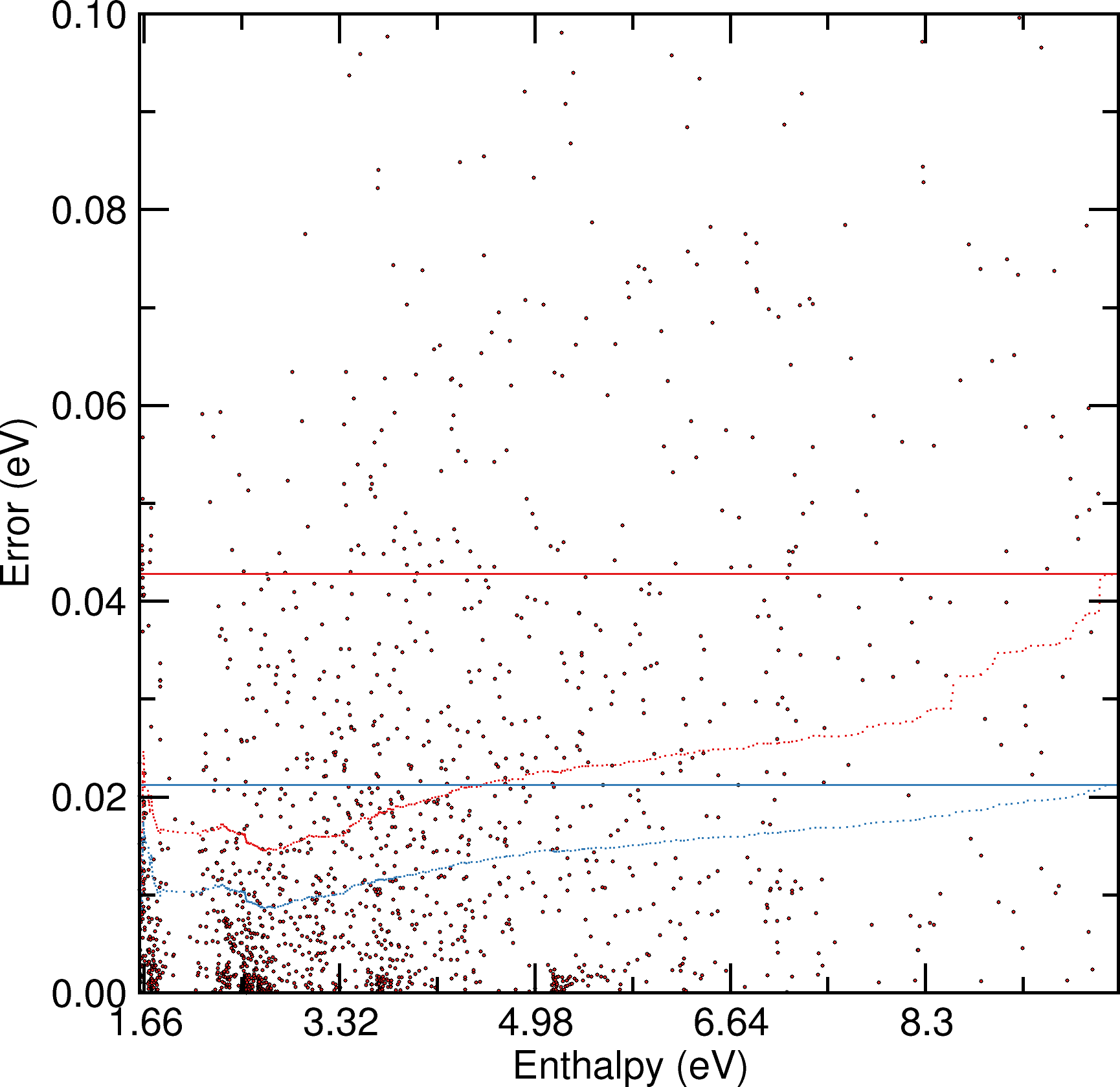}
\centering\begin{verbatim}
training    RMSE/MAE:  21.52  13.40  meV  Spearman  :  0.99983
validation  RMSE/MAE:  29.41  18.70  meV  Spearman  :  0.99981
testing     RMSE/MAE:  42.82  21.25  meV  Spearman  :  0.99975
\end{verbatim}
\clearpage

\flushleft{
\subsection{Mg-H}}
\subsubsection*{Searching}
\centering
\includegraphics[width=0.4\textwidth]{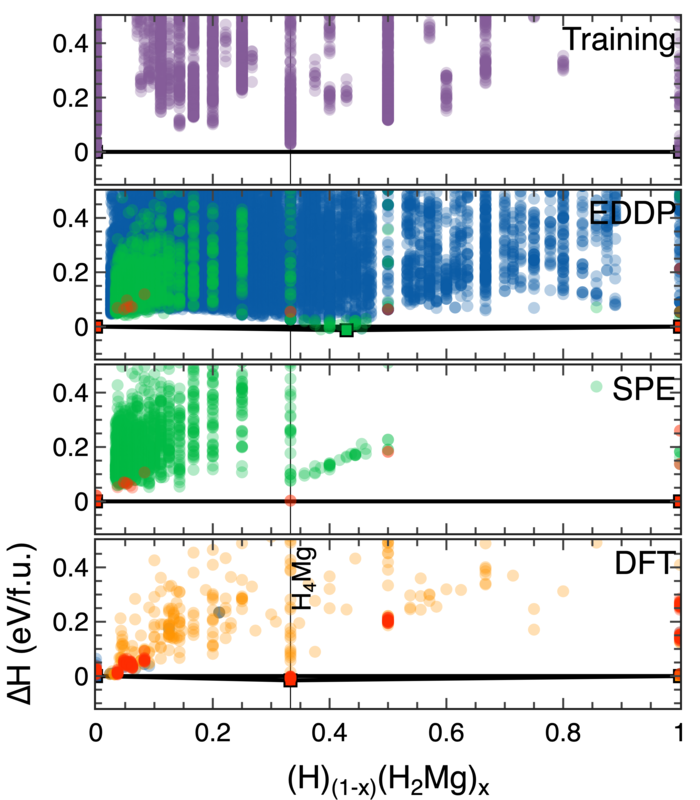}
\footnotesize


\flushleft{
\subsubsection*{\textsc{EDDP}}}
\centering
\includegraphics[width=0.3\textwidth]{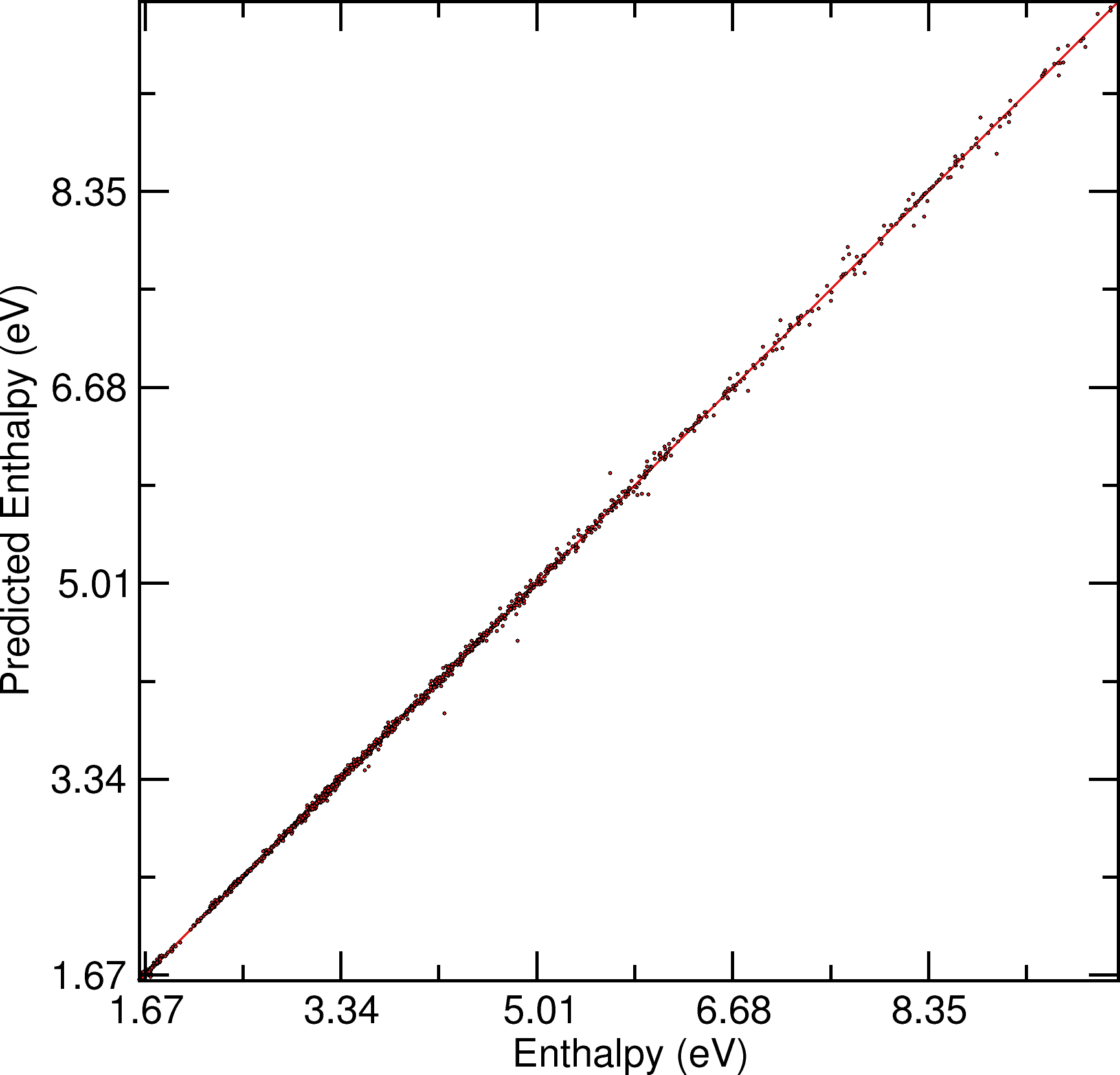}
\includegraphics[width=0.3\textwidth]{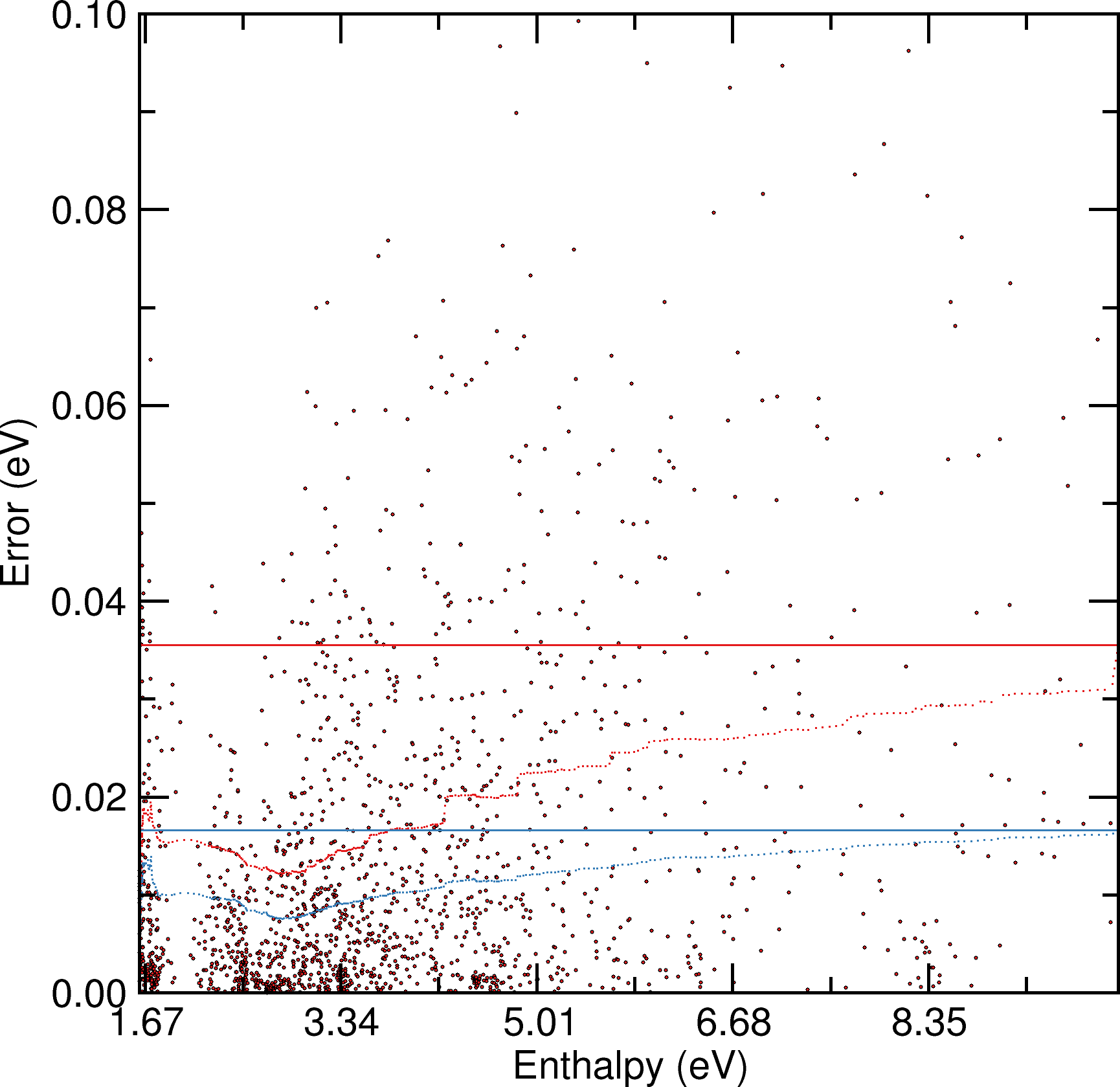}
\centering\begin{verbatim}
training    RMSE/MAE:  15.76  9.46   meV  Spearman  :  0.99988
validation  RMSE/MAE:  26.76  14.83  meV  Spearman  :  0.99986
testing     RMSE/MAE:  35.53  16.57  meV  Spearman  :  0.99980
\end{verbatim}
\clearpage

\flushleft{
\subsection{Mn-H}}
\subsubsection*{Searching}
\centering
\includegraphics[width=0.4\textwidth]{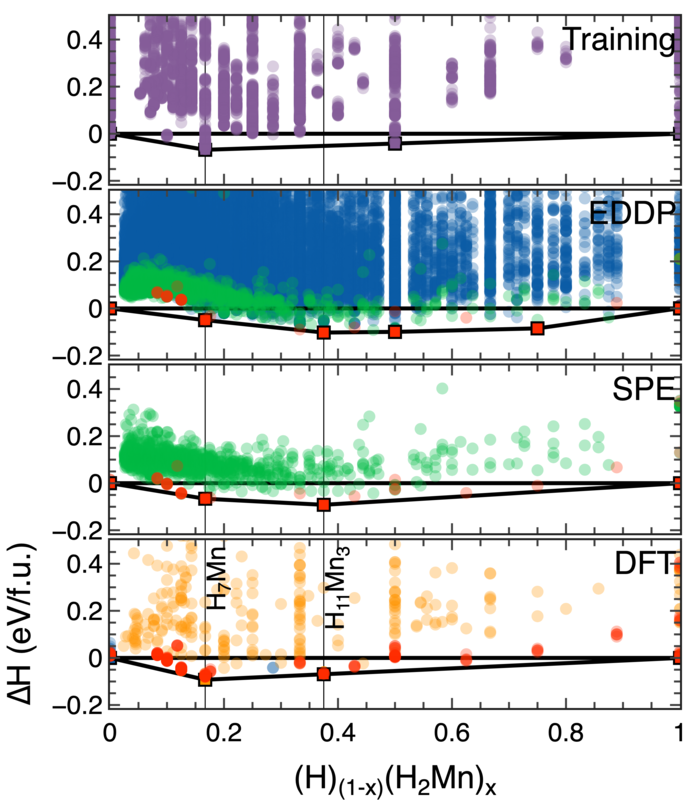}
\footnotesize


\flushleft{
\subsubsection*{\textsc{EDDP}}}
\centering
\includegraphics[width=0.3\textwidth]{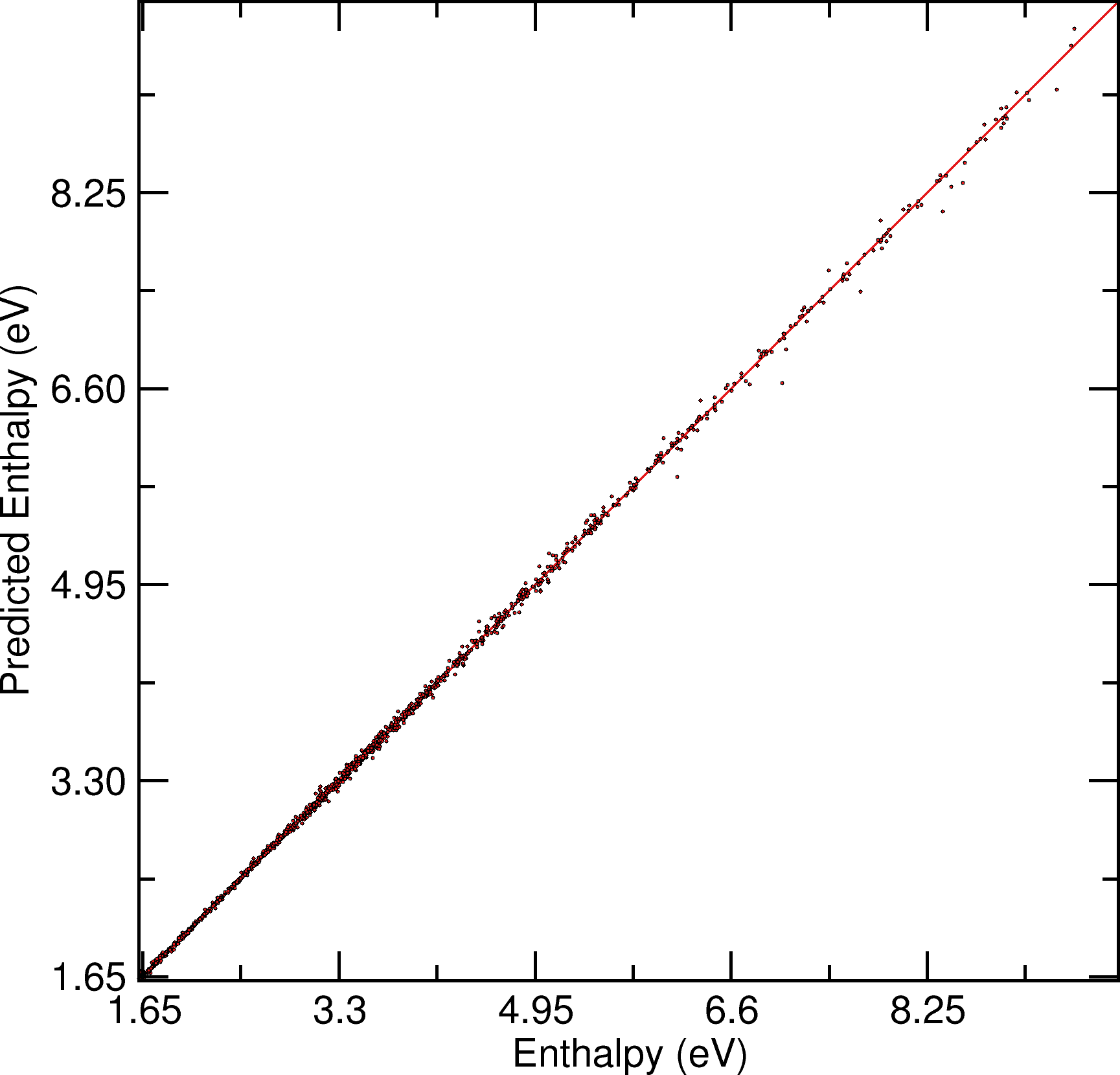}
\includegraphics[width=0.3\textwidth]{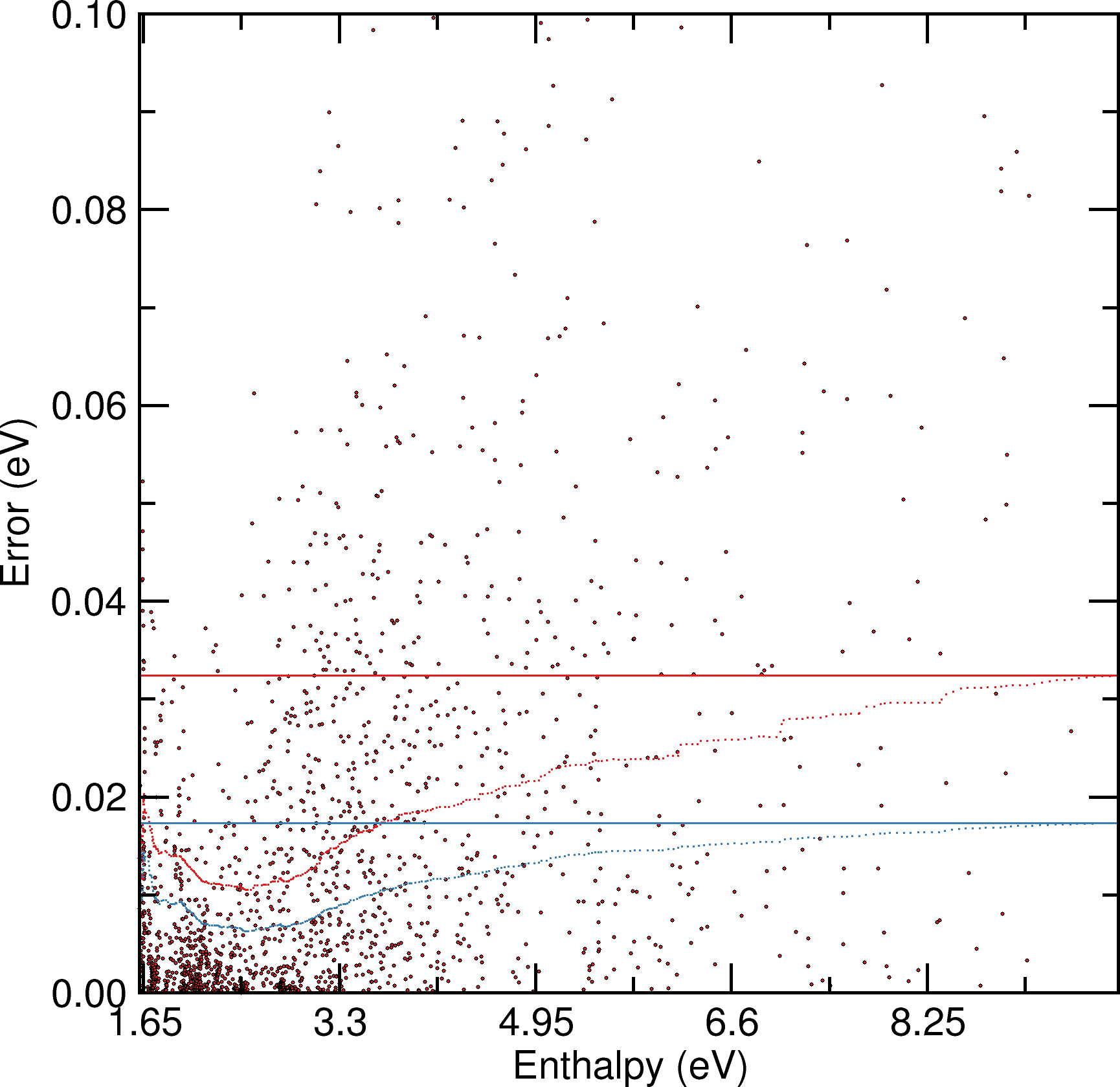}
\centering\begin{verbatim}
training    RMSE/MAE:  18.48  11.10  meV  Spearman  :  0.99985
validation  RMSE/MAE:  24.53  14.76  meV  Spearman  :  0.99978
testing     RMSE/MAE:  32.43  17.36  meV  Spearman  :  0.99979
\end{verbatim}
\clearpage

\flushleft{
\subsection{Mo-H}}
\subsubsection*{Searching}
\centering
\includegraphics[width=0.4\textwidth]{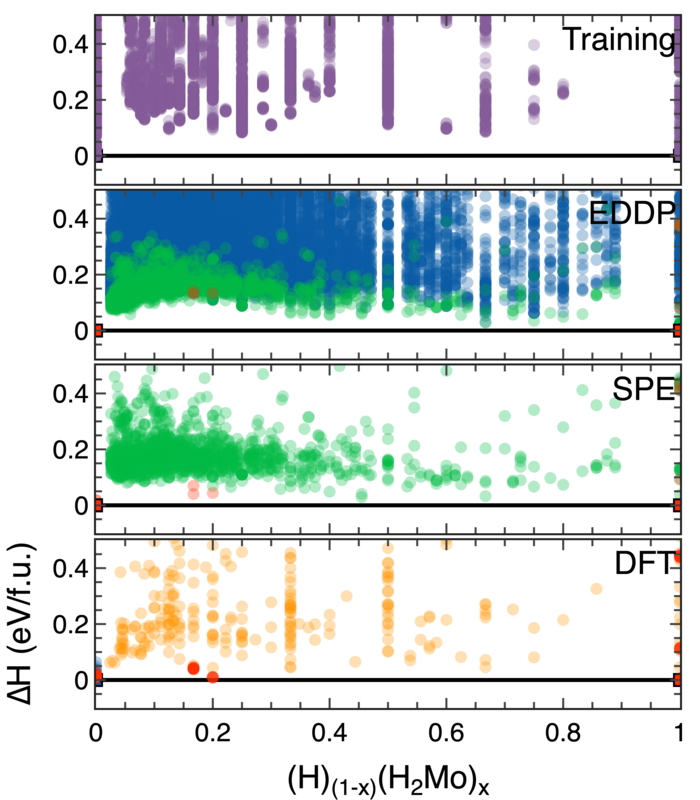}
\footnotesize


\flushleft{
\subsubsection*{\textsc{EDDP}}}
\centering
\includegraphics[width=0.3\textwidth]{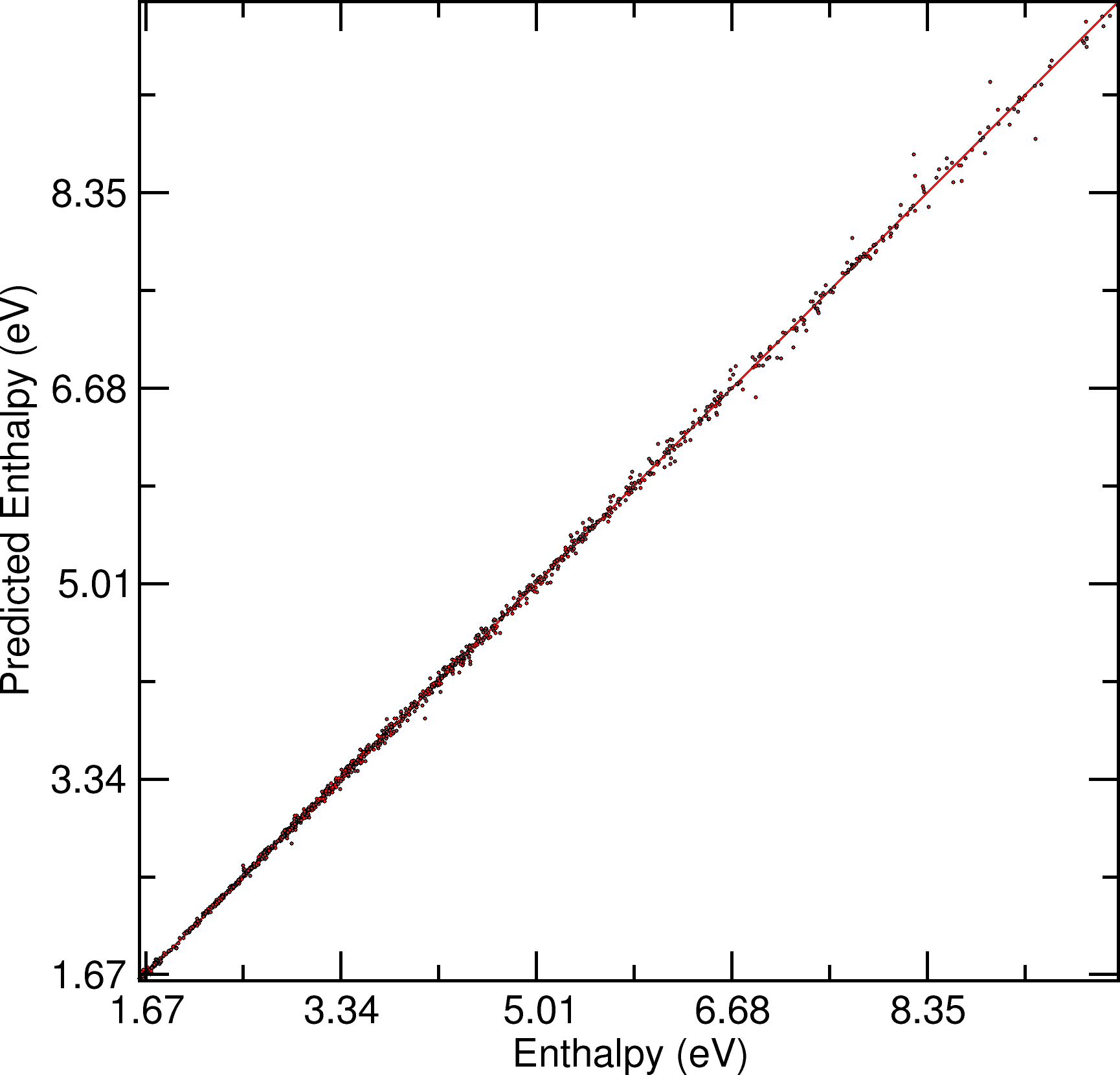}
\includegraphics[width=0.3\textwidth]{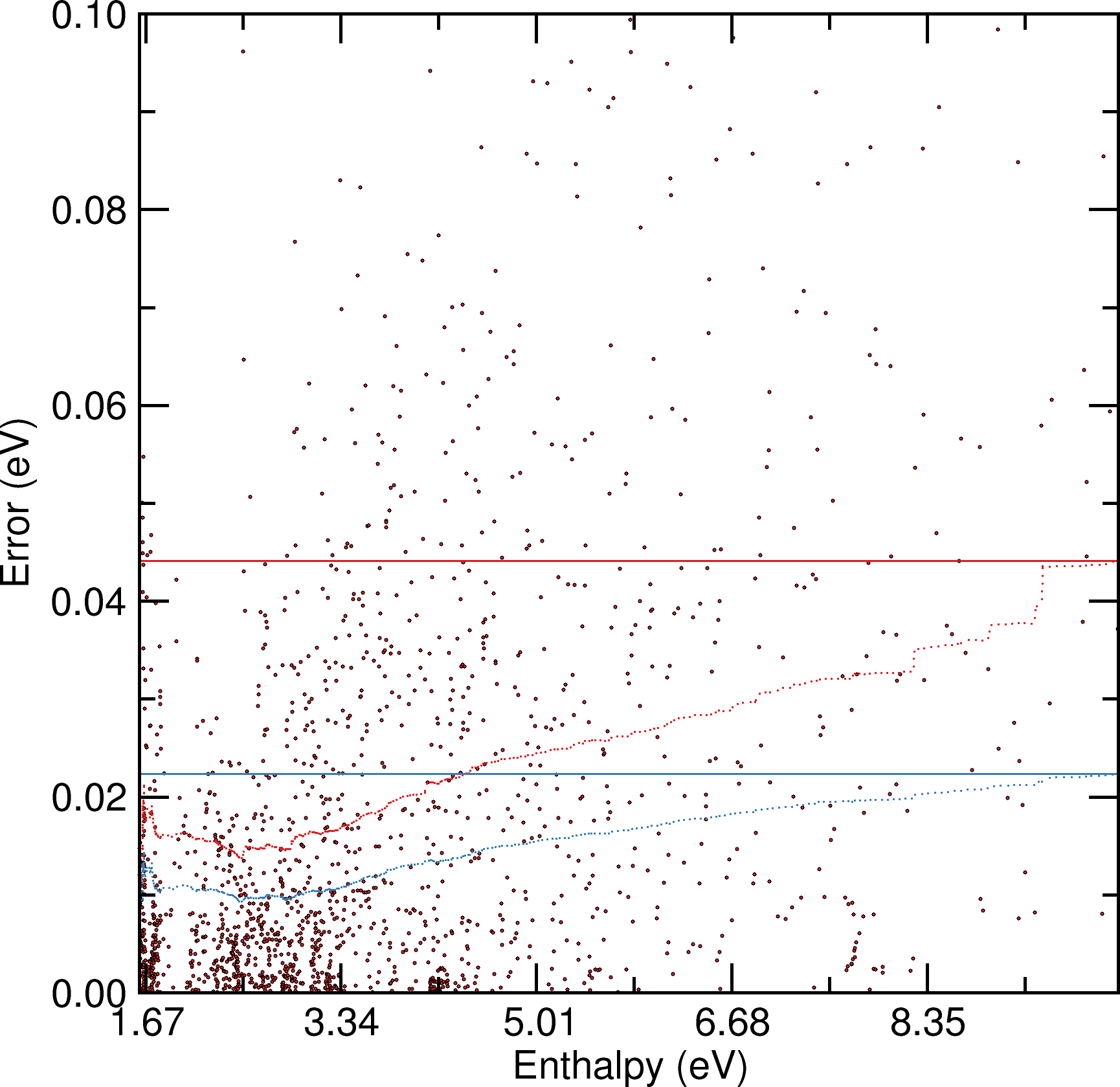}
\centering\begin{verbatim}
training    RMSE/MAE:  21.99  13.52  meV  Spearman  :  0.99984
validation  RMSE/MAE:  35.95  21.57  meV  Spearman  :  0.99976
testing     RMSE/MAE:  44.11  22.37  meV  Spearman  :  0.99976
\end{verbatim}
\clearpage

\flushleft{
\subsection{Na-H}}
\subsubsection*{Searching}
\centering
\includegraphics[width=0.4\textwidth]{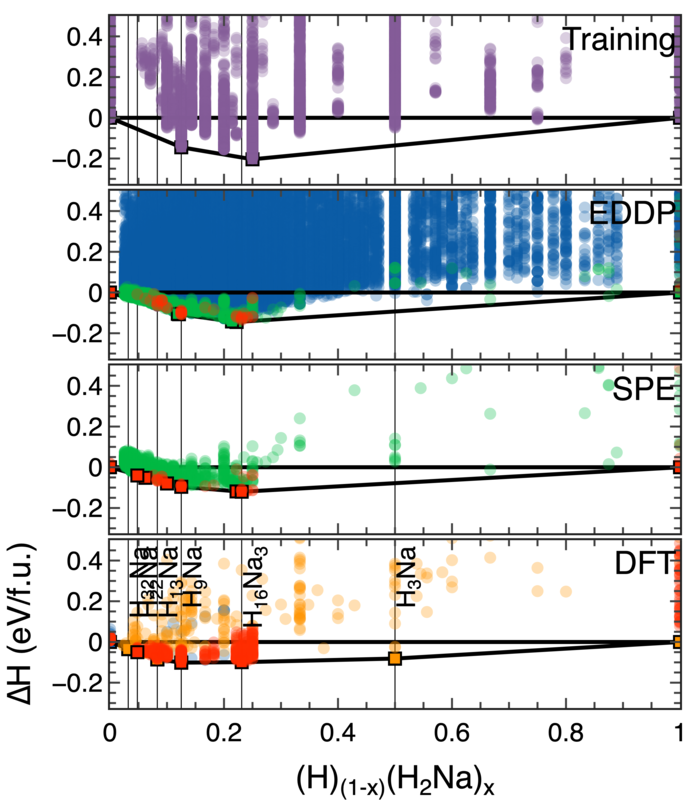}
\footnotesize


\flushleft{
\subsubsection*{\textsc{EDDP}}}
\centering
\includegraphics[width=0.3\textwidth]{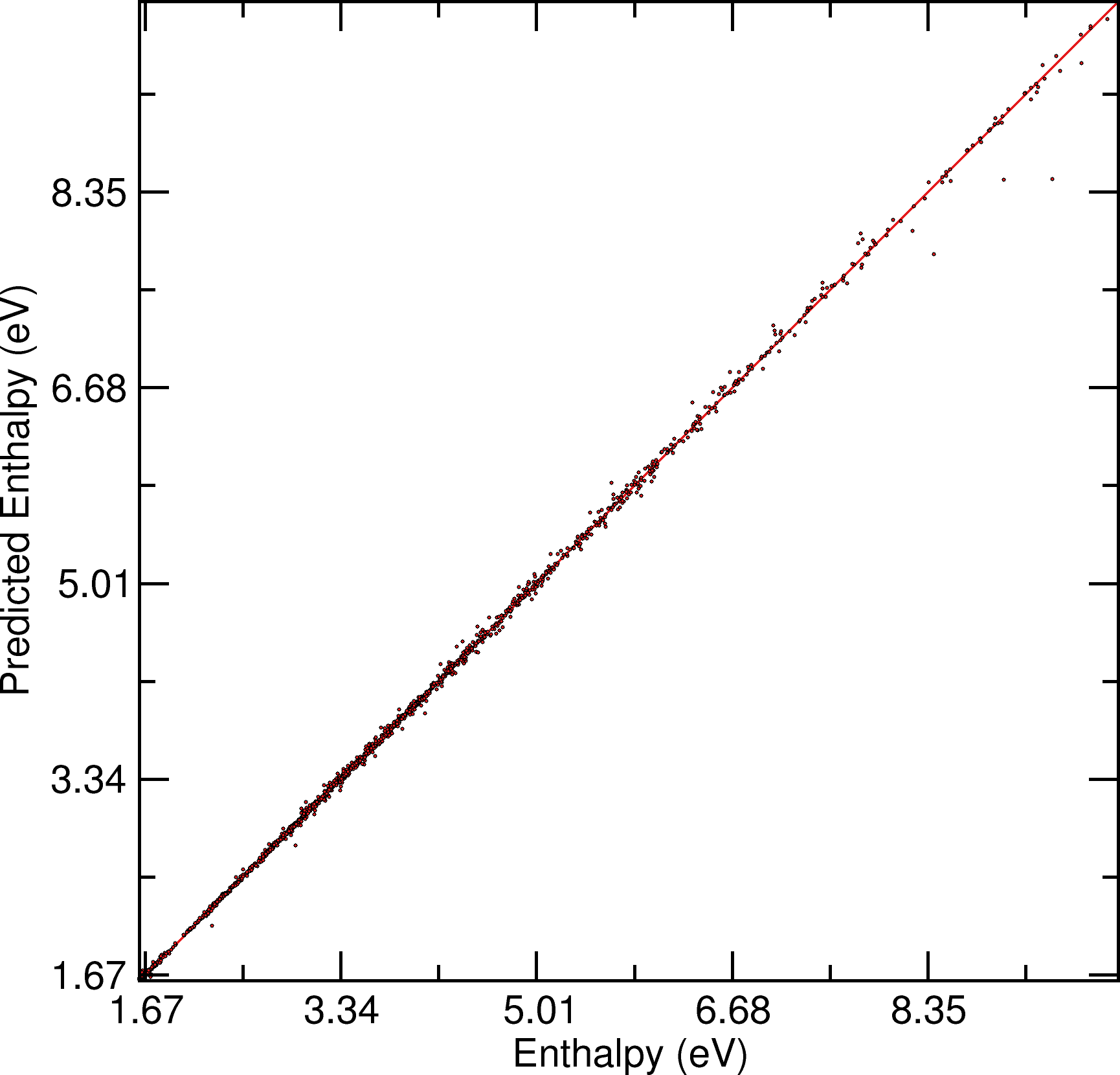}
\includegraphics[width=0.3\textwidth]{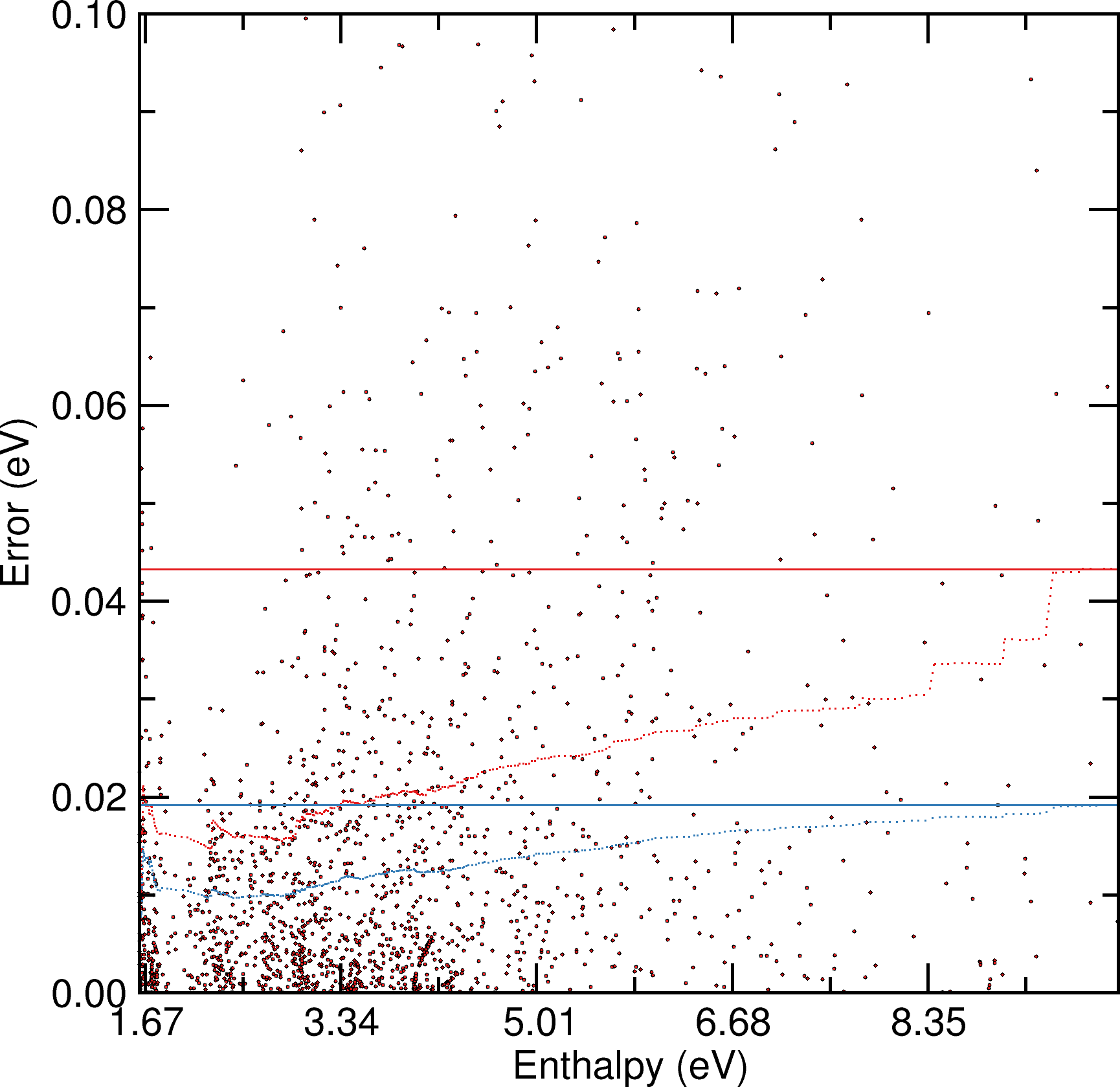}
\centering\begin{verbatim}
training    RMSE/MAE:  20.80  11.63  meV  Spearman  :  0.99985
validation  RMSE/MAE:  33.61  18.49  meV  Spearman  :  0.99976
testing     RMSE/MAE:  43.28  19.15  meV  Spearman  :  0.99976
\end{verbatim}
\clearpage

\flushleft{
\subsection{Nb-H}}
\subsubsection*{Searching}
\centering
\includegraphics[width=0.4\textwidth]{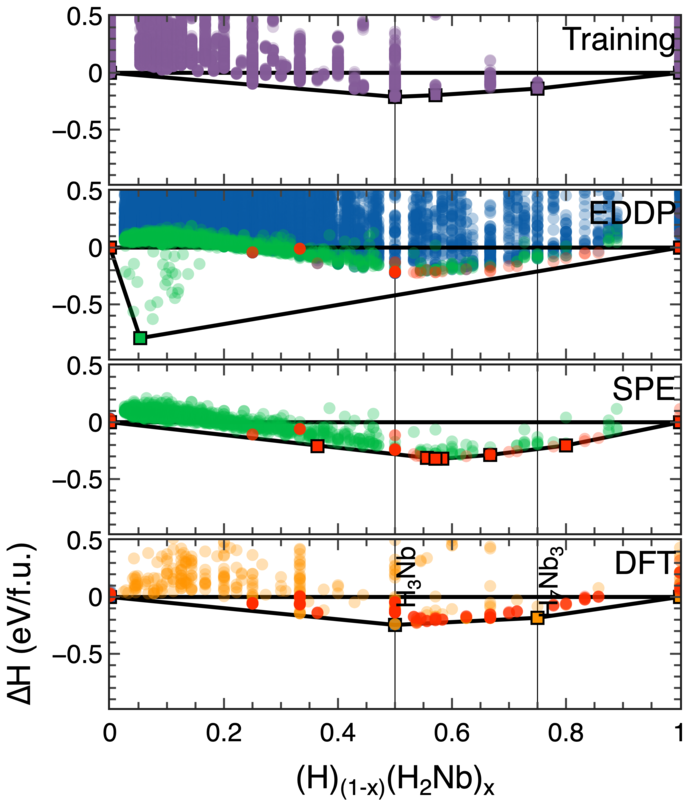}
\footnotesize


\flushleft{
\subsubsection*{\textsc{EDDP}}}
\centering
\includegraphics[width=0.3\textwidth]{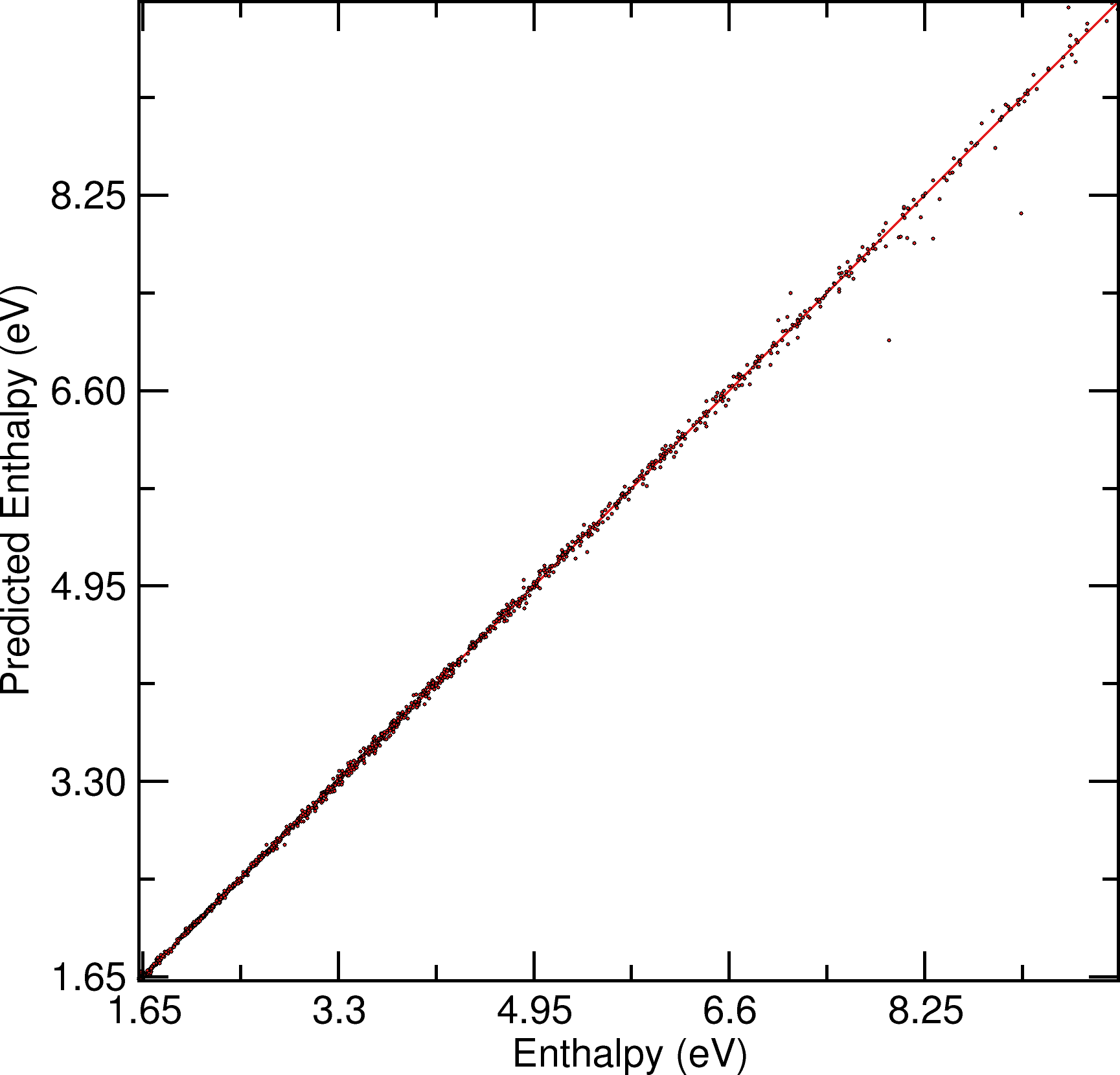}
\includegraphics[width=0.3\textwidth]{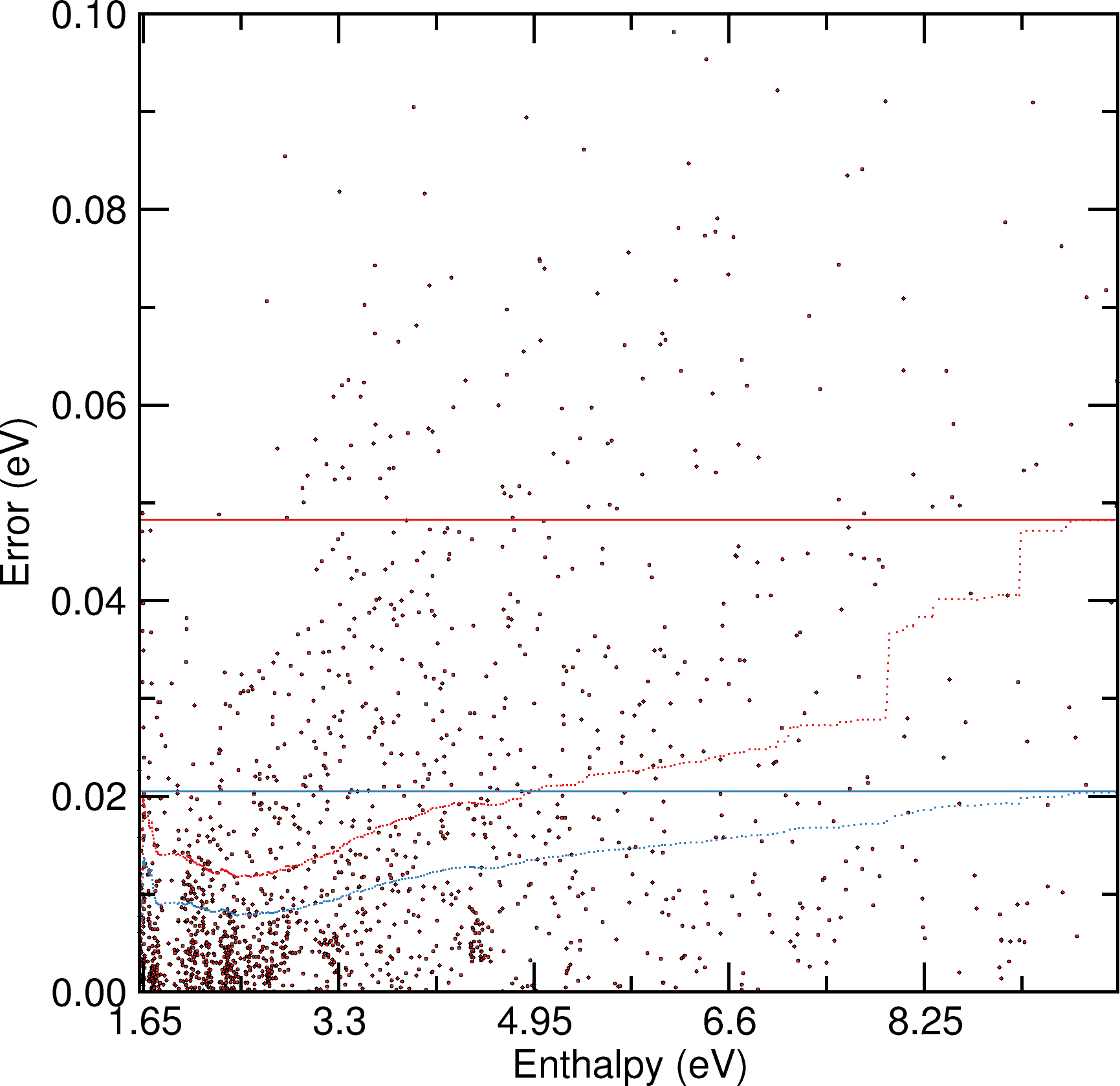}
\centering\begin{verbatim}
training    RMSE/MAE:  21.55  13.64  meV  Spearman  :  0.99986
validation  RMSE/MAE:  29.94  18.60  meV  Spearman  :  0.99978
testing     RMSE/MAE:  48.31  20.51  meV  Spearman  :  0.99983
\end{verbatim}
\clearpage

\flushleft{
\subsection{Nd-H}}
\subsubsection*{Searching}
\centering
\includegraphics[width=0.4\textwidth]{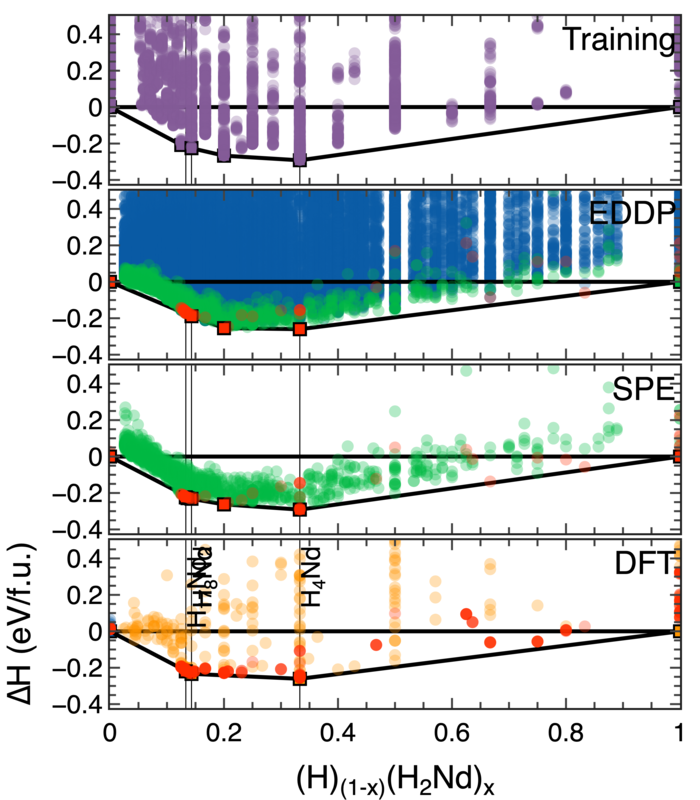}
\footnotesize


\flushleft{
\subsubsection*{\textsc{EDDP}}}
\centering
\includegraphics[width=0.3\textwidth]{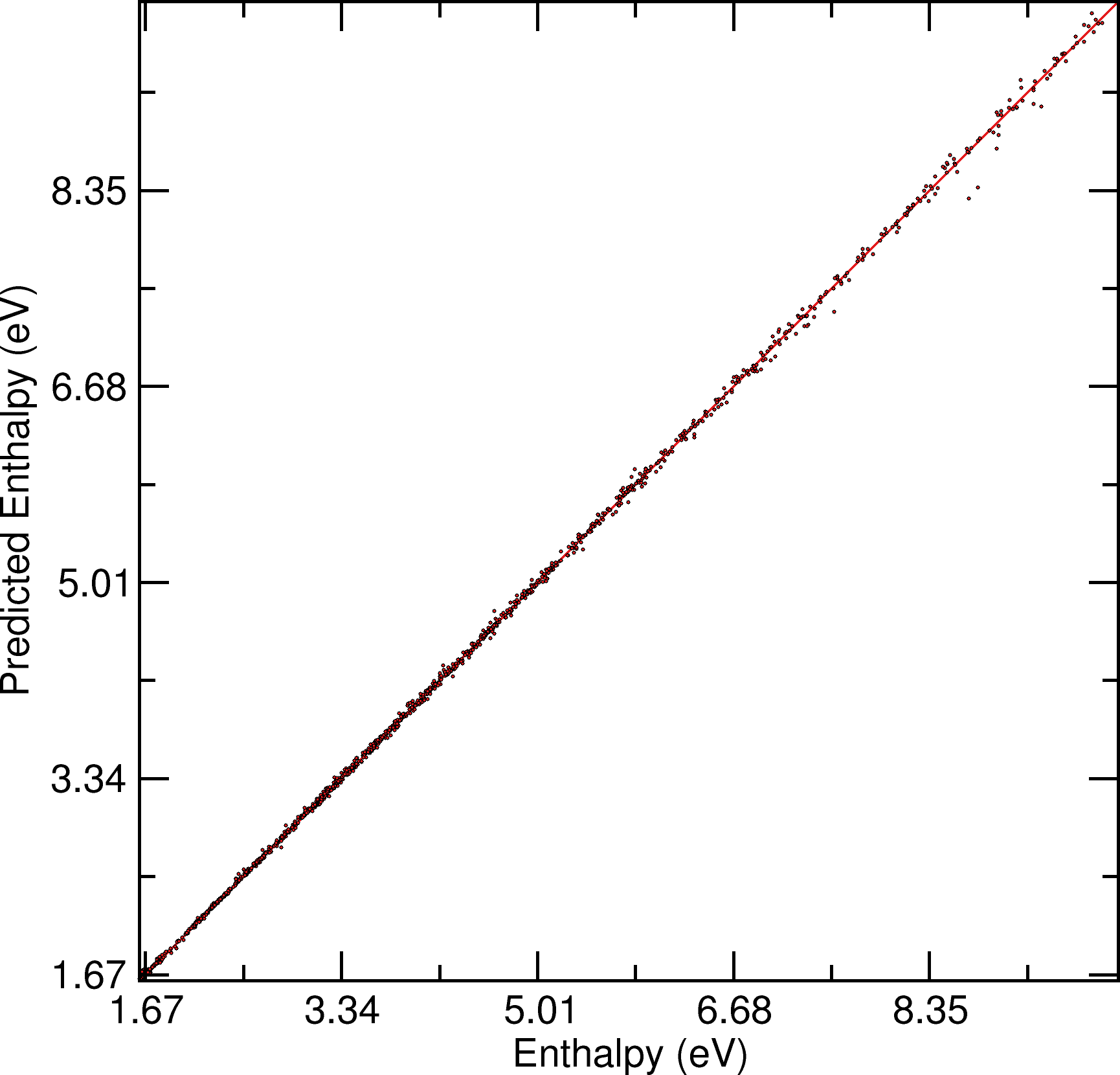}
\includegraphics[width=0.3\textwidth]{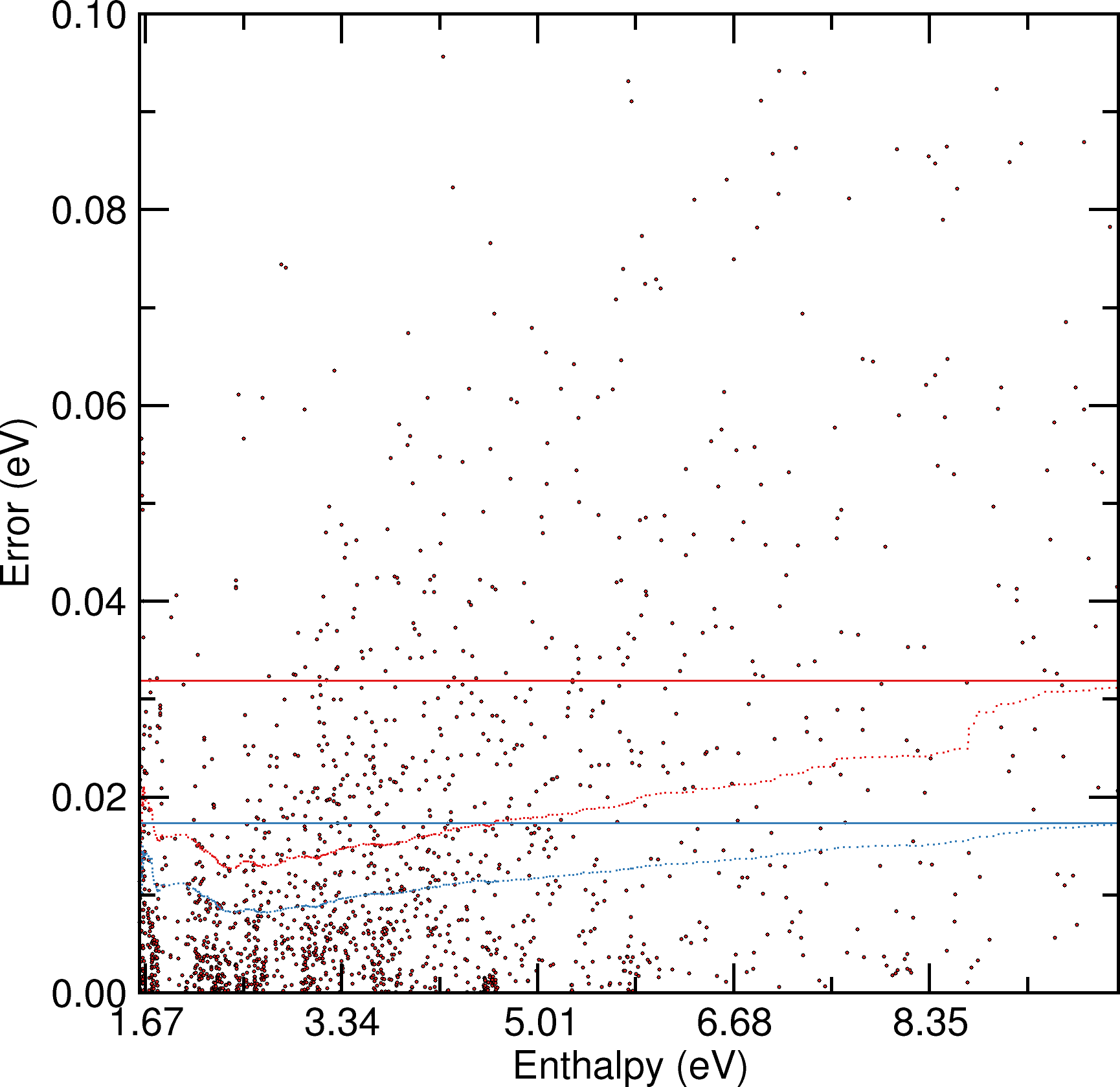}
\centering\begin{verbatim}
training    RMSE/MAE:  17.40  11.03  meV  Spearman  :  0.99988
validation  RMSE/MAE:  24.49  15.80  meV  Spearman  :  0.99984
testing     RMSE/MAE:  31.87  17.33  meV  Spearman  :  0.99986
\end{verbatim}
\clearpage

\flushleft{
\subsection{Ne-H}}
\subsubsection*{Searching}
\centering
\includegraphics[width=0.4\textwidth]{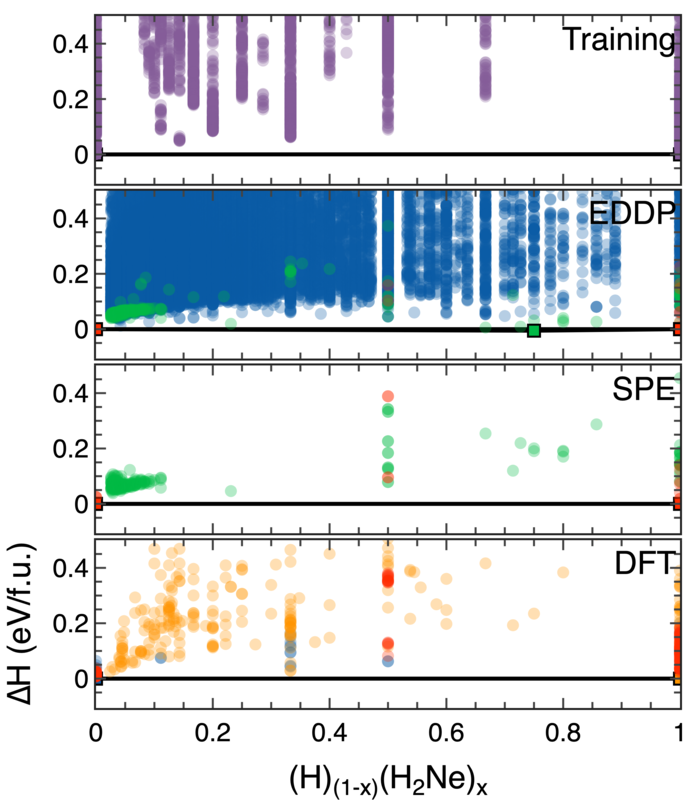}
\footnotesize


\flushleft{
\subsubsection*{\textsc{EDDP}}}
\centering
\includegraphics[width=0.3\textwidth]{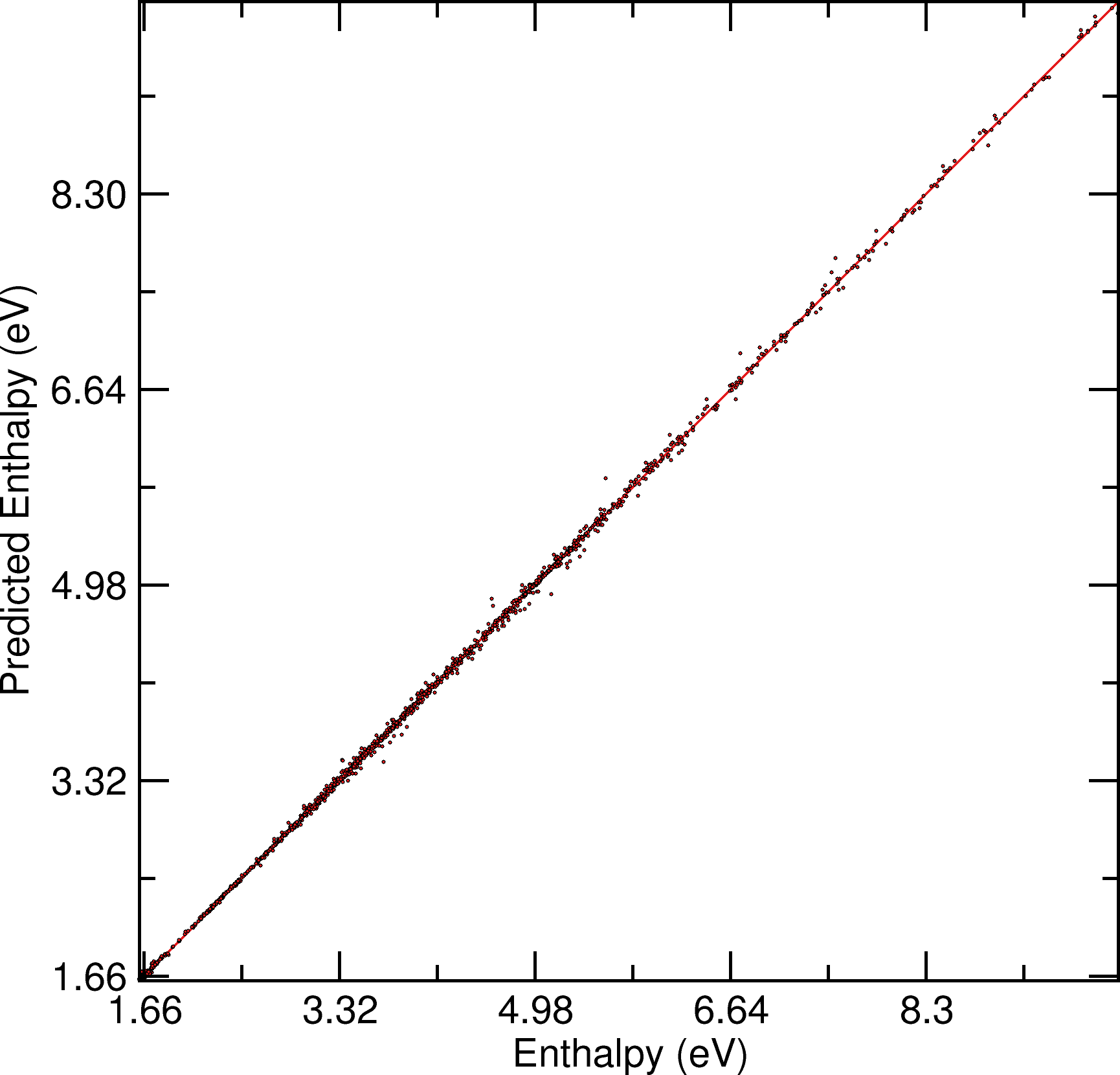}
\includegraphics[width=0.3\textwidth]{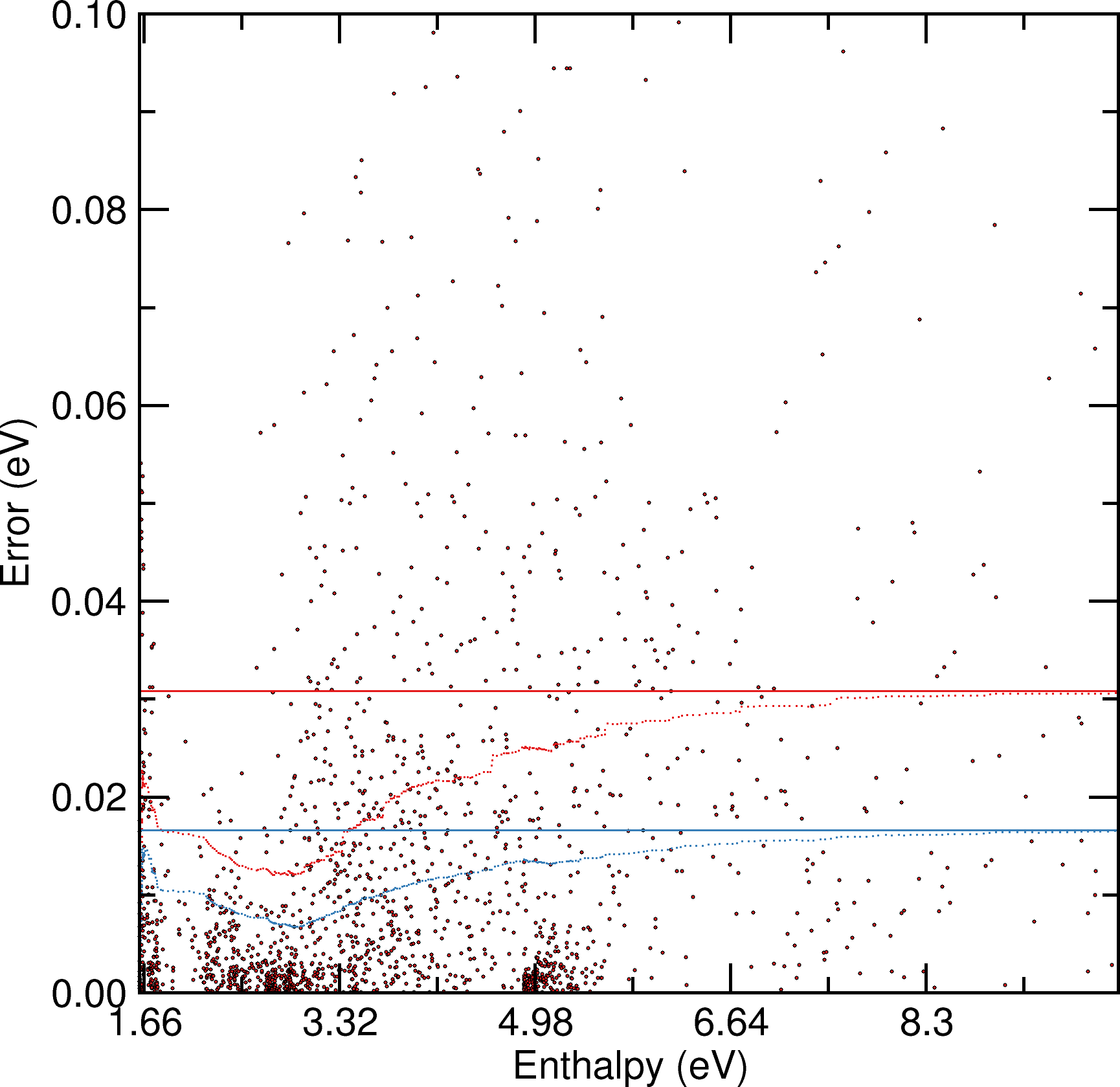}
\centering\begin{verbatim}
training    RMSE/MAE:  19.49  10.95  meV  Spearman  :  0.99986
validation  RMSE/MAE:  26.60  15.11  meV  Spearman  :  0.99977
testing     RMSE/MAE:  30.84  16.58  meV  Spearman  :  0.99974
\end{verbatim}
\clearpage

\flushleft{
\subsection{Ni-H}}
\subsubsection*{Searching}
\centering
\includegraphics[width=0.4\textwidth]{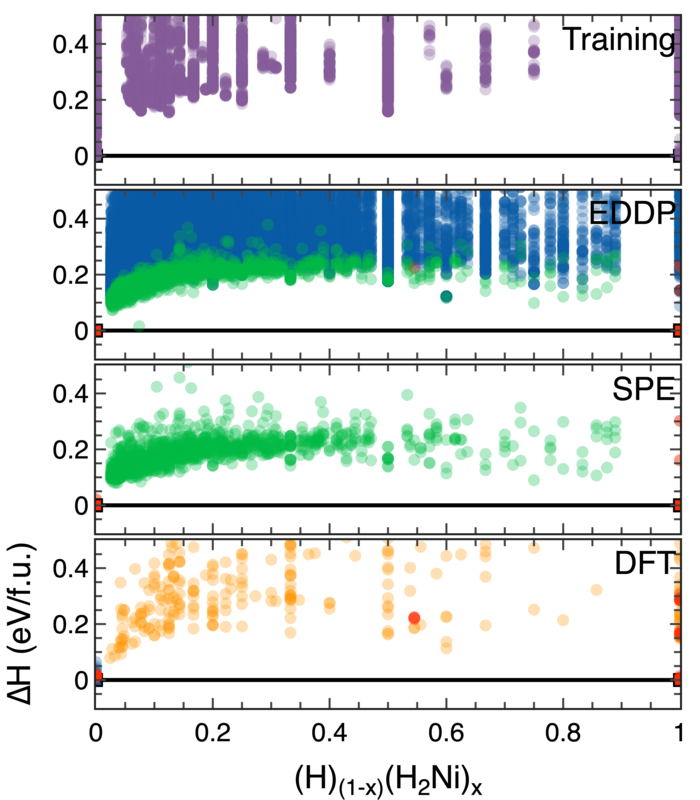}
\footnotesize


\flushleft{
\subsubsection*{\textsc{EDDP}}}
\centering
\includegraphics[width=0.3\textwidth]{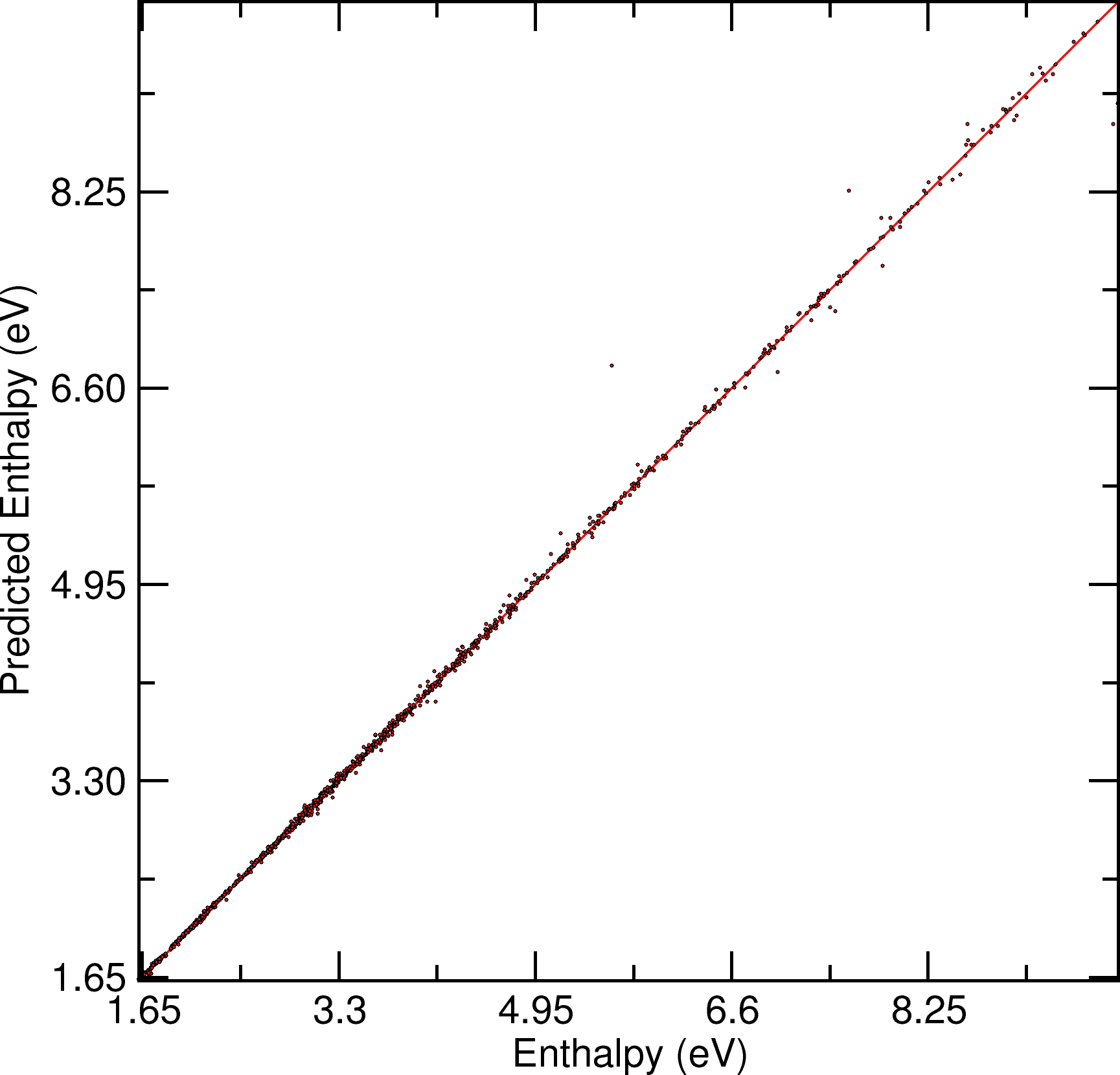}
\includegraphics[width=0.3\textwidth]{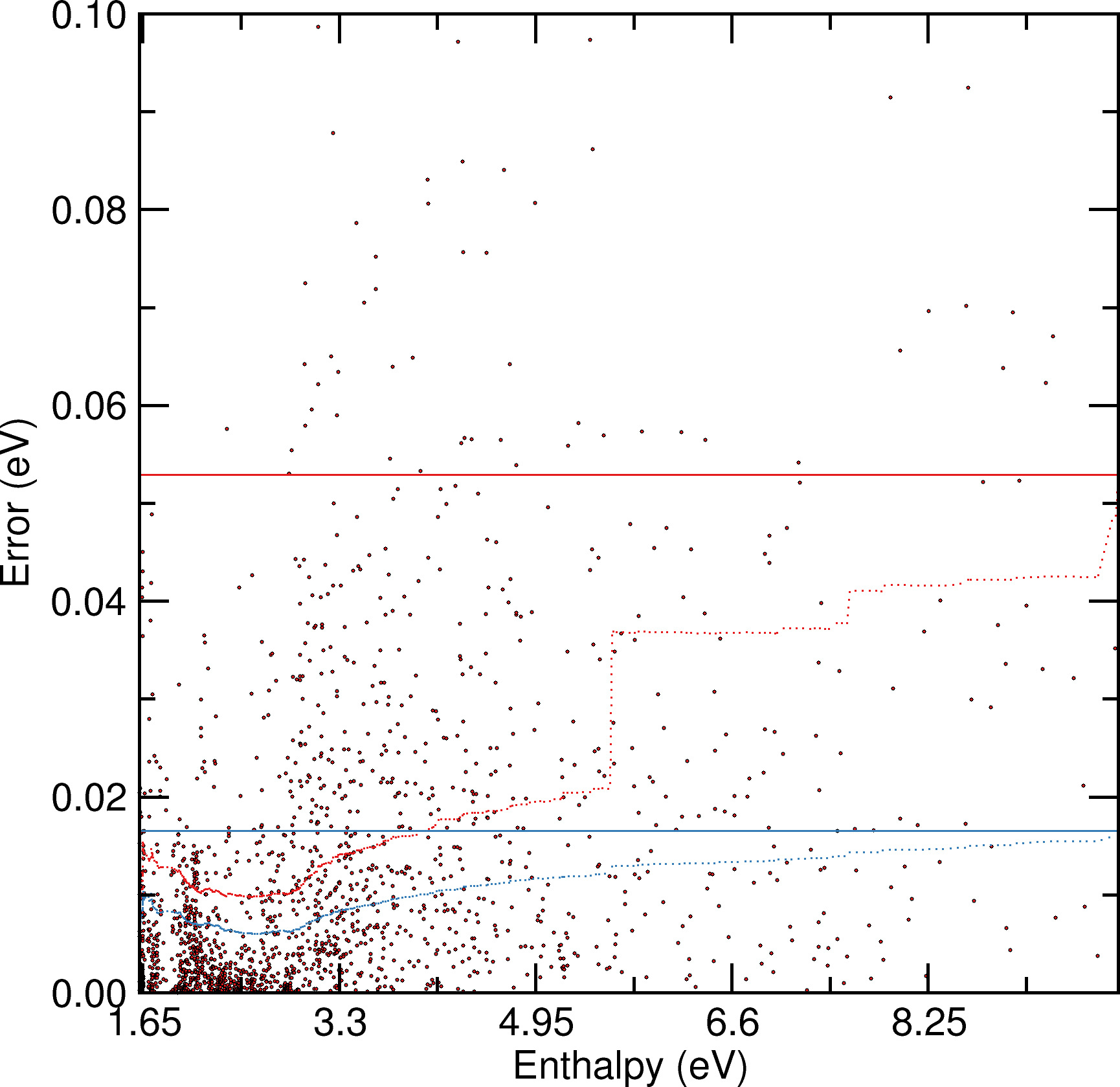}
\centering\begin{verbatim}
training    RMSE/MAE:  13.79  8.75   meV  Spearman  :  0.99989
validation  RMSE/MAE:  21.86  13.10  meV  Spearman  :  0.99982
testing     RMSE/MAE:  52.89  16.52  meV  Spearman  :  0.99984
\end{verbatim}
\clearpage

\flushleft{
\subsection{Os-H}}
\subsubsection*{Searching}
\centering
\includegraphics[width=0.4\textwidth]{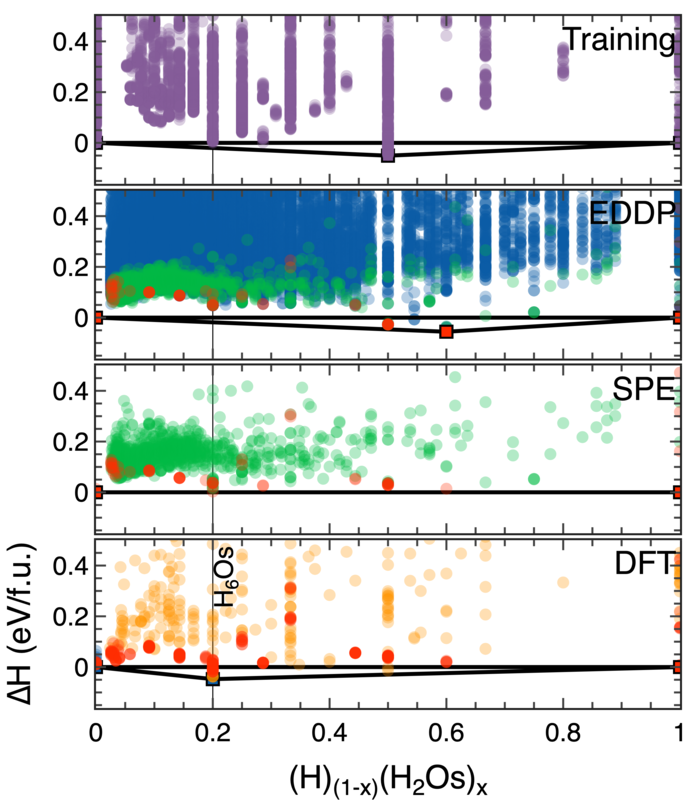}
\footnotesize


\flushleft{
\subsubsection*{\textsc{EDDP}}}
\centering
\includegraphics[width=0.3\textwidth]{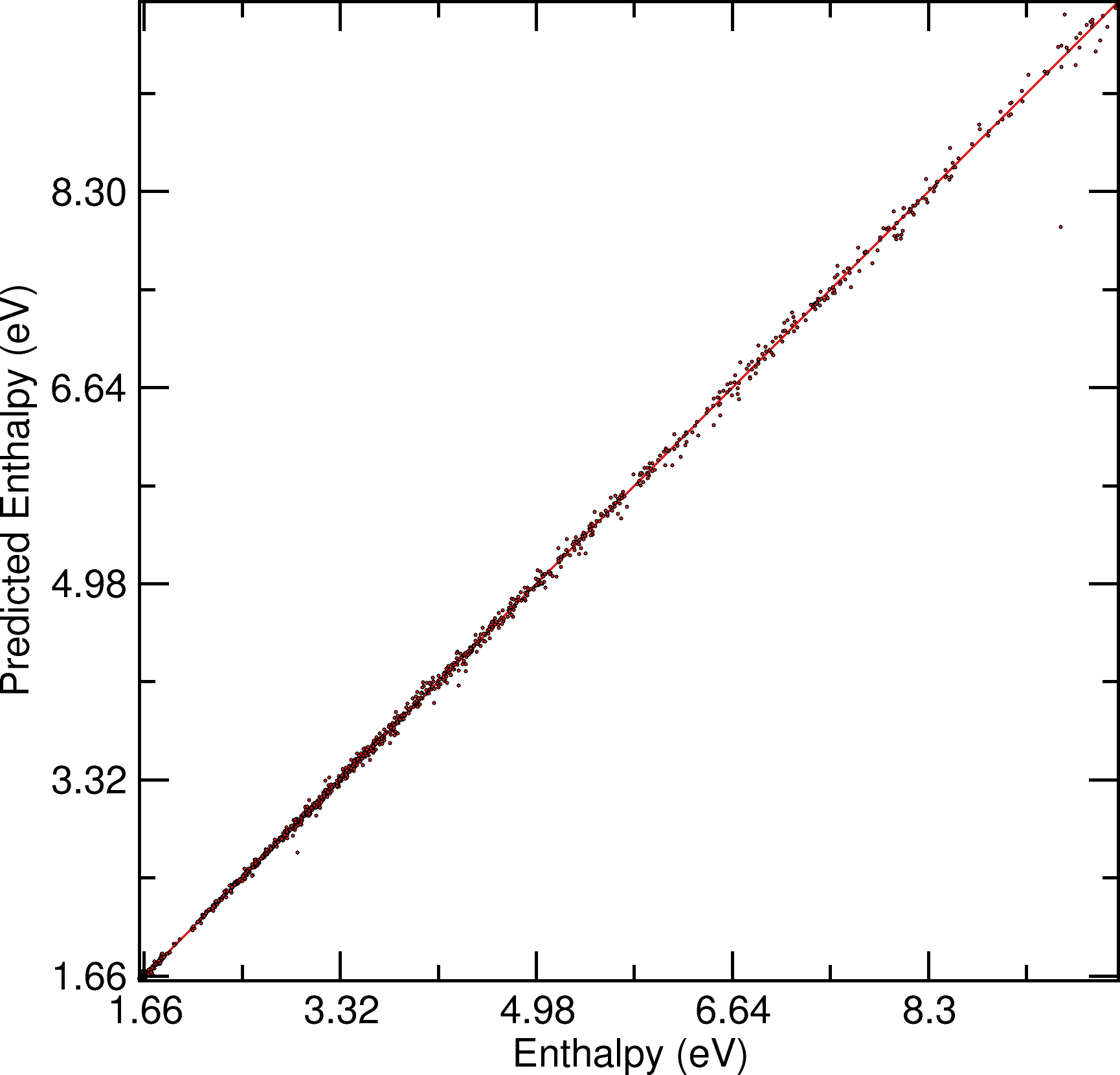}
\includegraphics[width=0.3\textwidth]{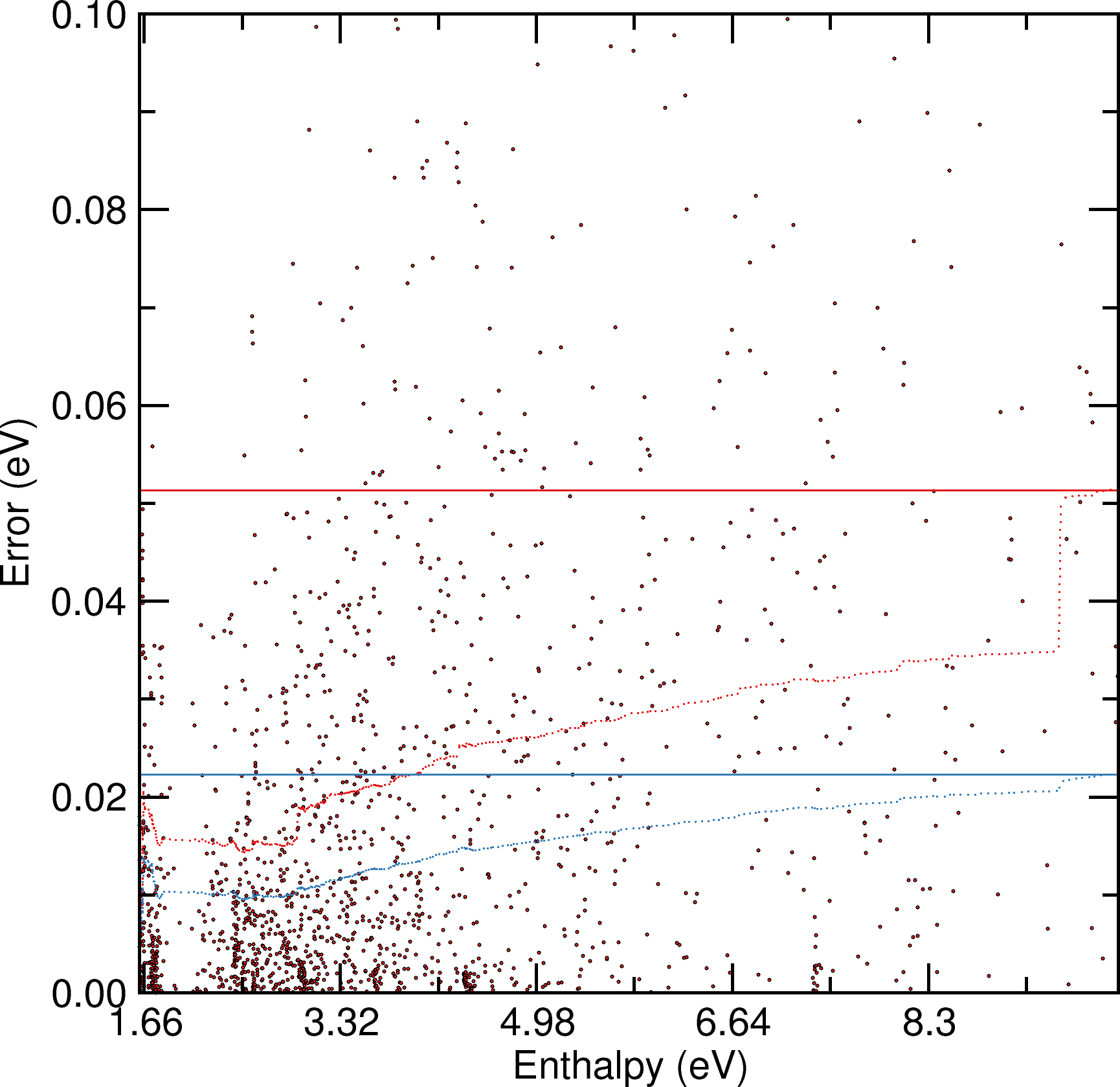}
\centering\begin{verbatim}
training    RMSE/MAE:  22.94  13.52  meV  Spearman  :  0.99982
validation  RMSE/MAE:  32.47  19.64  meV  Spearman  :  0.99978
testing     RMSE/MAE:  51.33  22.31  meV  Spearman  :  0.99973
\end{verbatim}
\clearpage

\flushleft{
\subsection{P-H}}
\subsubsection*{Searching}
\centering
\includegraphics[width=0.4\textwidth]{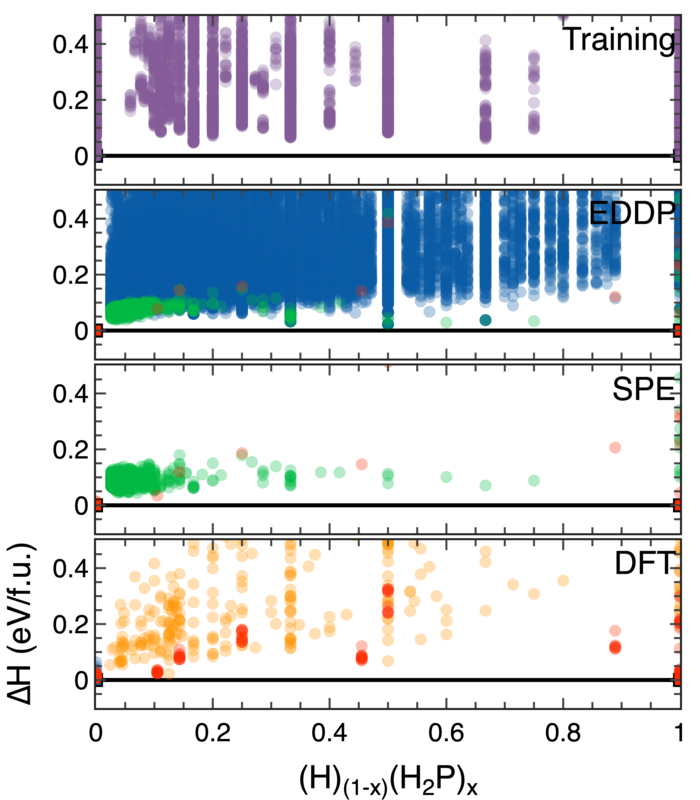}
\footnotesize


\flushleft{
\subsubsection*{\textsc{EDDP}}}
\centering
\includegraphics[width=0.3\textwidth]{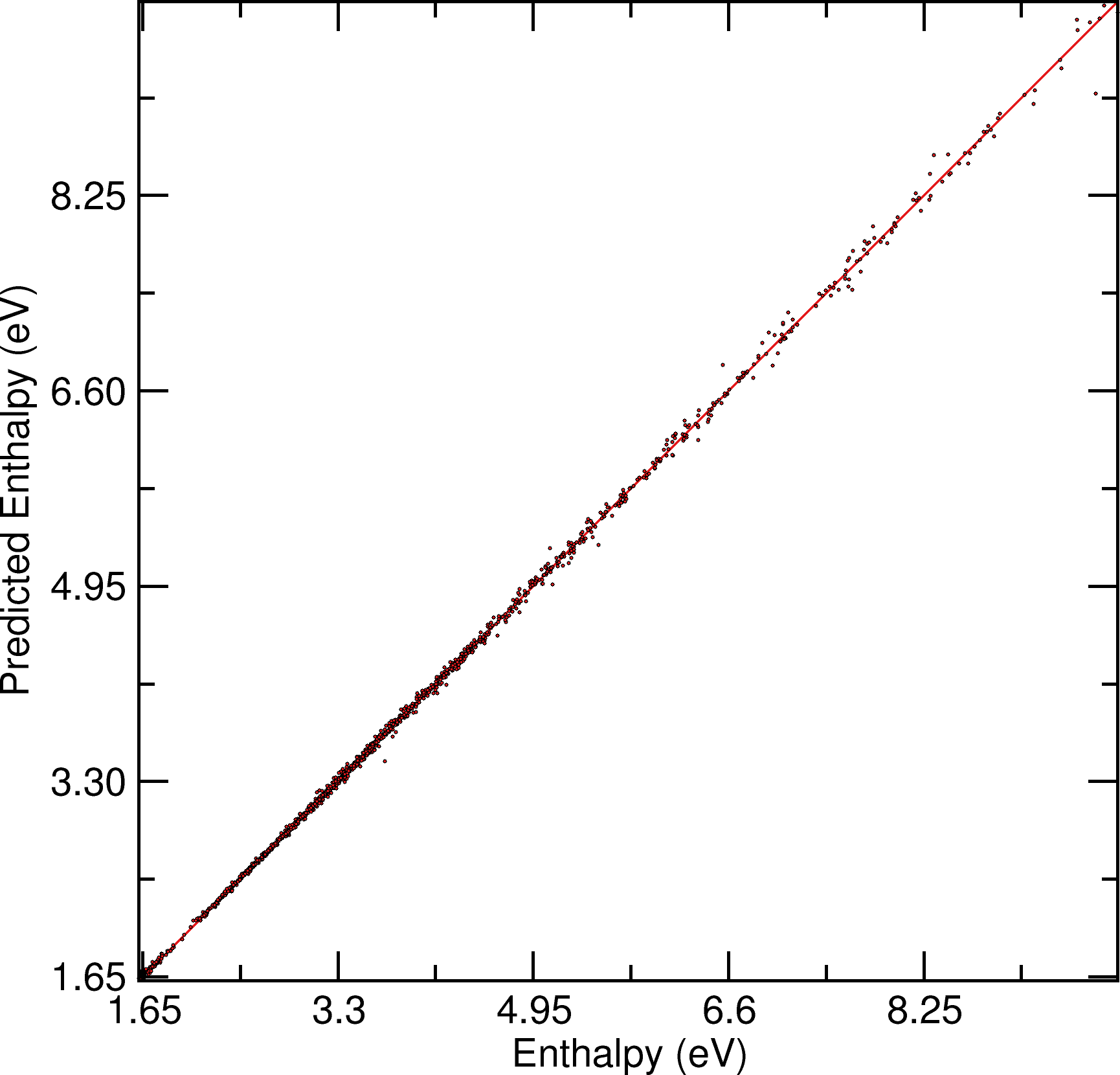}
\includegraphics[width=0.3\textwidth]{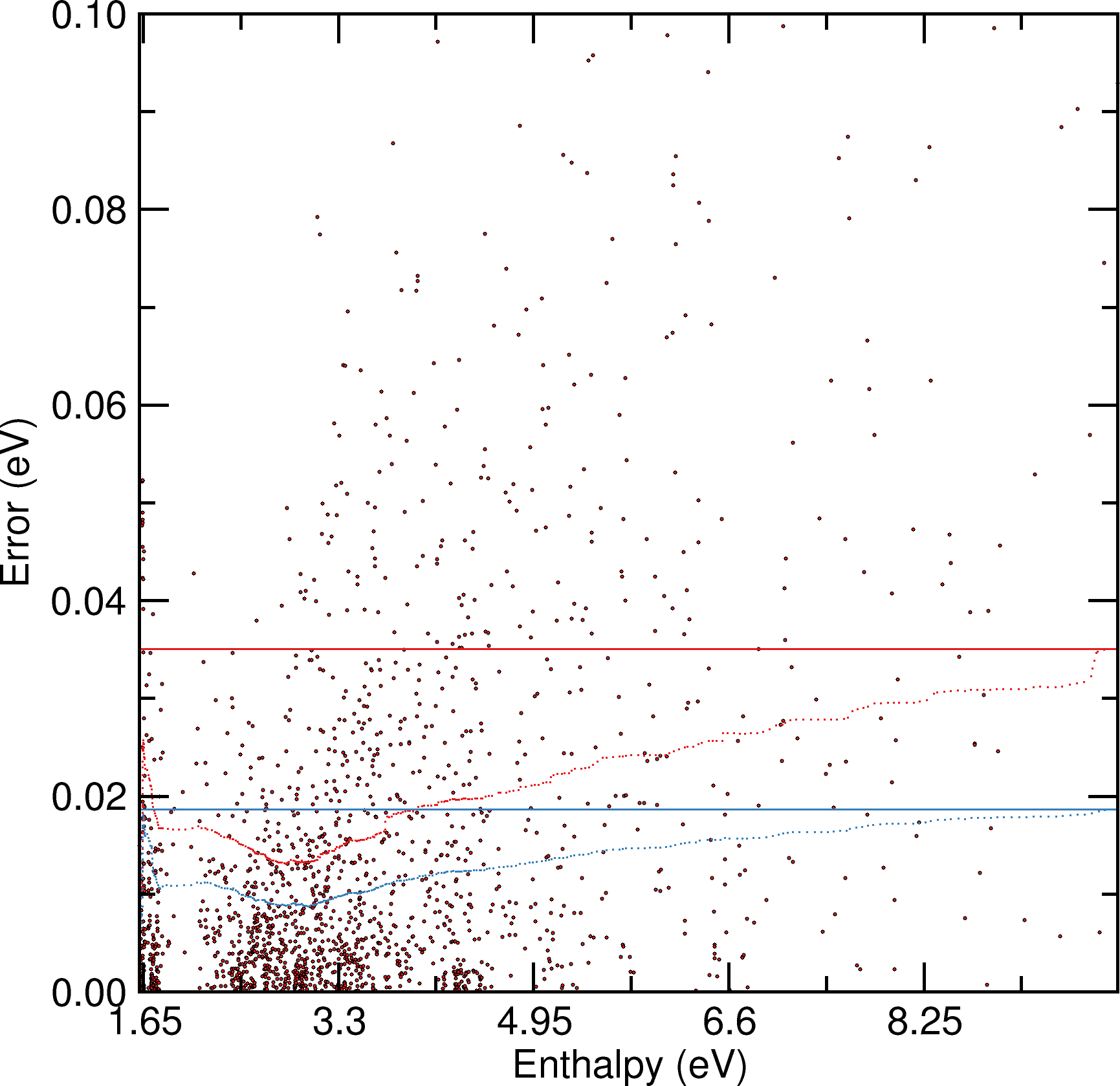}
\centering\begin{verbatim}
training    RMSE/MAE:  19.90  12.15  meV  Spearman  :  0.99984
validation  RMSE/MAE:  30.96  17.36  meV  Spearman  :  0.99980
testing     RMSE/MAE:  35.05  18.62  meV  Spearman  :  0.99976
\end{verbatim}
\clearpage

\flushleft{
\subsection{Pb-H}}
\subsubsection*{Searching}
\centering
\includegraphics[width=0.4\textwidth]{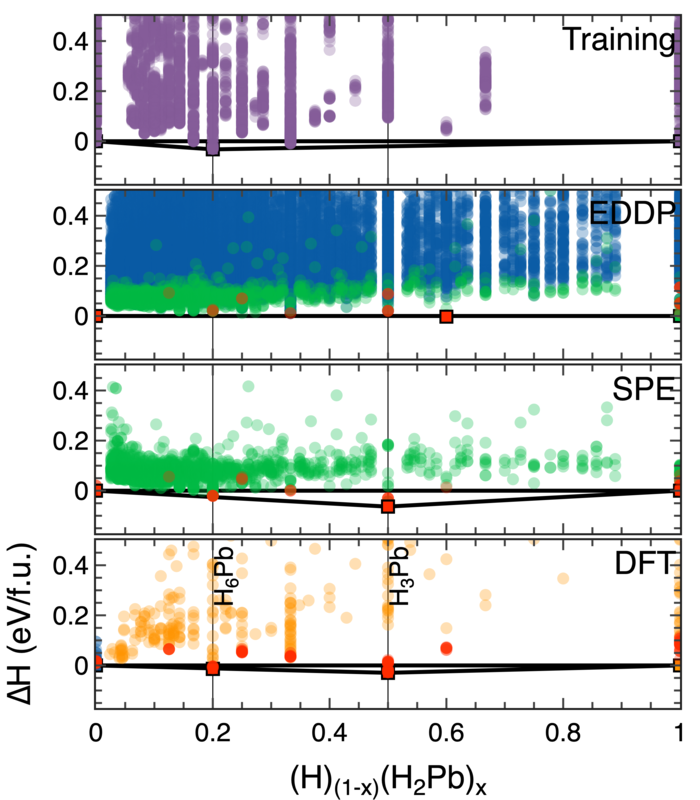}
\footnotesize


\flushleft{
\subsubsection*{\textsc{EDDP}}}
\centering
\includegraphics[width=0.3\textwidth]{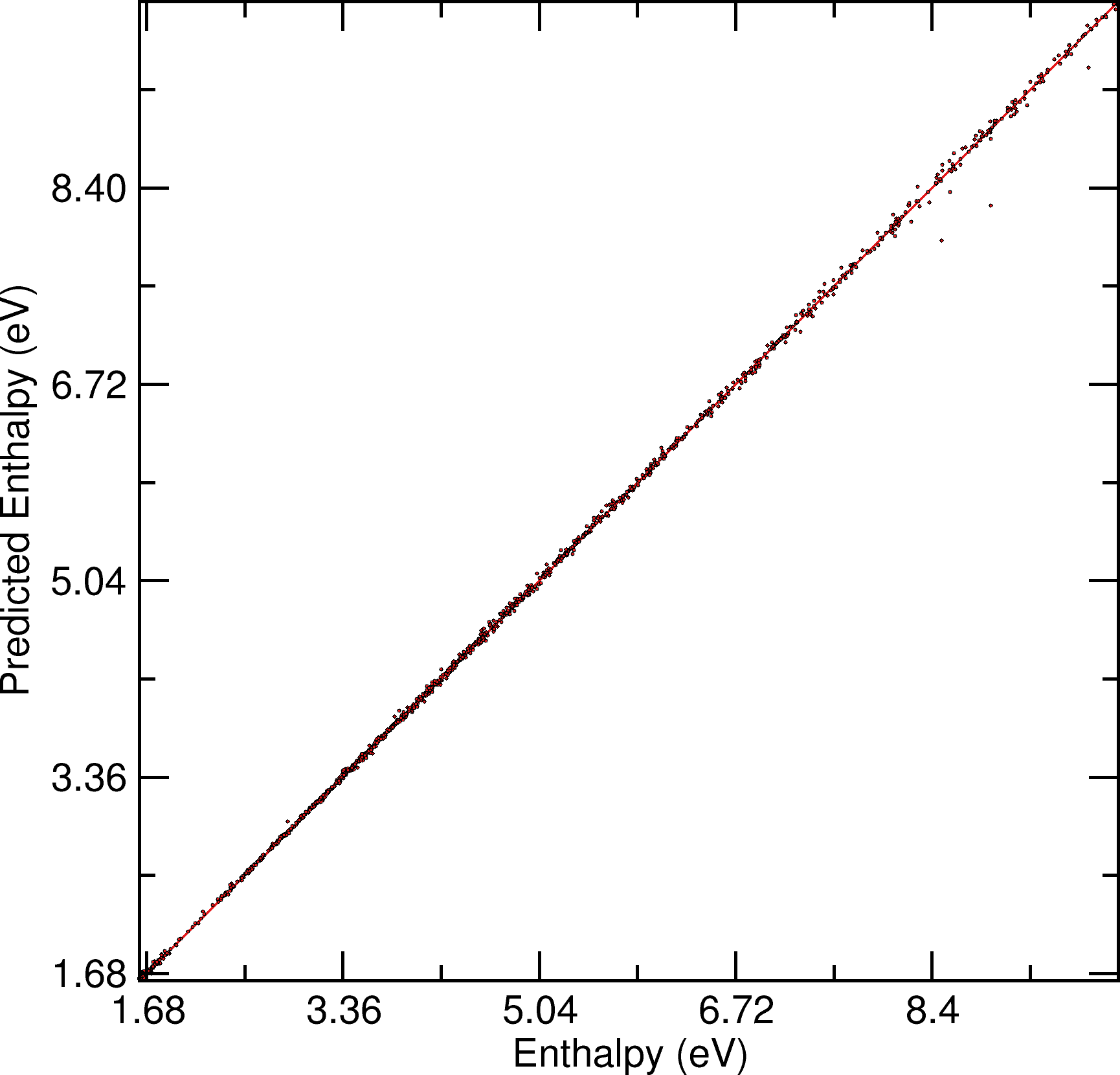}
\includegraphics[width=0.3\textwidth]{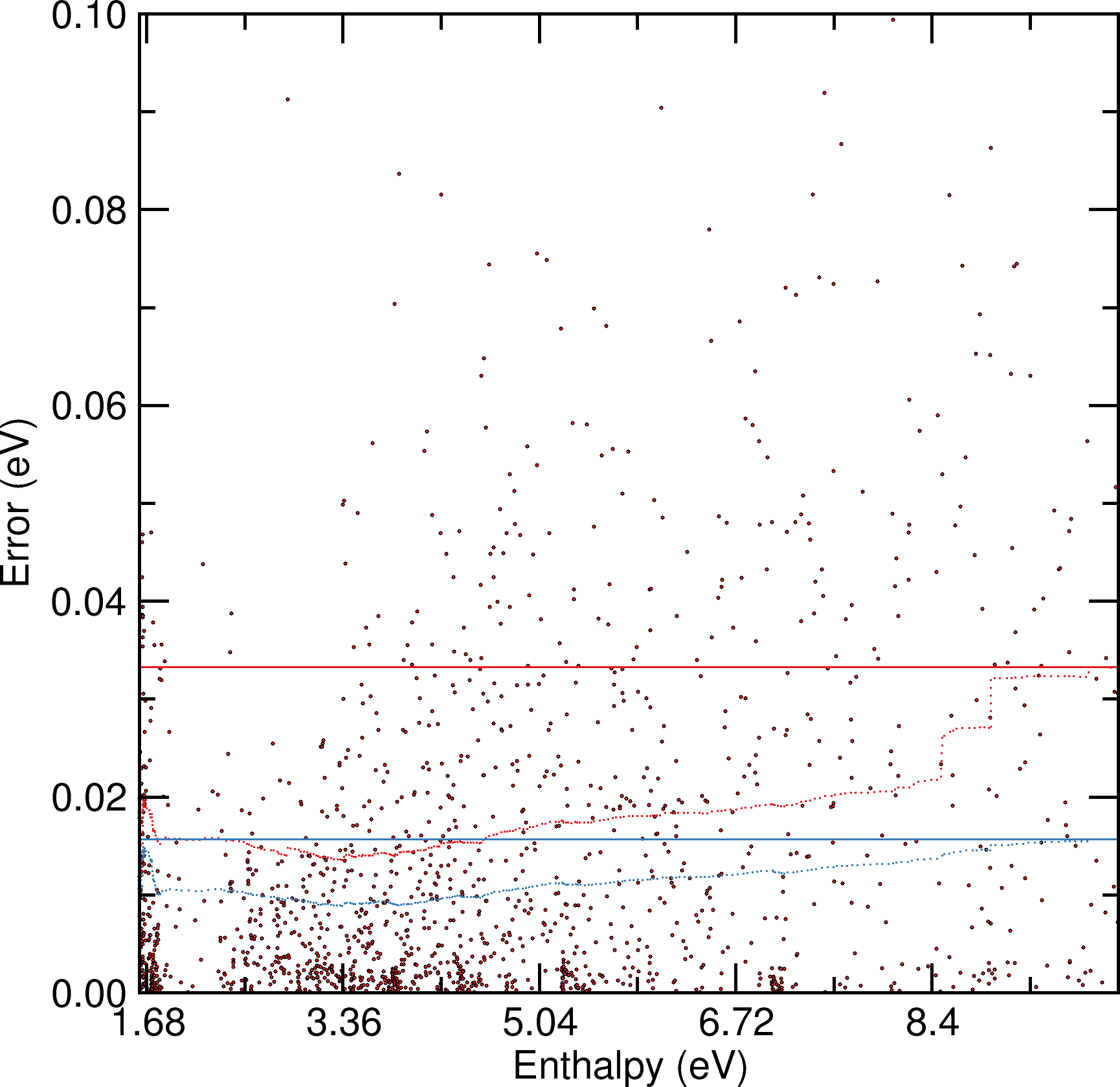}
\centering\begin{verbatim}
training    RMSE/MAE:  15.31  9.41   meV  Spearman  :  0.99989
validation  RMSE/MAE:  22.47  13.89  meV  Spearman  :  0.99987
testing     RMSE/MAE:  33.24  15.70  meV  Spearman  :  0.99982
\end{verbatim}
\clearpage

\flushleft{
\subsection{Pd-H}}
\subsubsection*{Searching}
\centering
\includegraphics[width=0.4\textwidth]{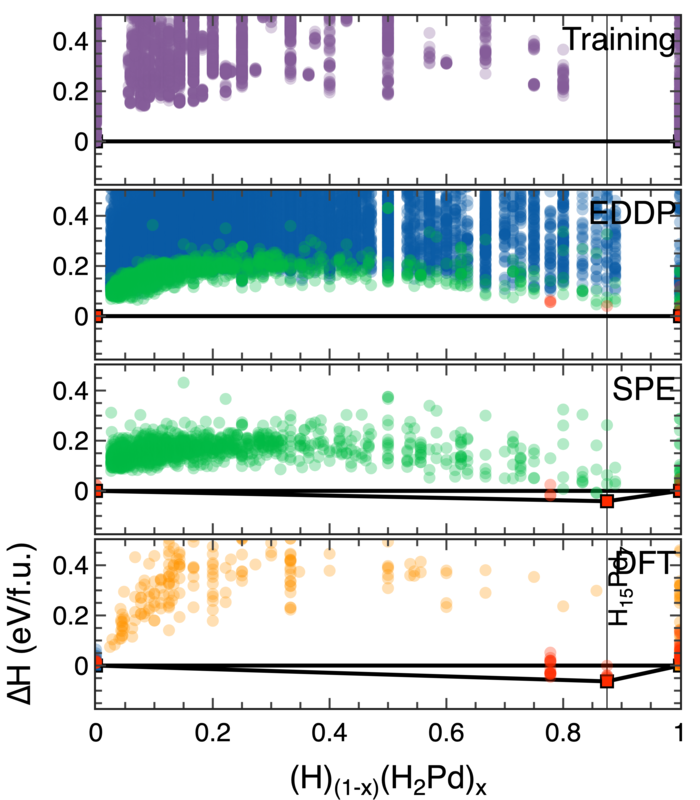}
\footnotesize


\flushleft{
\subsubsection*{\textsc{EDDP}}}
\centering
\includegraphics[width=0.3\textwidth]{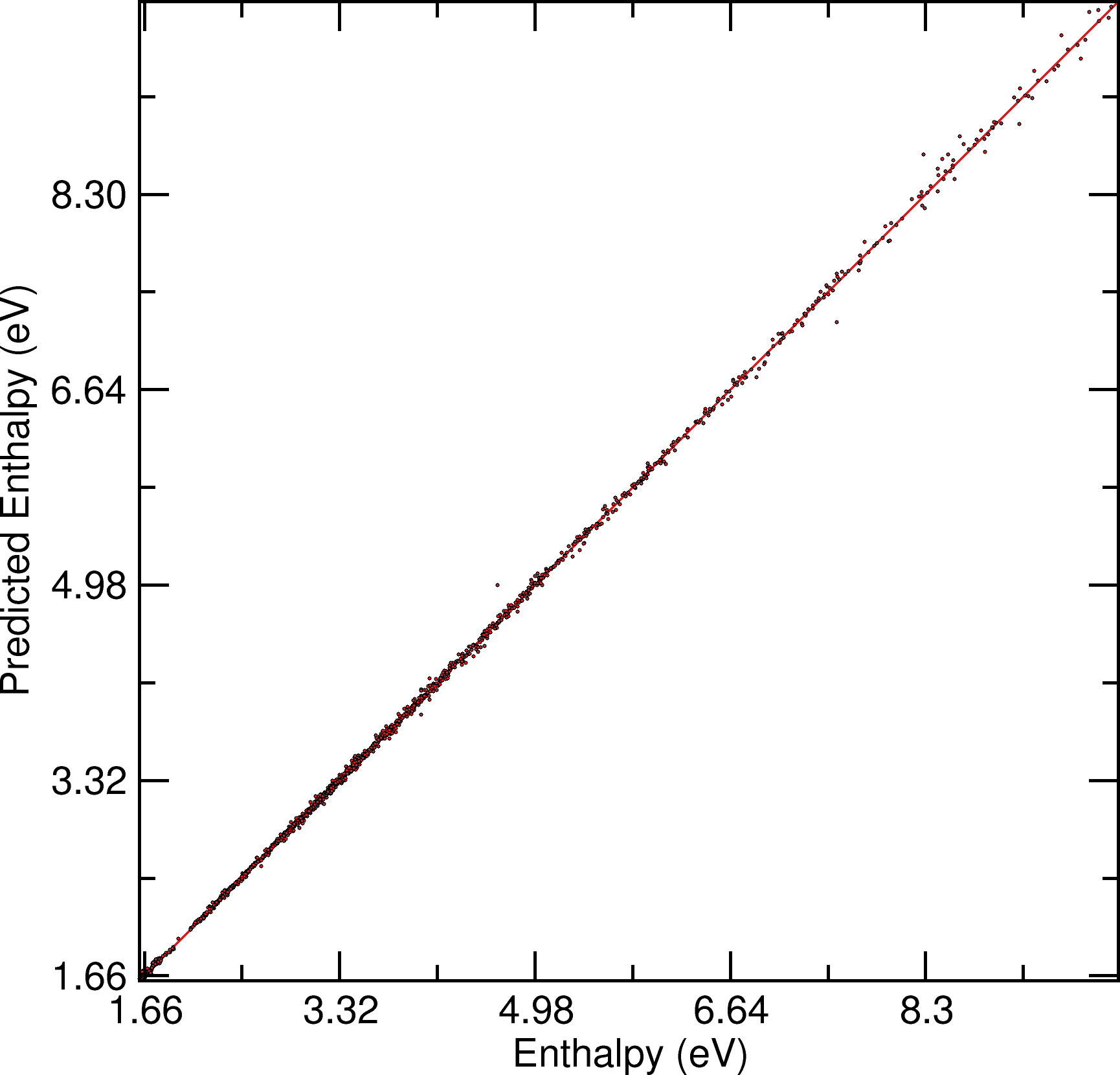}
\includegraphics[width=0.3\textwidth]{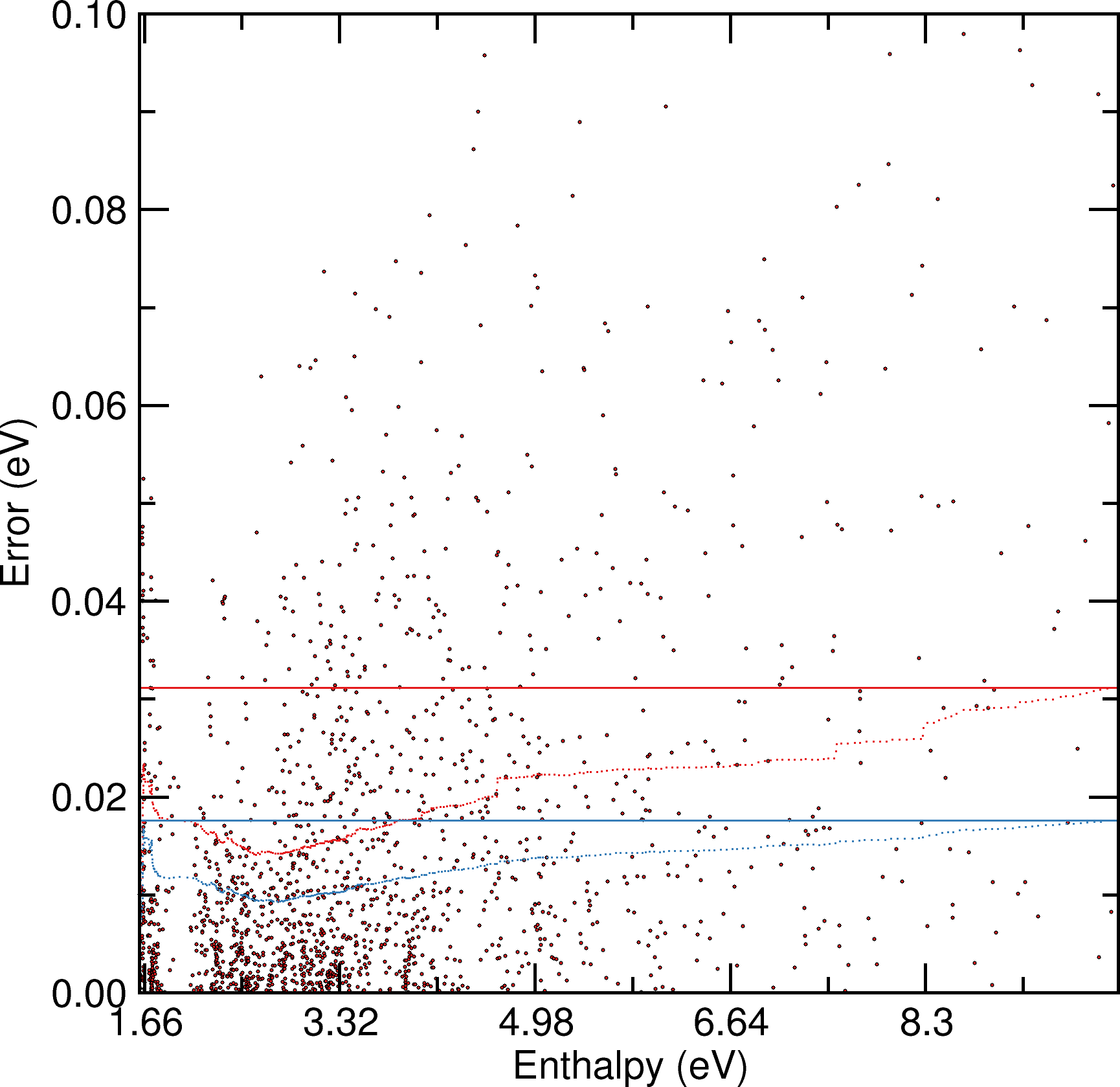}
\centering\begin{verbatim}
training    RMSE/MAE:  16.91  11.07  meV  Spearman  :  0.99985
validation  RMSE/MAE:  24.60  15.14  meV  Spearman  :  0.99985
testing     RMSE/MAE:  31.17  17.61  meV  Spearman  :  0.99978
\end{verbatim}
\clearpage

\flushleft{
\subsection{Pm-H}}
\subsubsection*{Searching}
\centering
\includegraphics[width=0.4\textwidth]{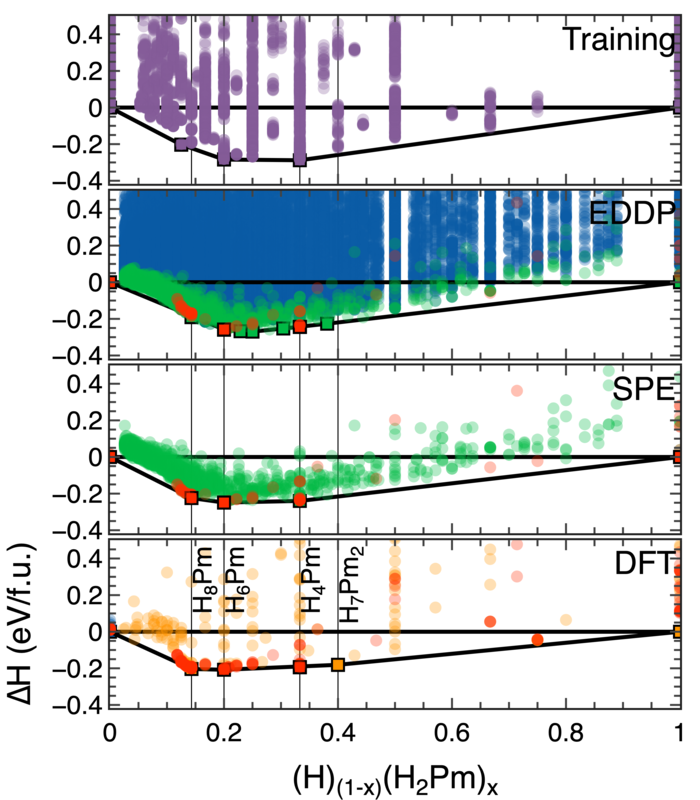}
\footnotesize


\flushleft{
\subsubsection*{\textsc{EDDP}}}
\centering
\includegraphics[width=0.3\textwidth]{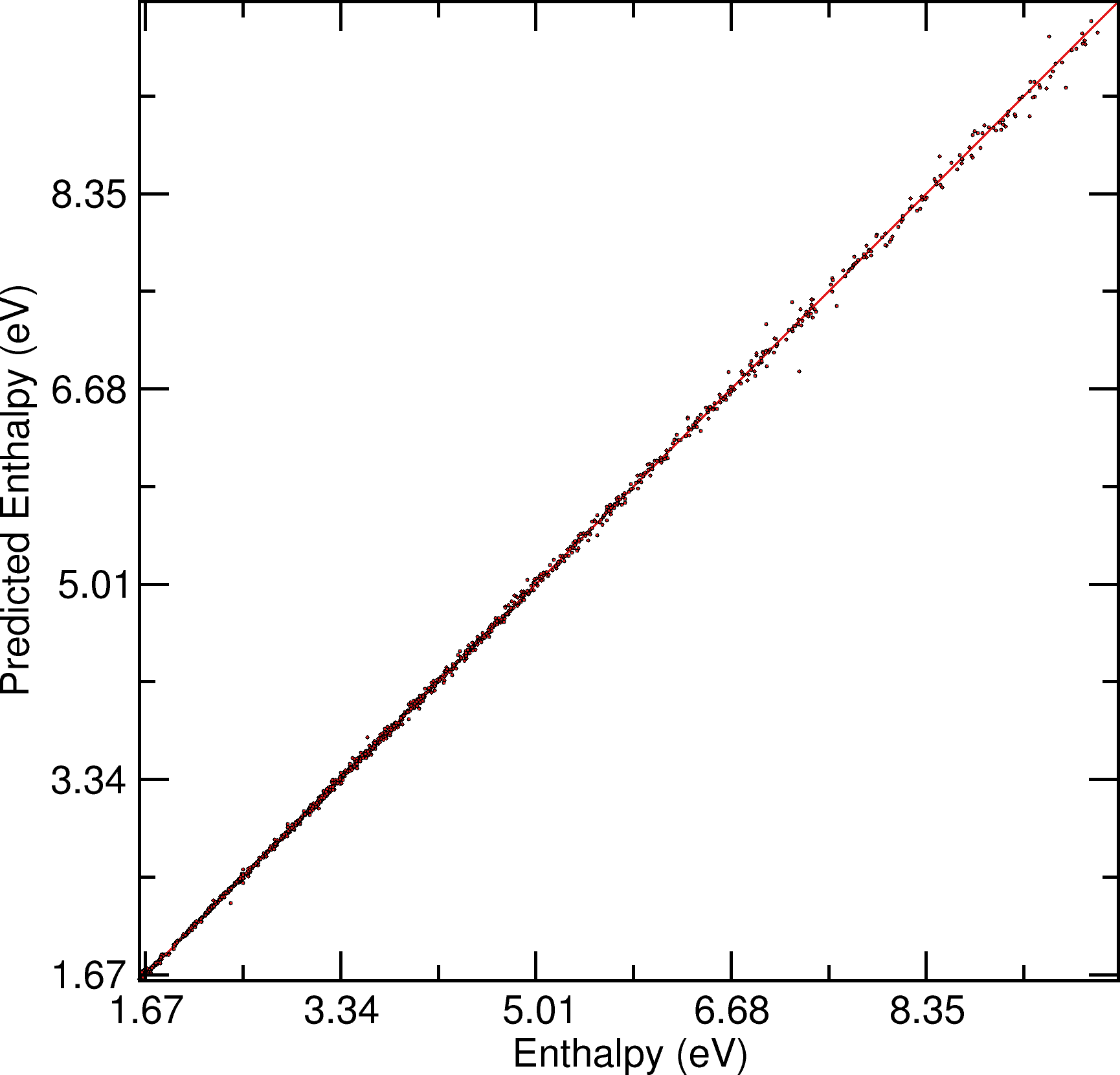}
\includegraphics[width=0.3\textwidth]{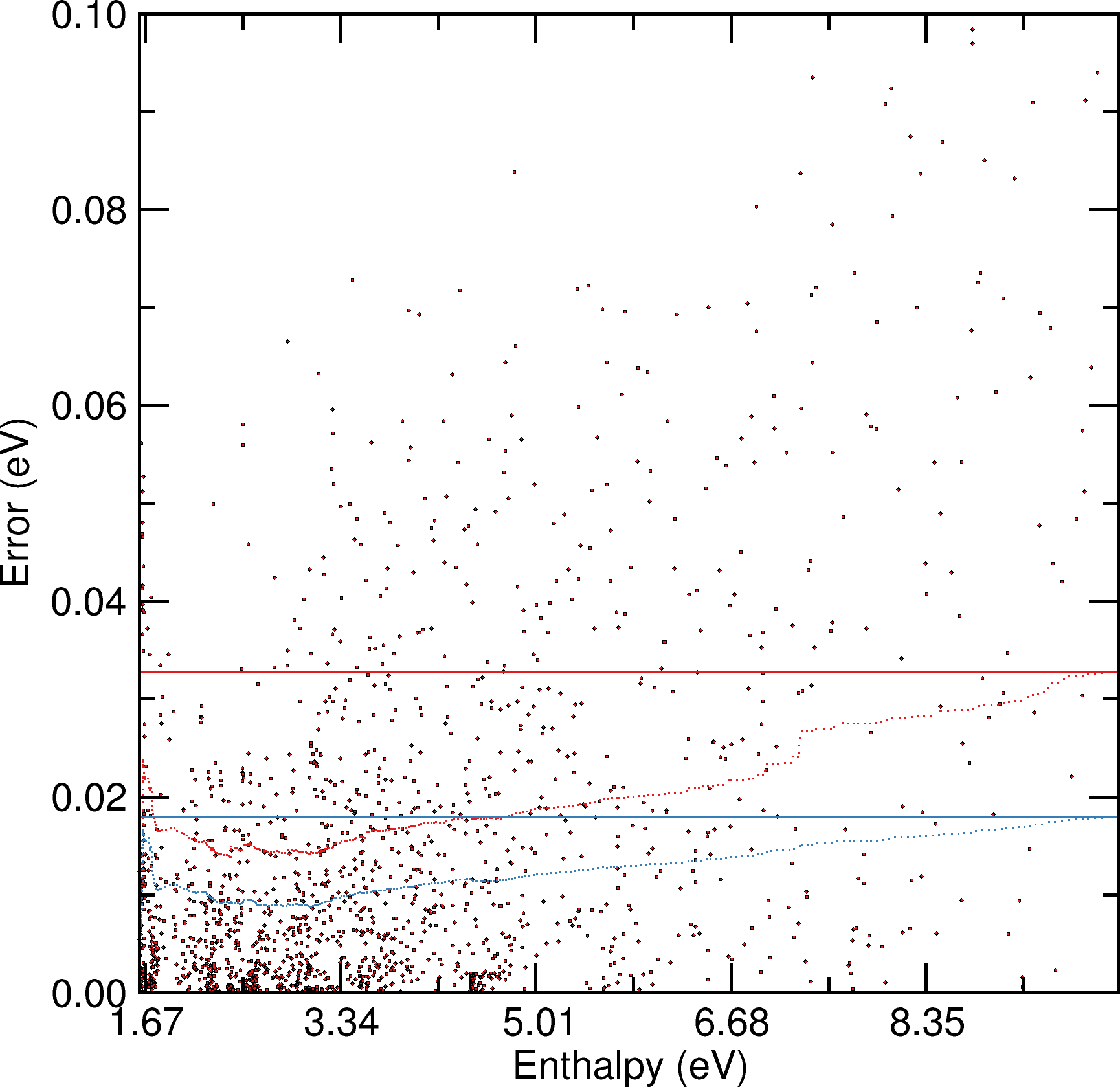}
\centering\begin{verbatim}
training    RMSE/MAE:  17.34  10.95  meV  Spearman  :  0.99989
validation  RMSE/MAE:  29.13  17.52  meV  Spearman  :  0.99982
testing     RMSE/MAE:  32.81  17.96  meV  Spearman  :  0.99985
\end{verbatim}
\clearpage

\flushleft{
\subsection{Pr-H}}
\subsubsection*{Searching}
\centering
\includegraphics[width=0.4\textwidth]{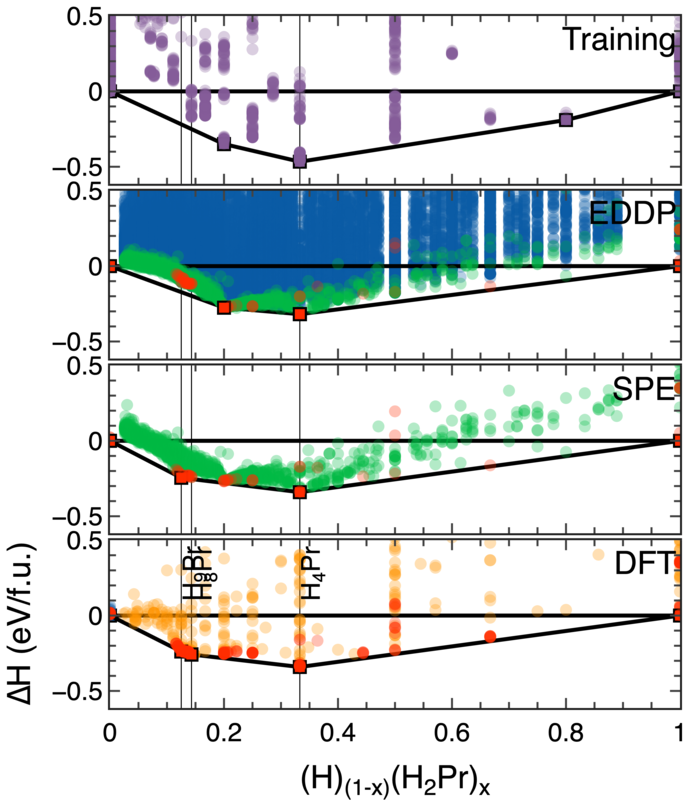}
\footnotesize


\flushleft{
\subsubsection*{\textsc{EDDP}}}
\centering
\includegraphics[width=0.3\textwidth]{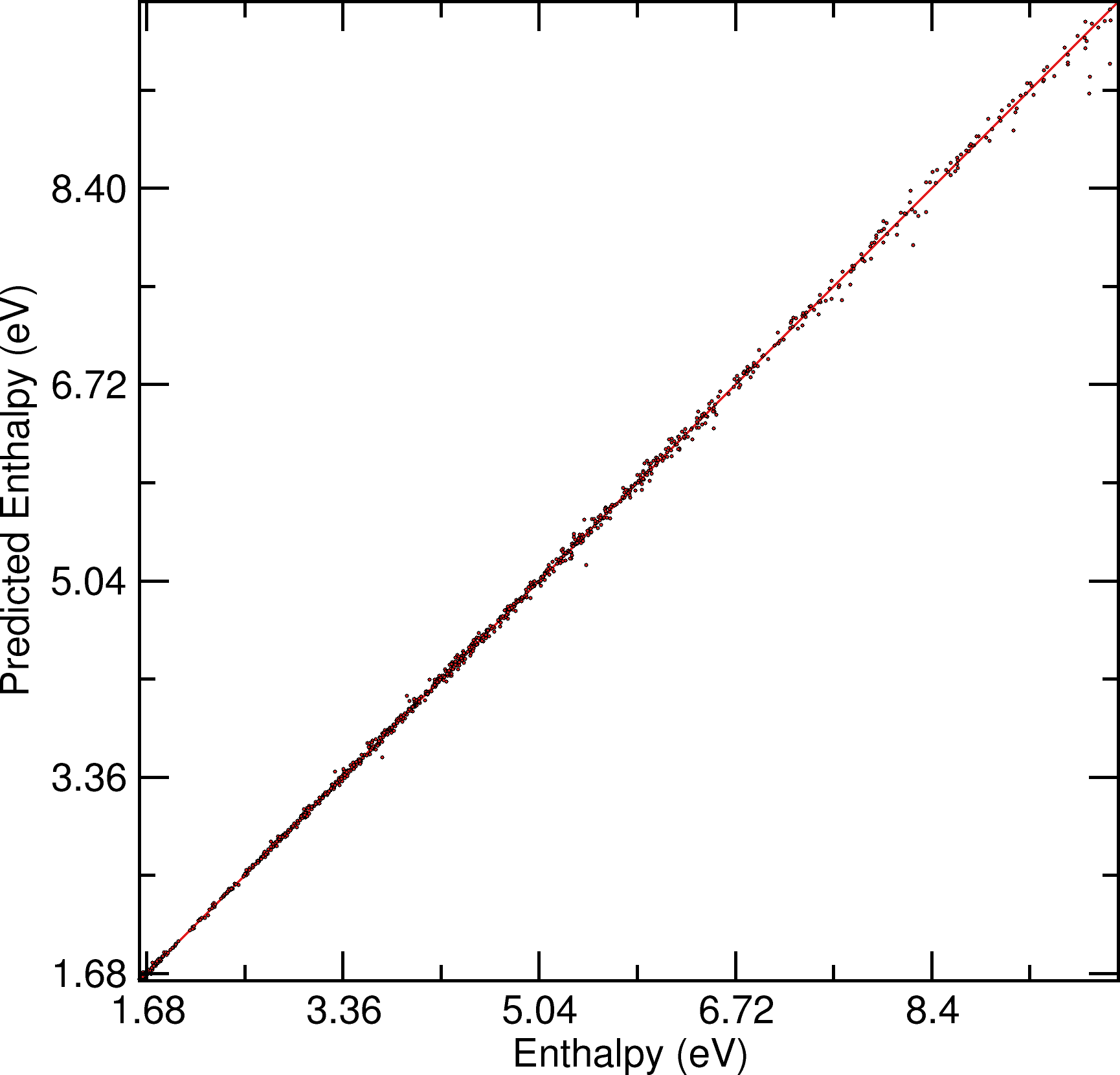}
\includegraphics[width=0.3\textwidth]{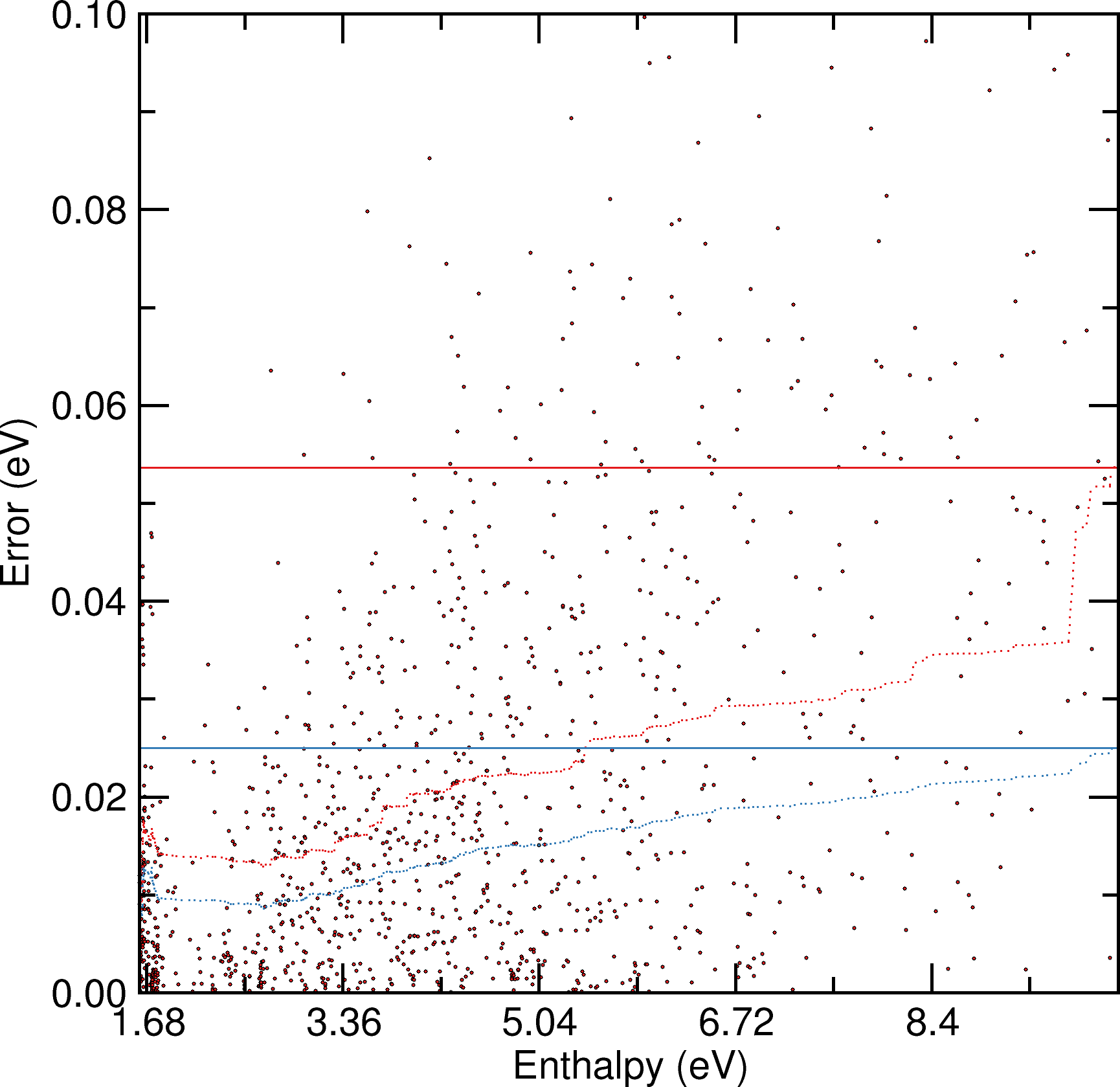}
\centering\begin{verbatim}
training    RMSE/MAE:  18.73  12.43  meV  Spearman  :  0.99978
validation  RMSE/MAE:  37.45  22.42  meV  Spearman  :  0.99968
testing     RMSE/MAE:  53.66  24.99  meV  Spearman  :  0.99972
\end{verbatim}
\clearpage

\flushleft{
\subsection{Pt-H}}
\subsubsection*{Searching}
\centering
\includegraphics[width=0.4\textwidth]{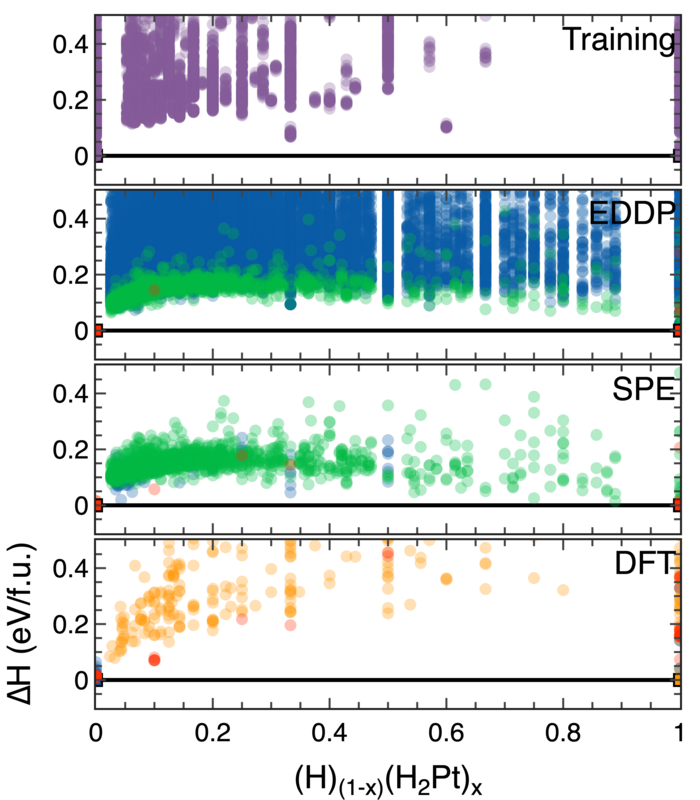}
\footnotesize


\flushleft{
\subsubsection*{\textsc{EDDP}}}
\centering
\includegraphics[width=0.3\textwidth]{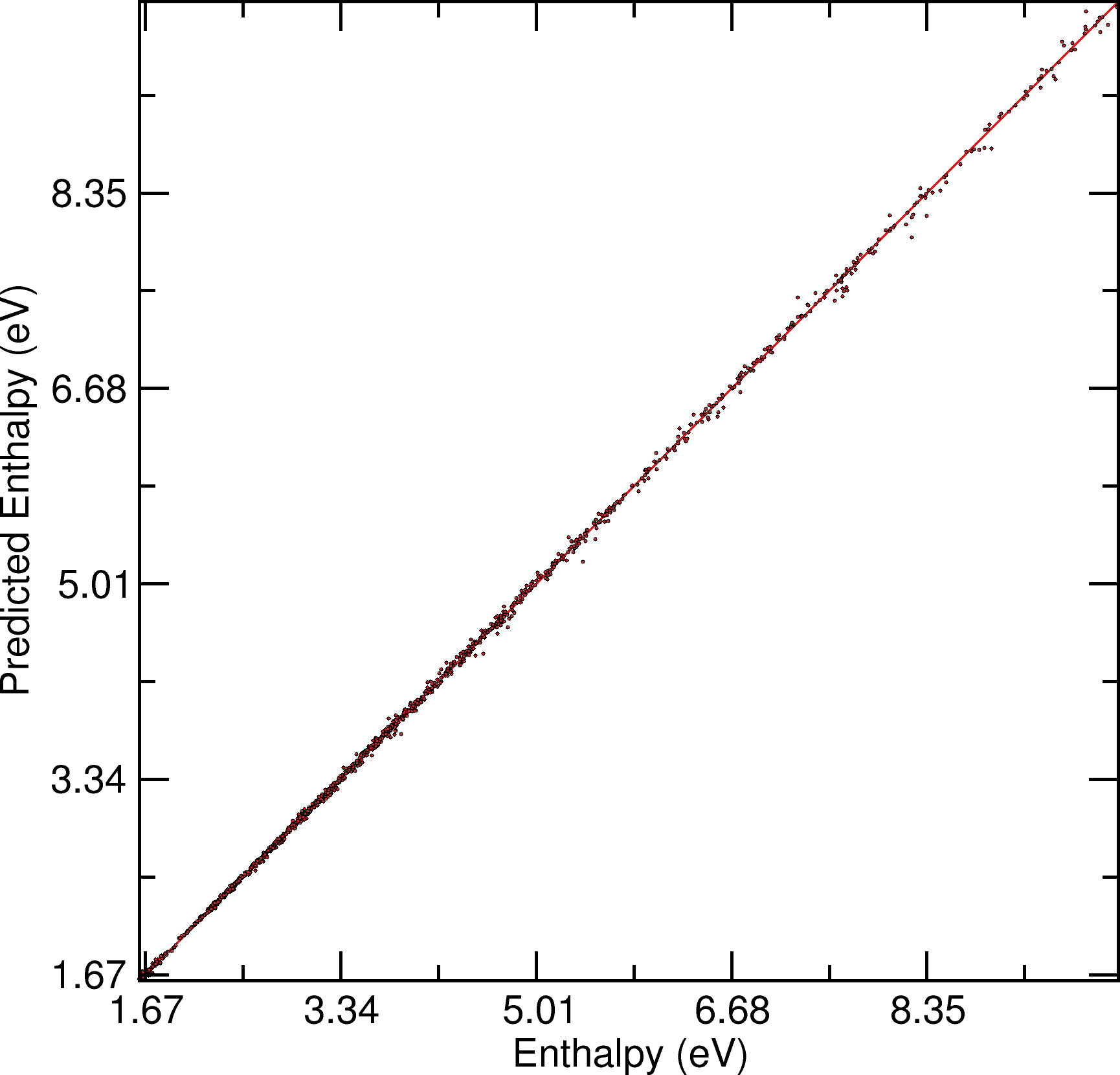}
\includegraphics[width=0.3\textwidth]{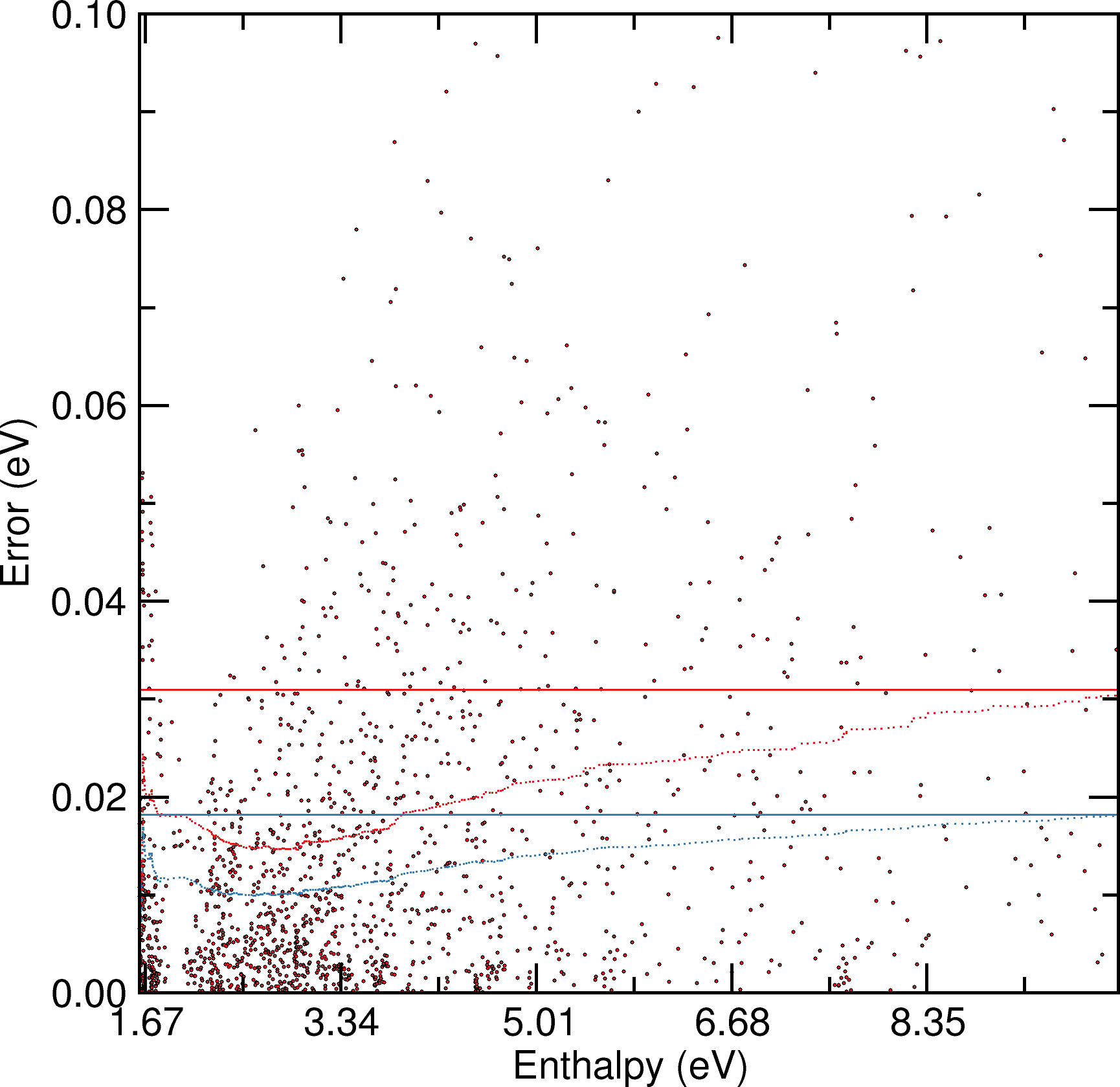}
\centering\begin{verbatim}
training    RMSE/MAE:  18.36  11.50  meV  Spearman  :  0.99987
validation  RMSE/MAE:  27.64  16.81  meV  Spearman  :  0.99978
testing     RMSE/MAE:  30.97  18.22  meV  Spearman  :  0.99973
\end{verbatim}
\clearpage

\flushleft{
\subsection{Rb-H}}
\subsubsection*{Searching}
\centering
\includegraphics[width=0.4\textwidth]{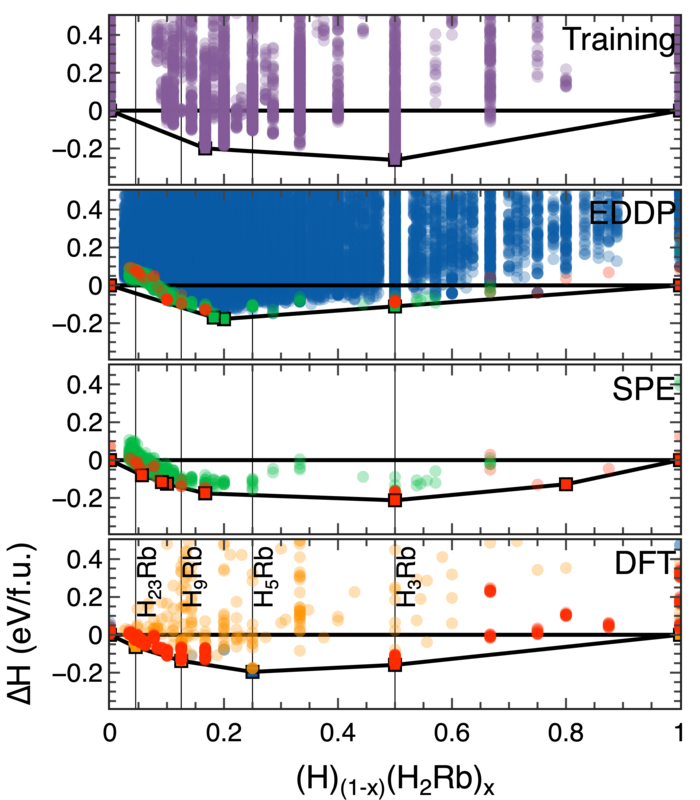}
\footnotesize


\flushleft{
\subsubsection*{\textsc{EDDP}}}
\centering
\includegraphics[width=0.3\textwidth]{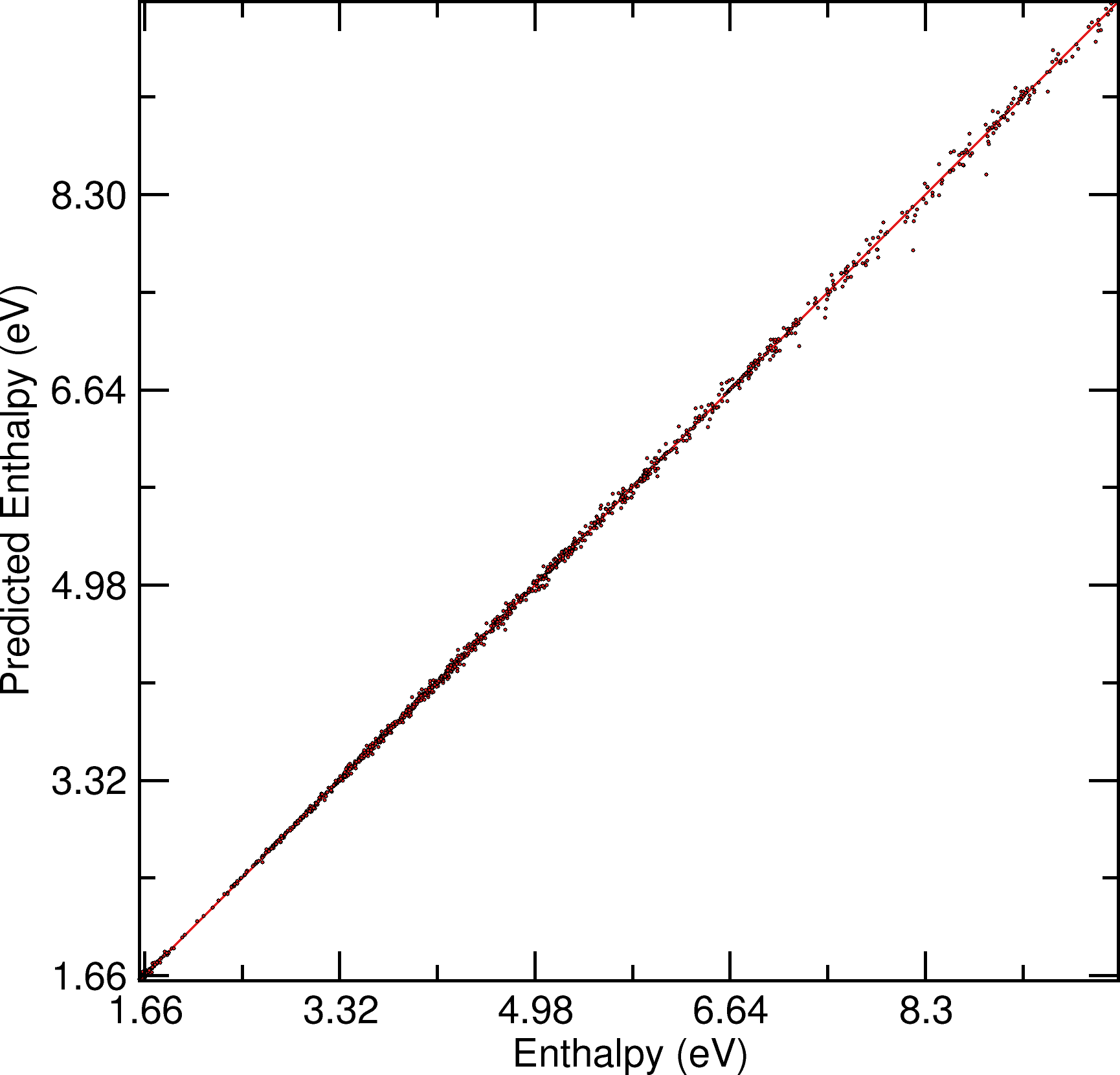}
\includegraphics[width=0.3\textwidth]{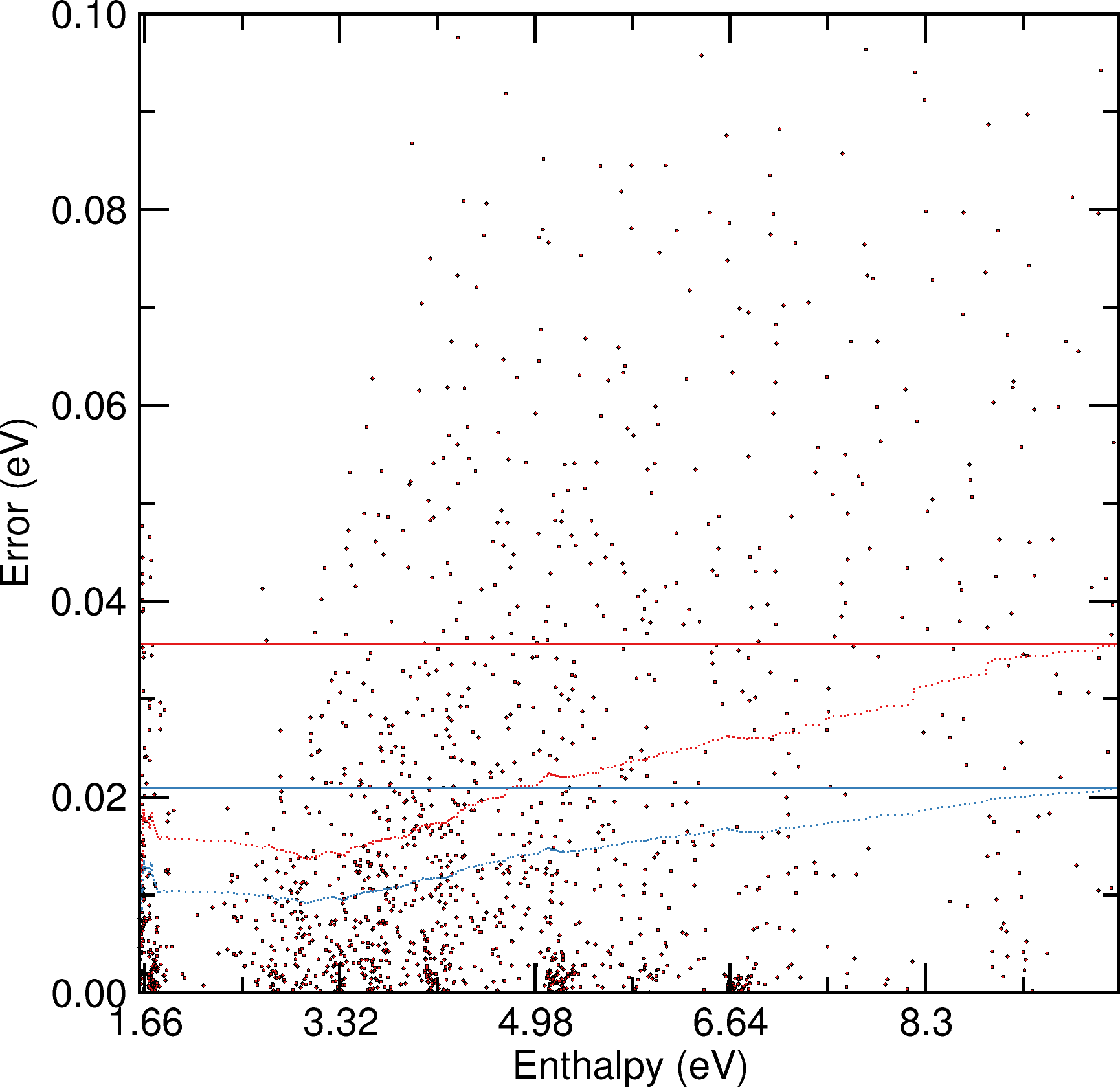}
\centering\begin{verbatim}
training    RMSE/MAE:  22.02  13.04  meV  Spearman  :  0.99986
validation  RMSE/MAE:  33.02  20.02  meV  Spearman  :  0.99978
testing     RMSE/MAE:  35.68  20.90  meV  Spearman  :  0.99977
\end{verbatim}
\clearpage

\flushleft{
\subsection{Re-H}}
\subsubsection*{Searching}
\centering
\includegraphics[width=0.4\textwidth]{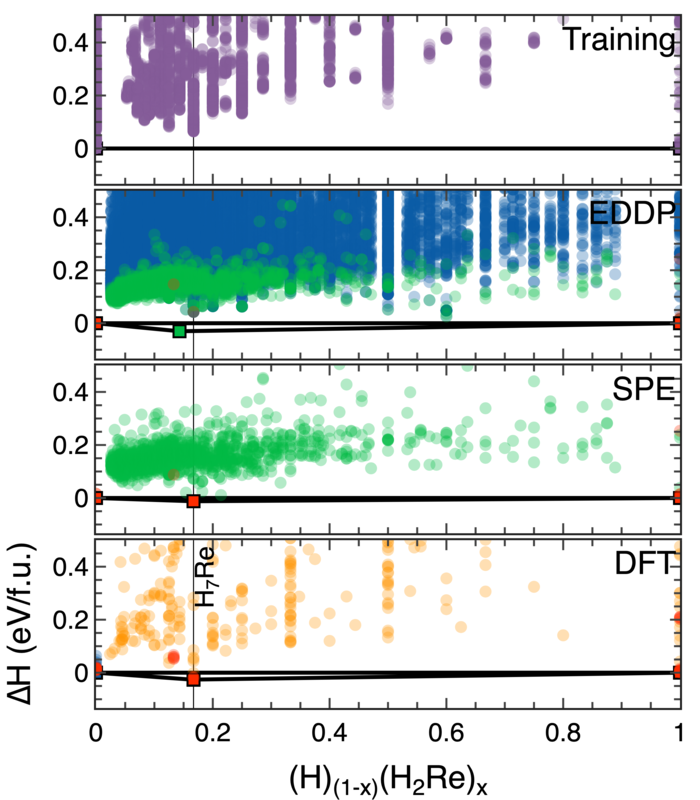}
\footnotesize


\flushleft{
\subsubsection*{\textsc{EDDP}}}
\centering
\includegraphics[width=0.3\textwidth]{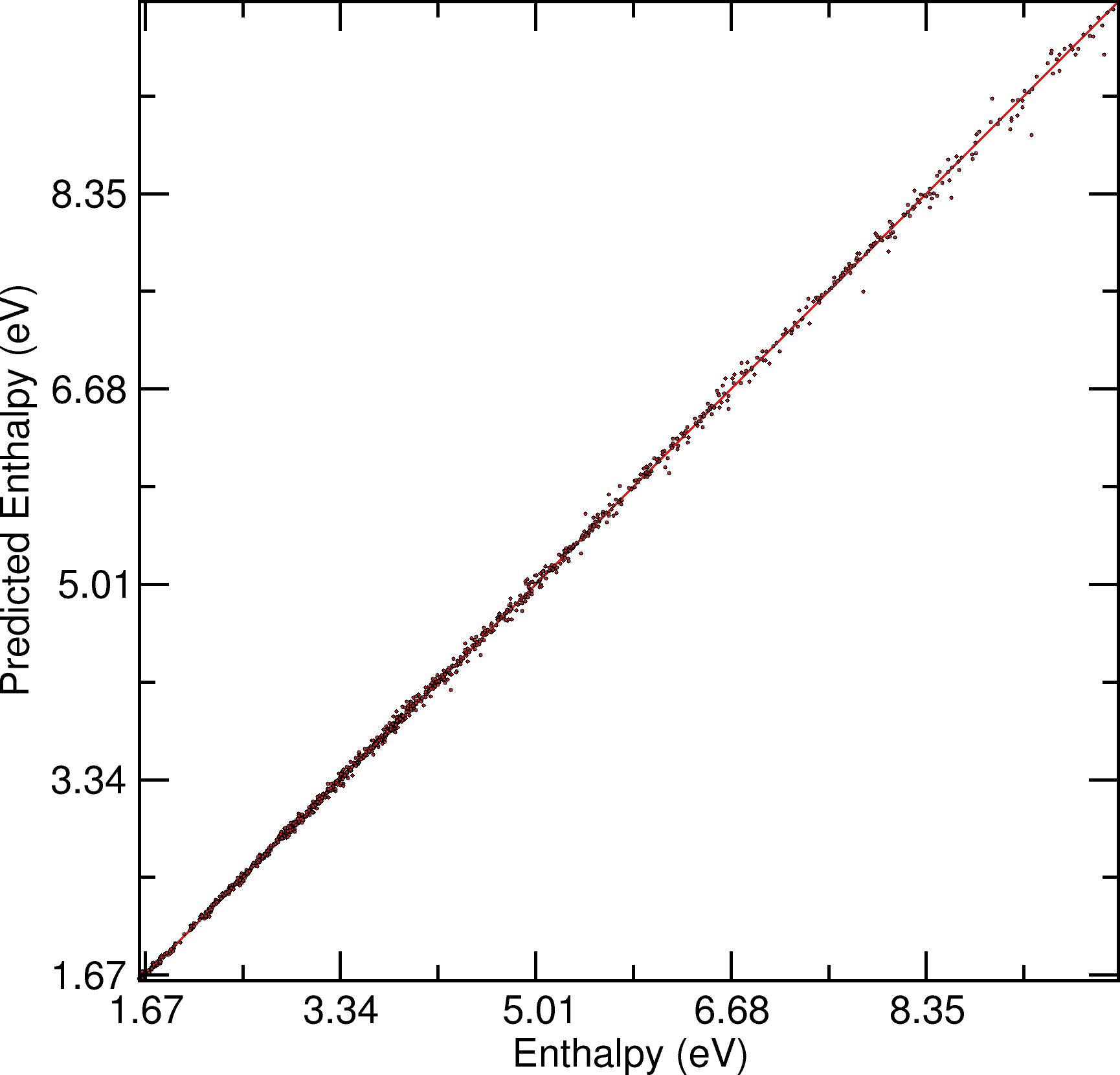}
\includegraphics[width=0.3\textwidth]{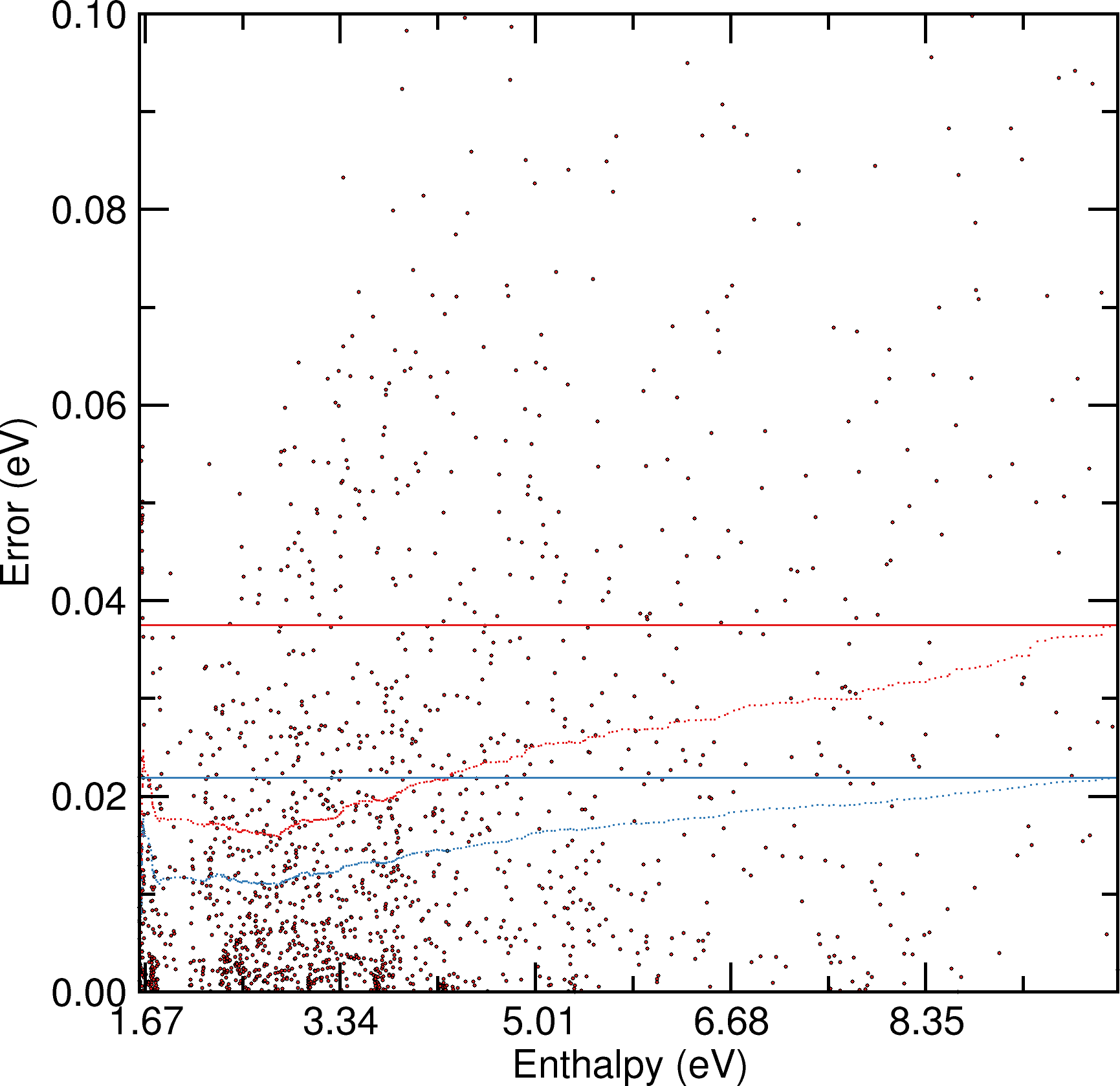}
\centering\begin{verbatim}
training    RMSE/MAE:  23.42  14.27  meV  Spearman  :  0.99981
validation  RMSE/MAE:  32.14  20.13  meV  Spearman  :  0.99977
testing     RMSE/MAE:  37.48  21.90  meV  Spearman  :  0.99977
\end{verbatim}
\clearpage

\flushleft{
\subsection{Rh-H}}
\subsubsection*{Searching}
\centering
\includegraphics[width=0.4\textwidth]{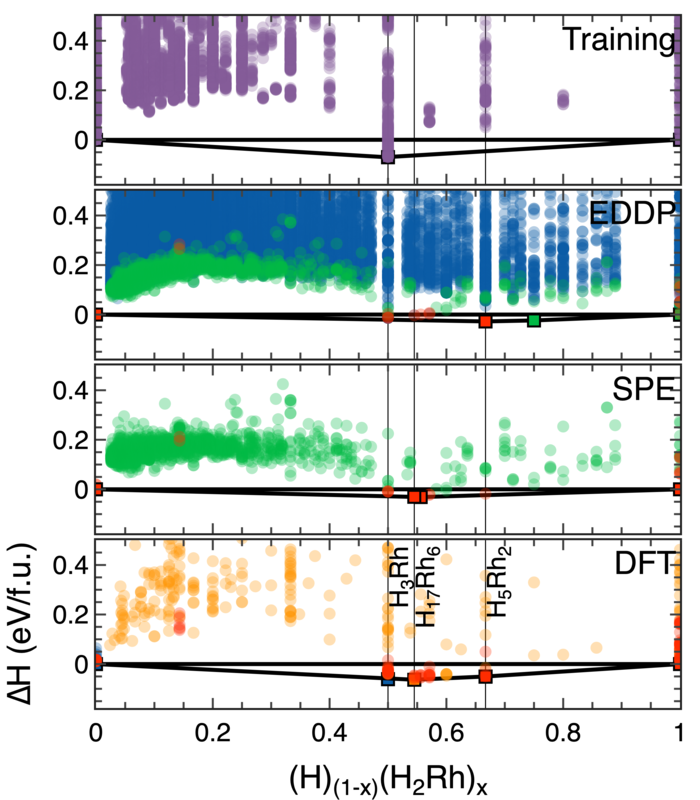}
\footnotesize


\flushleft{
\subsubsection*{\textsc{EDDP}}}
\centering
\includegraphics[width=0.3\textwidth]{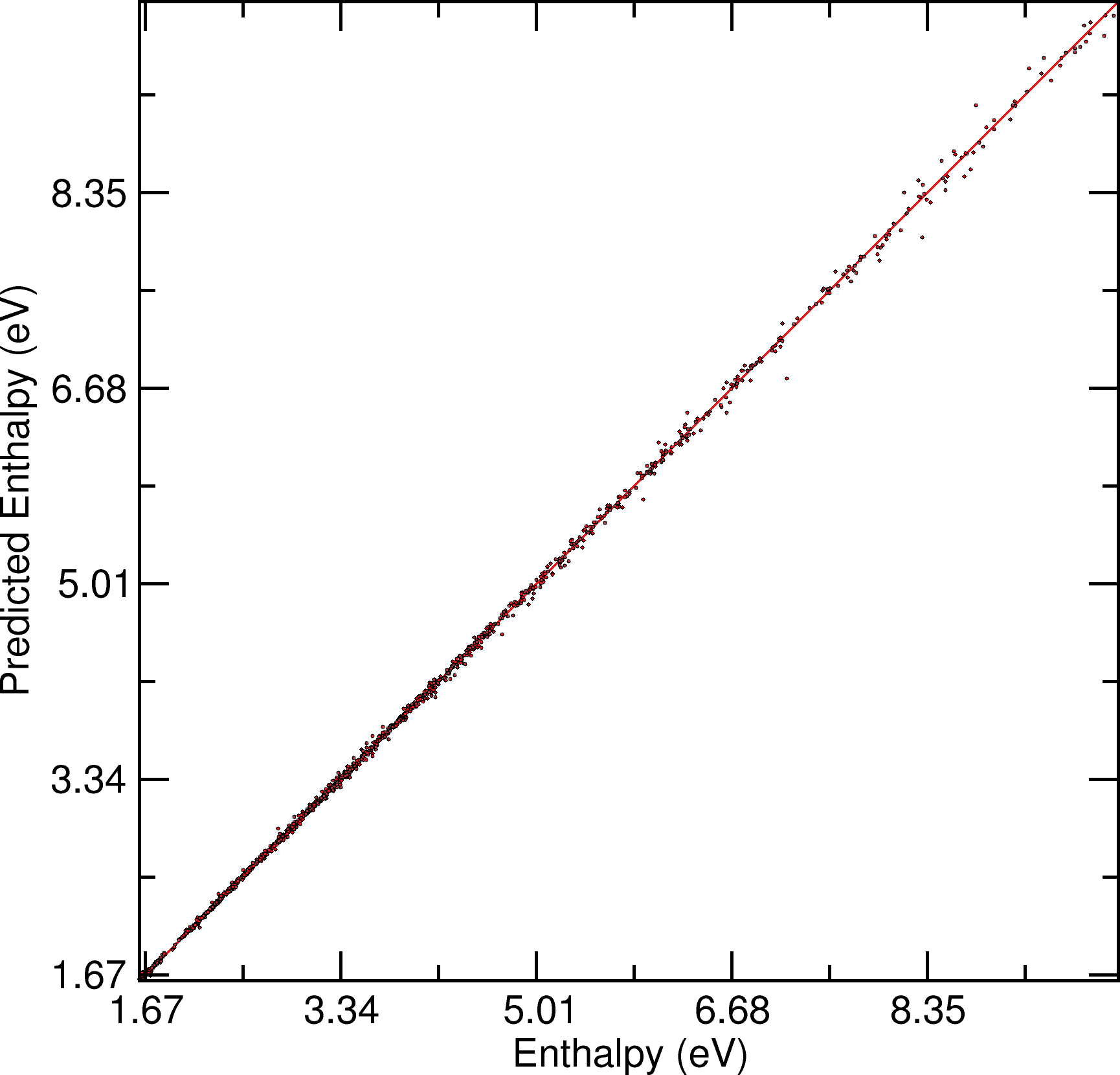}
\includegraphics[width=0.3\textwidth]{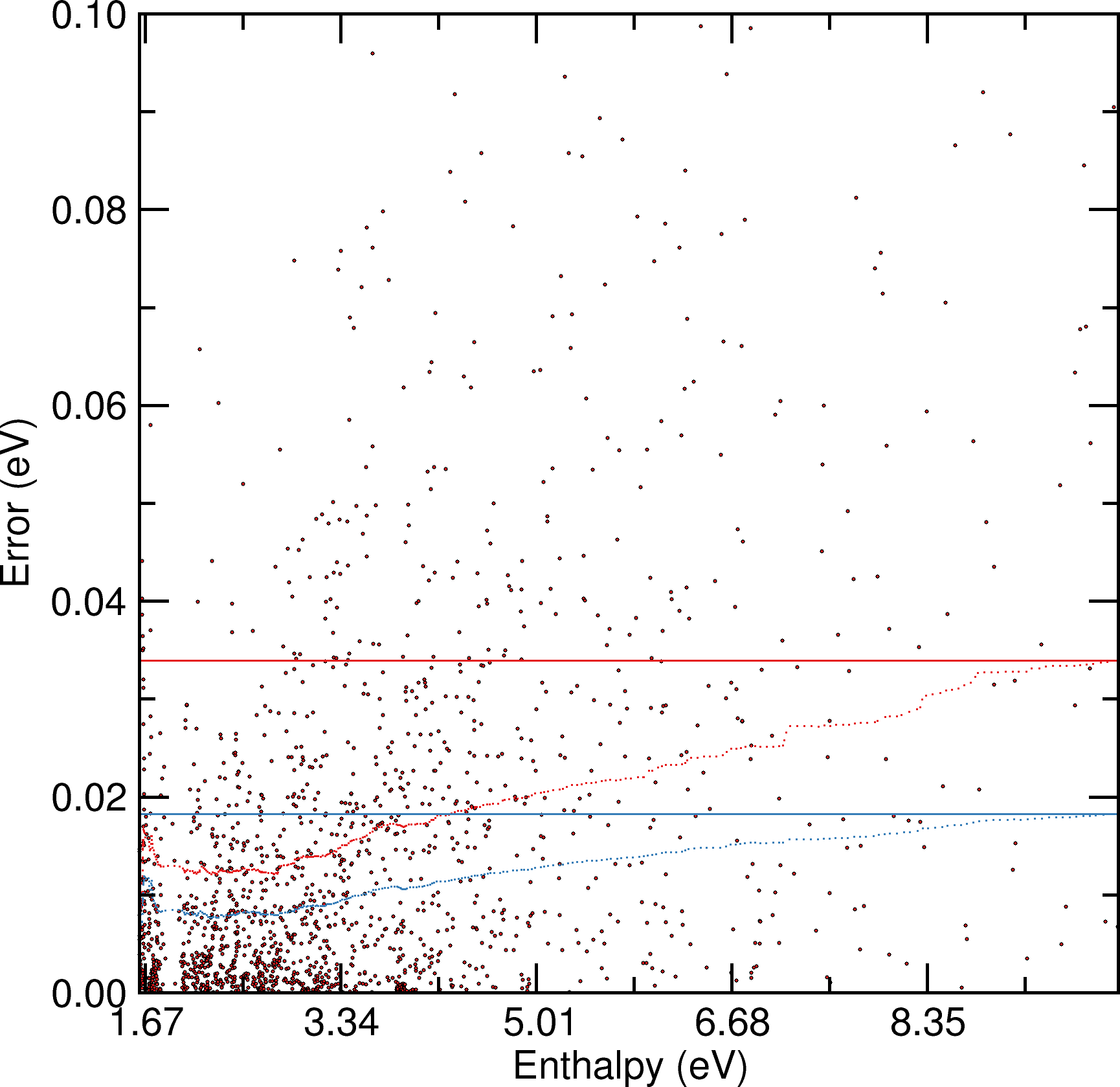}
\centering\begin{verbatim}
training    RMSE/MAE:  17.42  10.67  meV  Spearman  :  0.99988
validation  RMSE/MAE:  26.10  16.02  meV  Spearman  :  0.99976
testing     RMSE/MAE:  33.94  18.28  meV  Spearman  :  0.99984
\end{verbatim}
\clearpage

\flushleft{
\subsection{Ru-H}}
\subsubsection*{Searching}
\centering
\includegraphics[width=0.4\textwidth]{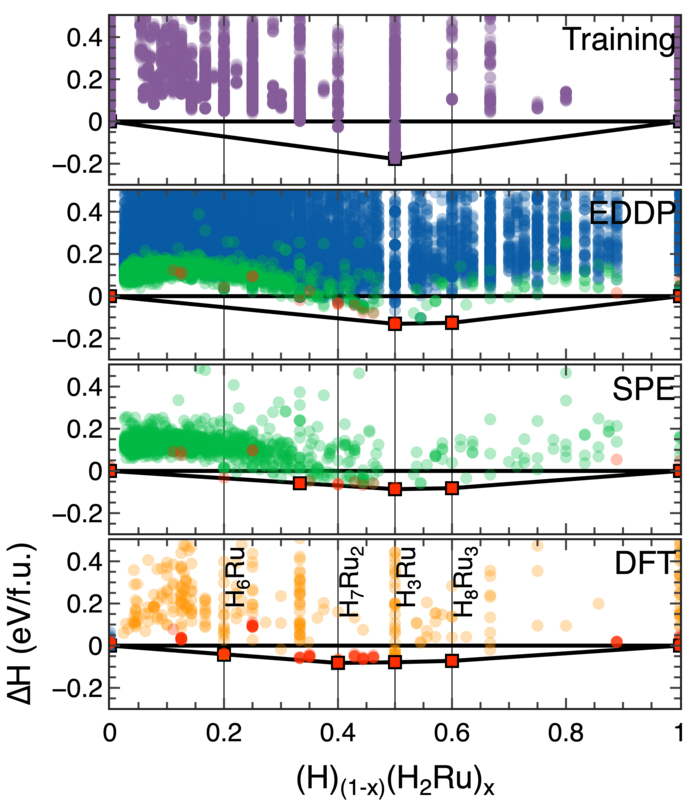}
\footnotesize


\flushleft{
\subsubsection*{\textsc{EDDP}}}
\centering
\includegraphics[width=0.3\textwidth]{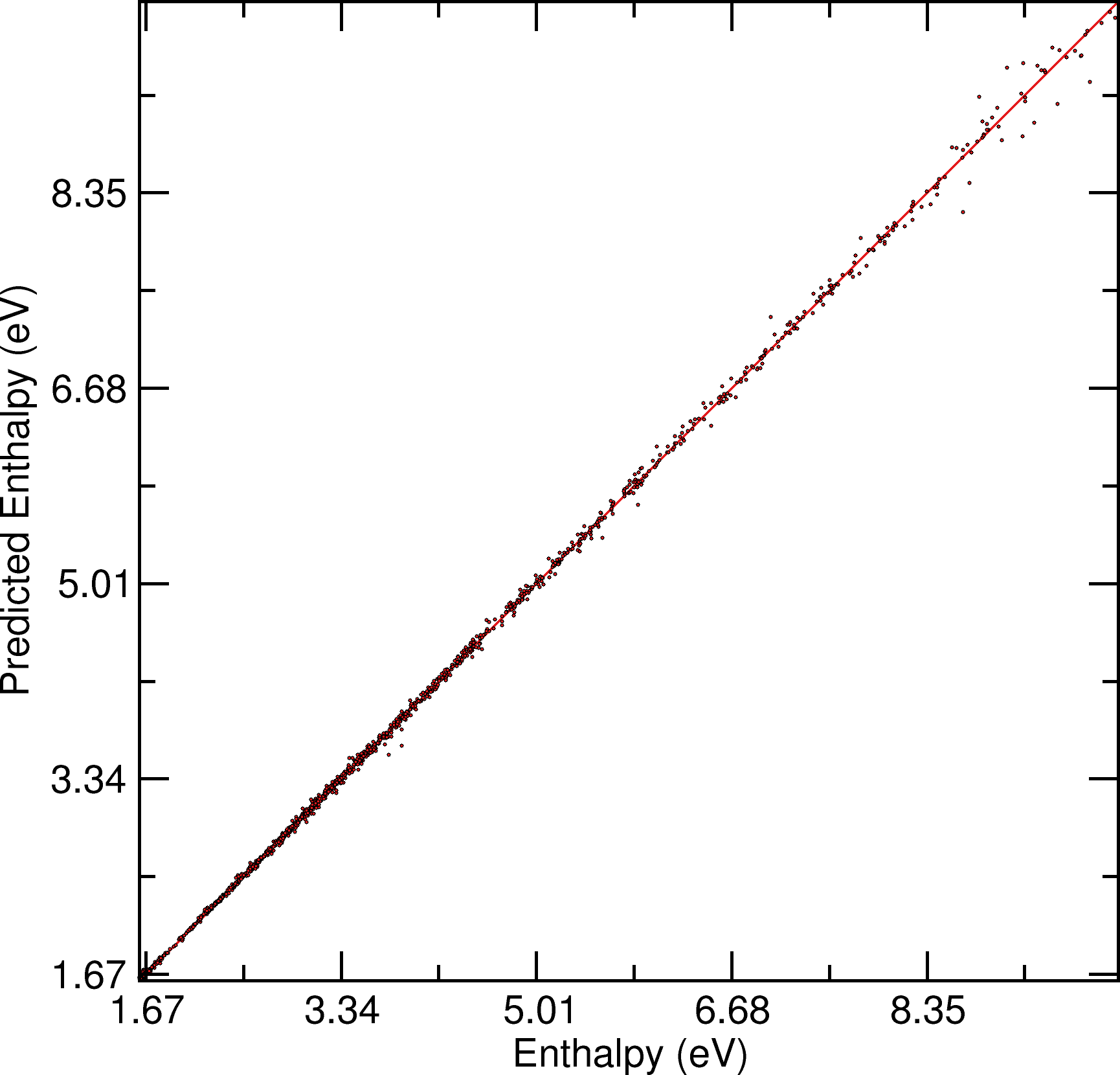}
\includegraphics[width=0.3\textwidth]{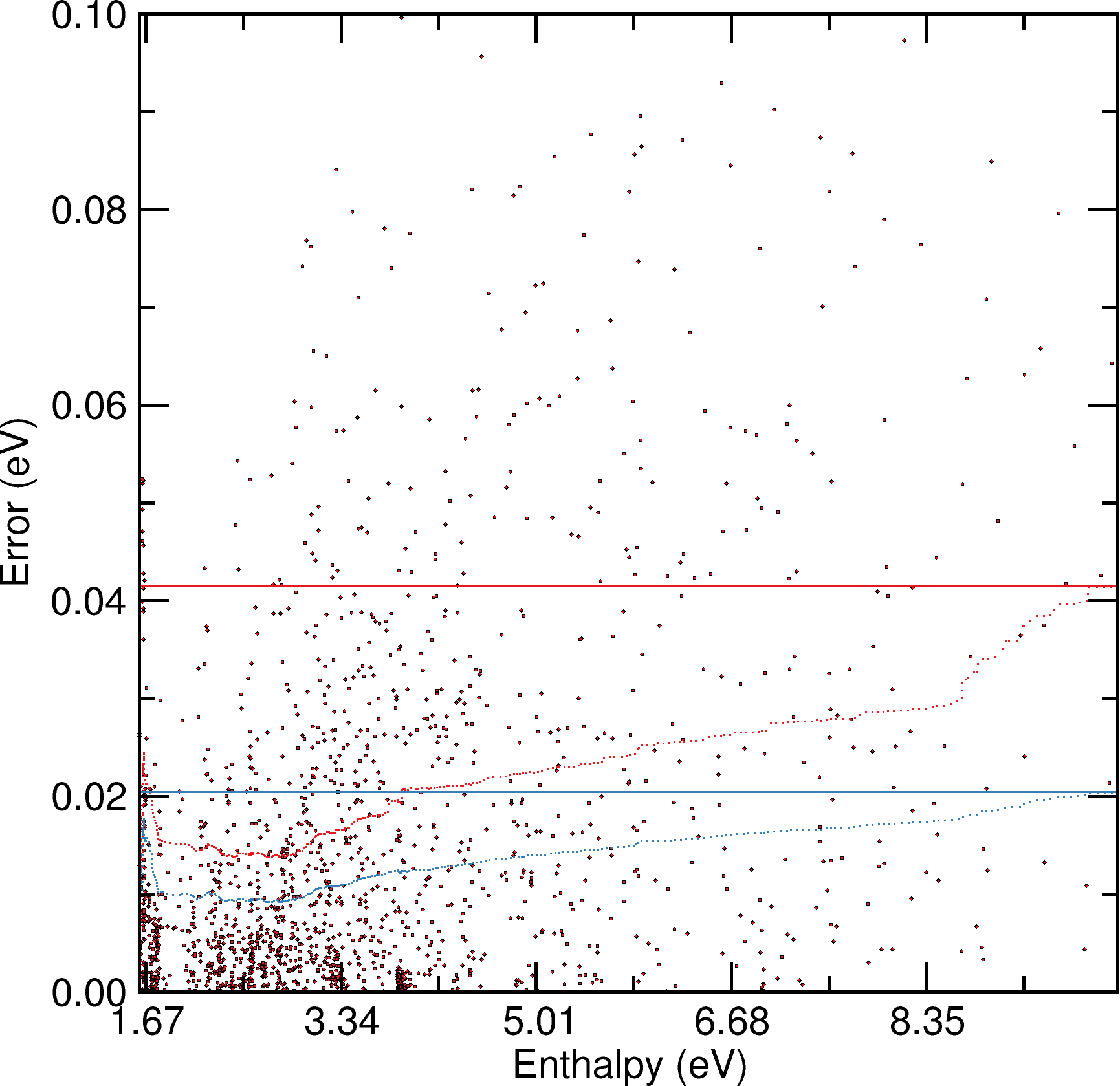}
\centering\begin{verbatim}
training    RMSE/MAE:  20.80  12.48  meV  Spearman  :  0.99983
validation  RMSE/MAE:  31.86  19.45  meV  Spearman  :  0.99976
testing     RMSE/MAE:  41.51  20.46  meV  Spearman  :  0.99973
\end{verbatim}
\clearpage

\flushleft{
\subsection{S-H}}
\subsubsection*{Searching}
\centering
\includegraphics[width=0.4\textwidth]{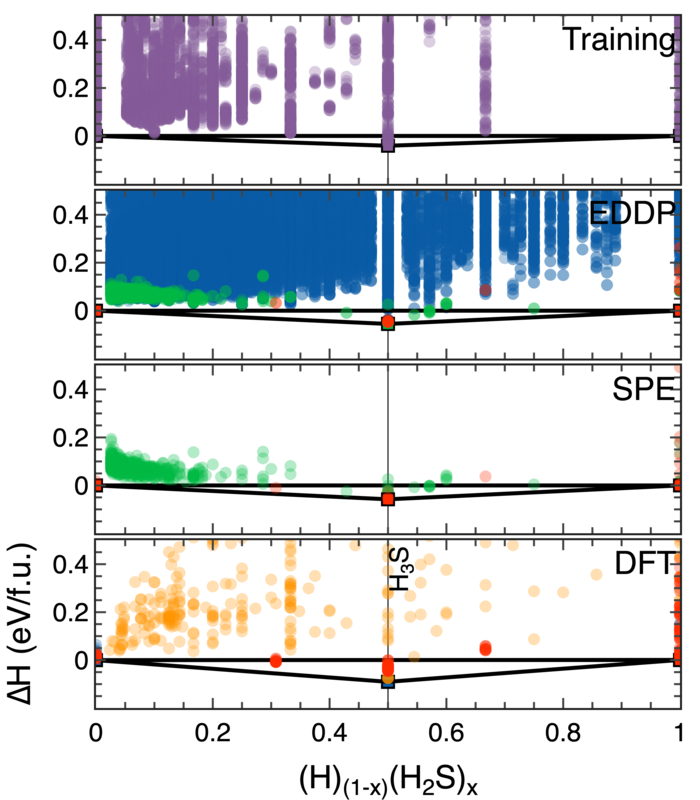}
\footnotesize


\flushleft{
\subsubsection*{\textsc{EDDP}}}
\centering
\includegraphics[width=0.3\textwidth]{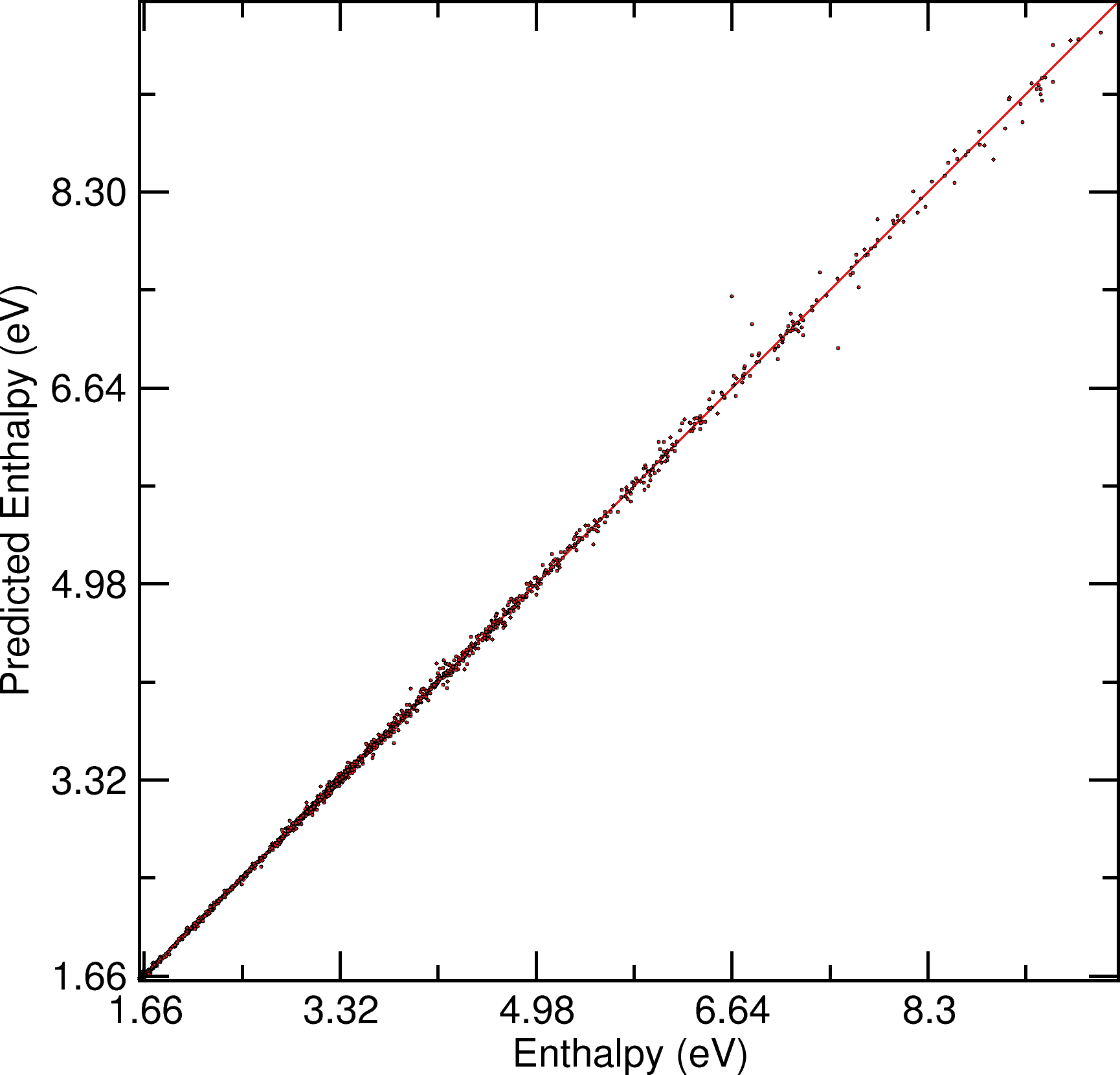}
\includegraphics[width=0.3\textwidth]{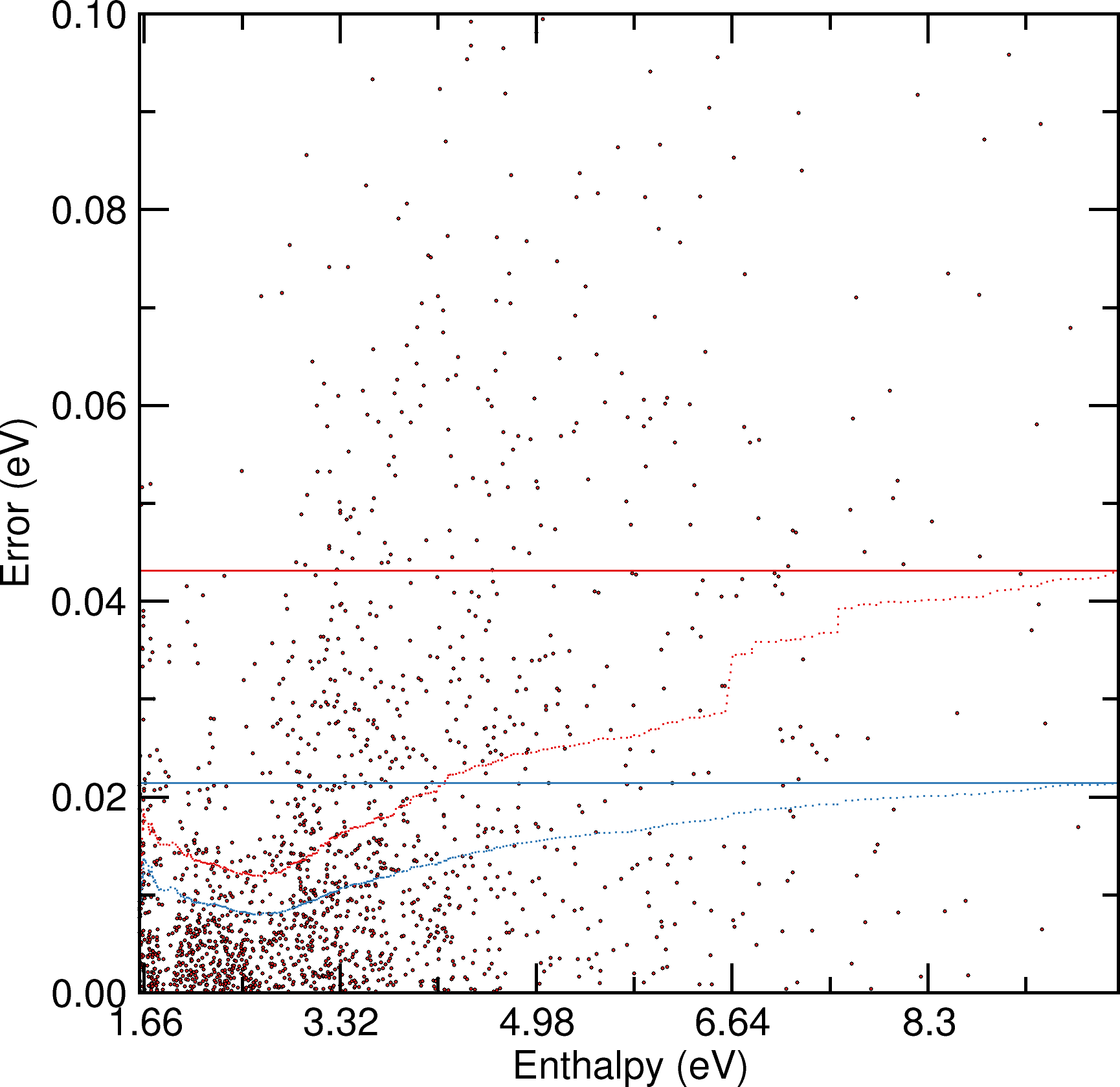}
\centering\begin{verbatim}
training    RMSE/MAE:  19.65  12.45  meV  Spearman  :  0.99985
validation  RMSE/MAE:  31.00  18.47  meV  Spearman  :  0.99977
testing     RMSE/MAE:  43.15  21.45  meV  Spearman  :  0.99976
\end{verbatim}
\clearpage

\flushleft{
\subsection{Sb-H}}
\subsubsection*{Searching}
\centering
\includegraphics[width=0.4\textwidth]{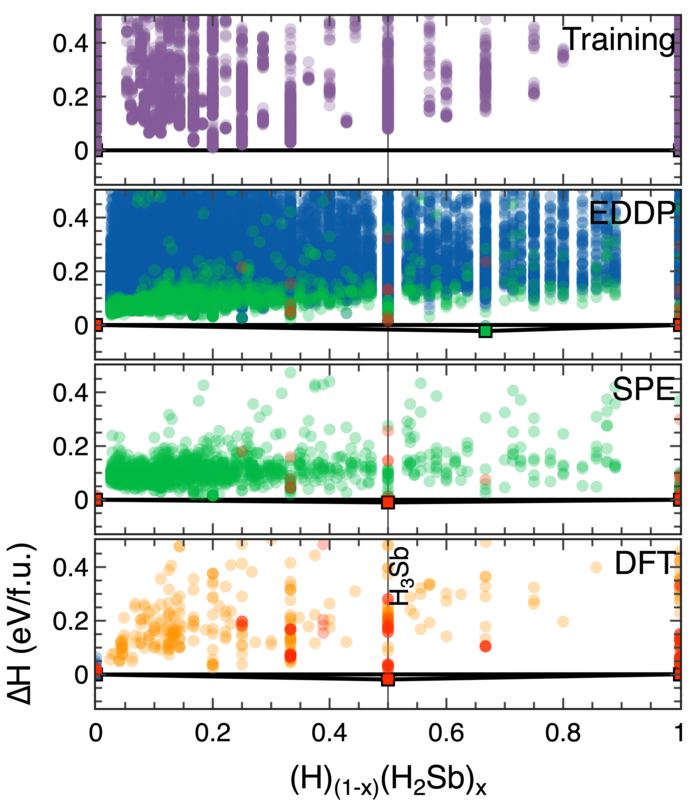}
\footnotesize


\flushleft{
\subsubsection*{\textsc{EDDP}}}
\centering
\includegraphics[width=0.3\textwidth]{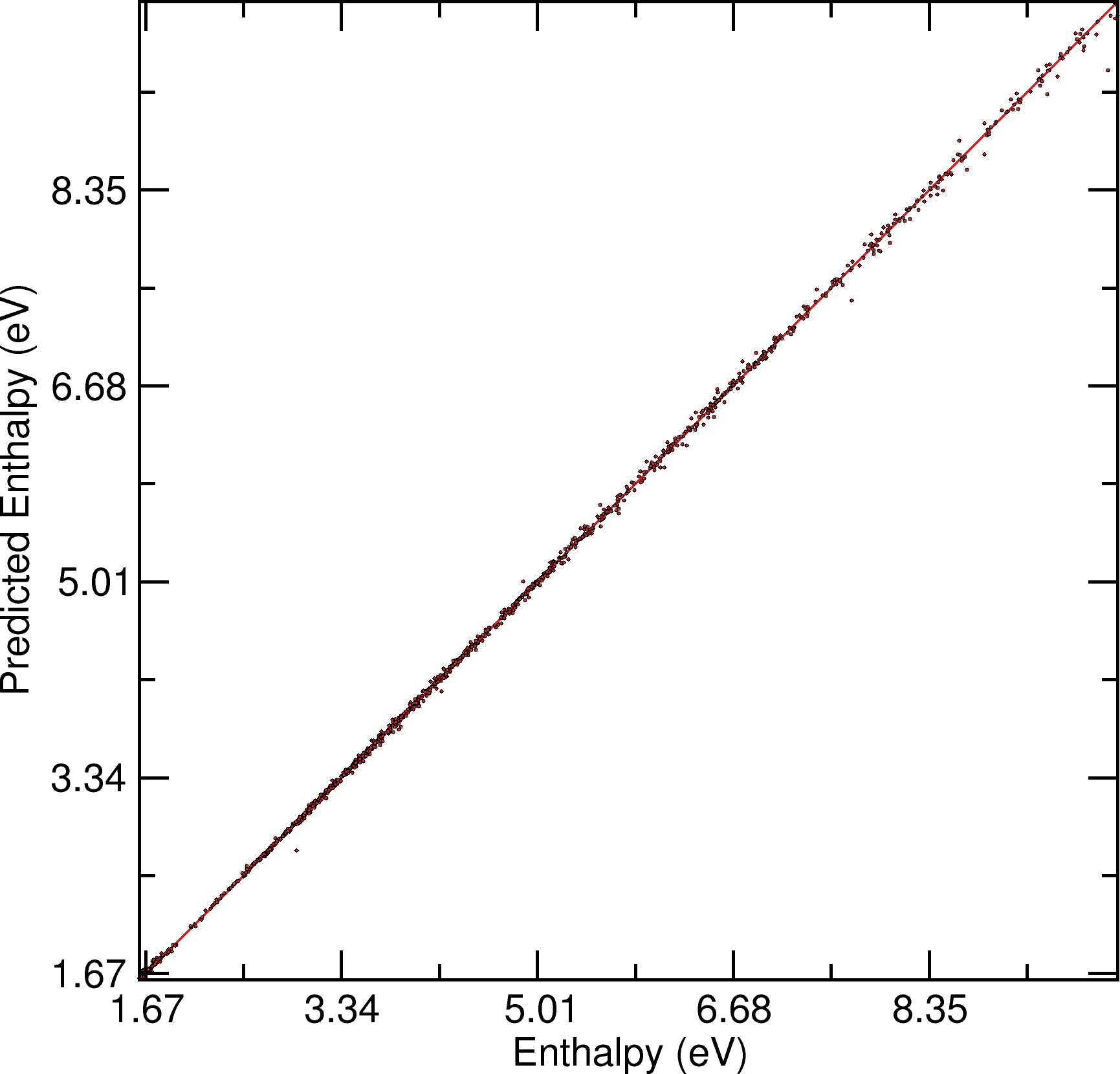}
\includegraphics[width=0.3\textwidth]{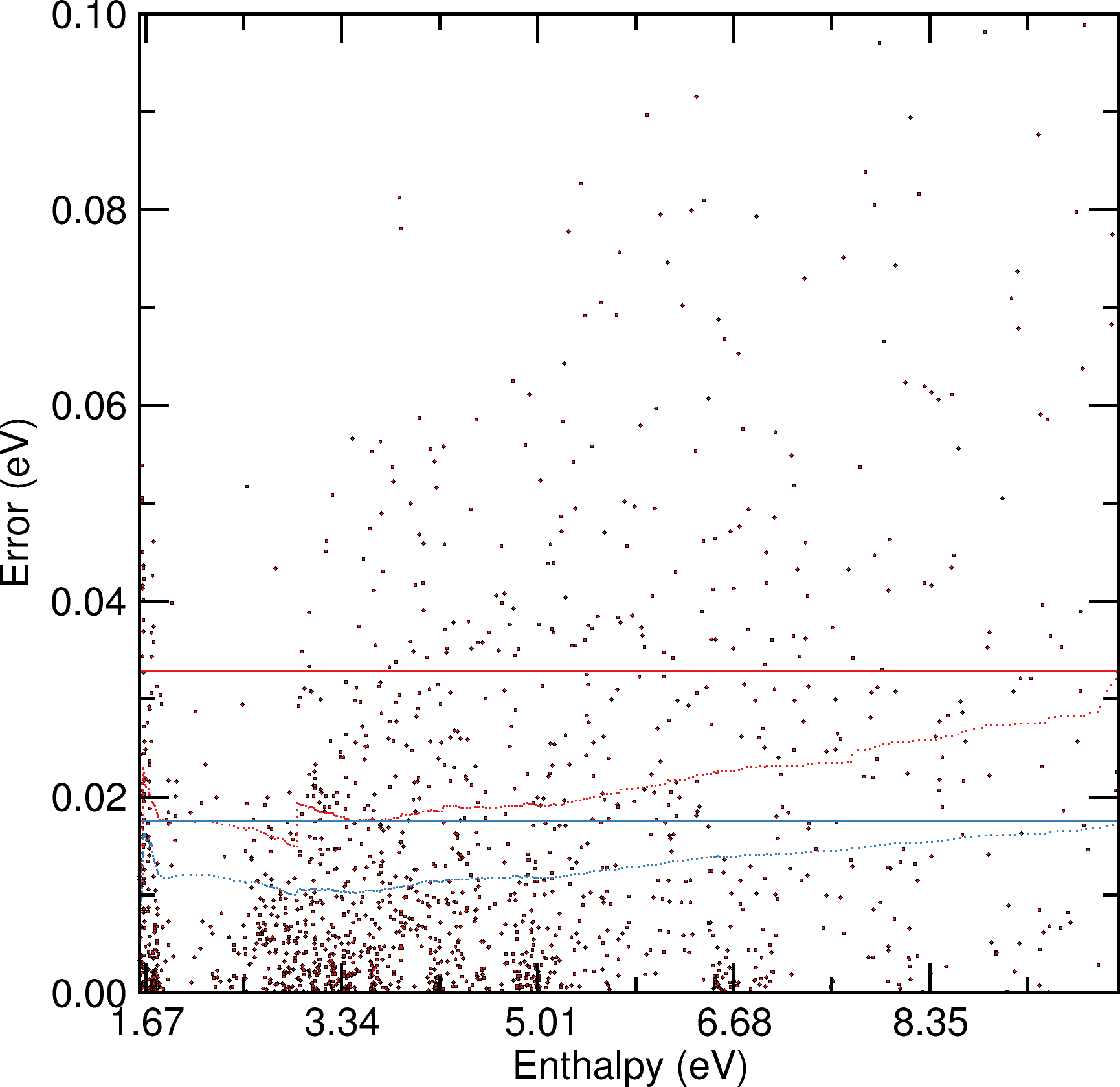}
\centering\begin{verbatim}
training    RMSE/MAE:  16.94  10.48  meV  Spearman  :  0.99989
validation  RMSE/MAE:  24.31  15.09  meV  Spearman  :  0.99983
testing     RMSE/MAE:  32.90  17.50  meV  Spearman  :  0.99978
\end{verbatim}
\clearpage

\flushleft{
\subsection{Sc-H}}
\subsubsection*{Searching}
\centering
\includegraphics[width=0.4\textwidth]{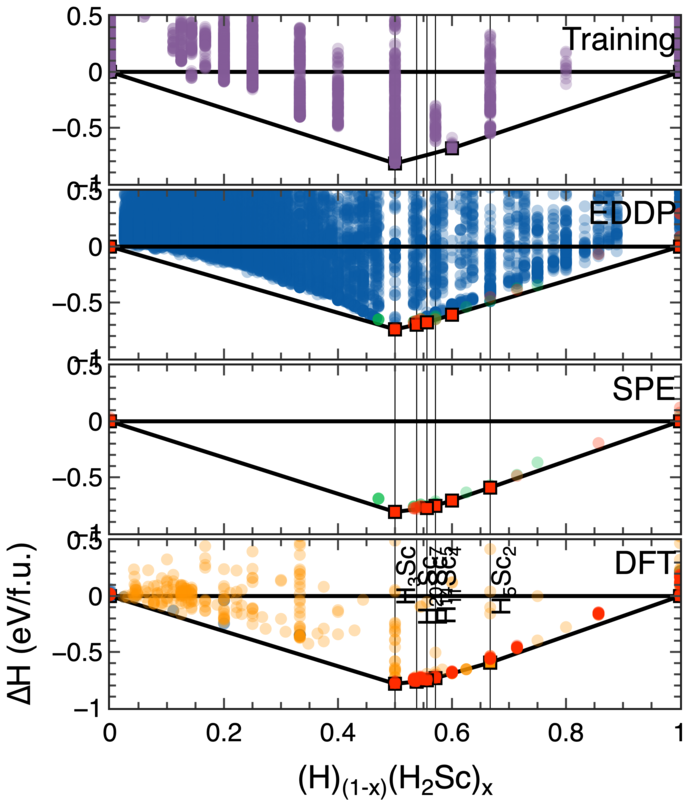}
\footnotesize


\flushleft{
\subsubsection*{\textsc{EDDP}}}
\centering
\includegraphics[width=0.3\textwidth]{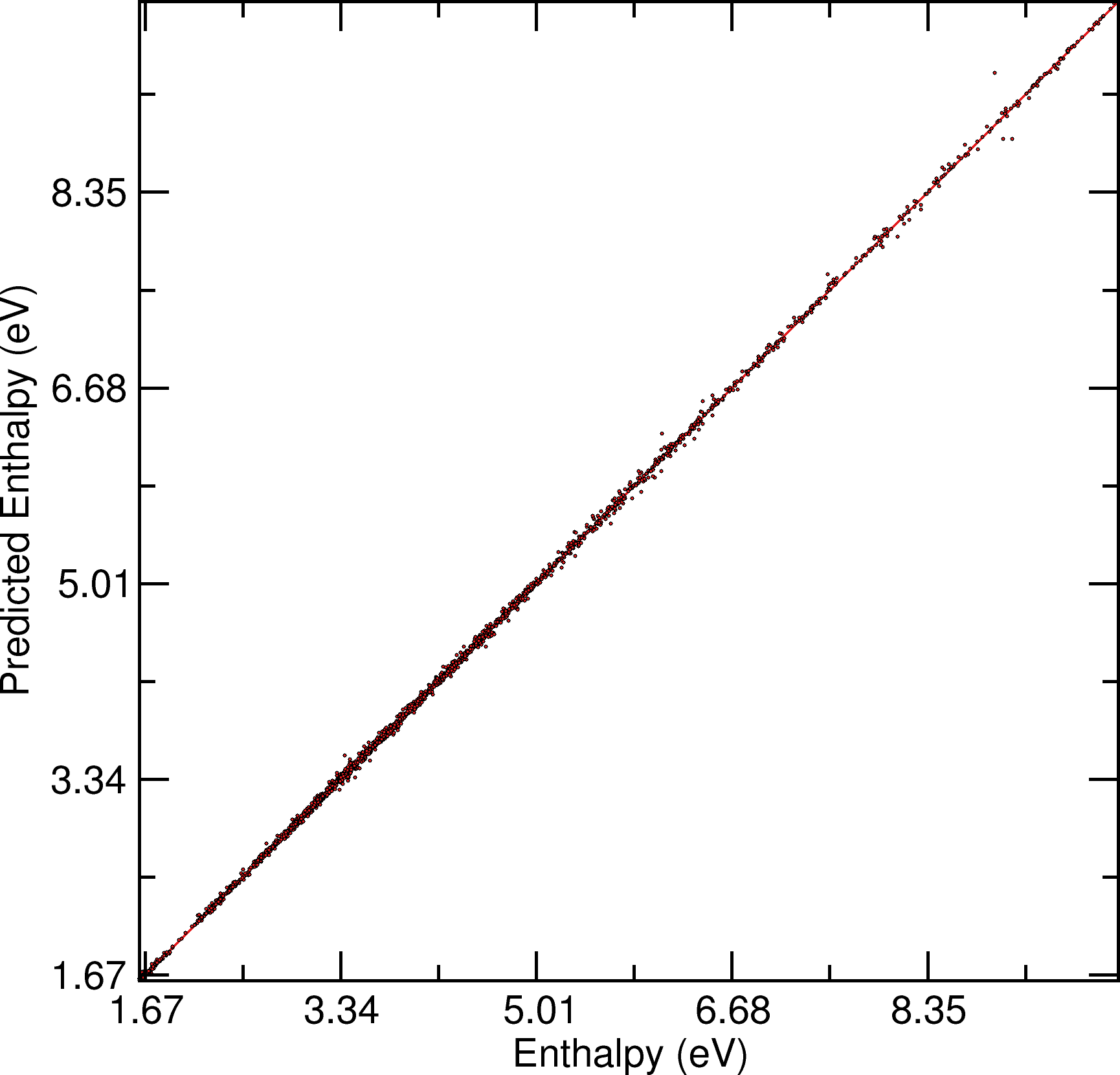}
\includegraphics[width=0.3\textwidth]{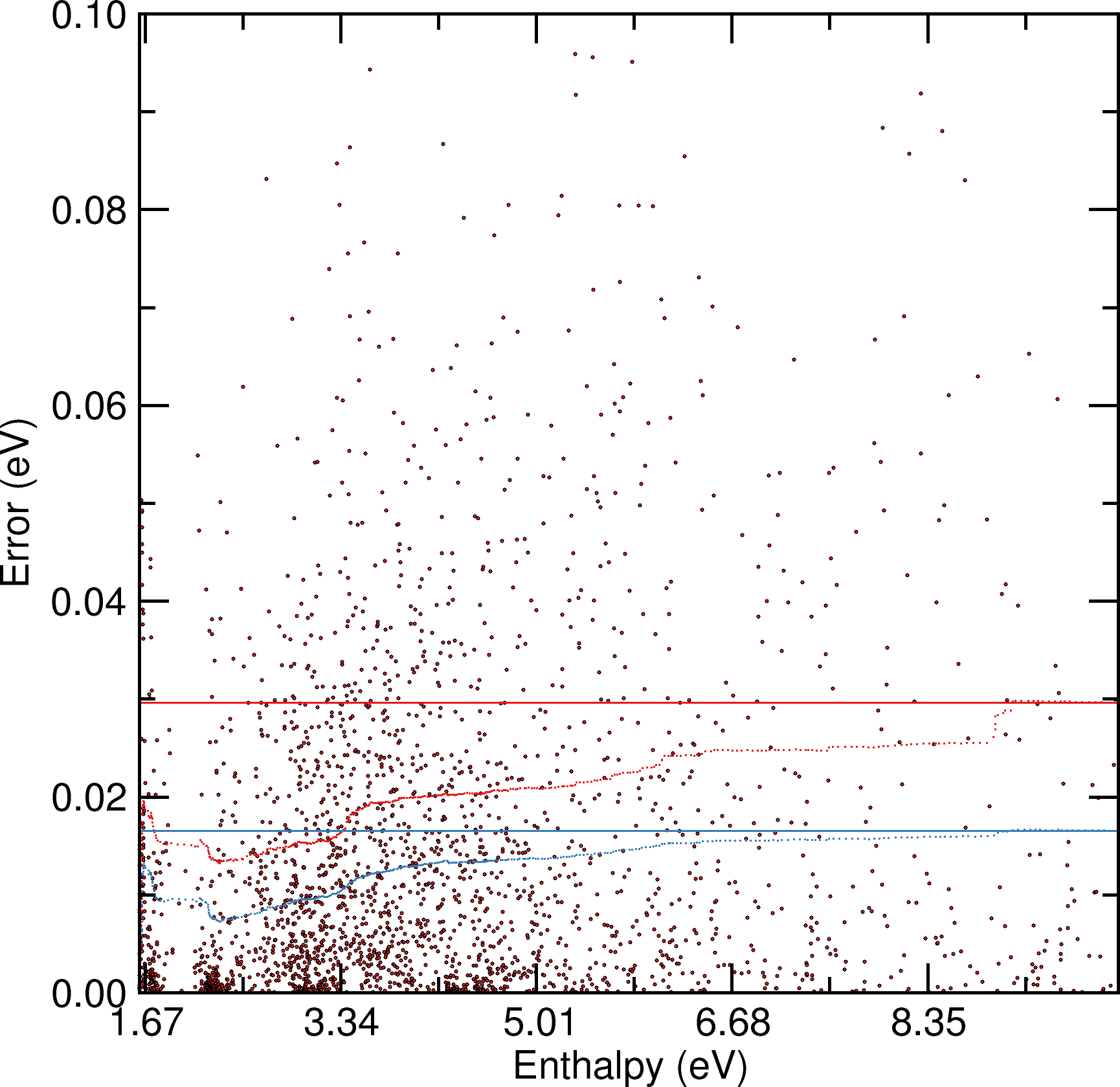}
\centering\begin{verbatim}
training    RMSE/MAE:  19.21  11.89  meV  Spearman  :  0.99989
validation  RMSE/MAE:  23.71  15.25  meV  Spearman  :  0.99985
testing     RMSE/MAE:  29.65  16.56  meV  Spearman  :  0.99983
\end{verbatim}
\clearpage

\flushleft{
\subsection{Se-H}}
\subsubsection*{Searching}
\centering
\includegraphics[width=0.4\textwidth]{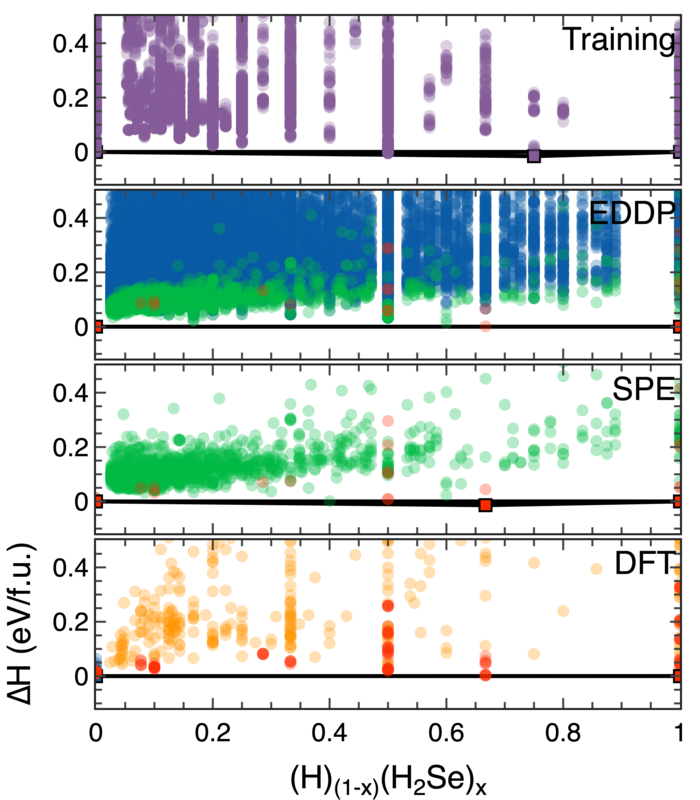}
\footnotesize


\flushleft{
\subsubsection*{\textsc{EDDP}}}
\centering
\includegraphics[width=0.3\textwidth]{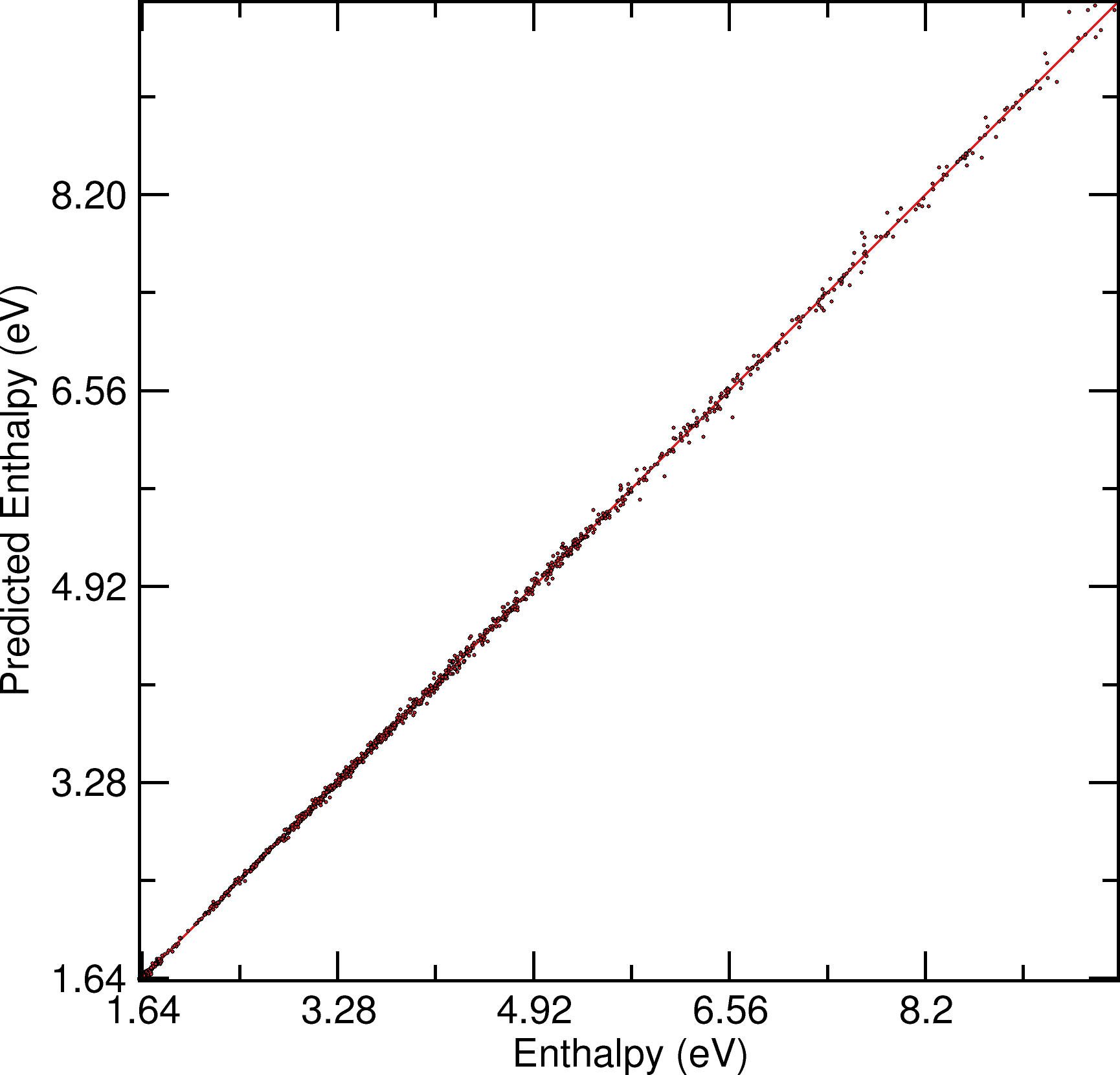}
\includegraphics[width=0.3\textwidth]{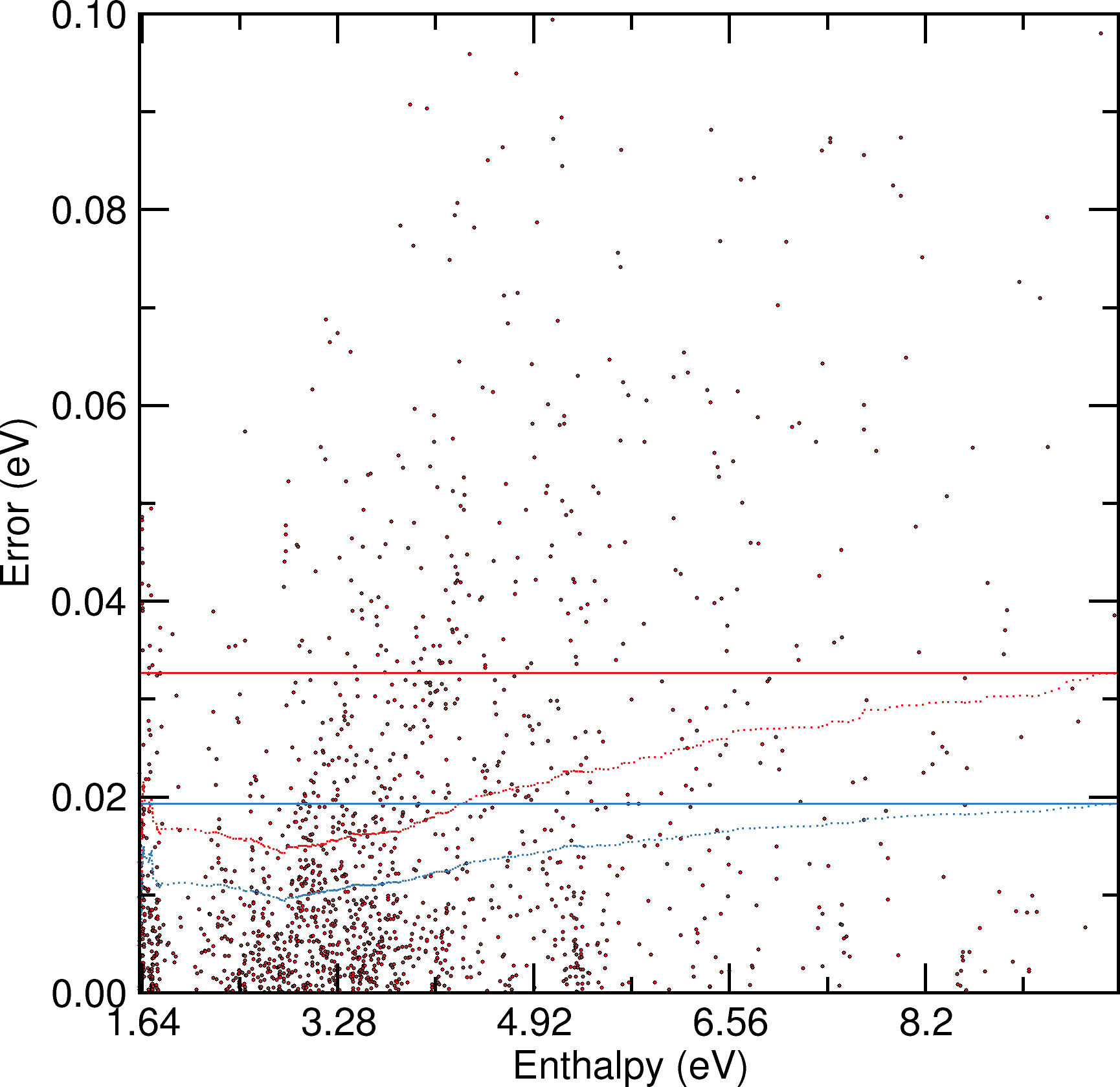}
\centering\begin{verbatim}
training    RMSE/MAE:  20.17  12.76  meV  Spearman  :  0.99985
validation  RMSE/MAE:  31.24  18.96  meV  Spearman  :  0.99975
testing     RMSE/MAE:  32.70  19.28  meV  Spearman  :  0.99978
\end{verbatim}
\clearpage

\flushleft{
\subsection{Si-H}}
\subsubsection*{Searching}
\centering
\includegraphics[width=0.4\textwidth]{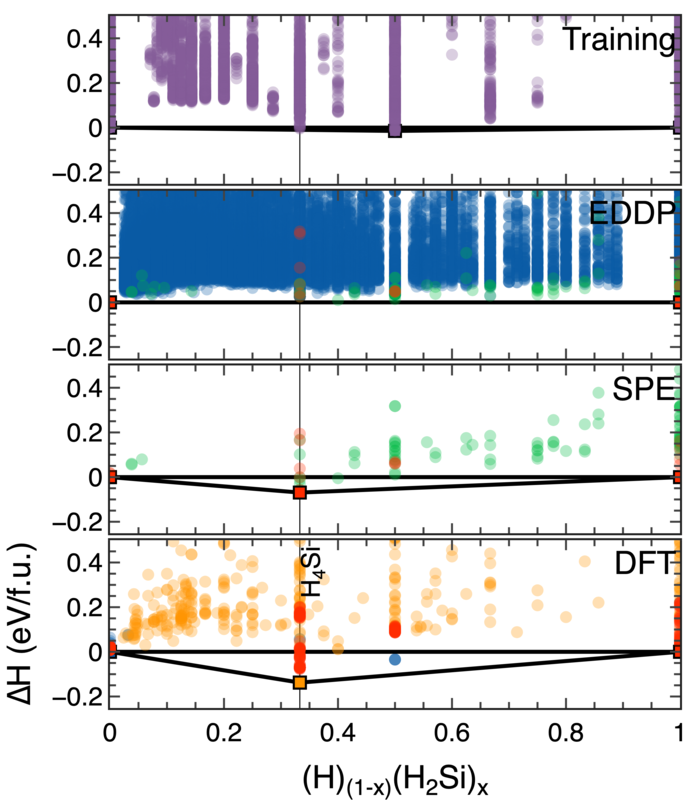}
\footnotesize


\flushleft{
\subsubsection*{\textsc{EDDP}}}
\centering
\includegraphics[width=0.3\textwidth]{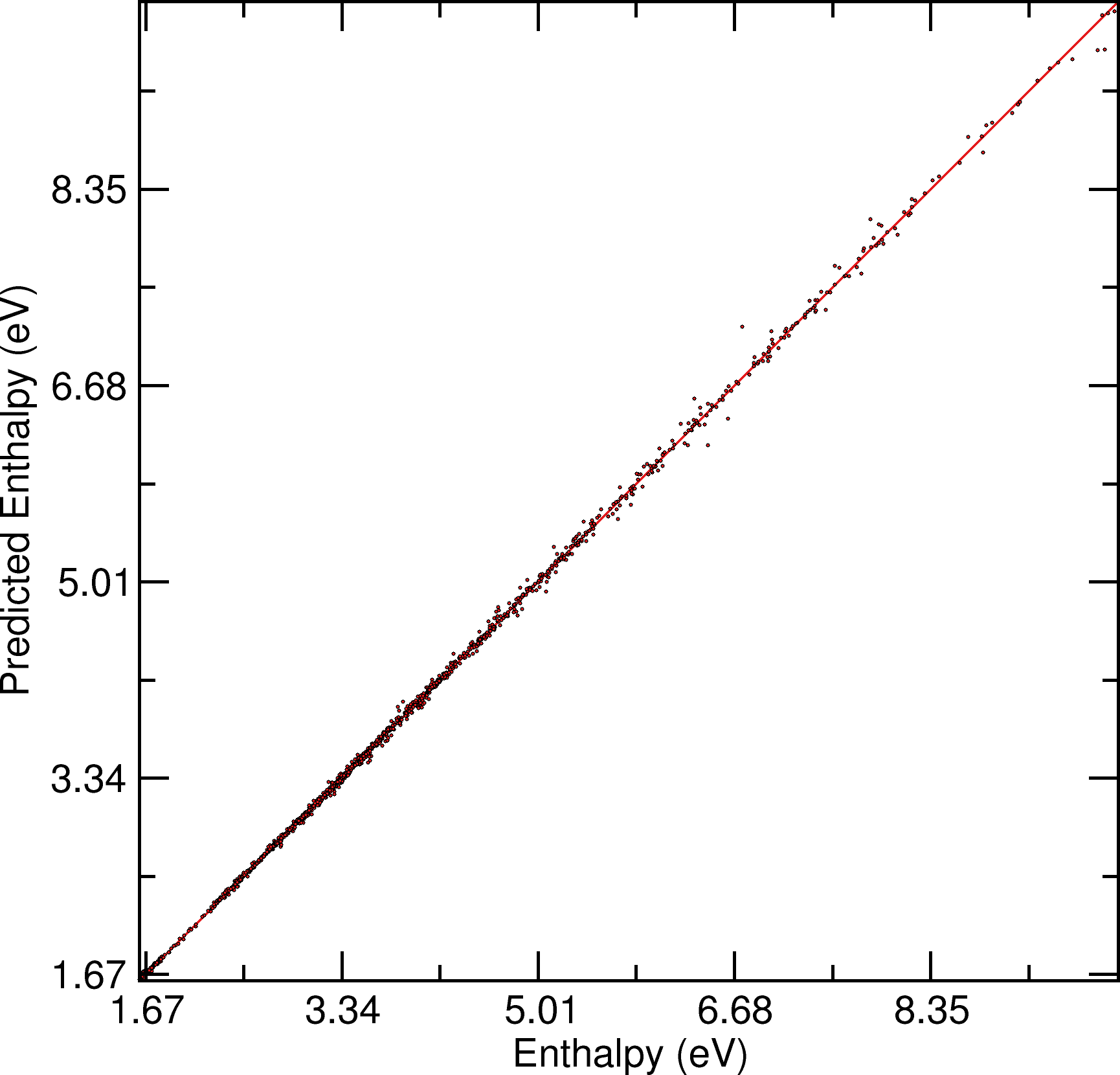}
\includegraphics[width=0.3\textwidth]{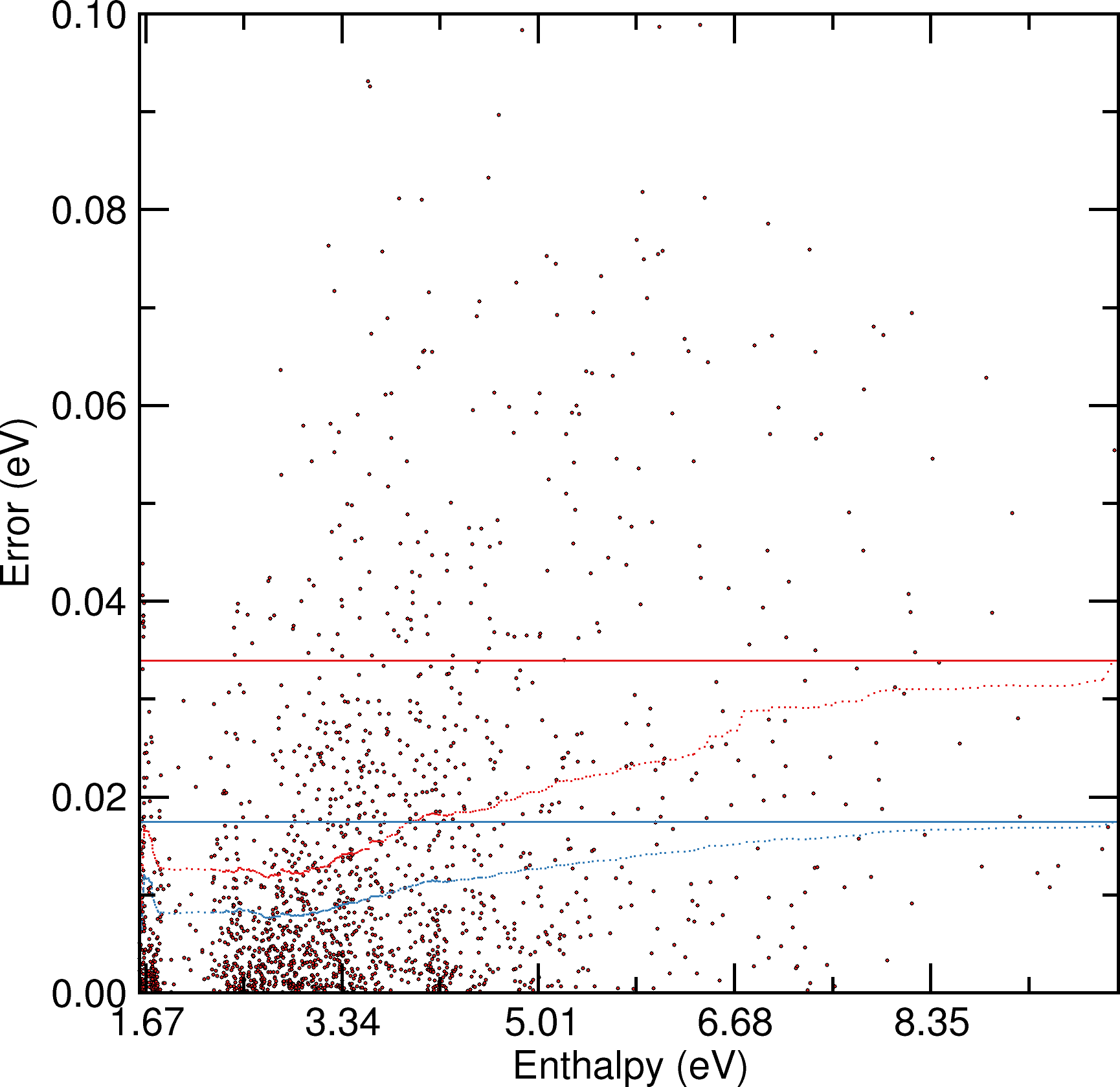}
\centering\begin{verbatim}
training    RMSE/MAE:  16.97  10.49  meV  Spearman  :  0.99986
validation  RMSE/MAE:  25.97  15.42  meV  Spearman  :  0.99979
testing     RMSE/MAE:  33.93  17.46  meV  Spearman  :  0.99980
\end{verbatim}
\clearpage

\flushleft{
\subsection{Sm-H}}
\subsubsection*{Searching}
\centering
\includegraphics[width=0.4\textwidth]{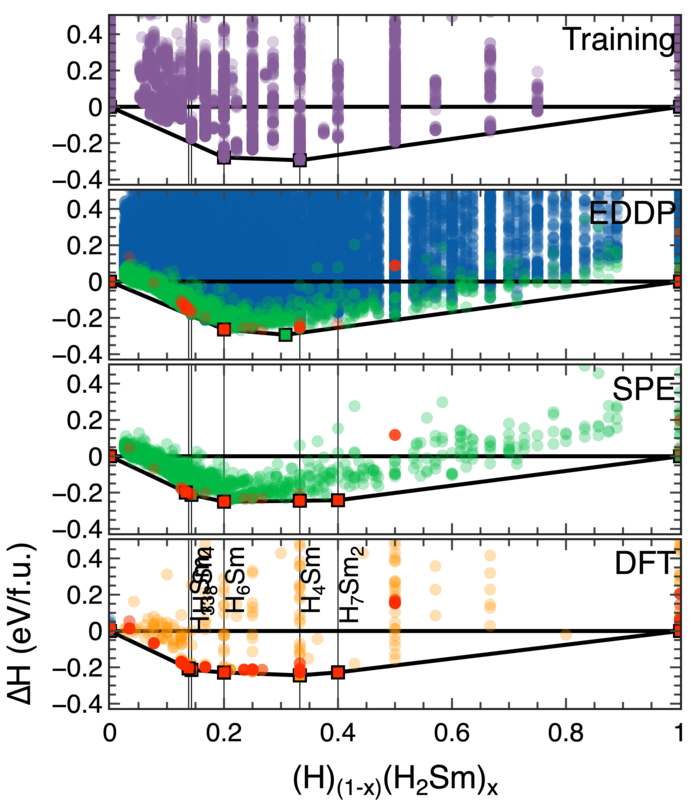}
\footnotesize


\flushleft{
\subsubsection*{\textsc{EDDP}}}
\centering
\includegraphics[width=0.3\textwidth]{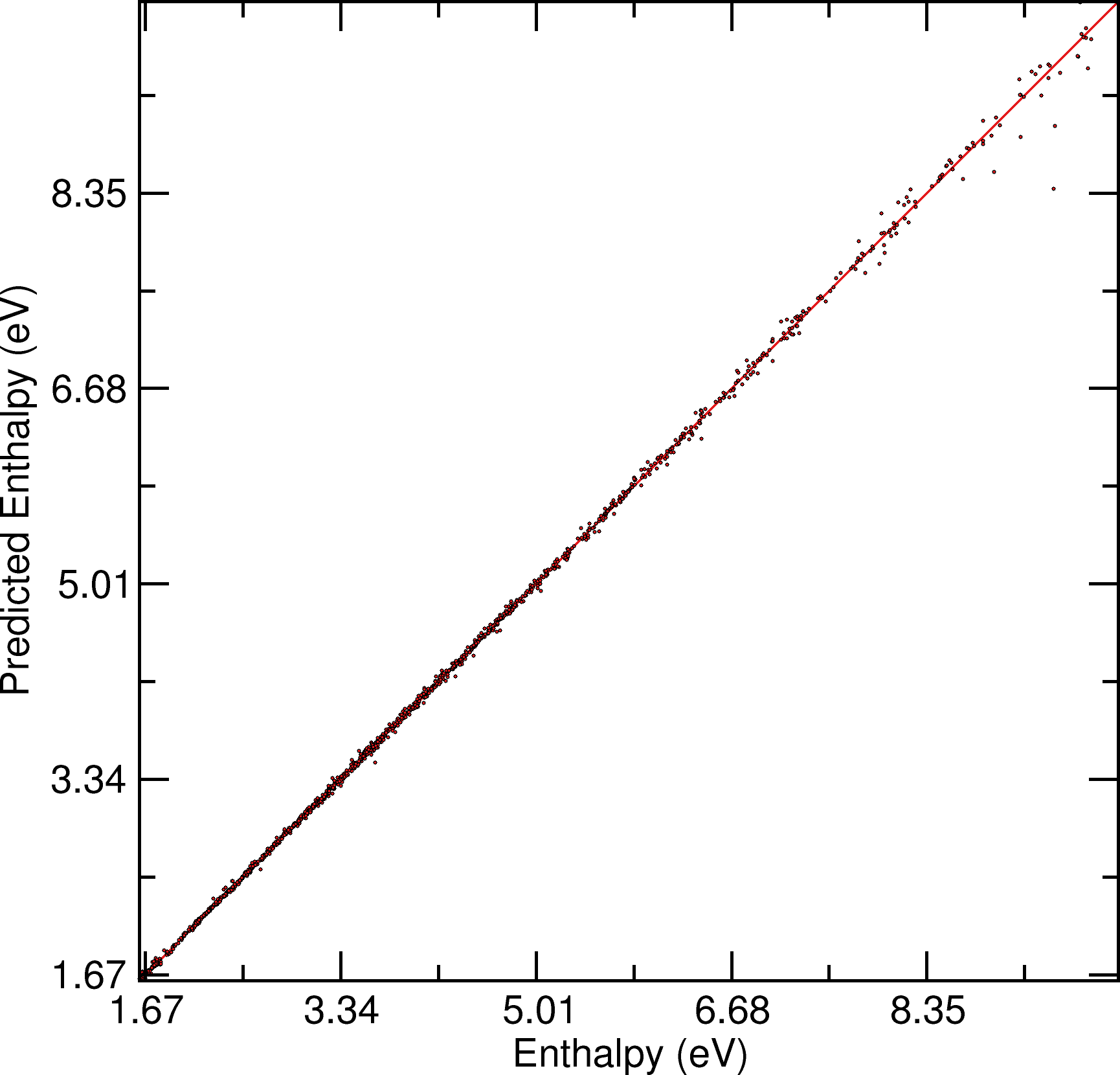}
\includegraphics[width=0.3\textwidth]{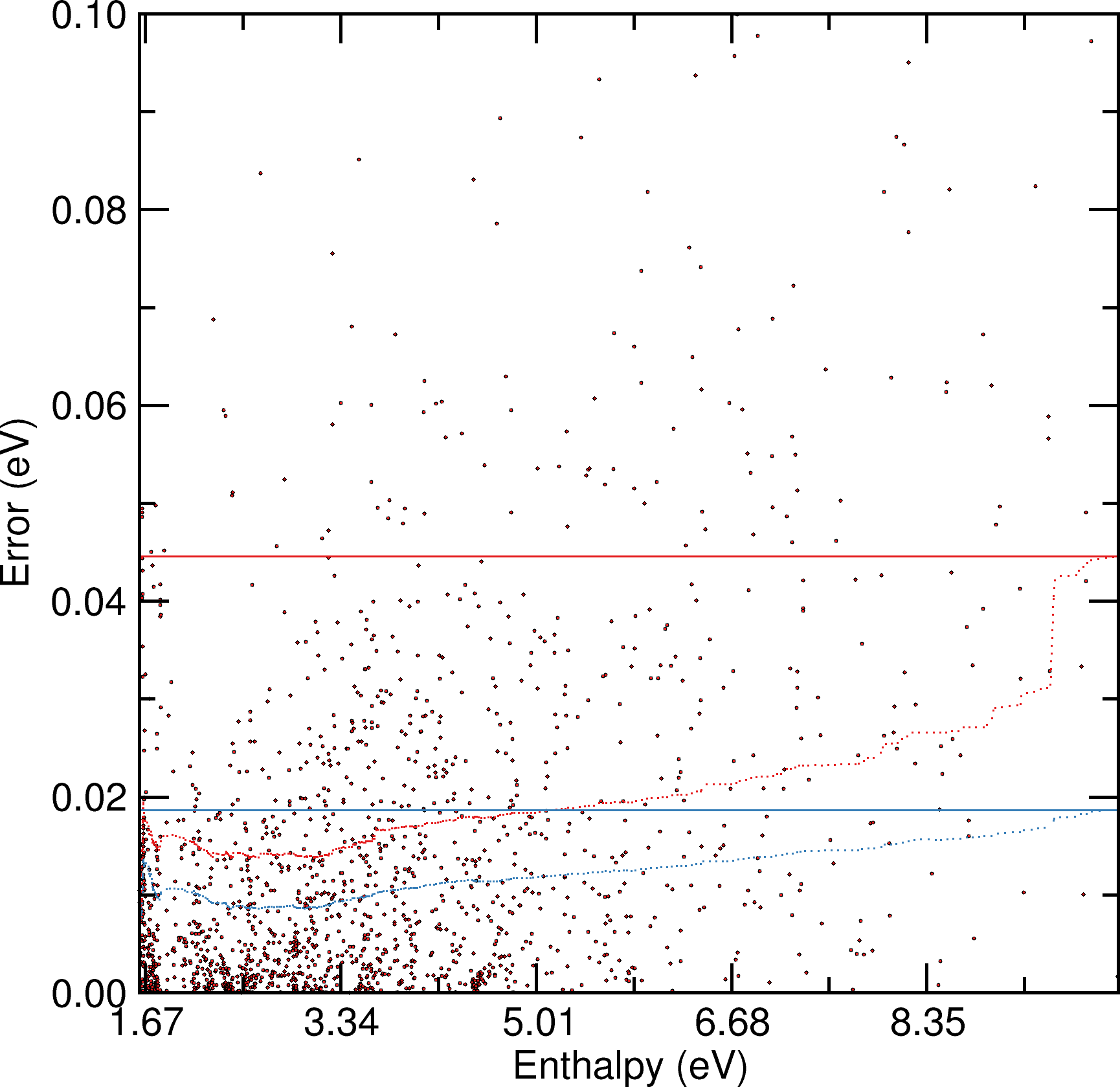}
\centering\begin{verbatim}
training    RMSE/MAE:  17.52  11.15  meV  Spearman  :  0.99988
validation  RMSE/MAE:  25.14  15.99  meV  Spearman  :  0.99986
testing     RMSE/MAE:  44.58  18.68  meV  Spearman  :  0.99980
\end{verbatim}
\clearpage

\flushleft{
\subsection{Sn-H}}
\subsubsection*{Searching}
\centering
\includegraphics[width=0.4\textwidth]{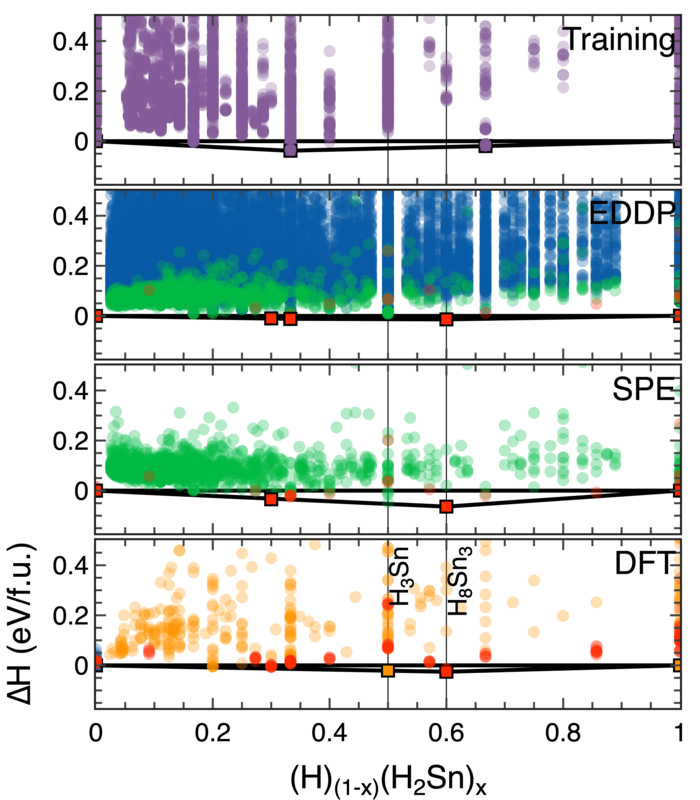}
\footnotesize


\flushleft{
\subsubsection*{\textsc{EDDP}}}
\centering
\includegraphics[width=0.3\textwidth]{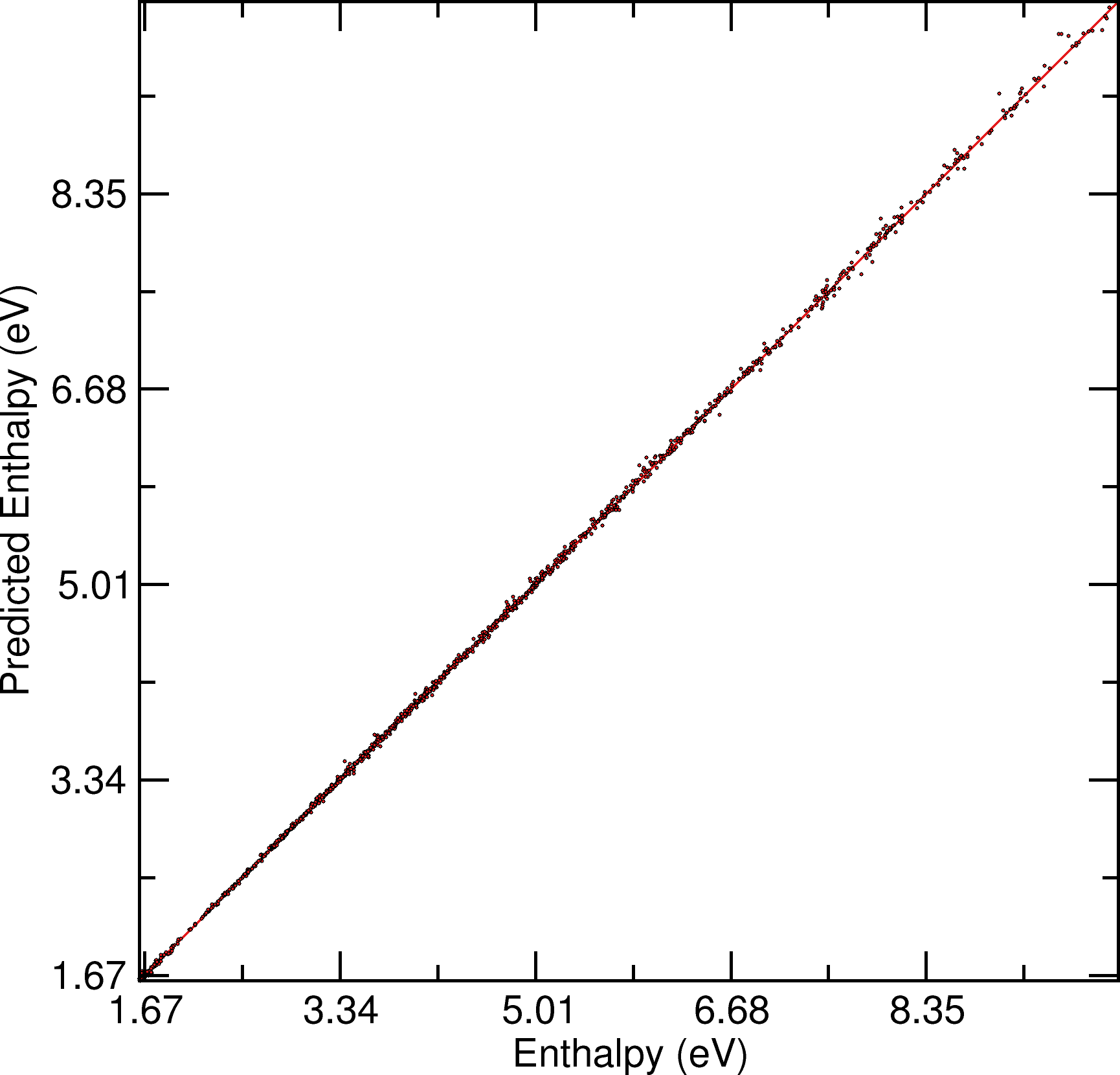}
\includegraphics[width=0.3\textwidth]{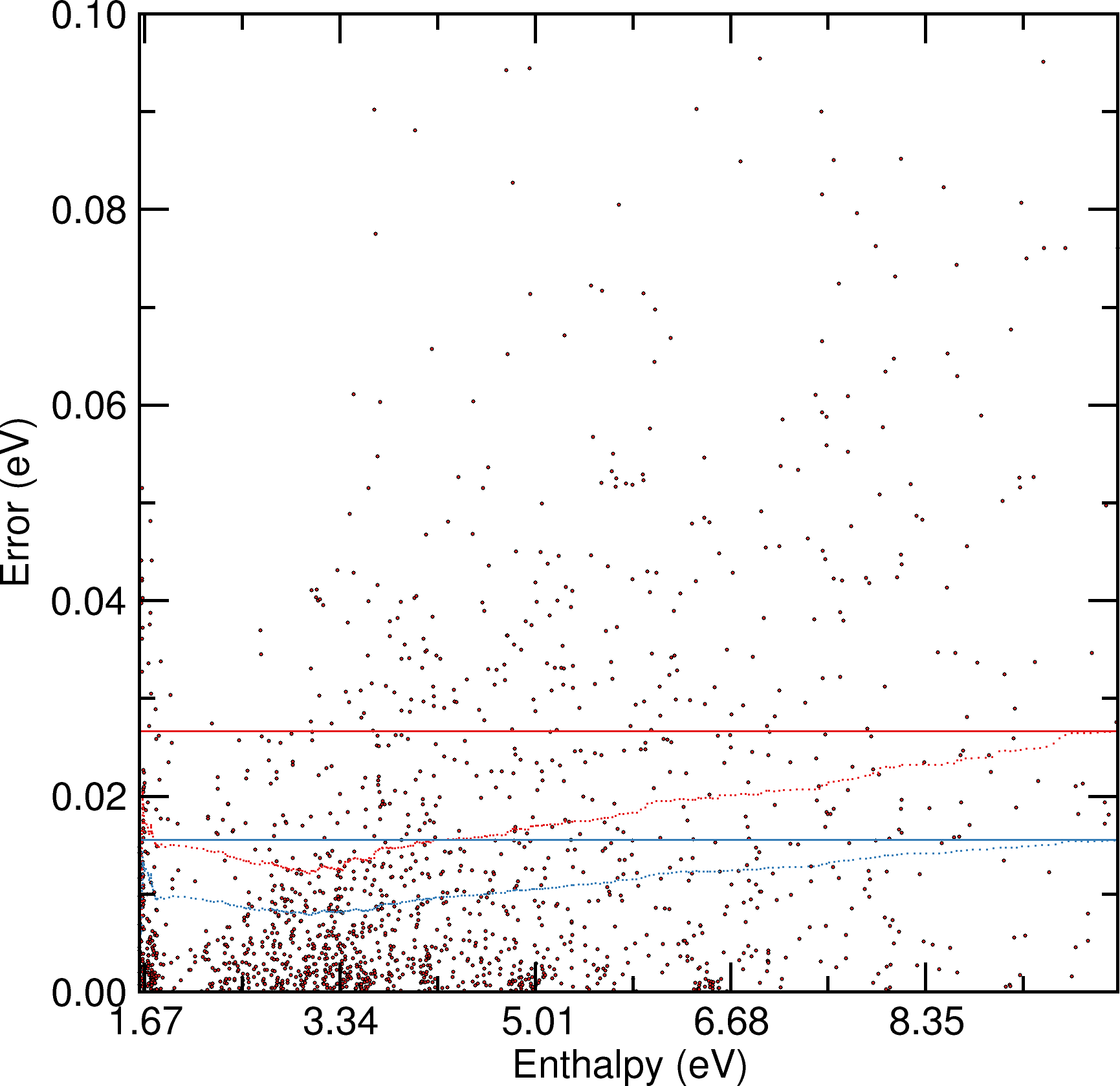}
\centering\begin{verbatim}
training    RMSE/MAE:  15.14  9.40   meV  Spearman  :  0.99990
validation  RMSE/MAE:  20.84  12.96  meV  Spearman  :  0.99987
testing     RMSE/MAE:  26.64  15.51  meV  Spearman  :  0.99984
\end{verbatim}
\clearpage

\flushleft{
\subsection{Sr-H}}
\subsubsection*{Searching}
\centering
\includegraphics[width=0.4\textwidth]{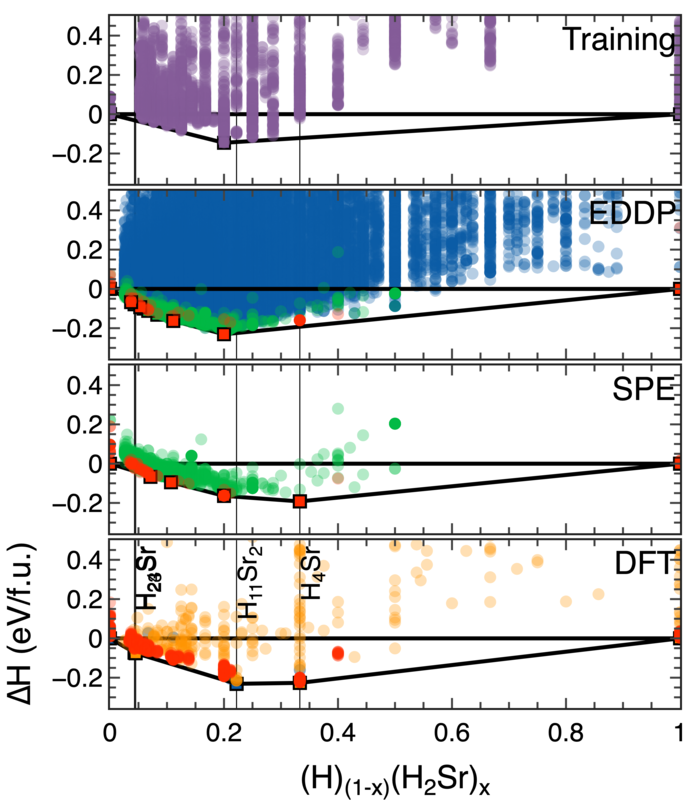}
\footnotesize


\flushleft{
\subsubsection*{\textsc{EDDP}}}
\centering
\includegraphics[width=0.3\textwidth]{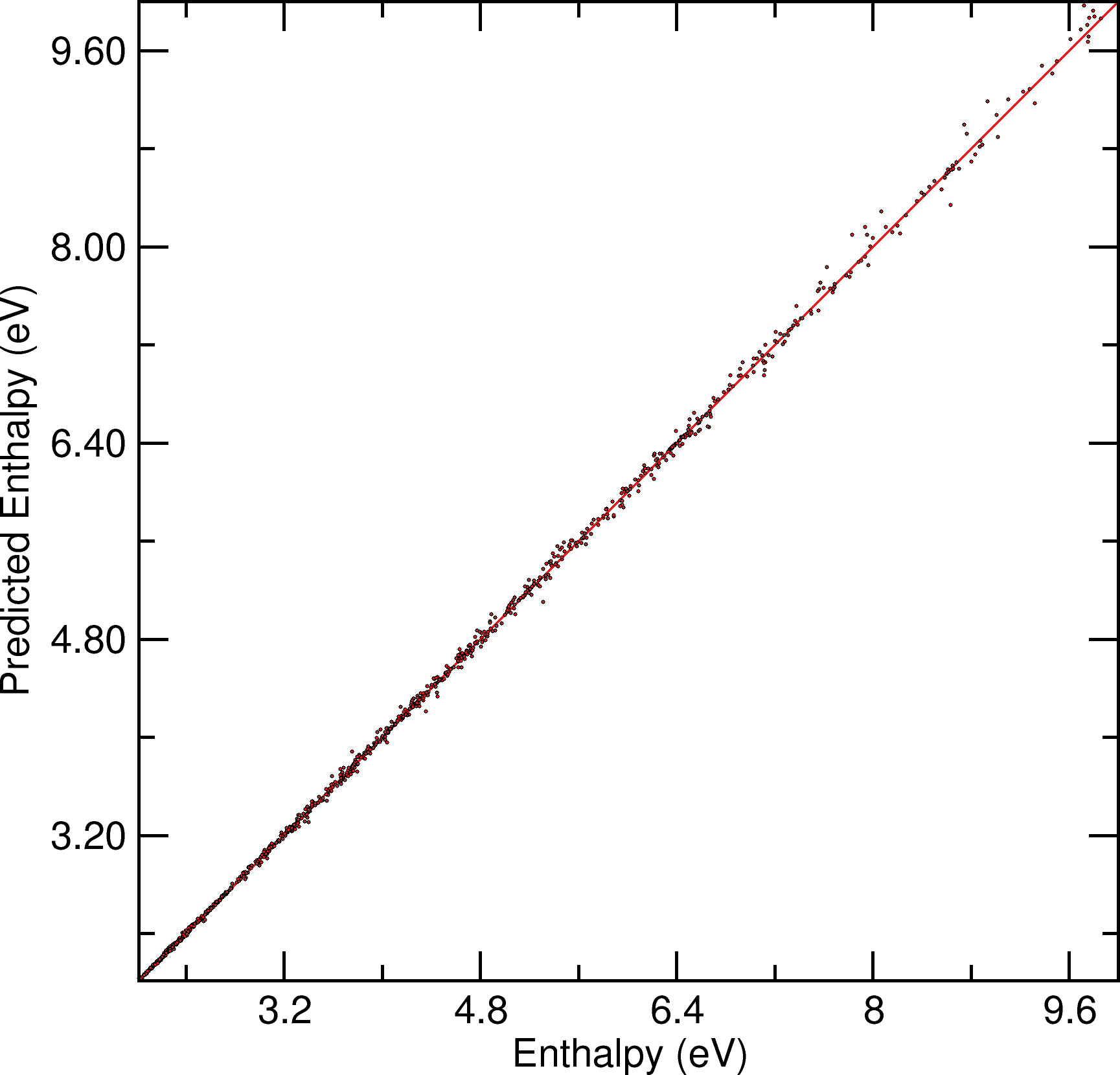}
\includegraphics[width=0.3\textwidth]{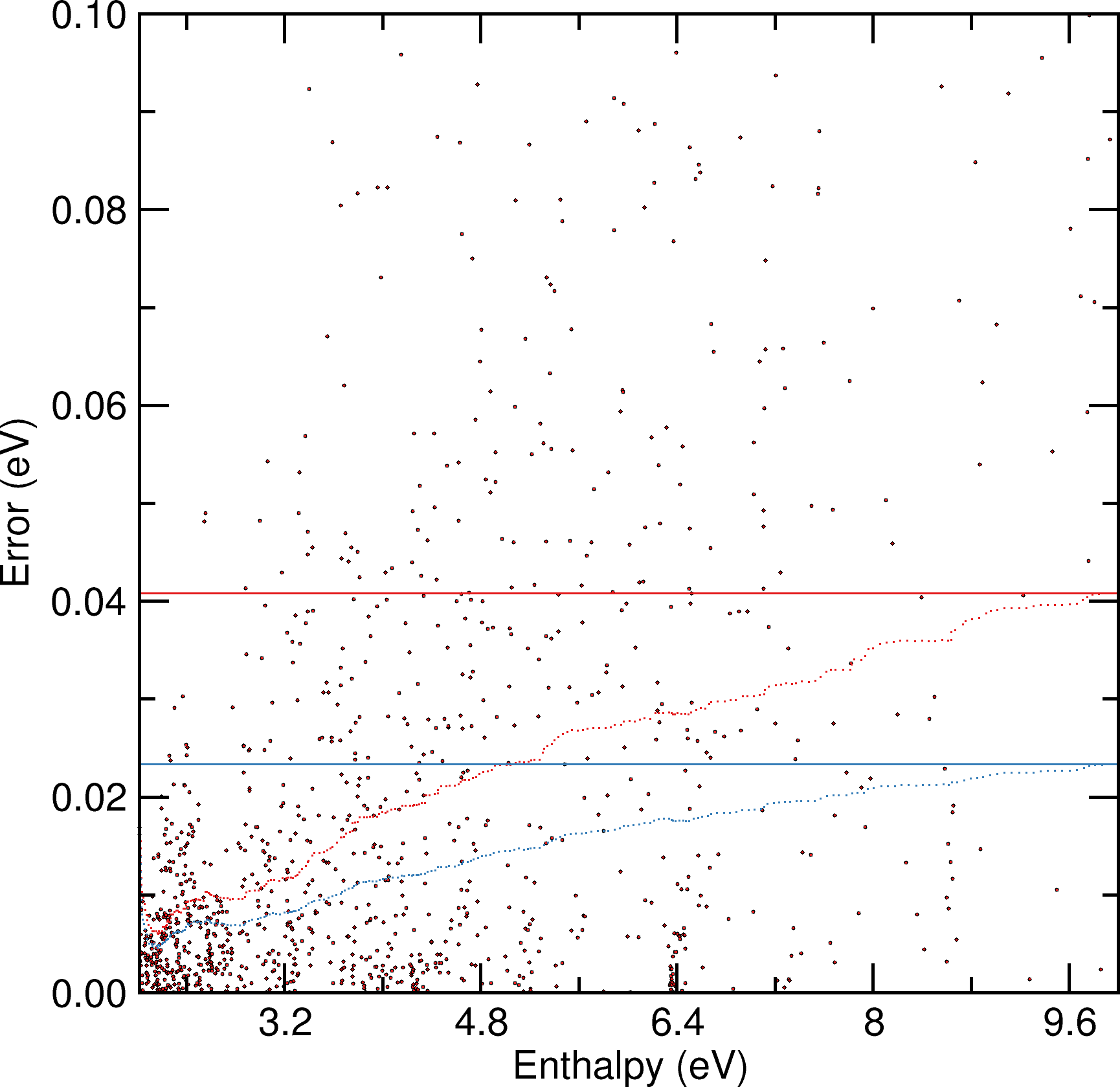}
\centering\begin{verbatim}
training    RMSE/MAE:  18.84  11.43  meV  Spearman  :  0.99993
validation  RMSE/MAE:  36.39  20.98  meV  Spearman  :  0.99985
testing     RMSE/MAE:  40.80  23.32  meV  Spearman  :  0.99984
\end{verbatim}
\clearpage

\flushleft{
\subsection{Ta-H}}
\subsubsection*{Searching}
\centering
\includegraphics[width=0.4\textwidth]{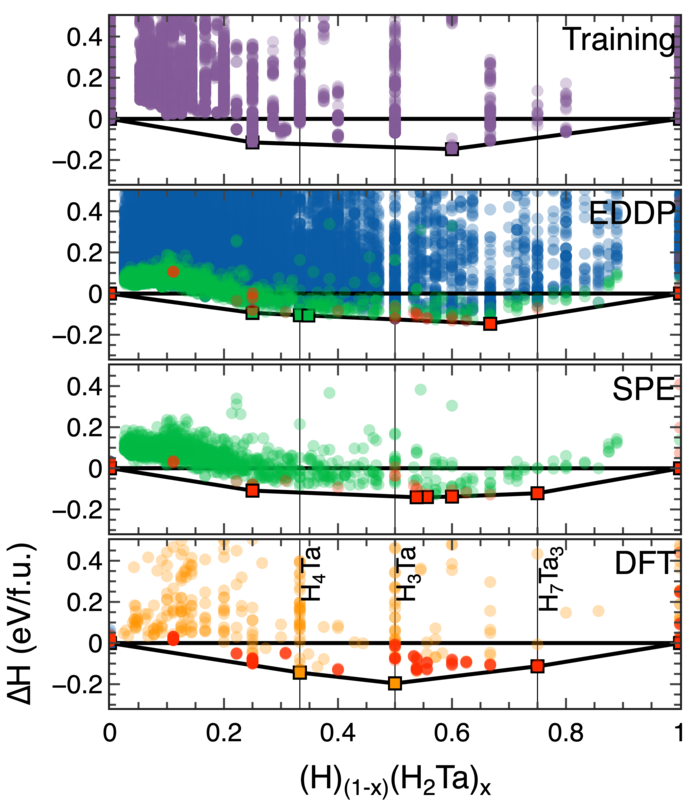}
\footnotesize


\flushleft{
\subsubsection*{\textsc{EDDP}}}
\centering
\includegraphics[width=0.3\textwidth]{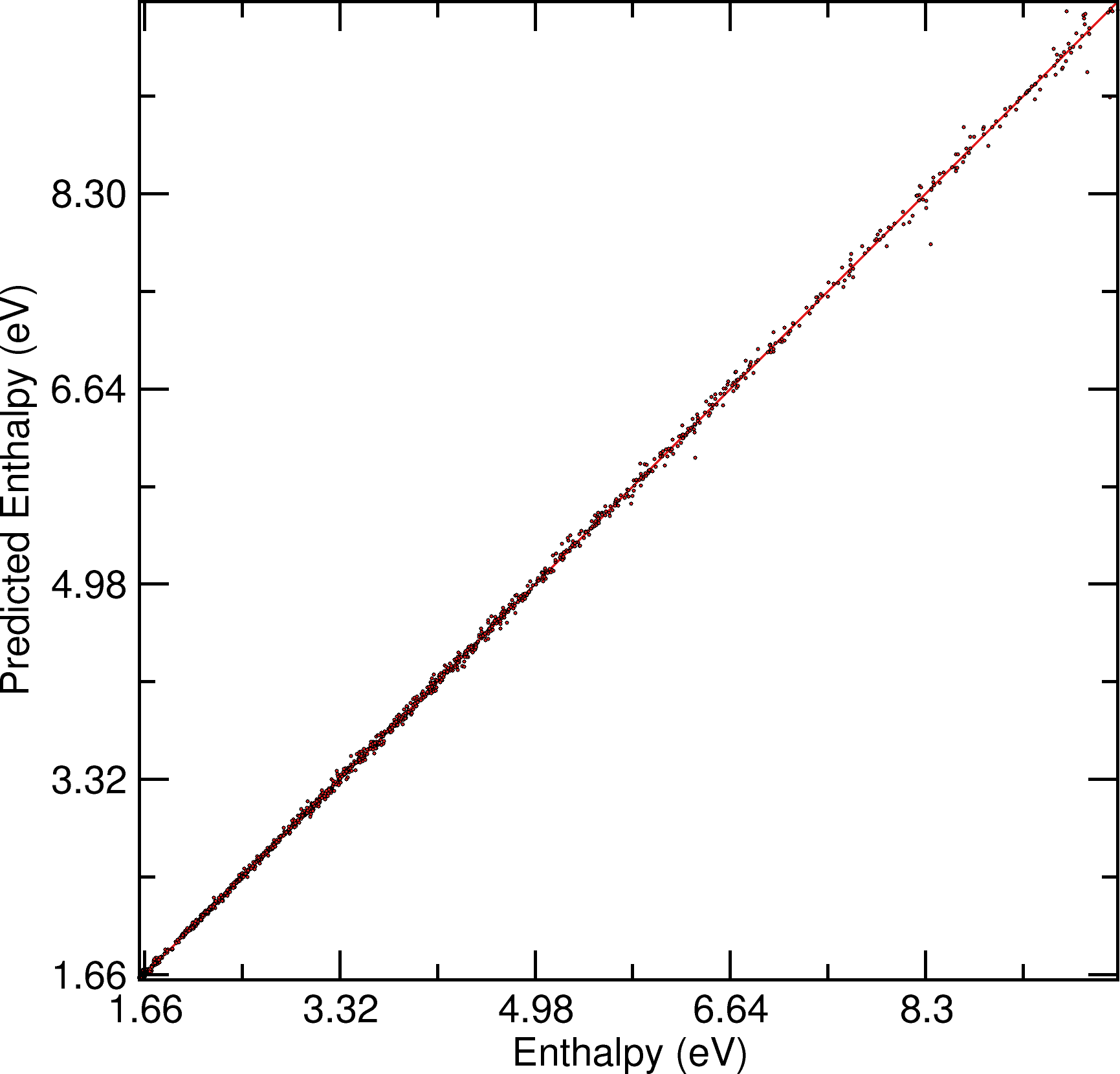}
\includegraphics[width=0.3\textwidth]{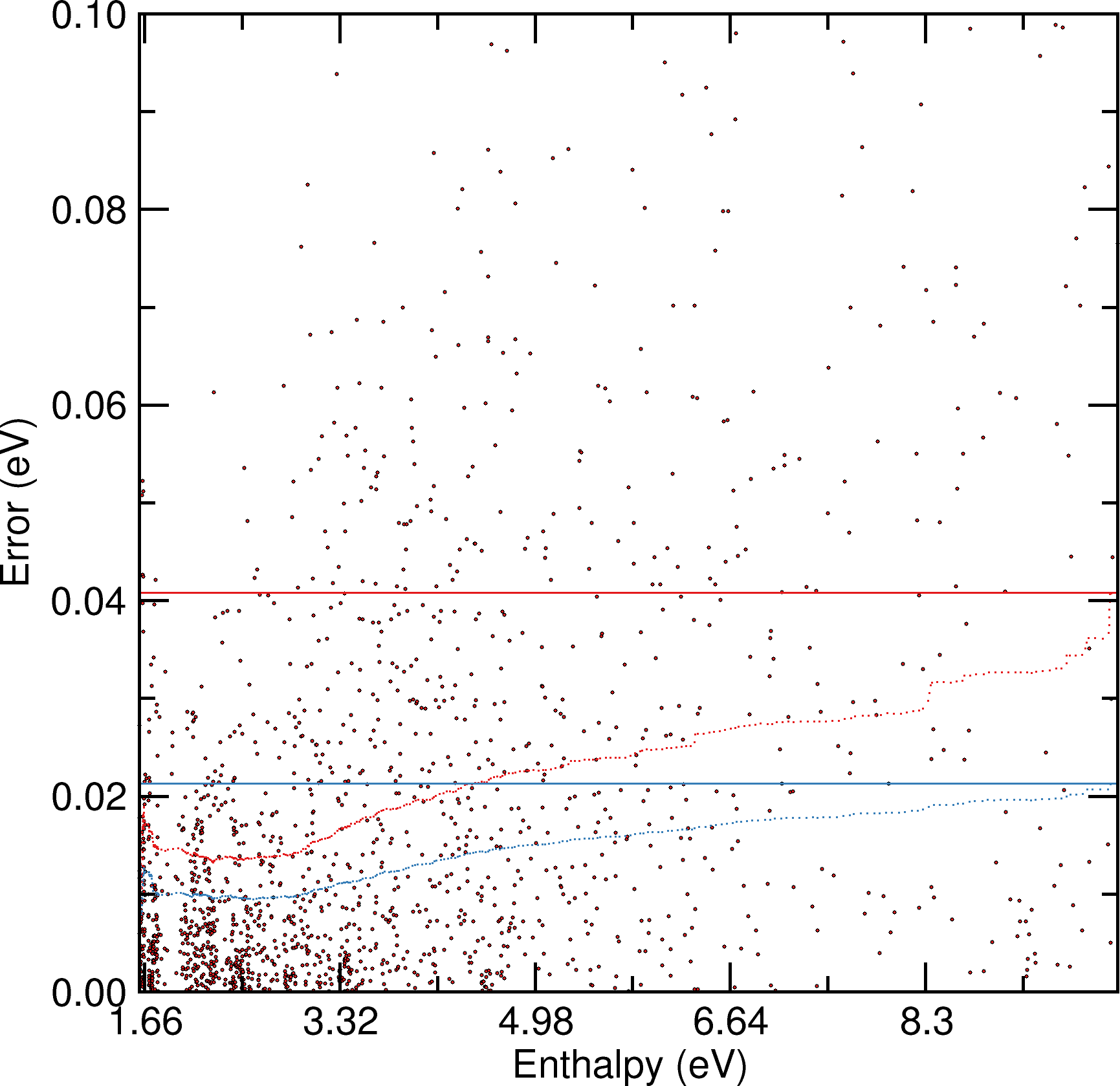}
\centering\begin{verbatim}
training    RMSE/MAE:  20.79  13.29  meV  Spearman  :  0.99985
validation  RMSE/MAE:  30.00  18.51  meV  Spearman  :  0.99981
testing     RMSE/MAE:  40.78  21.27  meV  Spearman  :  0.99980
\end{verbatim}
\clearpage

\flushleft{
\subsection{Tb-H}}
\subsubsection*{Searching}
\centering
\includegraphics[width=0.4\textwidth]{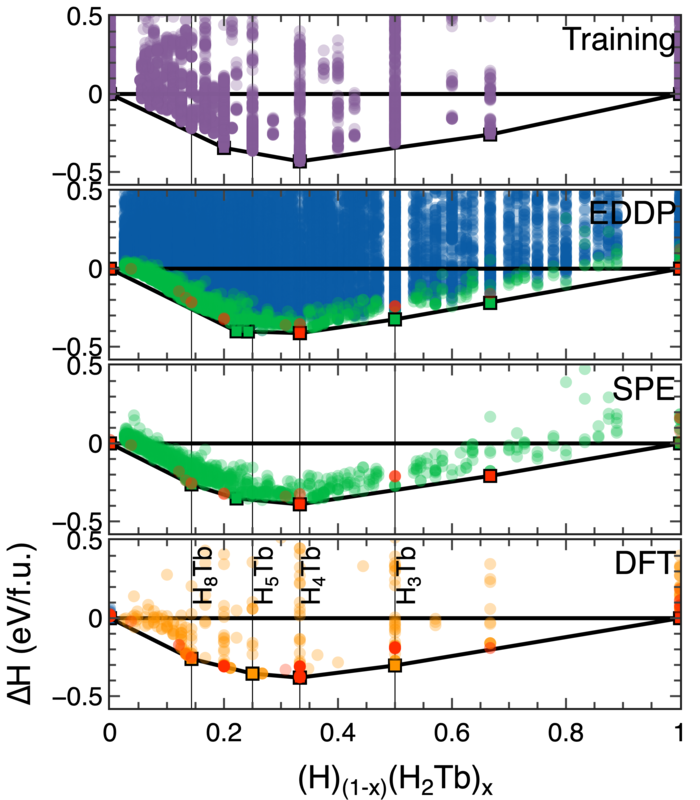}
\footnotesize


\flushleft{
\subsubsection*{\textsc{EDDP}}}
\centering
\includegraphics[width=0.3\textwidth]{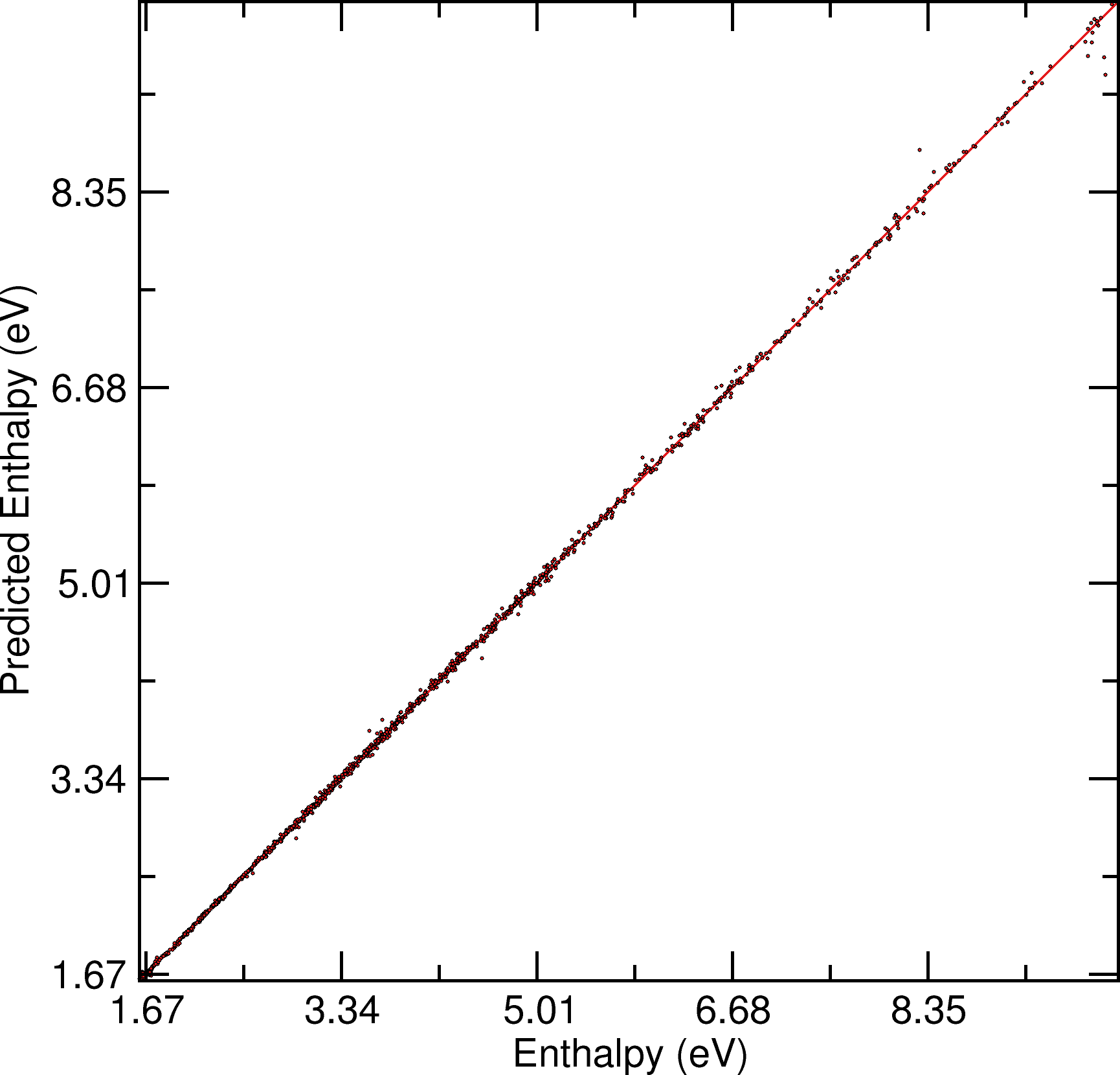}
\includegraphics[width=0.3\textwidth]{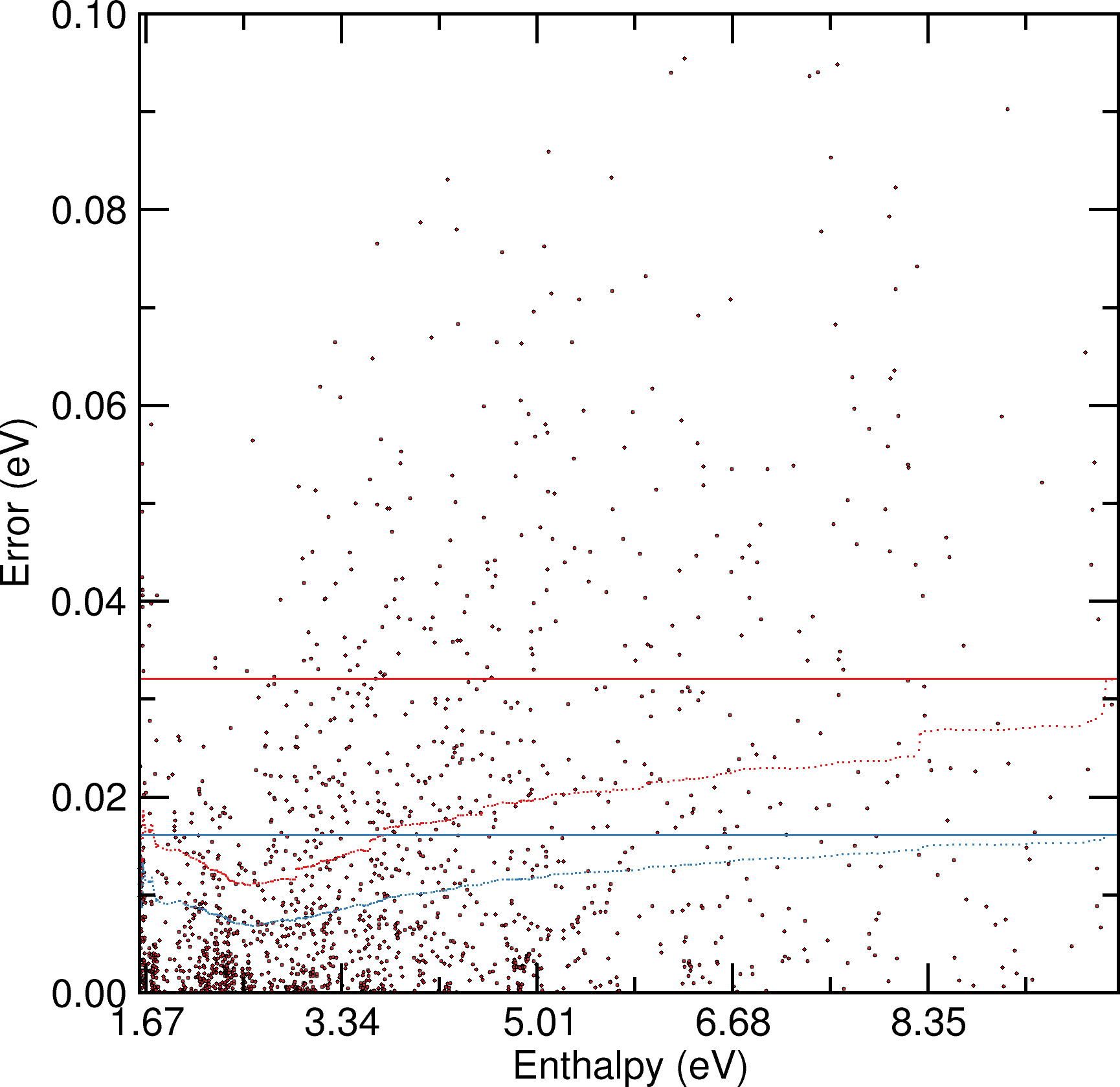}
\centering\begin{verbatim}
training    RMSE/MAE:  15.87  10.16  meV  Spearman  :  0.99988
validation  RMSE/MAE:  23.30  14.62  meV  Spearman  :  0.99985
testing     RMSE/MAE:  32.10  16.16  meV  Spearman  :  0.99984
\end{verbatim}
\clearpage

\flushleft{
\subsection{Tc-H}}
\subsubsection*{Searching}
\centering
\includegraphics[width=0.4\textwidth]{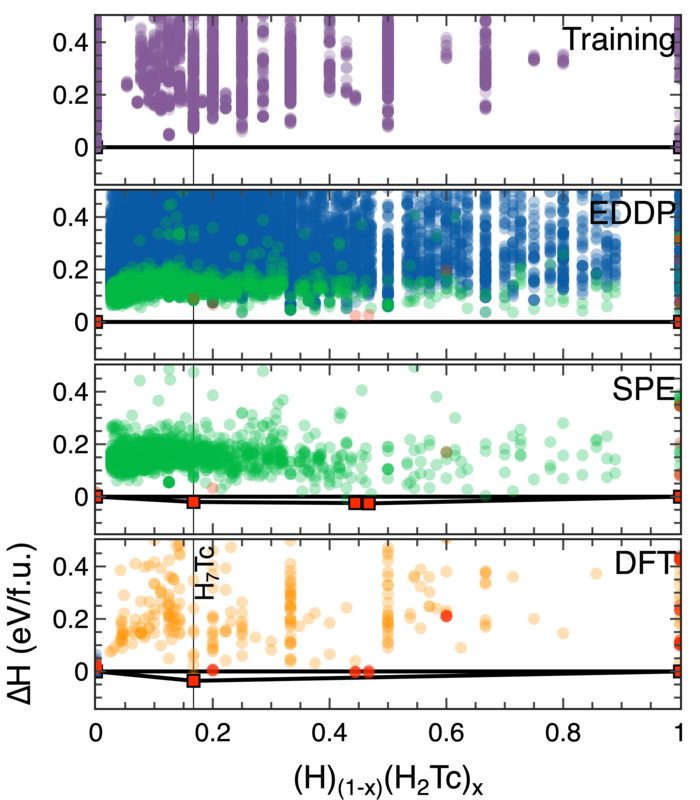}
\footnotesize


\flushleft{
\subsubsection*{\textsc{EDDP}}}
\centering
\includegraphics[width=0.3\textwidth]{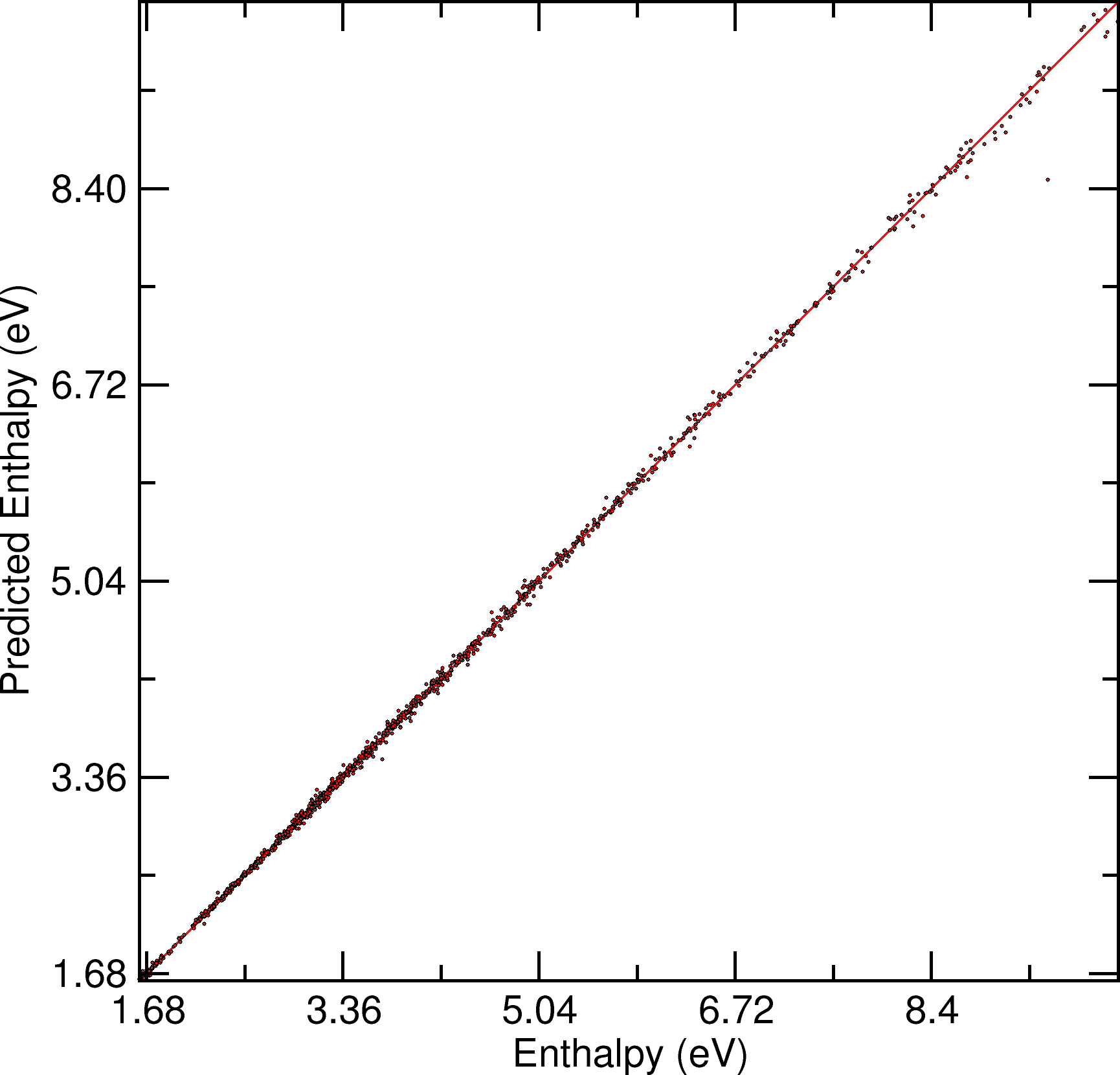}
\includegraphics[width=0.3\textwidth]{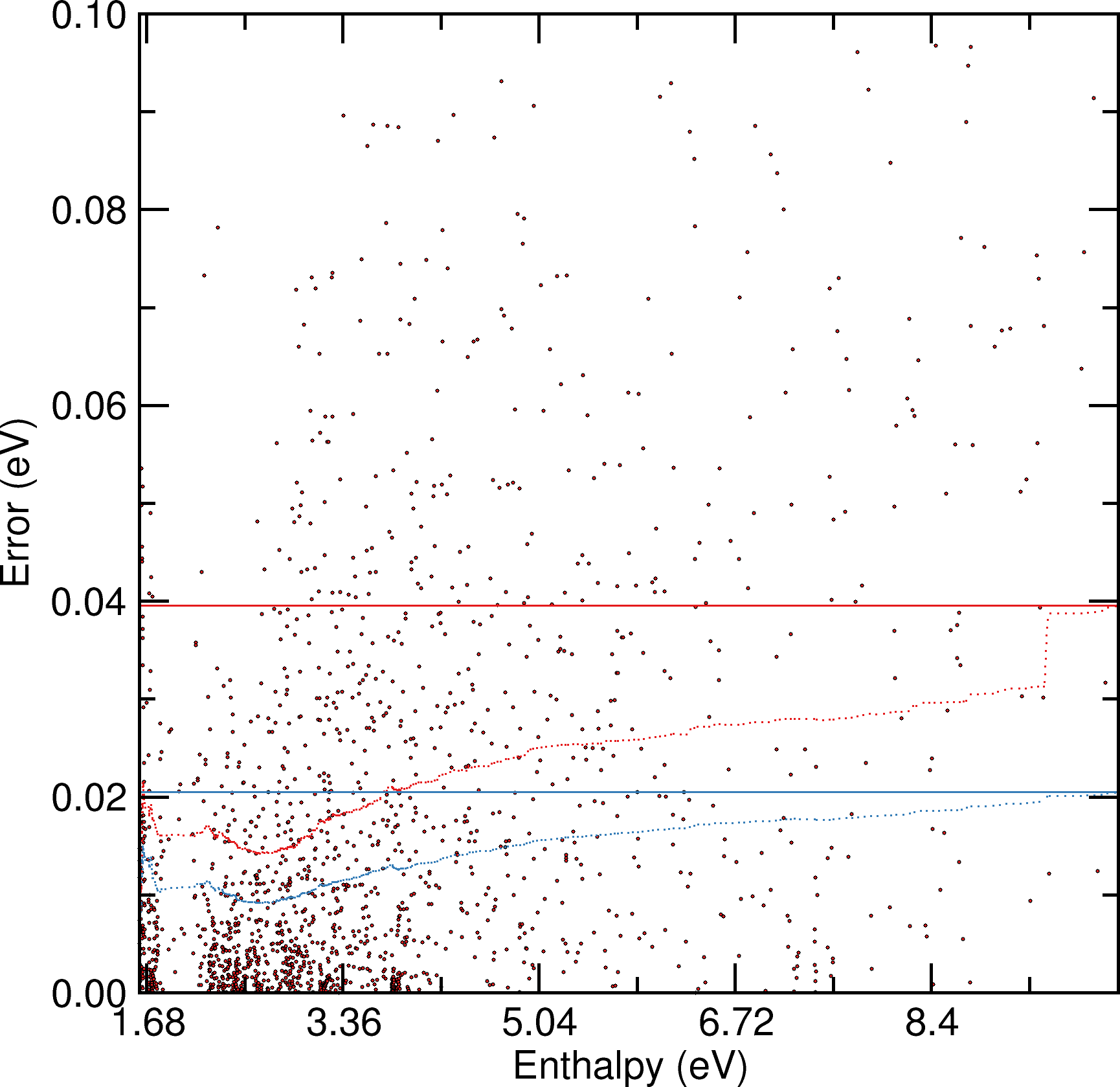}
\centering\begin{verbatim}
training    RMSE/MAE:  21.67  12.68  meV  Spearman  :  0.99983
validation  RMSE/MAE:  39.87  21.21  meV  Spearman  :  0.99975
testing     RMSE/MAE:  39.56  20.49  meV  Spearman  :  0.99971
\end{verbatim}
\clearpage

\flushleft{
\subsection{Te-H}}
\subsubsection*{Searching}
\centering
\includegraphics[width=0.4\textwidth]{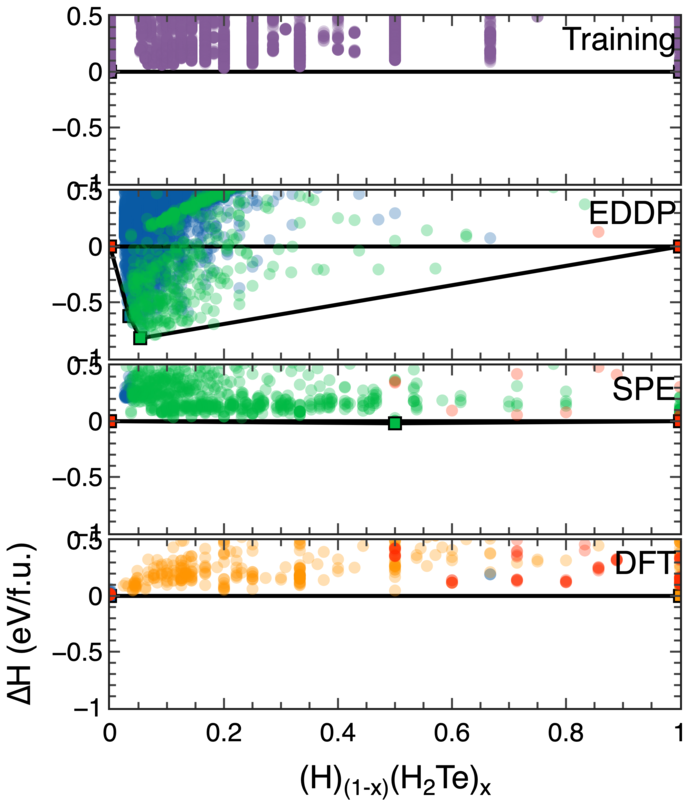}
\footnotesize


\flushleft{
\subsubsection*{\textsc{EDDP}}}
\centering
\includegraphics[width=0.3\textwidth]{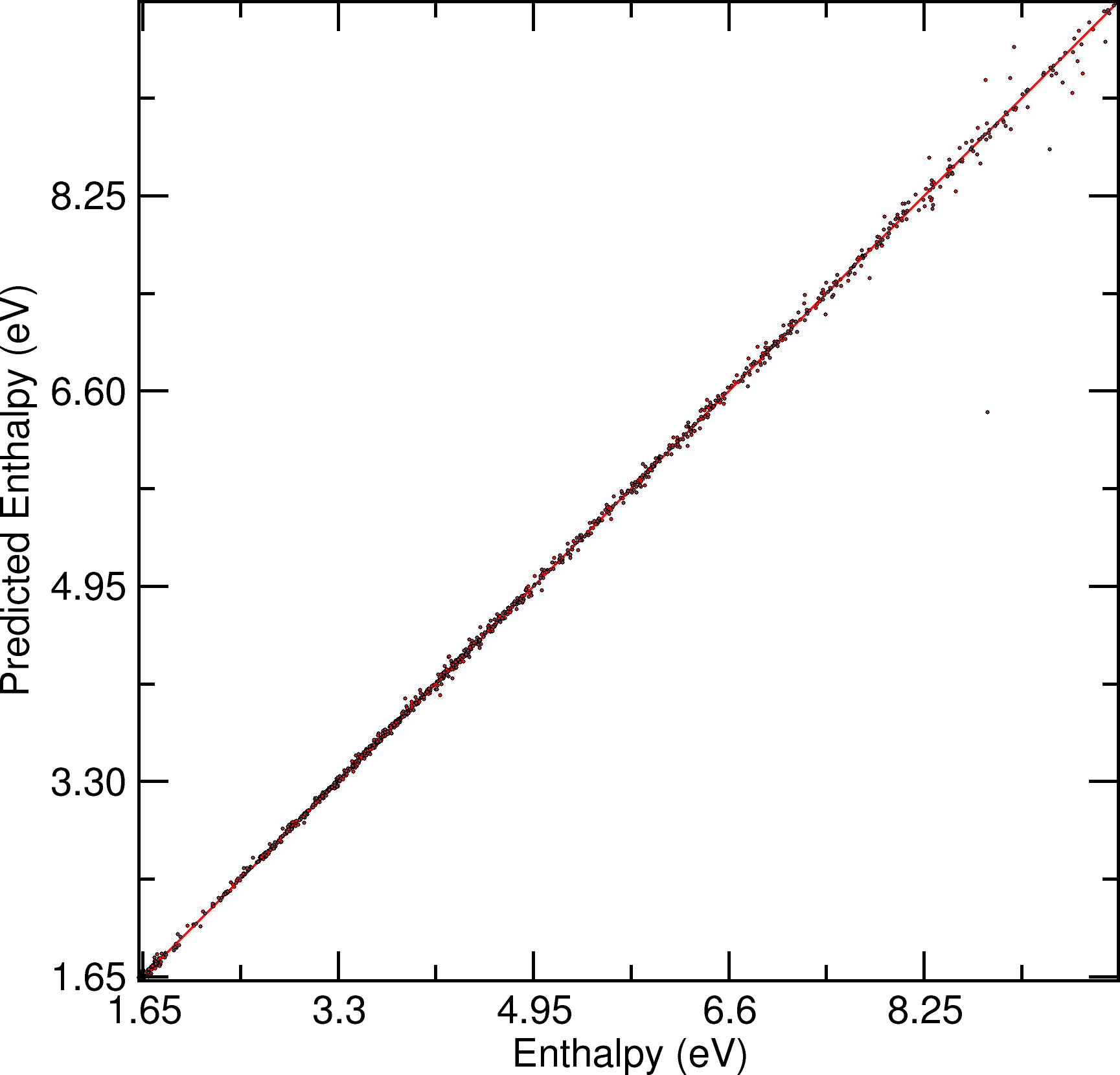}
\includegraphics[width=0.3\textwidth]{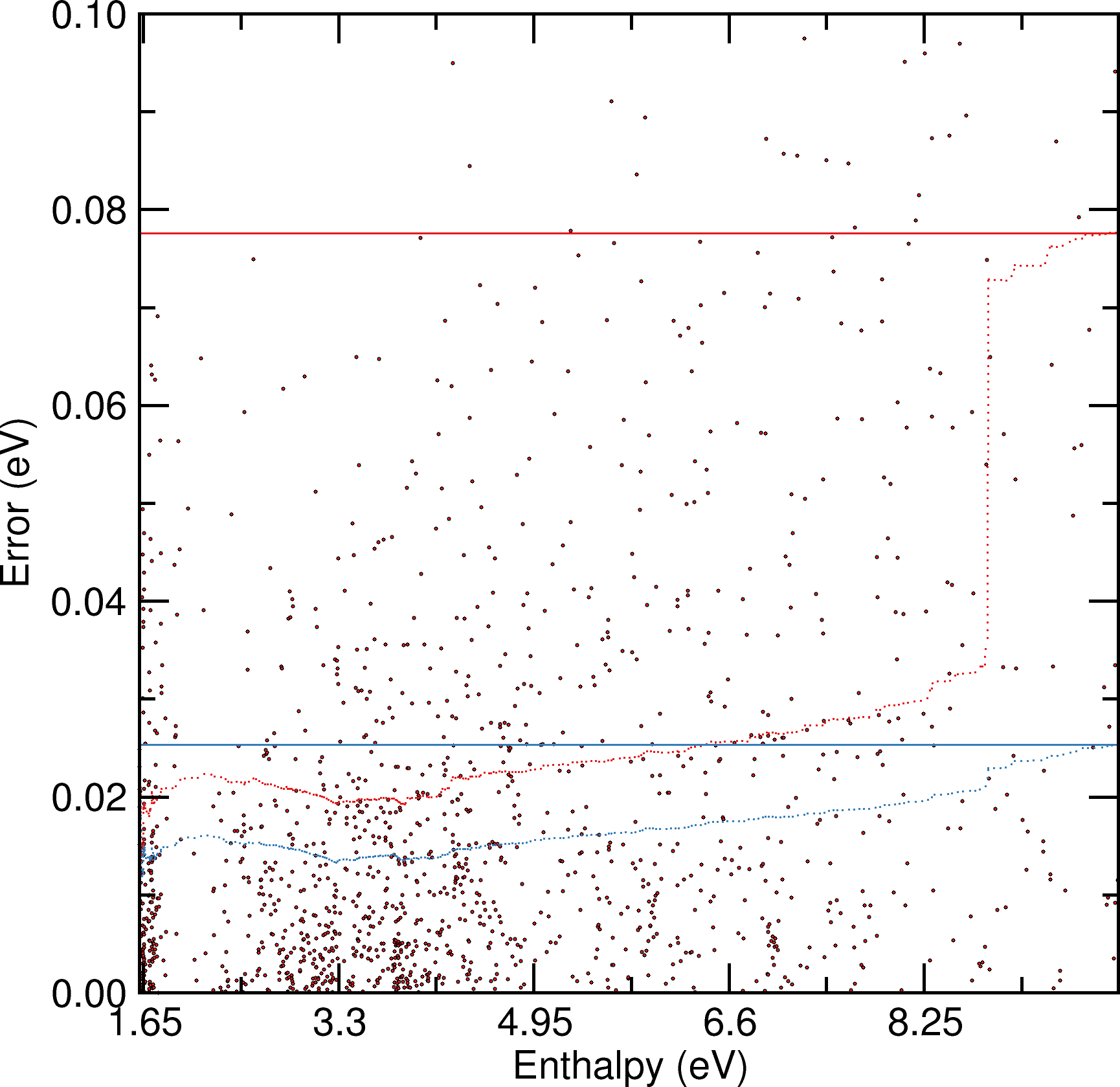}
\centering\begin{verbatim}
training    RMSE/MAE:  325.84  25.40  meV  Spearman  :  0.99865
validation  RMSE/MAE:  33.13   21.17  meV  Spearman  :  0.99975
testing     RMSE/MAE:  77.60   25.33  meV  Spearman  :  0.99962
\end{verbatim}
\clearpage

\flushleft{
\subsection{Ti-H}}
\subsubsection*{Searching}
\centering
\includegraphics[width=0.4\textwidth]{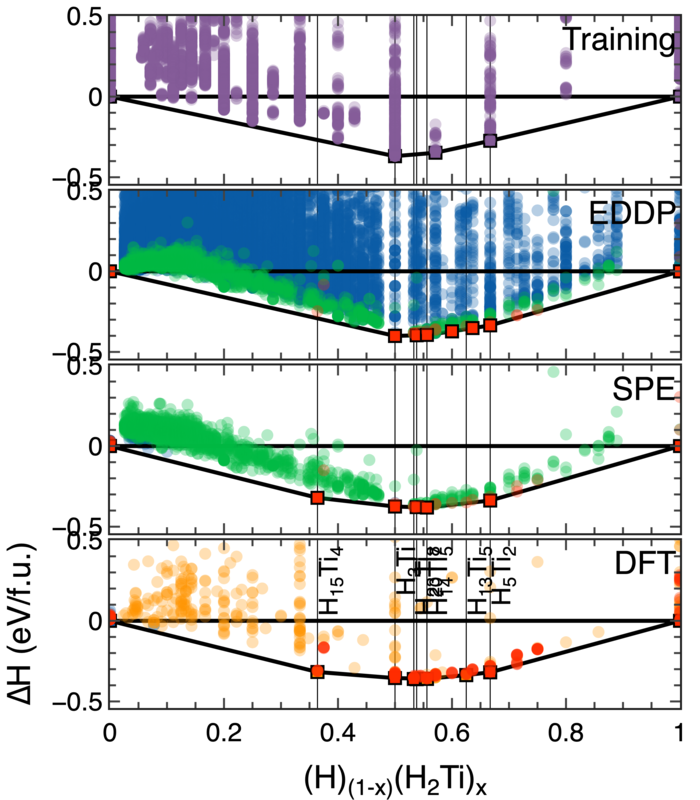}
\footnotesize


\flushleft{
\subsubsection*{\textsc{EDDP}}}
\centering
\includegraphics[width=0.3\textwidth]{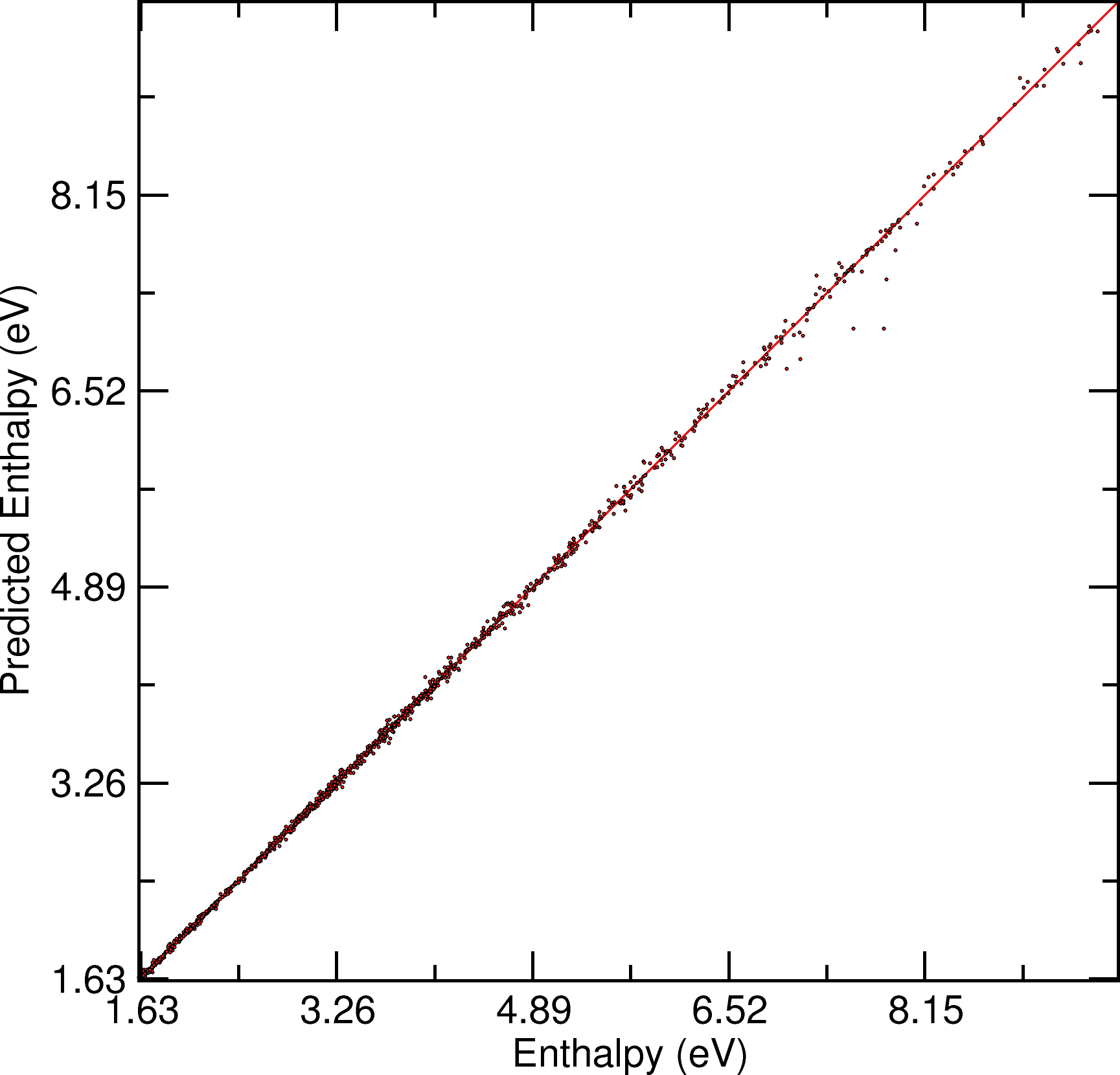}
\includegraphics[width=0.3\textwidth]{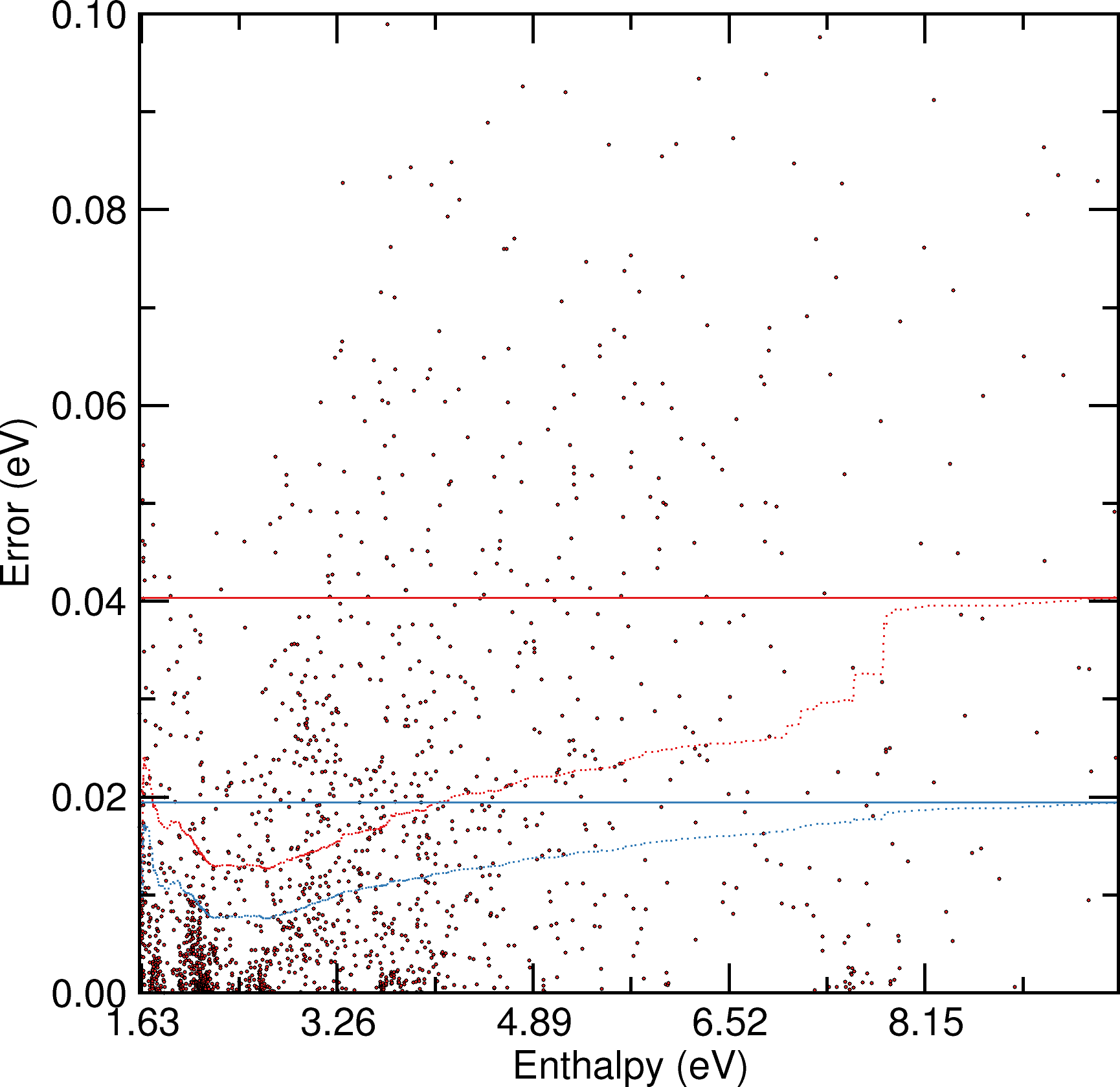}
\centering\begin{verbatim}
training    RMSE/MAE:  18.93  11.88  meV  Spearman  :  0.99983
validation  RMSE/MAE:  30.30  18.08  meV  Spearman  :  0.99982
testing     RMSE/MAE:  40.36  19.43  meV  Spearman  :  0.99977
\end{verbatim}
\clearpage

\flushleft{
\subsection{Tl-H}}
\subsubsection*{Searching}
\centering
\includegraphics[width=0.4\textwidth]{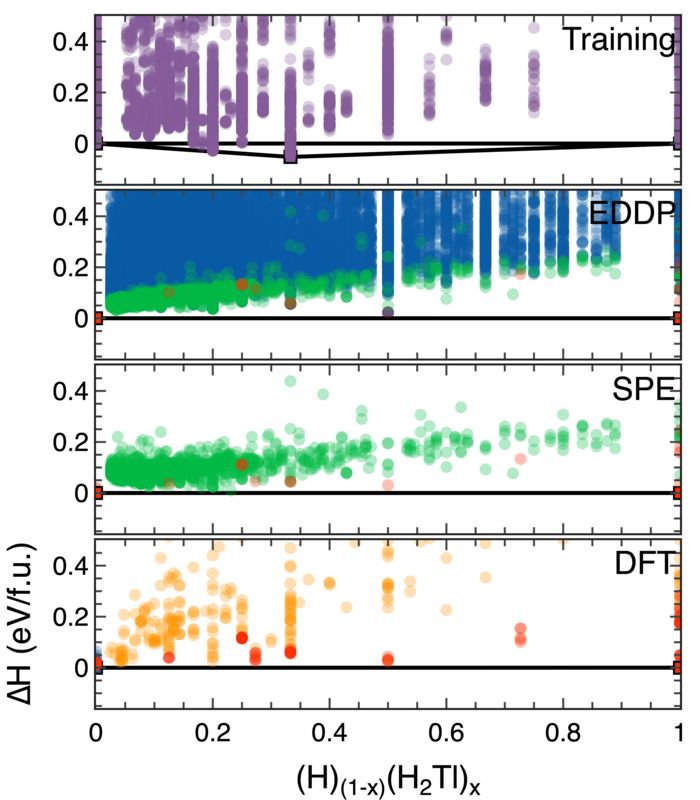}
\footnotesize


\flushleft{
\subsubsection*{\textsc{EDDP}}}
\centering
\includegraphics[width=0.3\textwidth]{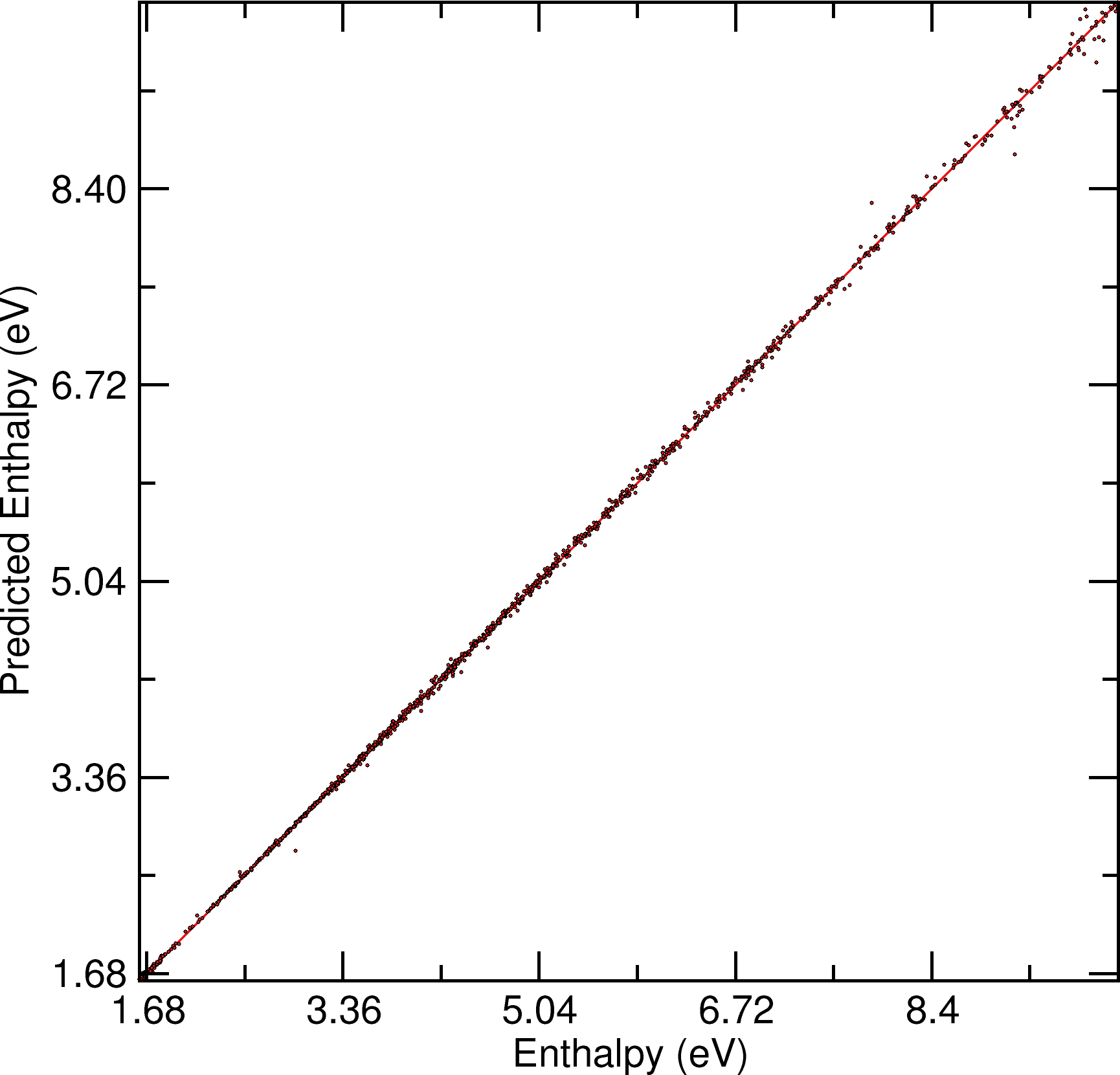}
\includegraphics[width=0.3\textwidth]{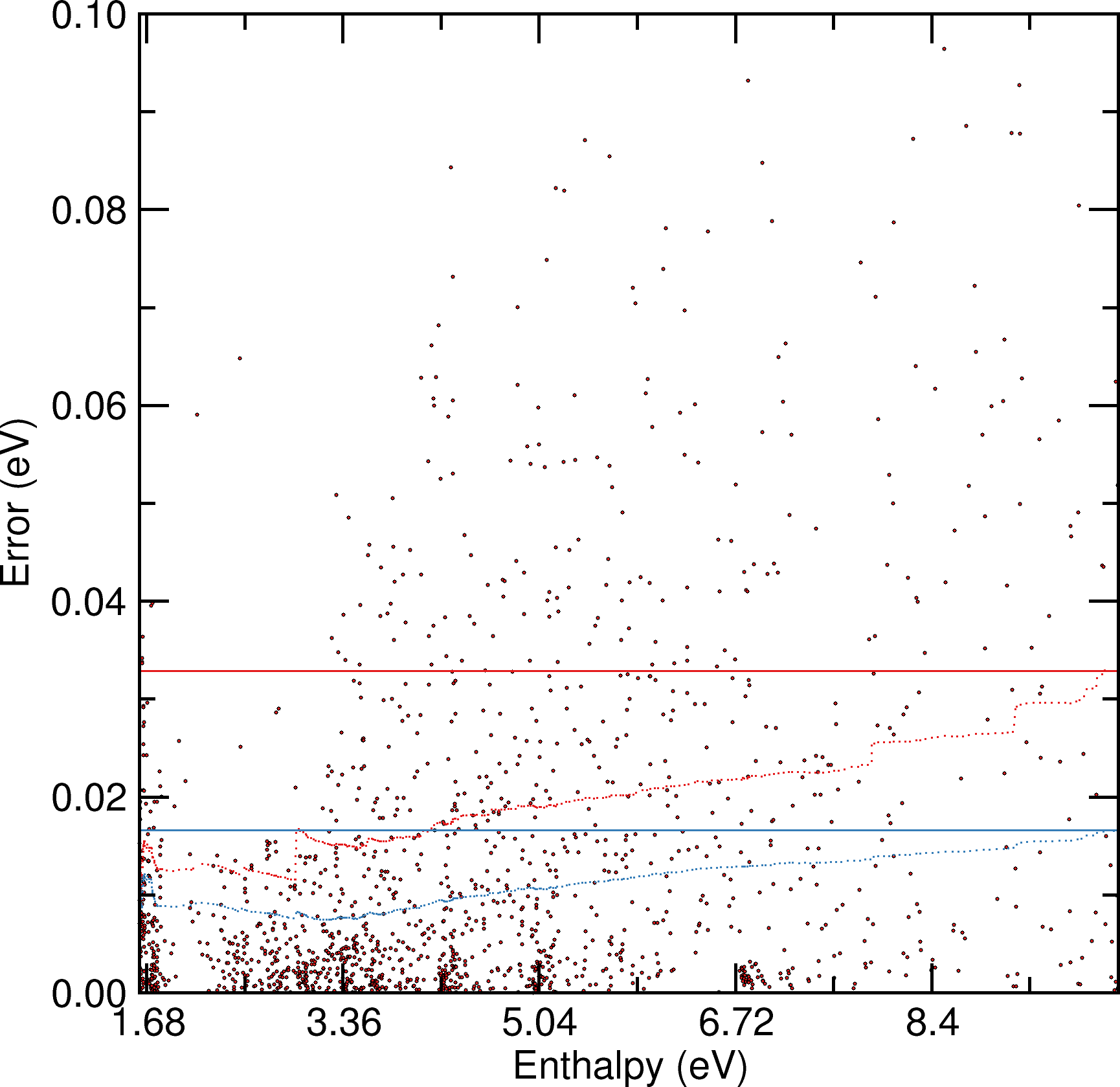}
\centering\begin{verbatim}
training    RMSE/MAE:  13.81  8.51   meV  Spearman  :  0.99991
validation  RMSE/MAE:  26.55  14.61  meV  Spearman  :  0.99985
testing     RMSE/MAE:  32.88  16.57  meV  Spearman  :  0.99987
\end{verbatim}
\clearpage

\flushleft{
\subsection{Tm-H}}
\subsubsection*{Searching}
\centering
\includegraphics[width=0.4\textwidth]{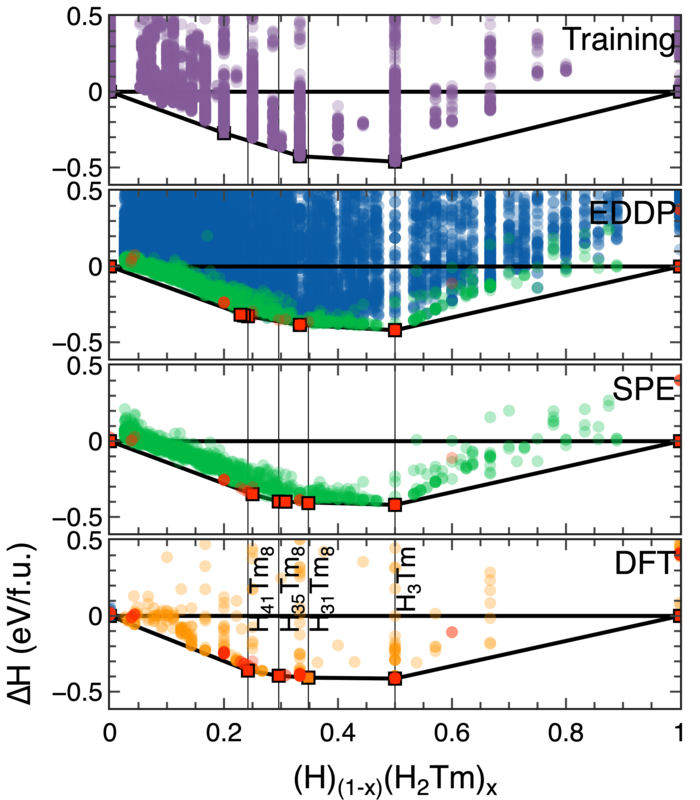}
\footnotesize


\flushleft{
\subsubsection*{\textsc{EDDP}}}
\centering
\includegraphics[width=0.3\textwidth]{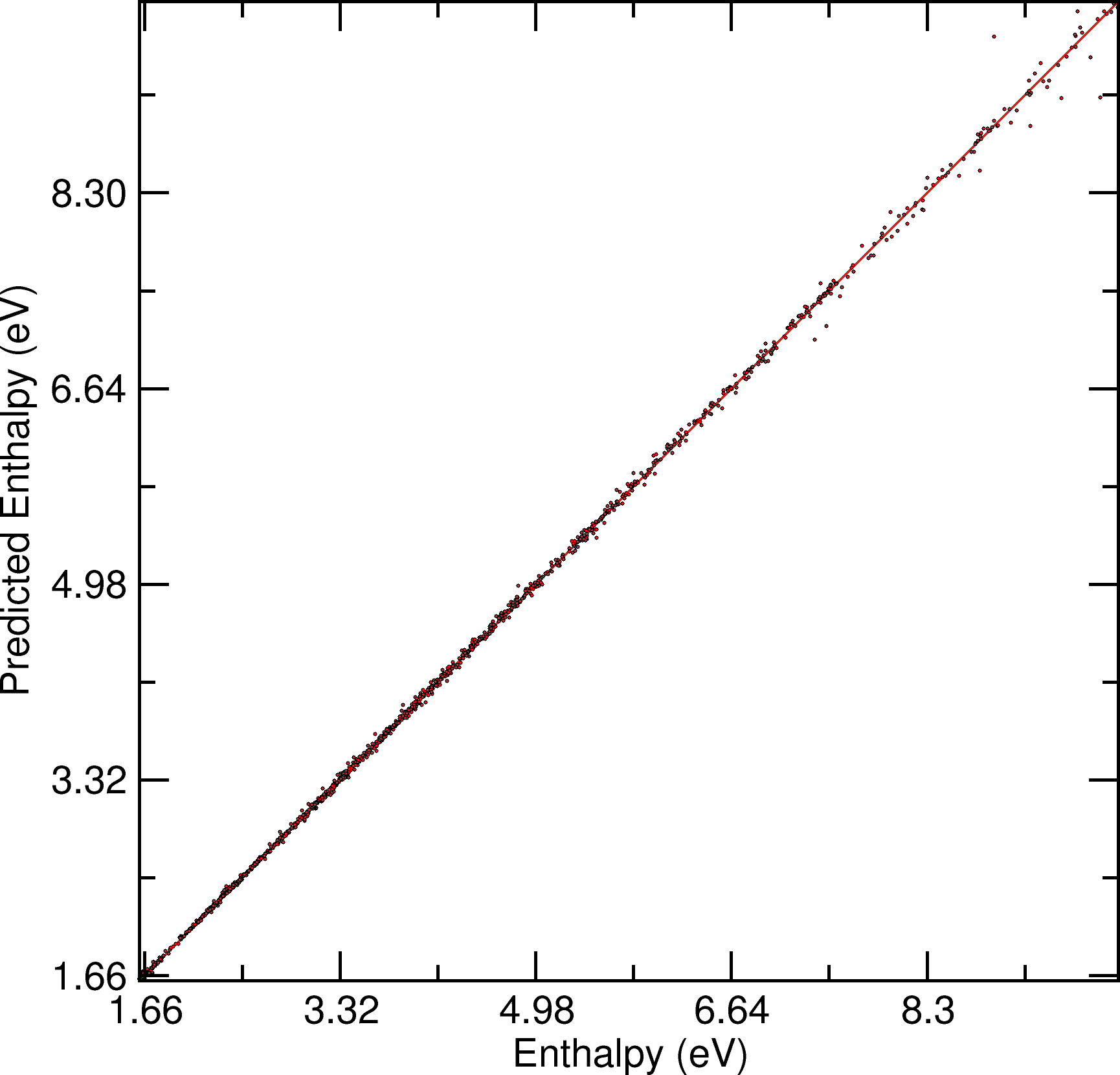}
\includegraphics[width=0.3\textwidth]{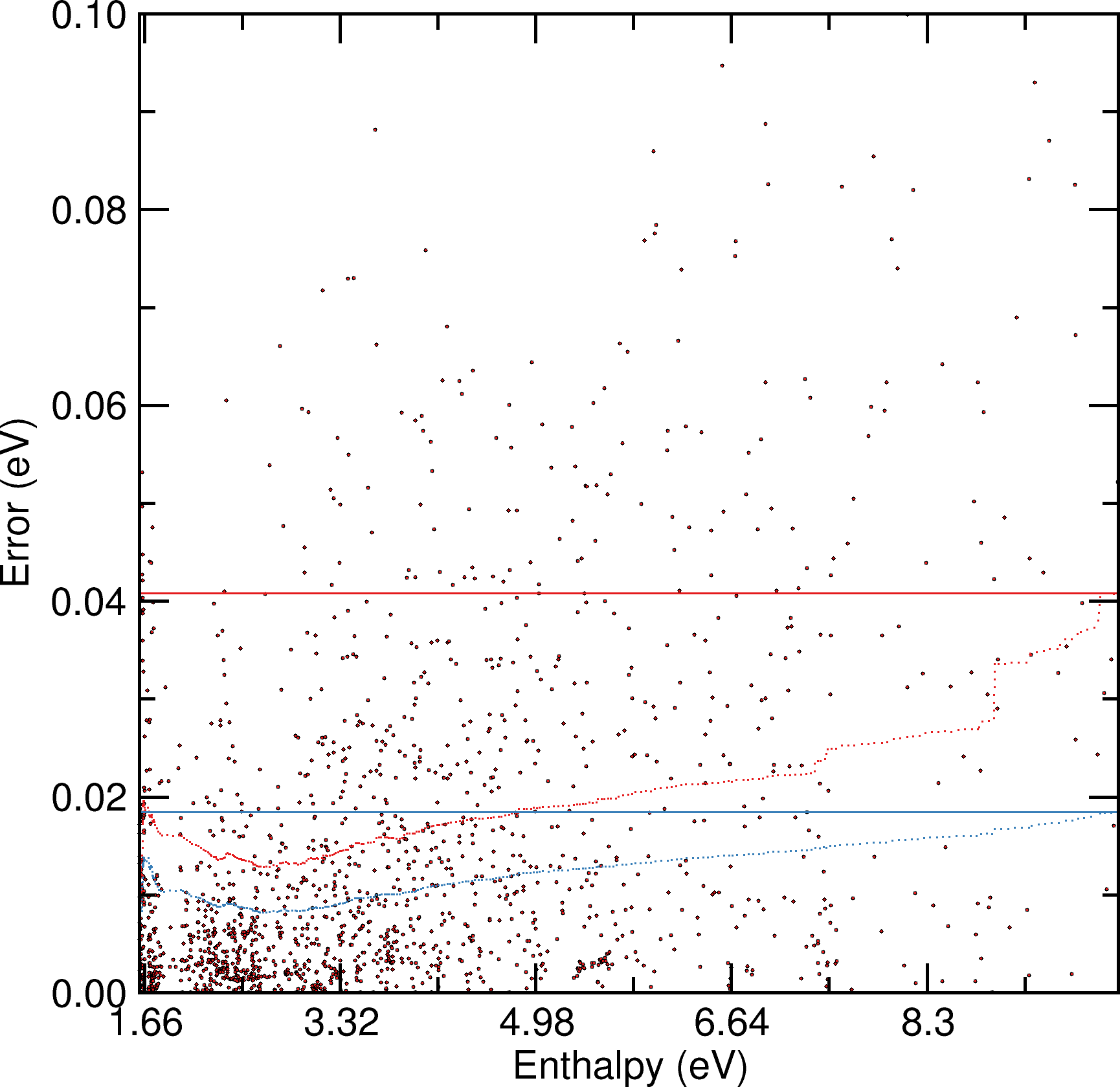}
\centering\begin{verbatim}
training    RMSE/MAE:  17.55  11.11  meV  Spearman  :  0.99989
validation  RMSE/MAE:  24.20  16.07  meV  Spearman  :  0.99982
testing     RMSE/MAE:  40.81  18.47  meV  Spearman  :  0.99983
\end{verbatim}
\clearpage

\flushleft{
\subsection{V-H}}
\subsubsection*{Searching}
\centering
\includegraphics[width=0.4\textwidth]{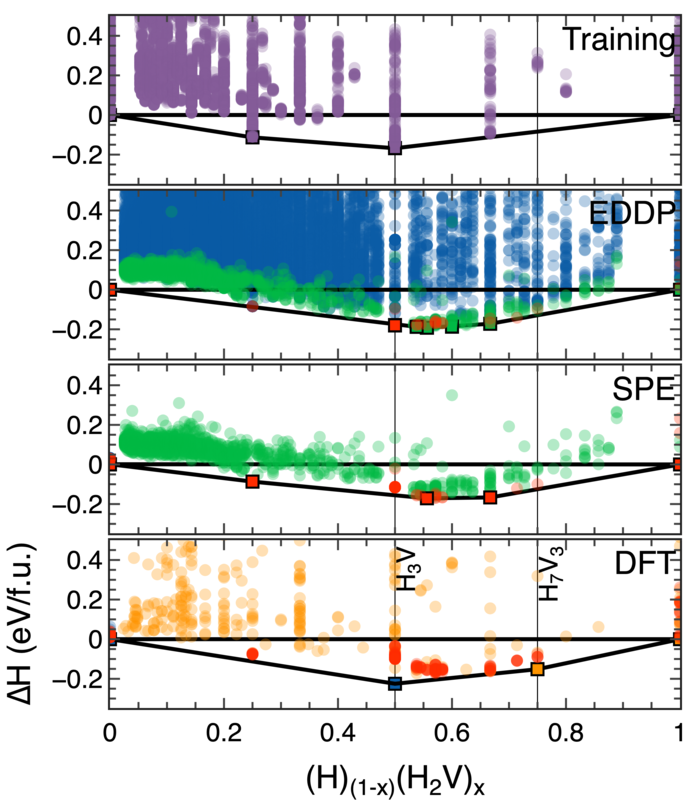}
\footnotesize


\flushleft{
\subsubsection*{\textsc{EDDP}}}
\centering
\includegraphics[width=0.3\textwidth]{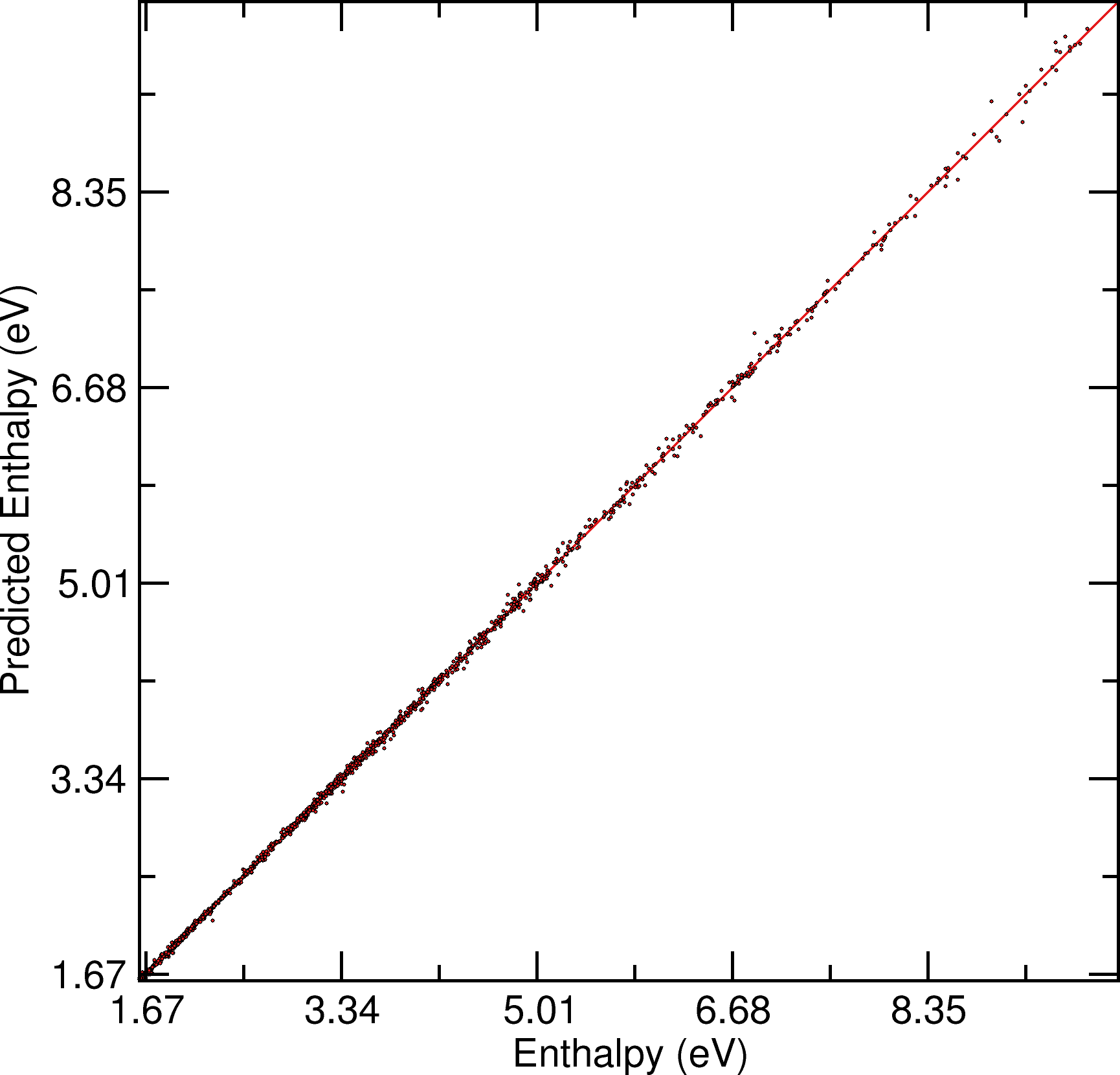}
\includegraphics[width=0.3\textwidth]{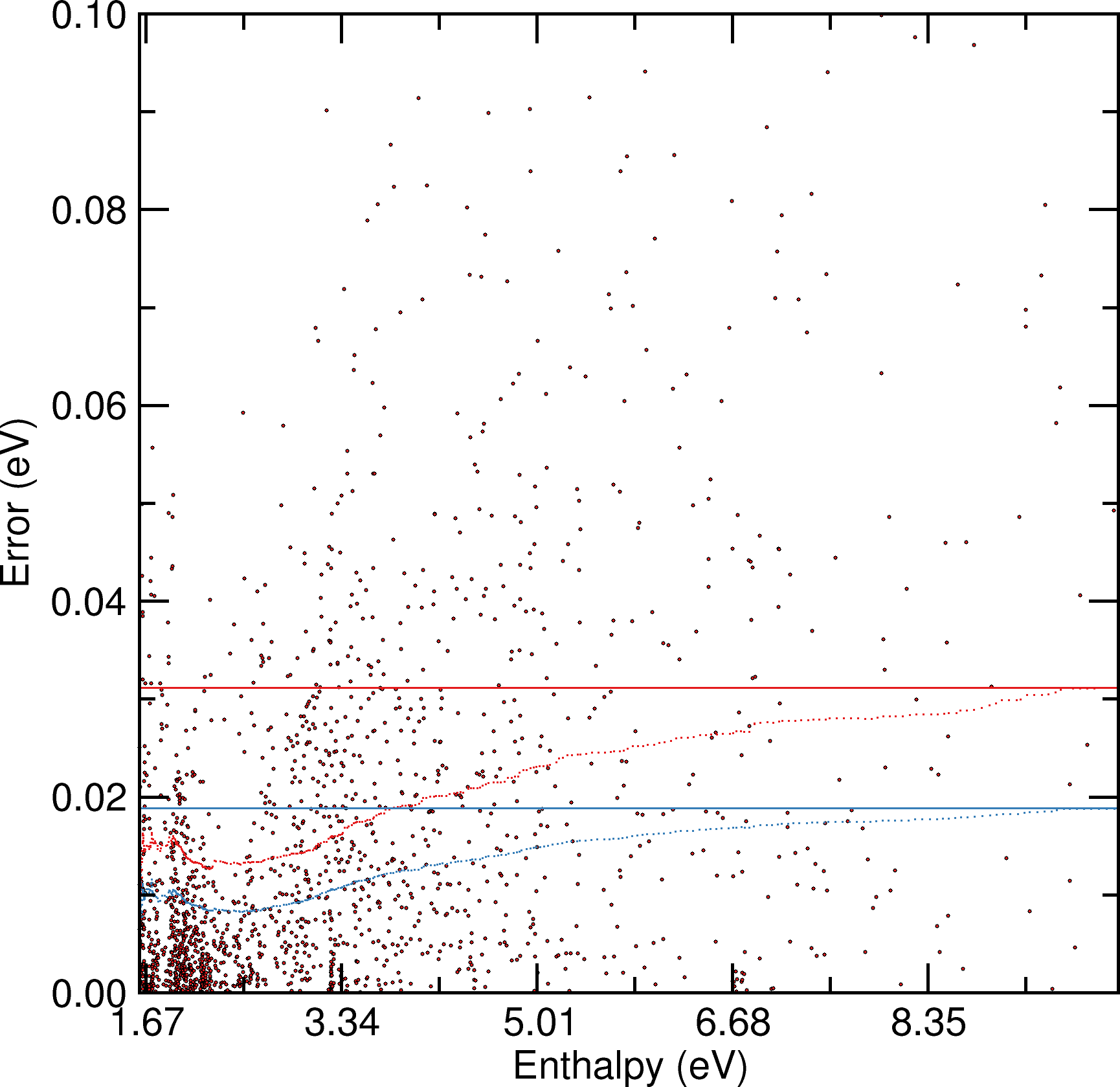}
\centering\begin{verbatim}
training    RMSE/MAE:  20.02  12.69  meV  Spearman  :  0.99982
validation  RMSE/MAE:  34.12  19.37  meV  Spearman  :  0.99973
testing     RMSE/MAE:  31.17  18.82  meV  Spearman  :  0.99973
\end{verbatim}
\clearpage

\flushleft{
\subsection{W-H}}
\subsubsection*{Searching}
\centering
\includegraphics[width=0.4\textwidth]{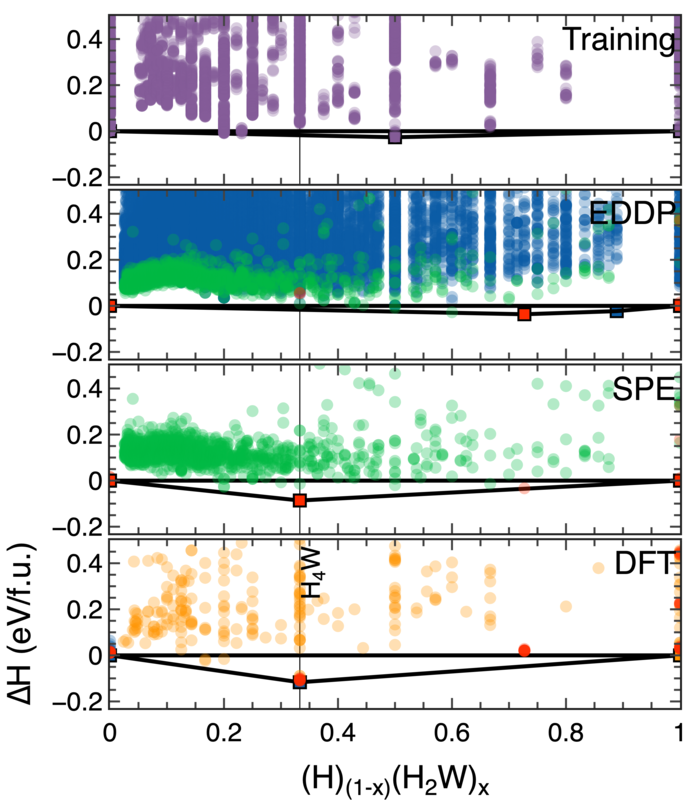}
\footnotesize


\flushleft{
\subsubsection*{\textsc{EDDP}}}
\centering
\includegraphics[width=0.3\textwidth]{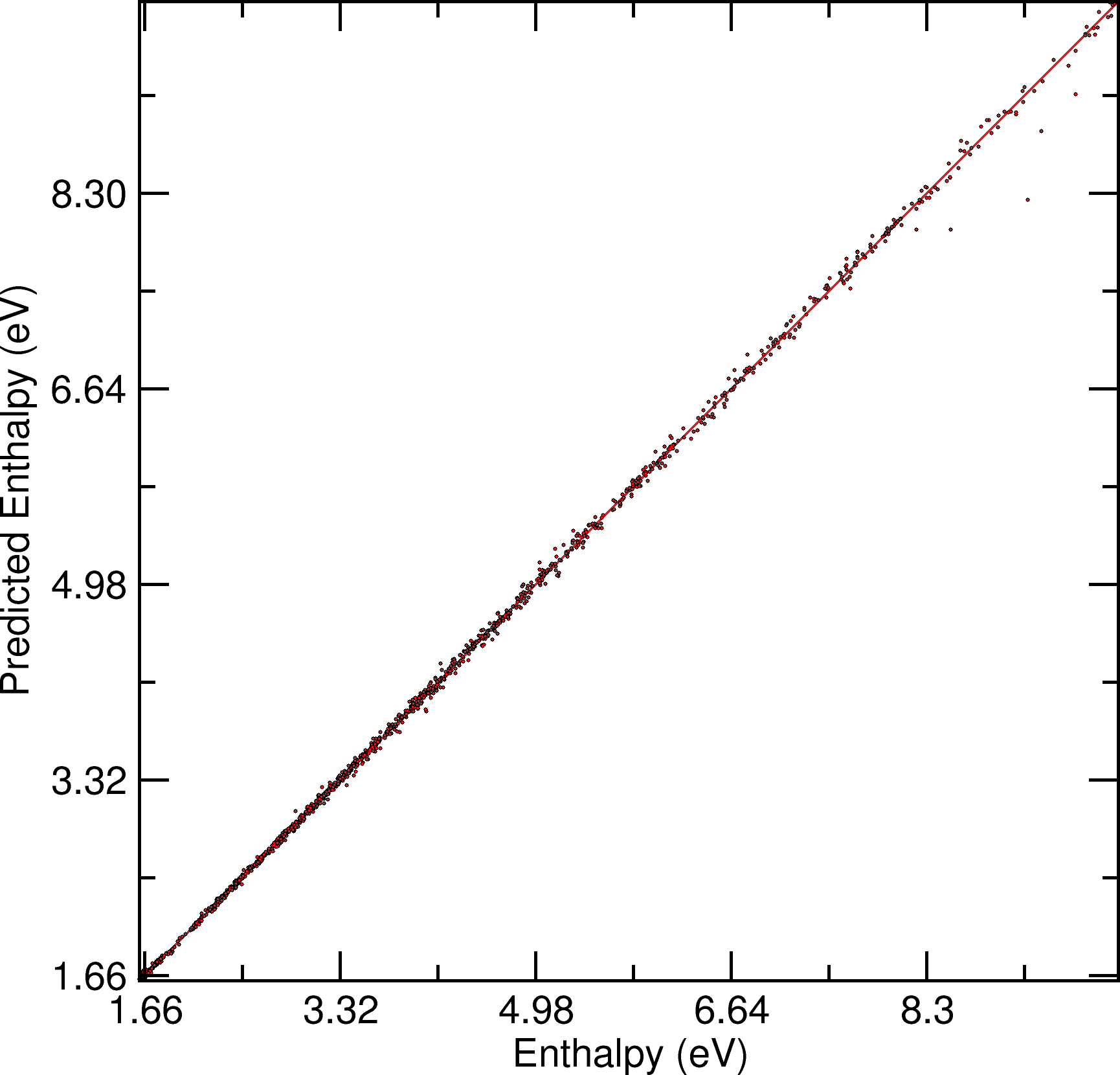}
\includegraphics[width=0.3\textwidth]{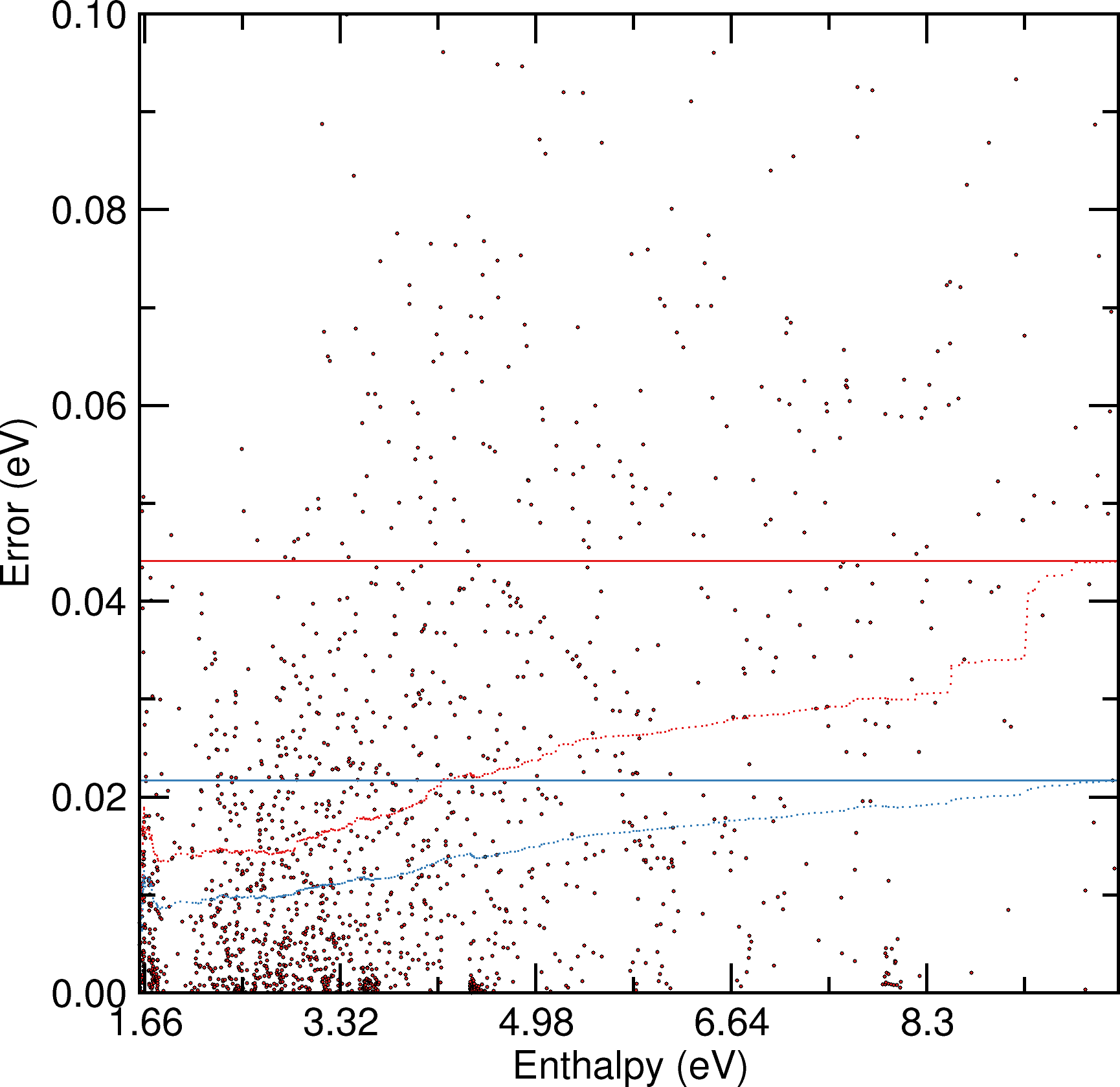}
\centering\begin{verbatim}
training    RMSE/MAE:  22.42  13.51  meV  Spearman  :  0.99983
validation  RMSE/MAE:  32.62  19.94  meV  Spearman  :  0.99974
testing     RMSE/MAE:  44.09  21.69  meV  Spearman  :  0.99980
\end{verbatim}
\clearpage

\flushleft{
\subsection{Xe-H}}
\subsubsection*{Searching}
\centering
\includegraphics[width=0.4\textwidth]{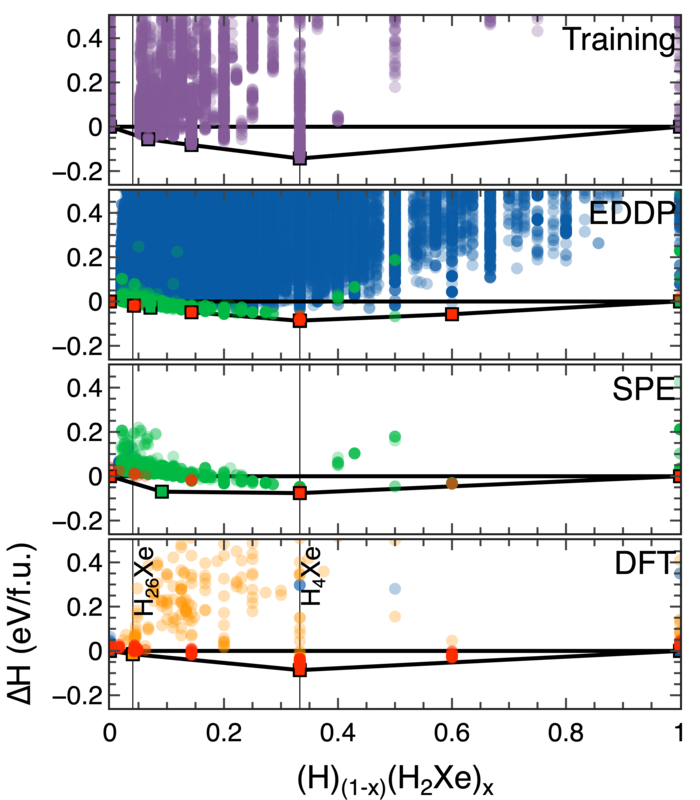}
\footnotesize


\flushleft{
\subsubsection*{\textsc{EDDP}}}
\centering
\includegraphics[width=0.3\textwidth]{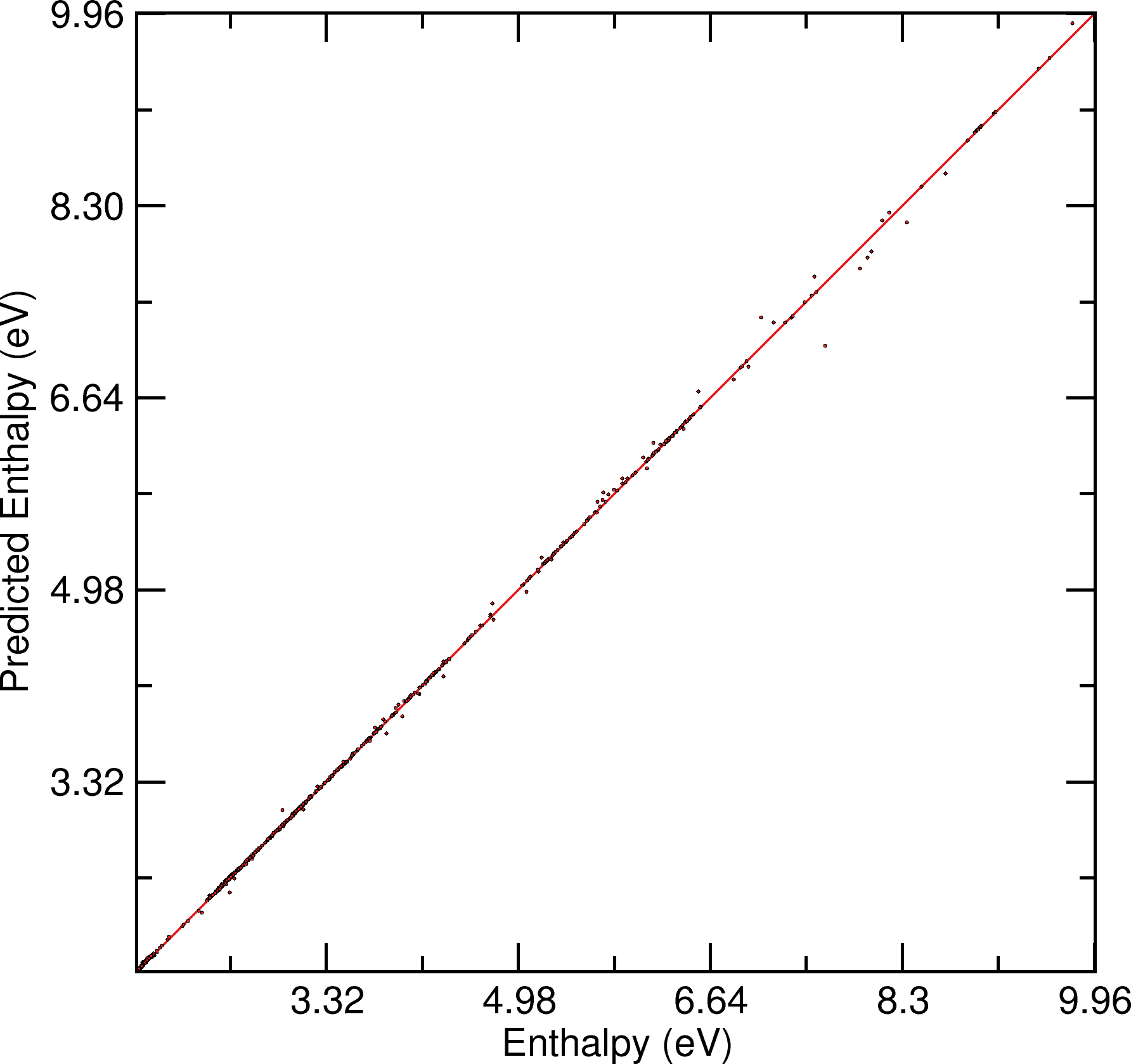}
\includegraphics[width=0.3\textwidth]{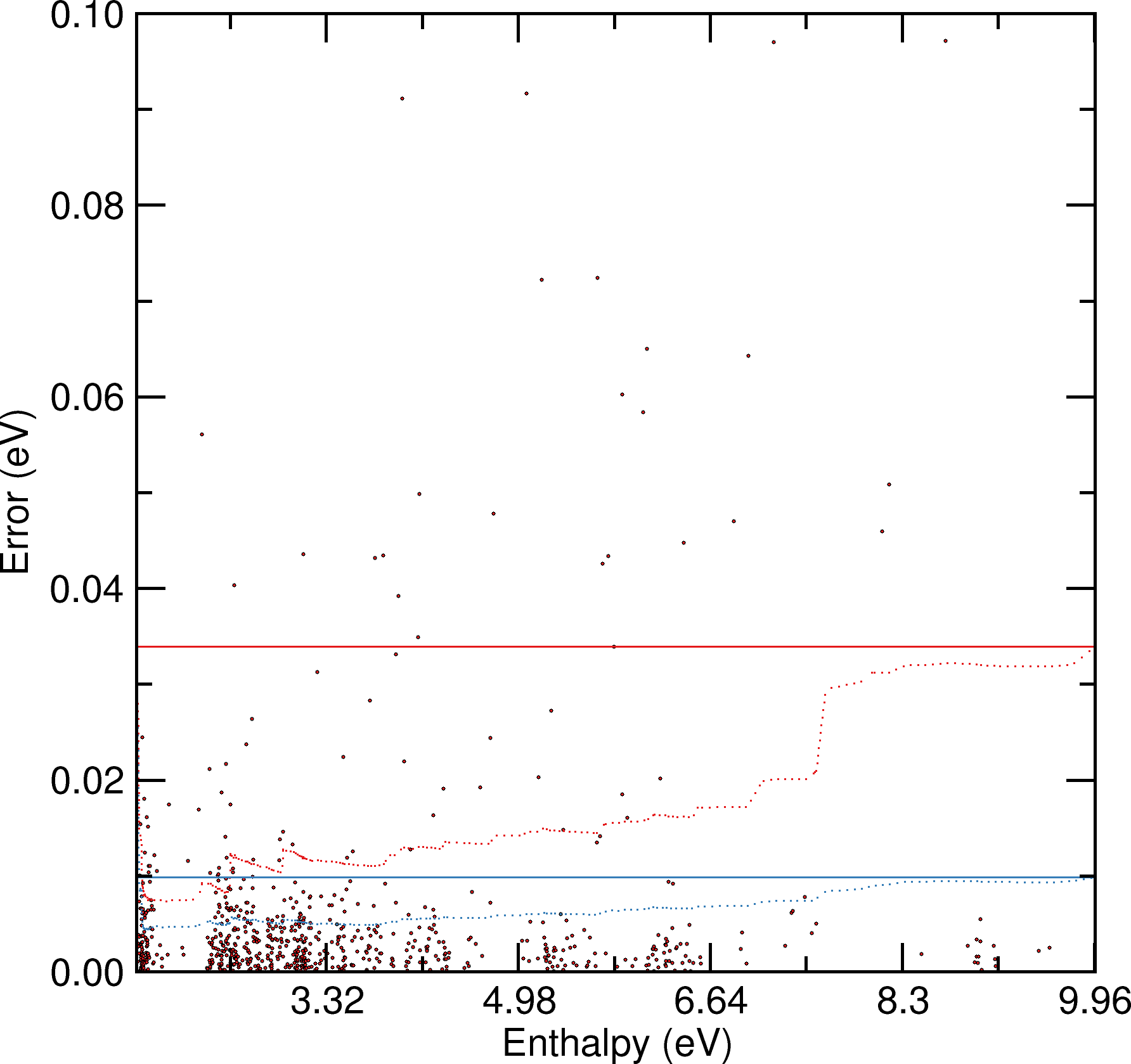}
\centering\begin{verbatim}
training    RMSE/MAE:  7.73   3.58   meV  Spearman  :  0.99995
validation  RMSE/MAE:  34.81  11.07  meV  Spearman  :  0.99980
testing     RMSE/MAE:  33.94  9.88   meV  Spearman  :  0.99983
\end{verbatim}
\clearpage

\flushleft{
\subsection{Y-H}}
\subsubsection*{Searching}
\centering
\includegraphics[width=0.4\textwidth]{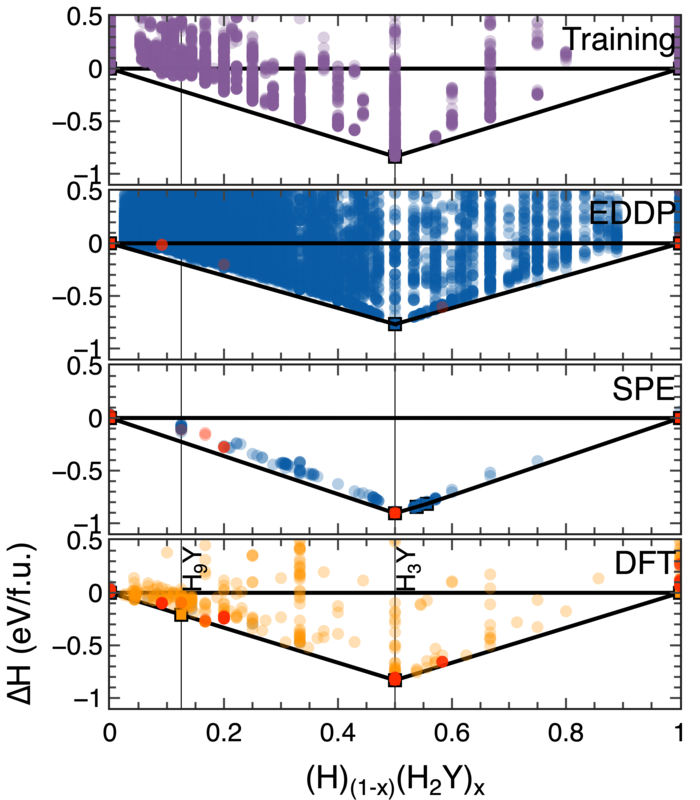}
\footnotesize


\flushleft{
\subsubsection*{\textsc{EDDP}}}
\centering
\includegraphics[width=0.3\textwidth]{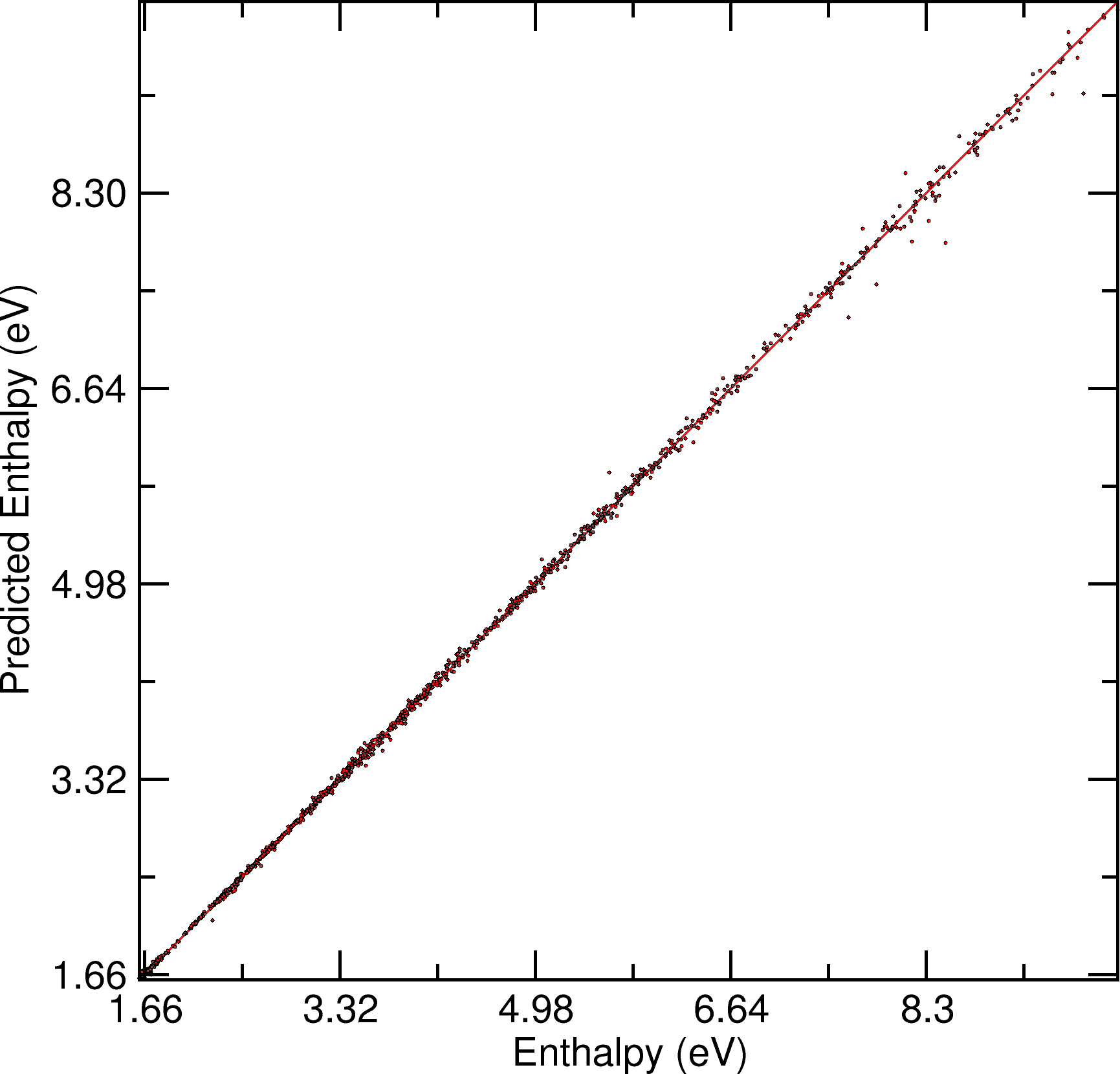}
\includegraphics[width=0.3\textwidth]{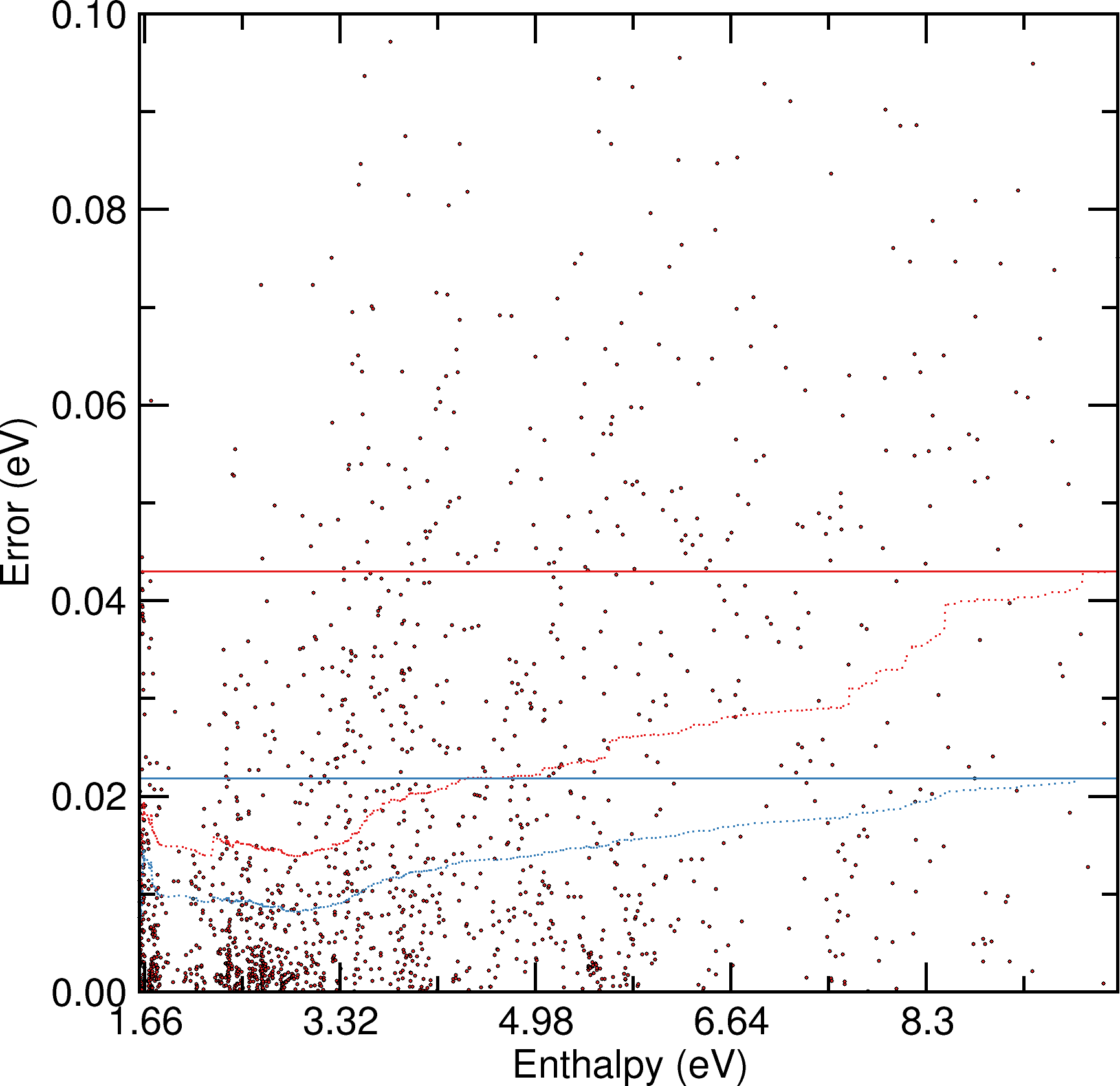}
\centering\begin{verbatim}
training    RMSE/MAE:  21.50  13.41  meV  Spearman  :  0.99987
validation  RMSE/MAE:  33.35  19.89  meV  Spearman  :  0.99985
testing     RMSE/MAE:  42.97  21.84  meV  Spearman  :  0.99976
\end{verbatim}
\clearpage

\flushleft{
\subsection{Yb-H}}
\subsubsection*{Searching}
\centering
\includegraphics[width=0.4\textwidth]{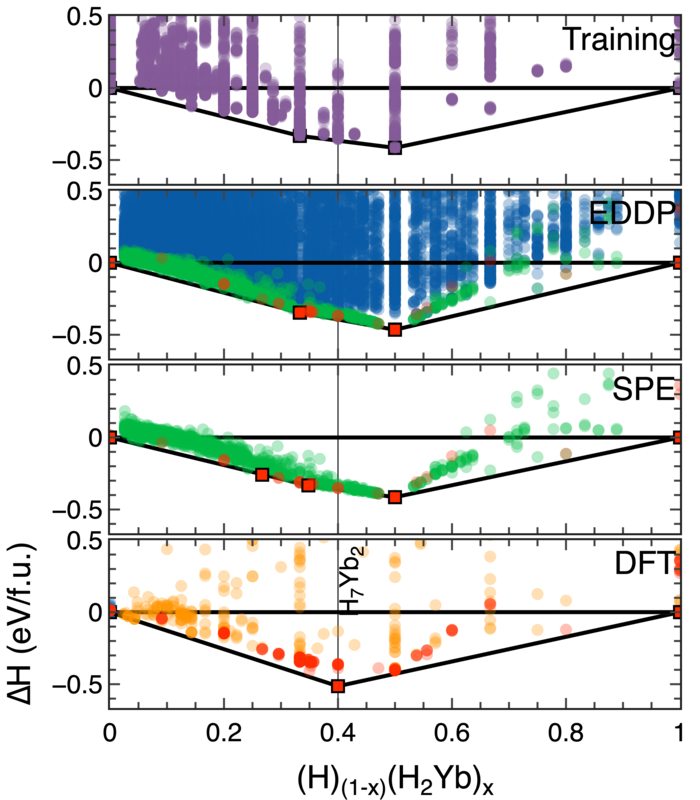}
\footnotesize


\flushleft{
\subsubsection*{\textsc{EDDP}}}
\centering
\includegraphics[width=0.3\textwidth]{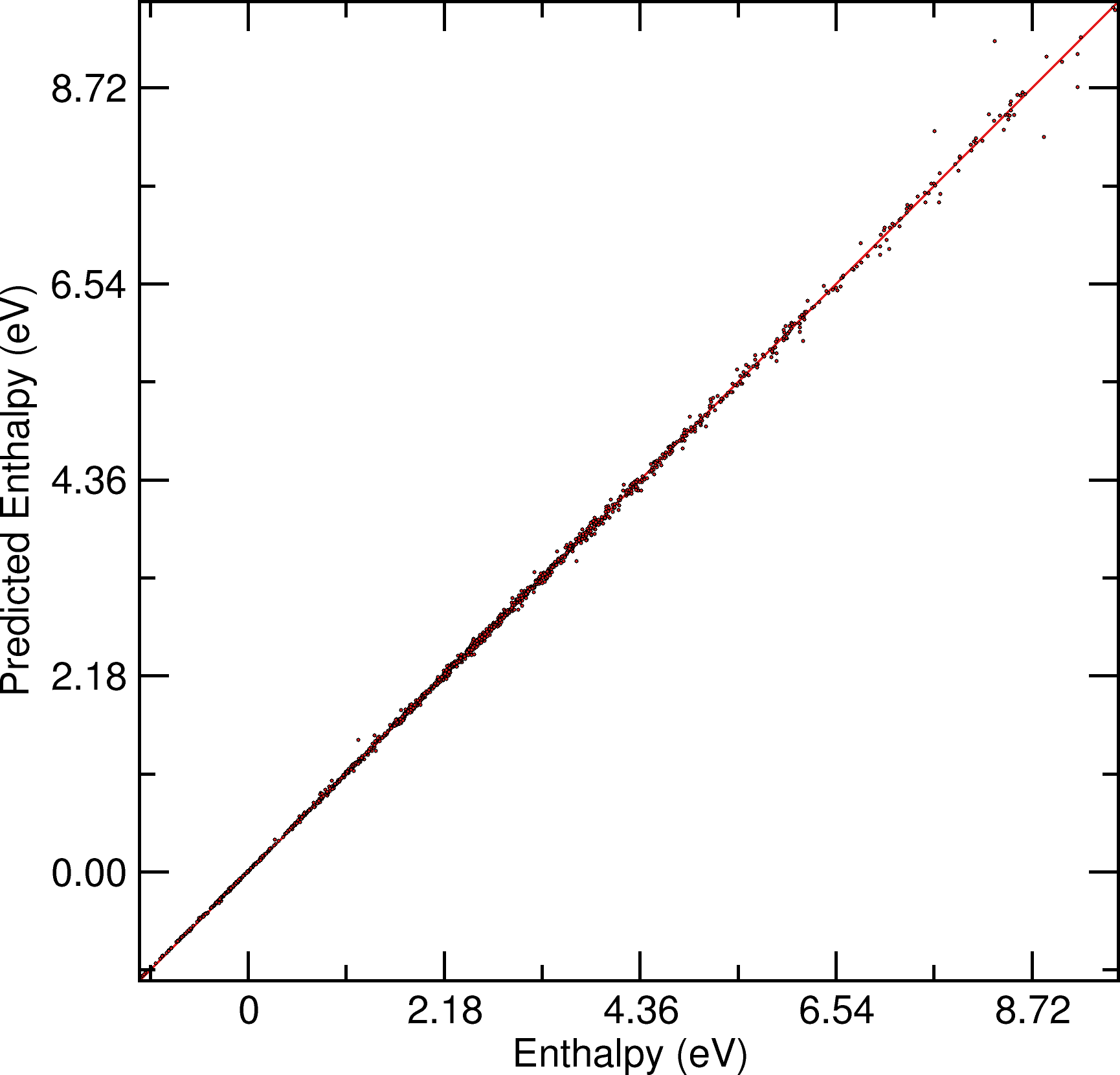}
\includegraphics[width=0.3\textwidth]{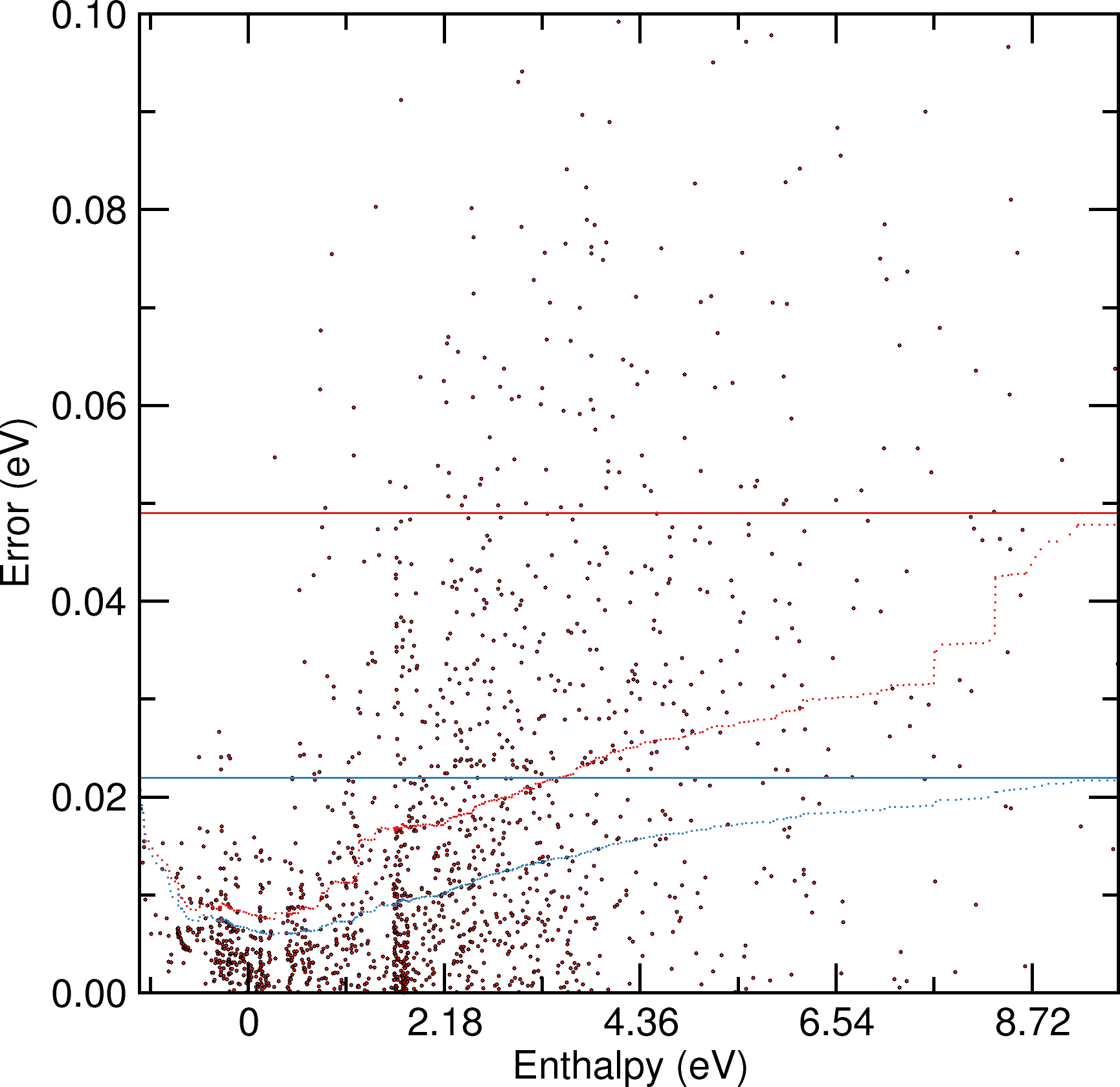}
\centering\begin{verbatim}
training    RMSE/MAE:  20.65  13.33  meV  Spearman  :  0.99985
validation  RMSE/MAE:  31.72  19.52  meV  Spearman  :  0.99981
testing     RMSE/MAE:  49.03  21.99  meV  Spearman  :  0.99982
\end{verbatim}
\clearpage

\flushleft{
\subsection{Zn-H}}
\subsubsection*{Searching}
\centering
\includegraphics[width=0.4\textwidth]{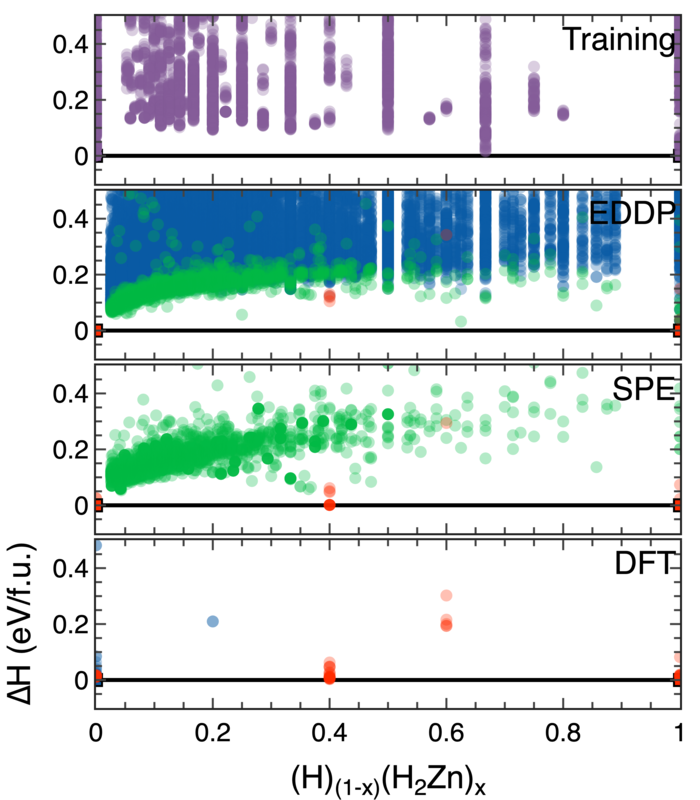}
\footnotesize\begin{verbatim}
Zn_100p0-2-HZn-P63mc                       100.02    12.551   -1792.950   0.496   0.496 -    2 HZn                P63mc          2
Zn_100p0-4-H6Zn-Pna21                       99.86    23.199   -1861.055   0.198   0.198 -    4 H6Zn               Pc             2
Zn_100p0-2500300-2337-59-XoOcLSc           100.00    48.378   -5448.348   0.180   0.180 -    2 H8Zn3              P1             5
Zn_100p0-1143057-5097-74-njBZqzx           100.01    56.540   -7229.223   0.112   0.112 -    4 H2Zn               Cmce          15
Zn_100p0-1143025-5087-34-ouKh5kH           100.03    70.194   -7311.390   0.076   0.076 -    2 H7Zn2              R3            20
Zn_100p0-2-Zn-P63mmc                       100.00     9.617   -1780.243   0.000   0.000 +    2 Zn                 P63/mmc        8
H-8-H-Pca21                                100.00    18.450    -109.594   0.000   0.000 +    8 H                  Pca21         48
\end{verbatim}

\flushleft{
\subsubsection*{\textsc{EDDP}}}
\centering
\includegraphics[width=0.3\textwidth]{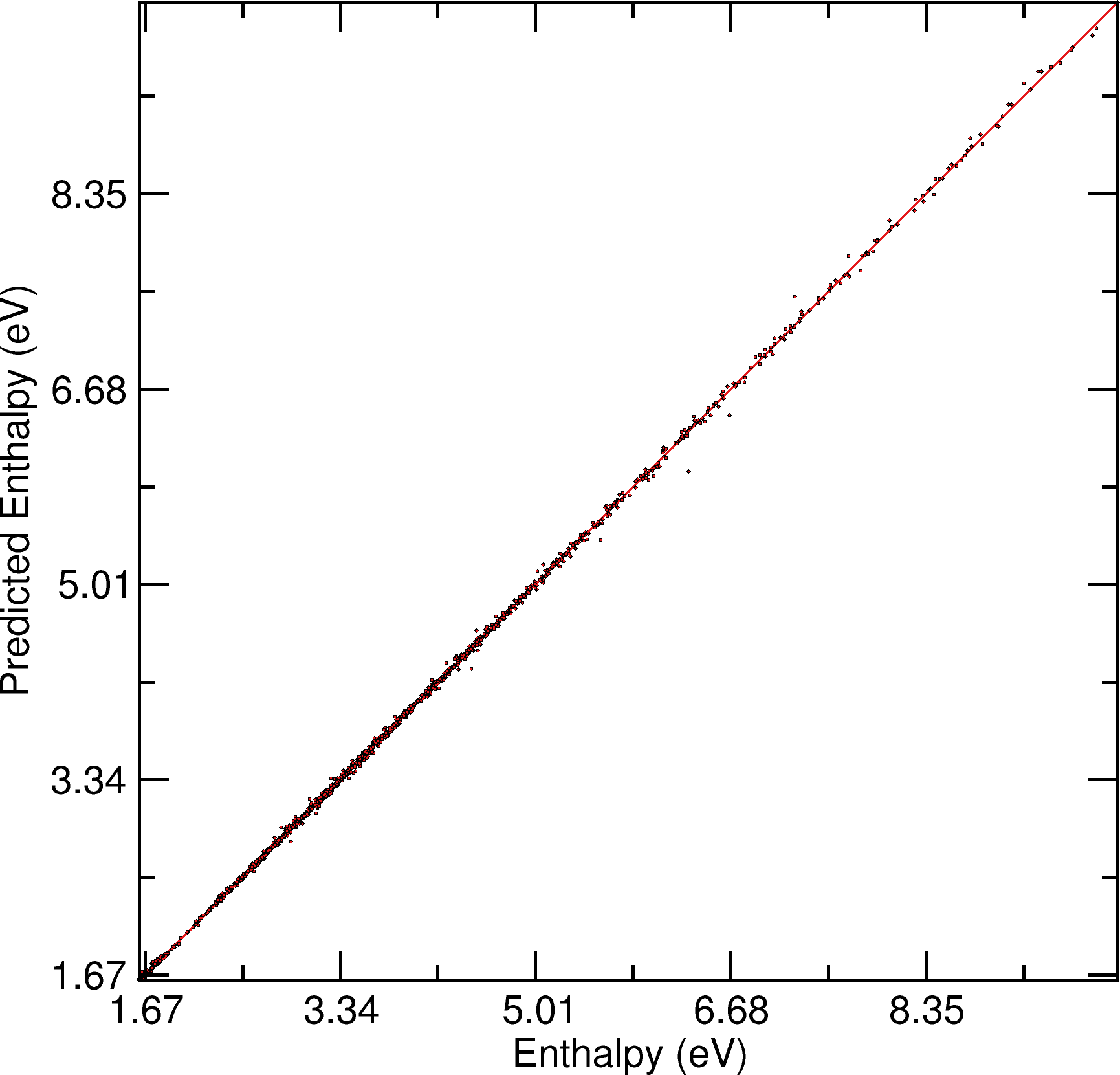}
\includegraphics[width=0.3\textwidth]{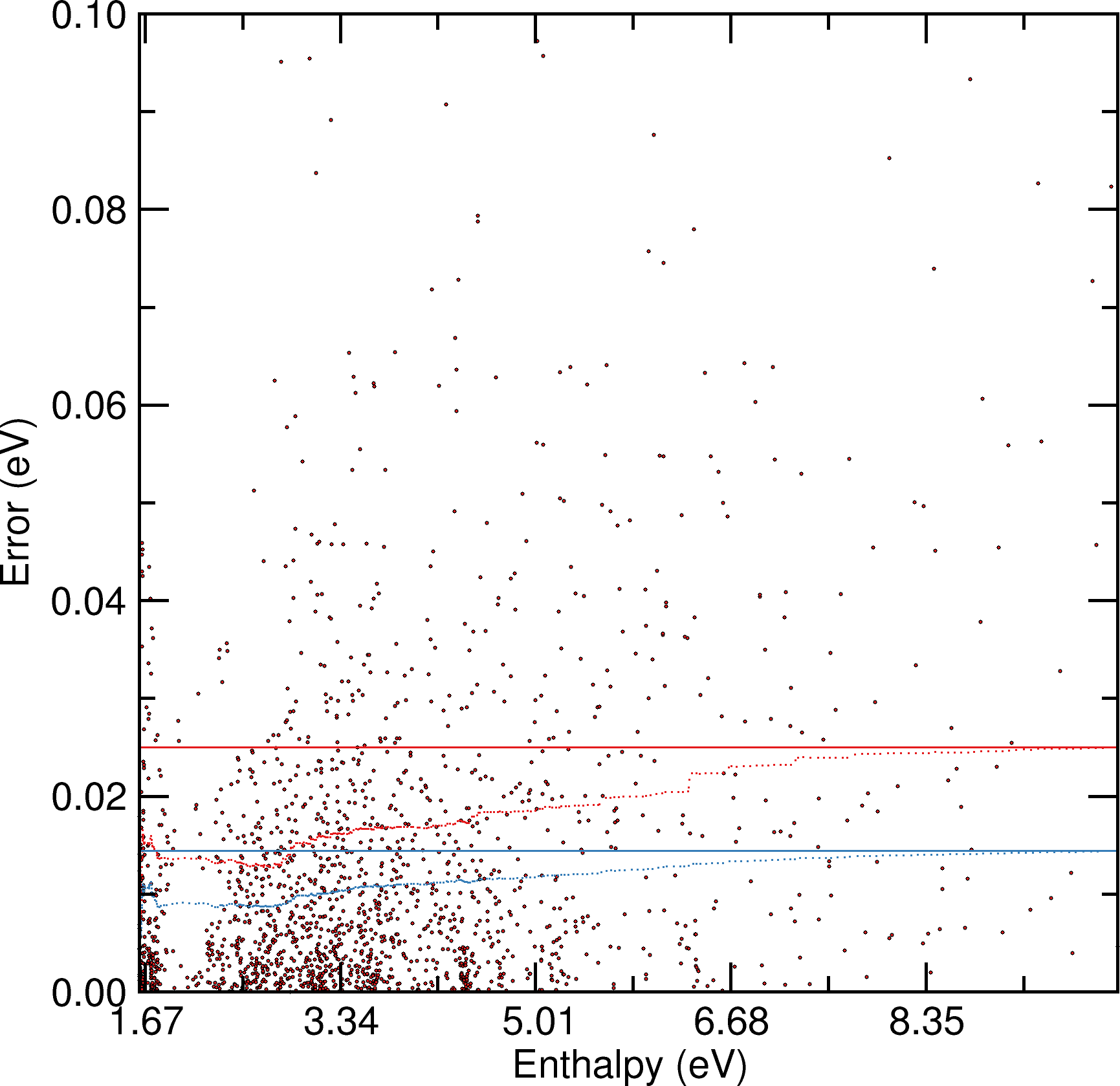}
\centering\begin{verbatim}
training    RMSE/MAE:  14.70  9.50   meV  Spearman  :  0.99987
validation  RMSE/MAE:  21.33  13.89  meV  Spearman  :  0.99981
testing     RMSE/MAE:  25.00  14.40  meV  Spearman  :  0.99979
\end{verbatim}
\clearpage

\flushleft{
\subsection{Zr-H}}
\subsubsection*{Searching}
\centering
\includegraphics[width=0.4\textwidth]{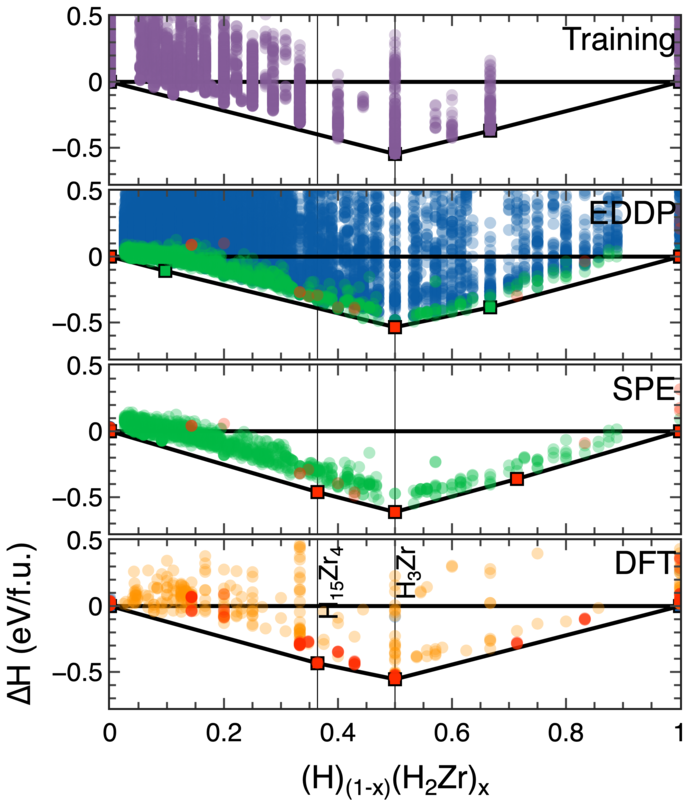}
\footnotesize


\flushleft{
\subsubsection*{\textsc{EDDP}}}
\centering
\includegraphics[width=0.3\textwidth]{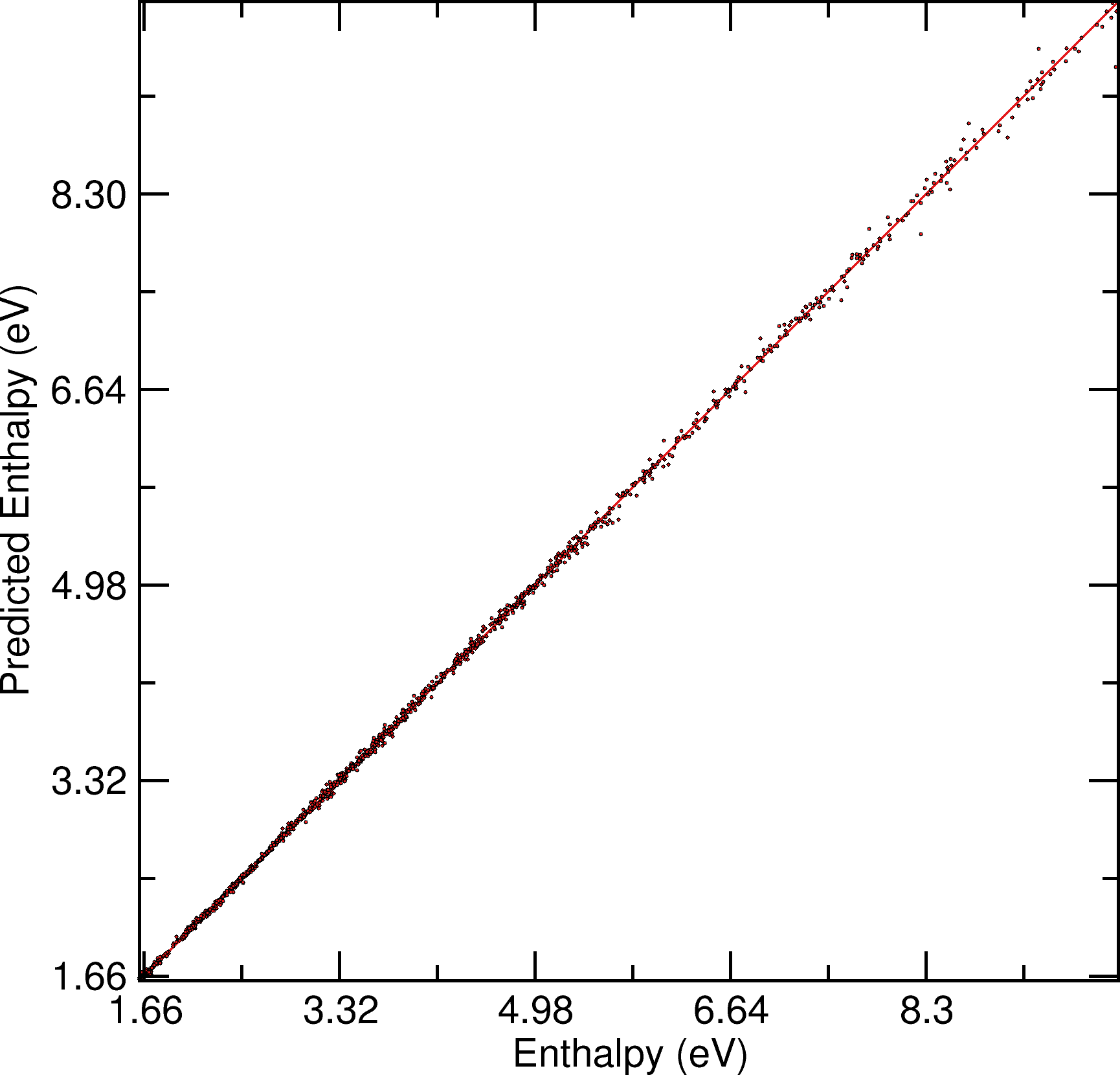}
\includegraphics[width=0.3\textwidth]{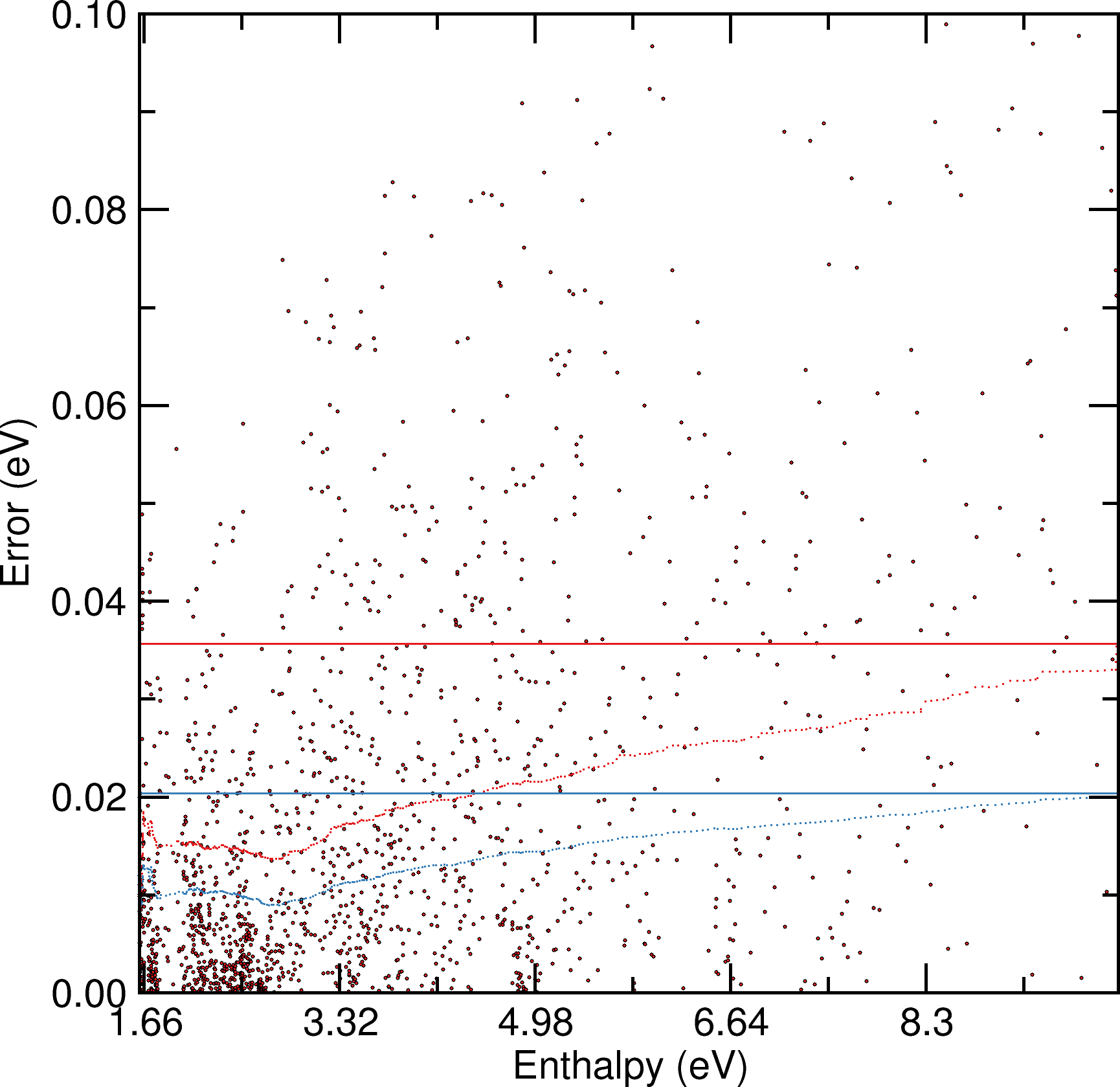}
\centering\begin{verbatim}
training    RMSE/MAE:  20.87  13.47  meV  Spearman  :  0.99985
validation  RMSE/MAE:  30.78  19.76  meV  Spearman  :  0.99982
testing     RMSE/MAE:  35.63  20.36  meV  Spearman  :  0.99982
\end{verbatim}
\clearpage

\end{document}